\documentclass[12pt,notitlepage]{report}
\usepackage{graphicx,vmargin,fancyhdr,amsfonts,amsmath,amsthm,amssymb,mathrsfs}
\usepackage{bibunits}
\usepackage{uwmthesis2}
\usepackage{times}

\usepackage{axodraw}

\makeatletter
\@addtoreset{equation}{section}
\makeatother

\usepackage{tensor}
\usepackage{color}
\usepackage{epsfig}
\usepackage{latexsym}

\setmarginsrb{1.0in}{1in}{1in}{0.65in}{0pt}{0.in}{0pt}{1cm}
\fancyhfoffset[]{0.25In}

\newcommand{\beq}{\begin{equation}}
\newcommand{\eeq}{\end{equation}}

\newcommand{\beqa}{\begin{eqnarray}}
\newcommand{\eeqa}{\end{eqnarray}}

\newcommand{\bea}{\begin{array}}
\newcommand{\eea}{\end{array}}
\newcommand{\iref}[2]{}
\newcommand{\com}[1]{}
\newcommand{\ci}[1]{}
\newcommand{\la}{\langle}
\newcommand{\ra}{\rangle}
\newcommand{\lpa}{\left(}
\newcommand{\rpa}{\right)}

\newcommand{\lsb}{\left[}
\newcommand{\rsb}{\right]}

\newcommand{\nn}{\nonumber \\}
\newcommand{\p}{\partial}
\newcommand{\pd}[1]{\frac {\partial} {\partial #1}}

\newcommand{\bay}[1]{\left(\begin{array}{#1}}
\newcommand{\eay}{\end{array}\right)}
\newcommand{\eg}{\textit{e.g.} }
\newcommand{\ie}{\textit{i.e.}, }

\def\xt{{\theta}}

\def\CA{{\cal A}}

\def\CD{{\cal D}}

\def\CF{{\cal F}}

\def\CH{{\cal H}}

\def\CL{{\cal L}}
\def\CM{{\cal M}}
\def\CN{{\cal N}}
\def\CO{{\cal O}}
\def\CP{{\cal P}}

\def\CS{{\cal S}}

\def\CV{{\cal V}}

\def\tit{{BEYOND \ THE \ STANDARD \  MODEL:\\LHC \ PHENOMENOLOGY,\\COSMOLOGY \ FROM \ POST-INFLATIONARY \ SOURCES,\\AND \ DARK \ MATTER \ PHYSICS}}
\def\aut{{\large Brian J. Vlcek}}

\newcommand{\postscript}[2]{\setlength{\epsfxsize}{#2\hsize}
   \centerline{\epsfbox{#1}}}

\newcommand{\sglspc}{\setstretch{1.1}}
\newcommand{\dblspc}{\setstretch{1.6}}

\begin{document}

\dblspc

\title{\tit}
\author{\aut}
\majorprof{Xavier Siemens}
\cmajorprof{Luis Anchordoqui}
\submitdate{August 2013}
\degree{Doctor of Philosophy}
\program{Physics}
\copyrightyear{2013}
\majordept{Physics}
\havededicationfalse
%\dedication
\haveminorfalse
%\haveminorfalse
\copyrightfalse
%\copyrighttrue
\doctoratetrue
\figurespagetrue
\tablespagetrue

% A the manuscript title page
\manuscriptp

\pagenumbering{roman}
\setcounter{page}{1}
\pagestyle{plain}

% B the Official Approval page
%\officialaprovp

\setcounter{page}{2}

%---------------------------------Abstract-----------------------------------------
\begin{center} 
\sglspc

\underline{ABSTRACT} \\
\ \\
{\uppercase \tit} \\		
\vskip 1em
\rm by\\
\vskip 1em	
\aut
\vskip 1em
The University of Wisconsin--Milwaukee, August \number\year\\
Under the Supervision of Professors Xavier Siemens and Luis Anchordoqui
\end{center}
\ \\
\dblspc %set double spacing
%--------Abstract Text-------------------------
\vskip -1cm
It is the goal of this dissertation to demonstrate that beyond the standard model, certain theories exist which solve conflicts between observation and theory -- conflicts such as massive neutrinos, dark matter, unstable Higgs vacuum, and recent Planck observations of excess relativistic degrees of freedom in the early universe.  Theories explored include a D-brane inspired construct of $U(3)\times Sp(1) \times U(1) \times U(1)$ extension of the standard model, in which we demonstrate several possible observables that may be detected at the LHC, and an ability to stabilize the Higgs mechanism. The extended model can also explain recent Planck data which, when added to HST data gives an excess of relativistic degrees of freedom of  $\Delta N = 0.574 \pm 0.25$ above the standard result.  Also explored is a possible non-thermal dark matter model for explanation of this result.  Recent observations of \emph{Fermi} bubble results indicate a signal of a $50~{\rm GeV}$ dark matter particle annihilating into $b \bar b$, with a thermally averaged annihilation cross section corresponding to $\langle \sigma_b v \rangle \sim 8 \times 10^{-27}~{\rm cm}^3/{\rm s}$, spurs interest in a Higgs portal model suggested by Steven Weinberg.  Other implications of this model are also explored such as its ability to explain dark matter direct detection results along with LHC Higgs data, and Planck data.  Particle physics is complimented by possible stochastic gravitational wave searches for which a model of second order global phase transitions is explored.  These transitions generate gravitational wave spectra with amplitudes of order $\Omega_{gw} h^2 \sim 10^{-24} - 10^{-15}$.  Furthermore, techniques into such calculations are investigated in hopes to improve the stability required in such lattice simulations.

\endabstract 
\newpage

% D the OPTIONAL copy right page
\ifcopyright\copyrightpage\fi
% E the OPTIONAL dedication page
\ifhavededication\dedicationpage\fi
% F the Table of Contents
\newpage
\renewcommand\contentsname{\begin{center}  \vspace{-2.5cm}TABLE \ OF \ CONTENTS\vspace{-1.5cm} \end{center}}        
\renewcommand\listfigurename{\vspace{-2.5cm}{\begin{center} LIST \ OF \ FIGURES \end{center}} \vspace{-0.5cm}}
\renewcommand\listtablename{\begin{center} \vspace{-2.5cm}LIST \ OF \ TABLES\vspace{-0.5cm} \end{center}}

\tableofcontents

\afterpreface

\newpage

%---------------------------------Acknowledgments-----------------------------------------
\begin{center}
{\LARGE \bf ACKNOWLEDGMENTS}
\end{center}
\ \\
I would like to acknowledge my co-authors in the various papers I've collaborated on in my graduate studies including, Ignatios Antoniadis, Haim Goldberg, Tom Giblin, Xing Huang, Dieter L\"{u}st, Larry Price, and Tomasz Taylor.  Also I thank the U.S. National Science Foundation (NSF) CAREER grant PHY-1053663, and Foundation grants  PHY-0970074, PHY-0955929, PHY-0758155, and the University of Wisconsin--Milwaukee Research Growth Initiative.  As well the Perimeter Institute for Theoretical Physics for the hospitality  during
the completion of part of this work enclosed.  I would especially like to thank Xavier Siemens for giving me an opportunity to start work on research in the field of physics and whose guidance was helpful throughout my studies.  I would also especially like to thank Luis Anchordoqui whose enthusiasm, excitement and encouragement helped me transition to a field of study I found more suiting, and for making me feel part of a team.  Furthermore, I extend my thanks to John Friedman, Dawn Erb, Philip Chang, Tom Paul, Peter Schwander, Buffey Hickling and Stefanie Pinnow for their participation in reviewing this dissertation.  I would also like to thank Kate Valerius for the innumerable times of helping me in tough situations throughout my graduate studies at the University of Wisconsin-Milwaukee.   I would also thank Tom Weiler for some valuable discussion regarding WIMPs. Lastly I thank my family, Carrie, Dave, Penny, Jimmy, Kevin who helped with editing, and especially my mother Laurie and my father James, whose sacrifices and support are the only reasons any of this is possible, Thank you. \\
\\
--Brian J. Vlcek,  August 2013

\newpage

%---------------------------------Units and Conventions-----------------------------------------
\begin{center}
{\LARGE \bf Units and Conventions}
\end{center}
\ \\
\sglspc %set single spacing 
\begin{itemize}
\item This dissertation makes use of the Einstein summing convention, where repeated indices are summed, \eg $A_\mu A^\mu = A_0 A^0 + A_1 A^1 + \dots + A_n A^n$ . \\
 
\item Unless otherwise specified, the Minkowski metric is given in the mostly minus form of $g_{\mu \nu} = {\rm diag}[1, -1, -1, \dots, -1]$ where there is $d$ entries of $-1$ along the diagonal for  $d+1$ space-time dimensions. \\
\item This dissertation also uses natural units where $c = 1 \ l_n t_n^{-1}, \ \hbar = 1 \ m_n l_n^2 t_n^{-1},$ and $k_b = 1 \ m_n l_n^2 t_n^{-2} T_n^{-1}$ where $m_n, \ l_n, \ t_n, \ T_n$ are the units of natural mass, length, time, and temperature respectively.  When using natural units, it is common practice to omit the natural unit symbols $(m_n,l_n,t_n,T_n)$ and solely represent units in terms of energy for which several common units may be represented by~\cite{Beringer:1900zz}
\beqa
&\hbar c& =  m_n l_n^3 t_n^{-2} \approx  197 \ {\rm MeV \ fm} \rightarrow 1 \ {\rm fm} = 5.08 \times 10^{-3} \ \rm MeV^{-1}, \nonumber \\
&c&  =  l_n t_n^{-1} \approx  3 \times10^{11} \ {\rm fm \ ps^{-1}}  \rightarrow 1 \ {\rm ps} = 1.52 \times 10^9 \ \rm MeV^{-1} ,  \nonumber \\
&k_b& = m_n l_n^2 t_n^{-2} T_n^{-1} \approx 8.62 \times 10^{-11} \ {\rm MeV \ K^{-1}} \rightarrow 1{\rm K} = 8.62 \times 10^{-11} \ {\rm MeV} ,  \nonumber \\
&M_{pl}& = \sqrt{\hbar c/G} \approx 1.22 \times 10^{19} \ {\rm  GeV \ c^{-2}} \rightarrow G^{-1/2} = M_{pl} = 1.22 \times 10^{19} \ {\rm  GeV} , \nonumber
\eeqa
where the arrows show the omission of natural units, $G$ is Newton's gravitational constant, and $M_{pl}$ the Planck mass. Natural units allow representation of units in terms of energy.
\end{itemize}

\newpage

%---------------------------------Preface-----------------------------------------
\begin{center}
{\LARGE \bf PREFACE}
\end{center}
\ \\
\indent Part I on BSM particle physics phenomenology is based on material from:
\begin{itemize}
\item  Luis A.~Anchordoqui, Ignatios Antoniadis, Haim Goldberg, Xing Huang, Dieter L\"{u}st, Tomasz R. Taylor, and {\bf Brian Vlcek} \\
{\it LHC Phenomenology and Cosmology of String-Inspired Intersecting D-Brane Models},
 Phys.\ Rev.\  D {\bf 86}, 066004 (2012)
  [arXiv:1206.2537 [hep-ph]]. 

\item  Luis A.~Anchordoqui, Ignatios Antoniadis, Haim Goldberg, Xing Huang, Dieter L\"{u}st, Tomasz R. Taylor, and {\bf Brian Vlcek} \\
{\it Vacuum Stability of Standard Model$^{++}$},
 \\ JHEP {\bf 1302}, 074 (2013)
  [arXiv:1208.2821 [hep-ph]]. 
\end{itemize}

Part II on gravitational waves is based on the following paper: 
\begin{itemize}
\item  John T. Giblin, Jr., Larry R. Price, Xavier Siemens, and {\bf Brian Vlcek}\\
  {\it Gravitational Waves from Global Second Order Phase Transitions},
\\  JCAP {\bf 1211}, 006 (2012) [arXiv:1111.4014 [astro-ph.CO]].
\end{itemize}

Part III on dark radiation and dark matter is based on material from:
\begin{itemize}
\item Luis~A.~Anchordoqui, Haim~Goldberg, and {\bf Brian~Vlcek}
\\ {\it Tracing the Interplay between Non-Thermal Dark Matter and Right-Handed Dirac Neutrinos with LHC Data,} 
in Proceedings of the 33rd International Cosmic Ray Conference (ICRC 2013), Rio de Janeiro, Brazil, 2 - 9 July; \\ \ [arXiv:1305.0146 [astro-ph.CO.]].

\item Luis~A.~Anchordoqui and {\bf Brian~J.~Vlcek}
\\{\it W-WIMP Annihilation as a Source of the Fermi Bubbles,}\\ 
\   Phys.\ Rev.\ D {\bf 88}, 043513 (2013)  [arXiv:1305.4625 [hep-ph]]. 
\end{itemize}

\dblspc%set double spacing 
\newpage

\pagenumbering{arabic}

\pagestyle{uwmheadings}

%---------------------------------Ch 1: Introduction-----------------------------------------
\chapter{Introduction}
\pagenumbering{arabic}
\thispagestyle{fancy}
\pagestyle{fancy} 
\vspace{-1 cm} 
%---------------------------------Ch 1: Section 1-----------------------------------------
\quad 	Modern physics is built on two outstanding theories, the standard model (SM) of particle physics and the concordance model of cosmology ($\Lambda$CDM cosmology)~\cite{Beringer:1900zz}.  The SM~\cite{WeinbergSM} is built on a mathematical footing known as quantum field theory (QFT) to describe the notions of quantum uncertainty to fields rather than point like particles.  This, along with principles of symmetry, allows physicists to accurately predict the outcomes of experiments of elementary particles down near the range of approximately $2.5 \times 10^{-20} \ {\rm m}$. To give some perspective of that scale, the radius of a hydrogen atom is about $0.5 \times 10^{-10} \ {\rm m}$ (the Bohr radius). If this scale is magnified to the size of a major city (taken to be $20 \ {\rm km}$), we would have knowledge of what's going on at the atomic level ($10^{-10}\ {\rm m}$) of the city model!  On the other end of the length spectrum, we have the $\Lambda$CDM model of cosmology, built upon Einstein's theory of general relativity (GR)~\cite{Einstein:1916}.  The $\Lambda$CDM model predicts the large scale behavior of the cosmos out to $13.7 \times 10^9 \ {\rm lyr} = 1.3 \times 10^{26} \ {\rm m}$ ~\cite{Weinberg_cosmo}.  Both theories are considered standard for physicists, however they cannot be final theories of their respective subjects.  They are effective theories, in the same sense that Newton's laws effectively describe projectiles and many other things at the macroscopic-scale, but it would be incorrect to use it to predict motions of electrons, nor the motions of highly relativistic objects~\cite{Einstein:1905}, there is a scale at which you use the laws of quantum physics, or general relativity.

	The SM and the $\Lambda$CDM model must somehow agree in their respective predictions to give a consistent theory of our observable universe.  One would expect that the SM is somehow more elementary as the physics of the very small should be able to describe the physics of the very large if taken to the effective limit.  The very name $\Lambda$CDM tells of our ignorance; ``$\Lambda$" is the canonical symbol for the unknown term in the Einstein field equations of GR that creates a repulsive like effect of gravity~\cite{Eisntein:1917}.  Currently this energy is associated with the non-zero energy of free space and is known as dark energy and accounts for $68.6\%$ of the energy density of the universe.  The CDM is cold dark matter, which accounts for $26.5\%$ of the matter content of the universe~\cite{Ade:2013lta}.  The rest of the universe is ordinary matter, for which the SM seems to correctly describe.  Where is the dark matter and dark energy in our current understanding of particle physics?  One glaring problem of our model of cosmology is that it is a classical theory and thus cannot correctly describe the state of the universe at the time of the big bang.
	
	As it stands today, the SM is incomplete.  One of the main reasons is that it does not include gravity as one of the interactions.  As well, after the recent (at the time of writing this) Higgs like particle discovery~\cite{ATLASnew, CMSnew}, with an apparent mass of $m_H = 125 \hbox{ GeV}$, the SM seems to have an unstable vacuum.  This means that the current formulation of the SM cannot be correct. There are many other reasons to suspect the SM is an effective theory.  These issues motivate the subject known as beyond the SM physics (BSM), where one attempts, through demanding new symmetries of the Lagrangian of the SM, to predict new particles or interactions.  Of the proposed theories, two are by far the most popular, supersymmetry (SUSY) and string theory, though they are related.  In Part I of this dissertation, we look at a possible extension of the SM through a string theoretical basis which imposes additional $U(N)$ symmetries on the Lagrangian of particle physics, rather than $SU(N)$ which is what the SM uses in its current formulation.  We will explore the phenomenology of this new model, including LHC signals, and cosmological observables.
		
\begin{figure}[tpbh]
\begin{center}
\postscript{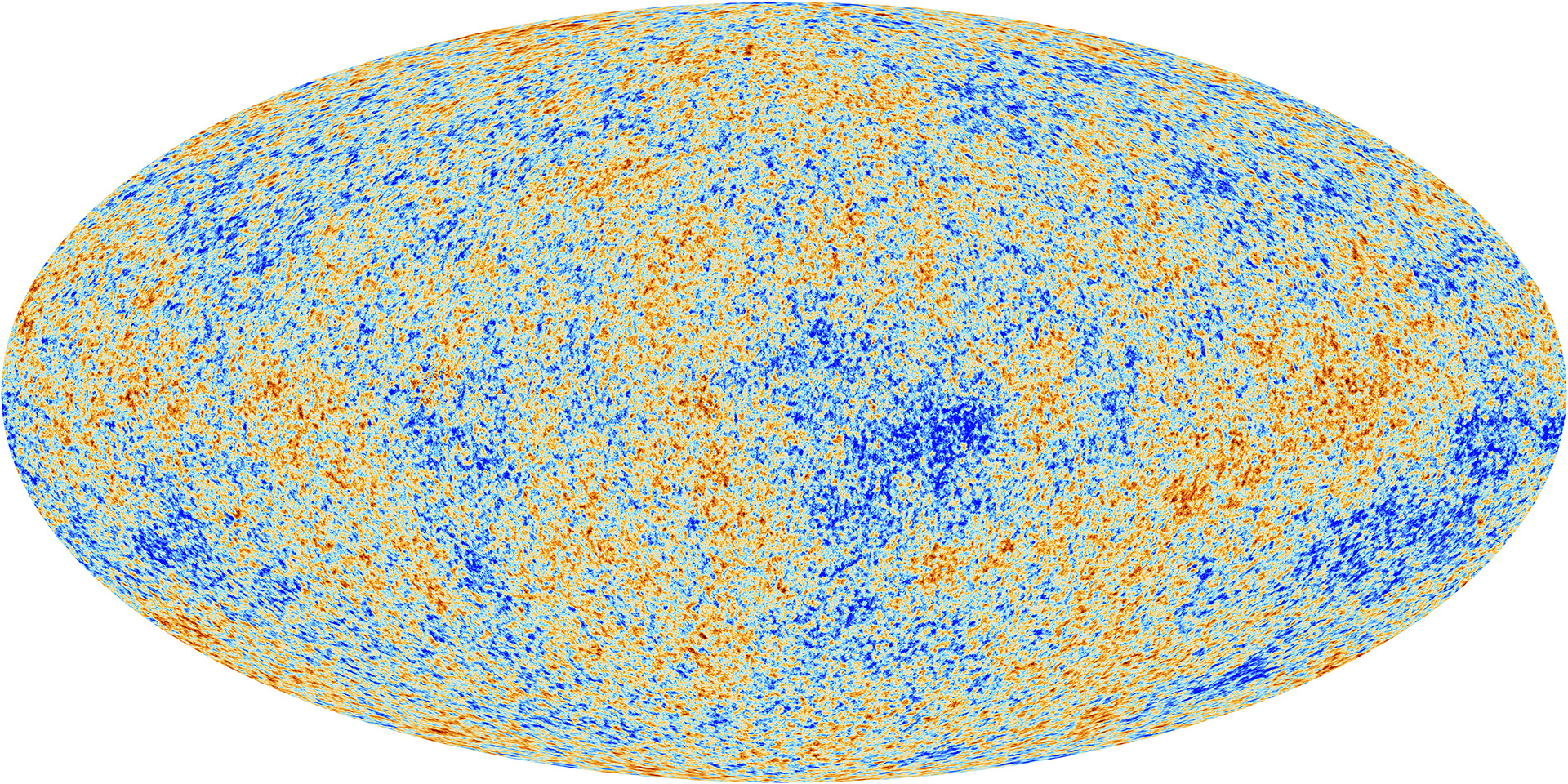}{0.8}
\caption[CMB Full Sky Anisotropy Map from Planck 2013 Results]{ \sglspc \small The highest resolution full sky anisotropy map from generated from the Planck 2013 results~\cite{PlanckANISO}.}
\end{center}
\label{fig:CMBanisoSky} 
\end{figure}

	One key concept of the $\Lambda$CDM model is the concept of inflation put forward by Guth in 1981~\cite{Guth}. It is widely accepted as the correct mechanism of flattening the universe (this will be expounded upon in Sec. \ref{sec:INFLATE}).  One exciting feature of inflation is that it takes whatever physics is occurring at the quantum level before the universe undergoes rapid expansion and magnifies it to the cosmic level!  This is considered as one of the main reasons for large scale anisotropies in the cosmic microwave background (CMB) as seen in Fig.~\ref{fig:CMBanisoSky}.
%----------------Figure------------------
%-----------------------------------------
	We might be able to exploit this magnification and search for physics that goes beyond what the standard model offers by observing effects in CMB anisotropies and theoretically a CMB-like background of gravitons.  We would expect to observe these gravitons in future interferometric gravitational wave observatories such as the laser interferometer gravitational wave observatory (LIGO) (admittedly a much more advanced detector would be needed but the concept is the same).  My work in Sec. \ref{sec:GWStart} explains the methods used to calculate what the observed background would be, given a model for the relevant fields at the time immediately after inflation.
	
	It is also interesting to look at CMB anisotropies as probes of BSM physics, specifically in the form of dark radiation.  Dark radiation is the term used to describe the excess relativistic degrees of freedom (r.d.o.f.) recently reported by the Planck Collaboration~\cite{Ade:2013lta}. This excess results from combing data from the Planck, and the Hubble Space Telescope result of $h = 0.738 \pm 0.024$~\cite{Riess:2011yx}, yielding $\Delta N = 0.574 \pm 0.25$.  Planck data suggests the excess is statistically significant, but is as of yet unexplained by SM physics, this along with other observations will be reviewed in Sec.~\ref{sec:DM} of this dissertation.  One possible solution to this problem is the inclusion of right chiral counterparts of the SM neutrinos which solves both neutrino masses(to be examined in the next section) and the possible r.d.o.f. excess.  In Sec.~\ref{sec:DM} we will explore how adding right chiral neutrinos, as well as relativistic dark matter into the $\Lambda$CDM model, may explain an excess in r.d.o.f., as well as setting limits on possible observables of an additional gauge boson at the LHC.
	
	CMB anisotropies observations are also complimented by observations of merging galaxy clusters~\cite{Markevitch} and rotation curves~\cite{VRubin} in suggesting dark matter.  The current theories suggest weakly interacting massive particles (WIMPs) are fermionic fields that explain all the observations associated with dark matter.  In the last section we will discuss a Higgs portal mechanism into exploring interactions of WIMPs with visible matter, specifically by exploring the implications of a Higgs portal model suggested by Weinberg~\cite{Weinberg:2013kea}, and possible detections in LHC data.

%---------------------------------Ch 1: Section 2-----------------------------------------
\section{Review of the SM of Particle Physics}
\subsection{States of Definite Chirality and Spinors}

\begin{figure}[h]
\begin{center}
\postscript{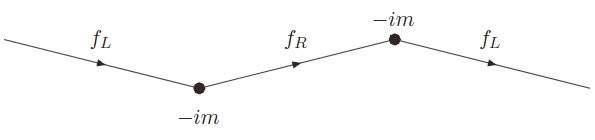}{0.7}
\caption[Chiral Fermion Propagation]{ \sglspc \small A fermion is viewed as a chiral states exchanging helicities by interactions with the vacuum as indicated by the dark points in the figure above. \dblspc }
\label{fig:ChiFer} 
\end{center}
\end{figure}

Before we try to expand on the SM, we review how it is currently formulated, while exposing its weakness along the way.  We start with the assumption that the reader has some knowledge of the mathematics of quantum field theory so that we may focus on the Lagrangian formulation of the SM, as well as knowledge with the workings of relativistic quantum mechanics.  When we construct theories, we start with a symmetry we think nature has and see what that forces the Lagrangian to look like.  One symmetry group that any theory must posses is invariance of operators of the Poincar\'{e} group, which consists of translations,  Lorentz boosts of special relativity and rotations in space. All of which seem to be symmetries of all the known laws of physics.  Another experimental fact we must consider is that there are two different types of elementary particles, bosons having integer spin, and fermions having half-integer spins.  When constructing representations of the Lorentz group it turns out that you essentially can have two different spin $1/2$ particles that furnish the representation, which are called chiral states. If the particles have no mass then their helicities (spin handedness along their directions of motion) are the same as their chirality. The two chiral states are the left handed spin $1/2$ fermions $f_L$ and the right handed spin $1/2$ fermions $f_R$.  However we do not observe massless fermions in the universe, so how do we justify Lorentz invariance, spin $1/2$ particles, and massive states?  The currently accepted answer is that if a fermion has mass then it is viewed as changing from one chiral state to the other by interacting with the vacuum of QFT as depicted in Fig. \ref{fig:ChiFer}. 
%-------FIG-----------------
%-----------------------------
With the two different states we write a Lorentz invariant Lagrangian for ``free" fermions (free in the sense that we observe fermions in massive states and consider them free though the Lagrangian below is not technically free)
\beq
\CL_f = i f_L^\dagger \bar{\sigma}^\mu \partial_\mu f_L + i f_R^\dagger \sigma^\mu \partial_\mu f_R - m(f_R^\dagger f_L + f_L^\dagger f_R),
\label{eq:One}
\eeq
where $\sigma^\mu = \left(1,\vec{\sigma} \right)$ and $\bar{\sigma}^\mu = \left(1,-\vec{\sigma} \right)$ with $\vec{\sigma}$ the Pauli matrices that may be familiar from ordinary quantum mechanics. We have also used the Einstein summation procedure for repeated indices.

We compact this notation from 2-component spinors $f_L,f_R$ to a 4-component Dirac spinor,
\beq
f = \left(\begin{array}{c} f_R \\ f_L \end{array} \right).
\label{eq:Two}
\eeq
This allows us to write the Dirac Lagrangian for free massive fermions
\beq
\CL_f = \bar{f}(i \gamma^\mu \partial_\mu - m)f, \hbox{ where } \gamma^\mu = \left( \begin{array}{cc} 0 & \bar{\sigma}^\mu \\ \sigma^\mu & 0 \end{array} \right),
\label{eq:Three}
\eeq
and with  $\bar{f} = f^\dagger \gamma^0$. The important property of the gamma matrices $\{\gamma^\mu,\gamma^\nu\} =2 g^{\mu \nu} I_{4\times 4}$, where $g^{\mu \nu}$ is the Minkowski flat metric and $I_{4\times4}$ a four dimensional identity matrix should be noted. To demonstrate this does in fact describe relativistic fermions we do a quick calculation using the principle of least action which gives
\beqa
\p_\mu \lpa \frac{\p \CL_f}{\p \bar{f},_\mu} \rpa - \frac{\p \CL_f}{\p \bar{f}} = 0,  \nonumber
\\
\lpa i \gamma^\mu \p_\mu - m \rpa f = 0,\nonumber
 \\
\lpa -i \gamma^\nu \p_\nu - m \rpa \lpa i \gamma^\mu \p_\mu - m \rpa f = 0, \nonumber
 \\
\lpa \p^\mu \p_\mu + m^2 \rpa f = 0.
\label{eq:Four}
\eeqa
The last equation is the Klein-Gordon equation or the quantum version of the special relativistic energy relation $p_\mu p^\mu  = p^2 = m^2$. From this point forward we will use the notation $p_\mu p^\mu = p^2$,  it should be apparent when contraction of indices is used versus the squaring of a c-number. 

\subsection{Gauge Theories}
Guided by Lorentz invariance and observation of the spin $1/2$ nature of fermions, we were able to construct a relativistic Lagrangian,  which has a symmetry we did not intend to supply to it. Spinors undergoing the transformation
\beq
f \rightarrow e^{-i \theta} f,
\label{eq:Five}
\eeq
where $\theta$ is some real number, leave the Dirac Lagrangian invariant.  In fact if we have multiple species of fermions $f_i$ we can repeat the same Dirac Lagrangian for each species which will have a more complex but similar symmetry
\beq
\CL_{\rm fermion} = \bar{f_i}\lpa i \delta_{ij} \gamma^\mu \p_\mu - m_{i j} \rpa f_j,
\label{eq:Six}
\eeq
where $m_{ij}$ is a diagonal mass matrix and we sum over $i,j$ with a Euclidean metric. We perform a transformation on the collection of spinor species given by
\beq
f_i \rightarrow e^{-i \theta^a \lpa T^a \rpa_{ij}}f_j = \left( e^{-i \Theta}f \right)_i, 
\label{eq:Seven}
\eeq    
with $T^a$, a collection of matrices, and $\theta^a$ a collection of real numbers.   In the second equality, I compacted the notation with $\Theta =\theta^a T^a$ (summing over internal group indices with a Euclidean metric) and
\beq
f = \lpa \bea{c} f_1 \\ f_2 \\ \vdots \eea\rpa.
\label{eq:Eight}
\eeq 
The fermion Lagrangian will remain invariant under this transformation\footnote{ \sglspc \small Note that $T^a$ and $\gamma^\mu$ commute as one operates on the spinors, and one mixes the spinor species.} if $(T^a)^\dagger =  T^a$.  Operators constructed with this requirement such as $U = e^{i \Theta}$ will necessarily be unitary $U^\dagger U = 1$.  It is then said that the collection of $N$ relativistic massive fermions will be invariant under operation of the global symmetry of the $U(N)$ group (why it is called global will become apparent below).

Now we change from $\theta^a$ being real numbers to $\theta^a(x)$ now being real functions of space-time. The transformation $U(x)$ now depends on one's position in space-time; which are termed local $U(N)$ gauge transformations. The Lagrangian is no longer invariant under local $U(N)$ transformations, and the action of the gauge group operator leaves us with
\beq
\CL_f \rightarrow \CL_f + \bar{f} \lpa \gamma^\mu \p_\mu \Theta \rpa f.
\label{eq:Nine}
\eeq 
The problem is the derivative in the original Lagrangian.  If we construct a derivative operator $D_\mu$ that transforms like $D_\mu \rightarrow U D_\mu U^\dagger$ then we will be able to construct a Lagrangian that is invariant to local $U(N)$ transformations.  We accomplish this by introducing gauge fields so the derivative term becomes
\beq
D_\mu = \p_\mu - i g T^a W^a_\mu(x) = \p_\mu - i g W_\mu,
\label{eq:Ten}
\eeq
where under gauge transformations $W_\mu \rightarrow e^{i \Theta} W_\mu e^{-i \Theta} - \frac{1}{g} \p_\mu \Theta$. Note that I have introduced a compact notation $W_\mu = T^a W^a_\mu$.  Interestingly, if the gauge group is $U(1)$, then the gauge transformation rule is that of electrodynamics.  Requiring invariance of the Lagrangian to operators of $U(N)$ we are able to write our new Lagrangian as
\beq
\CL_f = \bar{f} \lpa i \gamma^\mu D_\mu - m \rpa f.
\label{eq:Eleven}
\eeq
However we did not write the most general local gauge invariant terms.  We left out a gauge field kinetic. We define the gauge field strength by
\beq
F_{\mu \nu} = \p_\mu W_\nu - \p_\nu W_\mu + \left[W_\mu,W_\nu\right].
\label{eq:Twelve}
\eeq
This transforms under local gauge transformations like $F_{\mu \nu} \rightarrow U F_{\mu \nu} U^\dagger$ so we may make a gauge invariant kinetic term from this by use of the permutative property of the trace operation
\beq
\frac{1}{2} \hbox{Tr}\left( F^2 \right) \rightarrow \frac{1}{2} \hbox{Tr} \left( U F U^\dagger U F U^\dagger\right) = \frac{1}{2} \hbox{Tr} \left(U^\dagger U F^2 \right) = \frac{1}{2} \hbox{Tr} \left(F^2\right).
\label{eq:Thirteen}
\eeq
If we restrict ourselves to using the sub group of $U(N)$ to that of $SU(N)$ then we have $\hbox{Tr}\left(T^a T^b\right) = \frac{1}{2} \delta^{ab}$ which allows us to explicitly write the kinetic term for the gauge fields
\beq
\frac{1}{2} \hbox{Tr} \left( F^2 \right) = \frac{1}{2} \hbox{Tr} \left( F^a_{\mu \nu} F_b^{\mu \nu} T^a T^b\right) = \frac{1}{4} F^a_{\mu \nu} F^{\mu \nu}_a.
\label{eq:Fourteen}
\eeq
All together we write out the local $SU(N)$ gauge invariant Lagrangian of $N$ types of relativistic fermions as  
\beqa
\CL &=& \bar{f} \lpa i \gamma^\mu D_\mu - m \rpa f - \frac{1}{2} \hbox{Tr}\left(F^2\right), \\
  &=& \bar{f_i} \lpa i \delta_{ij} \gamma^\mu \p_\mu - m_{ij} \rpa f_j + g \lpa\bar{f_i}\gamma^\mu \lpa T^a\rpa_{ij}  f_j\rpa W^a_\mu - \frac{1}{4}F^a_{\mu \nu} F_a^{\mu \nu}. \\
\label{eq:Fifteen}
\eeqa
It is important to notice that the Lagrangian cannot have a mass term $m_a^2 W_a^\mu W^a_\mu$ as it will break the local $SU(N)$ gauge invariance.

\subsection{Electroweak Model}

The final formulation of the SM didn't occur until 1967 when three seminal papers ~\cite{GWS1, GWS2, GWS3}  formed the Glashow-Weinberg-Salam (GWS) electroweak model of interactions.  This, combined with Higgs paper ~\cite{Higgs} on symmetry breaking with a non-zero vacuum expectation value of a scalar field, allowed one to reconcile data with theory. As well the work of Feynman, Bjorken, t'Hooft, among others culminated in adding in QCD to the SM for which the Nobel Prize of 2004 was awarded to Gross, Politzer and Wilczek for there asymptotic freedom work allowing perturbative calculations to pave the way for QCD analysis in particle collider experiments~\cite{Wilczek1,Wilczek2, Politzer1, Politzer2} .  The GWS model is built by imposing a $SU_L(2)\times U_Y(1)$ symmetry.  The subscript $L$ on $SU(2)$ indicates that left handed chiral states participate differently in the gauge group operations, than the right handed chiral states.  The right handed chiral states are said to be singlets of the $SU_L(2)$ symmetry; which means they transform under the symmetry operation as a scalar (no transformation) and effectively do not couple to the gauge bosons associated with that symmetry. The $Y$ for $U(1)$ is deemed hypercharge and is similar to the charge in electromagnetism, however as we will see via the Weinberg mixing of $SU_L(2)$ and $U_Y(1)$ gauge bosons, we will form the electromagnetic gauge field we are familiar with.  

Starting with the covariant derivative ($D_\mu$ operator), we have $4$ gauge bosons associated with the symmetry group
\beq
D_\mu = \p_\mu - i g_2 \tau^a W^a_\mu - i g Y B_\mu.
\label{eq:Sixteen}
\eeq
The operator $Y$ acting on multiplets has eigenvalues that are the hypercharge of that multiplet. This does not affect the gauge invariance so it is allowed in our formulation.  Further, we have $\tau^a = \sigma^a/2$ where $\sigma^a$ are the Pauli matrices. This allows $\tau^a$ to form the generators of $SU(2)$ in the fundamental representation.

  In the SM the left chiral the lepton sector doublet for one generation is,
\beq
E_L = \lpa \bea{c} \nu_L \\ e_L \eea \rpa,
\label{eq:Seventeen}
\eeq
the subscript $L$ on the Dirac spinors composing the doublet are to inform you that this spinor only contains left handed chiral components.   Typically this doublet is obtained by projecting out the right chiral state from the Dirac spinor used in quantum electrodynamics (QED) via the relation
\beq
e_L = P_L e = \frac{1}{2} \lpa 1- \gamma^5 \rpa  e.
\label{eq:Eighteen}
\eeq
Along with the left chiral doublet we have a $SU_L(2)$ singlet (which is the right chiral state $e_R$),  using the doublet and the singlet, we complete our leptonic sector for the SM for one generation, which results in
\beq
\CL_{\rm Lepton} = \bar{E}_L i \gamma^\mu D_\mu E_L + \bar{e}_R i \gamma^\mu D_\mu e_R - \frac{1}{4} W^a_{\mu \nu} W_a^{\mu \nu} - \frac{1}{4}B^{\mu \nu}B_{\mu \nu},
\label{eq:Nineteen}
\eeq
where $W^a_{\mu \nu}, \ B_{\mu \nu}$ are the field strengths for $SU_L(2), \ U_Y(1)$ gauge bosons respectively.

 Notice we do not have masses for any of the leptons involved; the SM in this form cannot be recognized as physically relevant.  To put it in a form that is physically meaningful, we make a change of basis for the gauge fields,
\beq
\lpa \bea{c} W^+_\mu \\ W^-_\mu \\ Z^0_\mu \\ A_\mu \eea \rpa = 
\lpa \bea{cccc} \frac{1}{\sqrt{2}} & -\frac{i}{\sqrt{2}} & 0 & 0 \\
              \frac{1}{\sqrt{2}} & \frac{i}{\sqrt{2}} & 0 & 0 \\
							 0 & 0 & \cos(\theta_w) & -\sin(\theta_w) \\
							 0 & 0 & \sin(\theta_w) & \cos(\theta_w) \\
			\eea
\rpa
\lpa \bea{c} W^1_\mu \\ W^2_\mu \\ W^3_\mu \\ B_\mu \eea \rpa.
\label{eq:Twenty}
\eeq
We also define a new operator $Q = \tau^3 + Y$ which we identify as electrical charge we are familiar with; as well we define  $\tau^\pm = \tau^1 \mp i \tau^2$.  In this new basis the covariant derivative takes the form
\beq
D_\mu = \p_\mu - i \frac{g_2}{\sqrt{2}}\tau^- W^+_\mu - i \frac{g_2}{\sqrt{2}}\tau^+ W^-_\mu - i \frac{g_2}{\cos(\theta_w)}\lpa \tau^3 - \sin^2(\theta_w) Q \rpa Z^0_\mu - i e Q A_\mu,
\label{eq:Twentyone}
\eeq
where $\tan(\theta_w) = g/g_2$ and $e = g_2 \sin(\theta_w)$, the elementary charge of an electron. Additionally we choose $Y E_L = -\frac{1}{2} E_L$ and $Y e_R = - e_R$, which allows us to expand the Lagrangian (with the gauge field kinetic terms not shown) as
\beqa
\CL_{\rm Lepton} &= & \bar{\nu}_L i \gamma^\mu \p_\mu \nu_L + \bar{e}_L i \gamma^\mu \p_\mu e_L + \bar{e}_R i \gamma^\mu \p_\mu e_R + \frac{g_2}{\sqrt{2}} \lpa \bar{\nu}_L \gamma^\mu e_L \rpa W^+_\mu \nonumber
 \\
& + & \frac{g_2}{\sqrt{2}} \lpa \bar{e}_L \gamma^\mu \nu_L \rpa W^-_\mu + \frac{g_2}{\cos(\theta_w)}\lpa\frac{1}{2} + \sin^2(\theta_w)\rpa \lpa \bar{\nu}_L \gamma^\mu \nu_L \rpa Z^0_\mu \nonumber
\\
&-& \frac{g_2}{\cos(\theta_w)}\lpa\frac{1}{2}-\sin^2(\theta_w)\rpa \lpa \bar{e}_L \gamma^\mu e_L \rpa Z^0_\mu \nonumber
\\
&-& e \lpa \bar{e}_L\gamma^\mu e_L \rpa A_\mu - e \lpa \bar{e}_R \gamma^\mu e_R \rpa A_\mu + \ldots
\label{eq:Twentytwo}
\eeqa  
\ \\
If we put this in terms of Dirac spinors of QED, we have
\beqa
\CL_{\rm Lepton} &=& \bar{\nu} \ i \gamma^\mu \p_\mu P_L \nu + \bar{e} \ i \gamma^\mu \p_\mu e \nonumber
\\
&+& \frac{g_2}{\sqrt{2}} \lpa \bar{\nu} \gamma^\mu P_L e \rpa W^+_\mu + \frac{g_2}{\sqrt{2}} \lpa \bar{e} \gamma^\mu P_L \nu \rpa W^-_\mu \nonumber 
\\
&+& \frac{g_2}{\cos(\theta_w)}\lpa\frac{1}{2} + \sin^2(\theta_w)\rpa \lpa \bar{\nu} \gamma^\mu P_L \nu \rpa Z^0_\mu \nonumber
\\
&-&  \frac{g_2}{\cos(\theta_w)}\lpa\frac{1}{2}-\sin^2(\theta_w)\rpa \lpa \bar{e} \gamma^\mu P_L e \rpa Z^0_\mu  \nonumber
 \\
&-& e \lpa \bar{e}\gamma^\mu e \rpa A_\mu+ \ldots
\label{eq:Twentythree}
\eeqa  
In this form we see we have charged vector-axial currents mediated by $W^\pm_\mu$ bosons.  As well we have the typical vector current of quantum electrodynamics $\bar{e} \gamma^\mu e$ mediated by the neutral photon $A_\mu$ boson. There is also a third vector-axial current that also has neutral currents and a neutral mediator (the $Z^0_\mu$ boson).  It is important to note that we have not included a right chiral state for the neutrino as it is consistently experimentally found in a left handed chiral state. 

We are able to include quarks and finish the SM gauge theory by repeating the process with the multiplets for the first generation of quarks
\beq
Q_L = \lpa \bea{c} u_L \\ d'_L \eea \rpa, \ u_R, \ d'_R.
\label{eq:Twentyfour}
\eeq
We use the symbol $d'_L$ and not $d_L$ because mass eigenstates of the quarks are not the same as the flavor eigenstates with the weak force interacts with.  The Cabibbo-Kobayashi-Maskawa (CKM) matrix relates the mass eigenstates to the flavor eigenstates by a mixing matrix
\beq
(d'_L)^i = V^{ij} (d_L)^j,
\label{eq:Twentyfive}
\eeq
where $(d_L)^j$ generically denotes $(d_L, s_L, b_L)$ for the down, strange, and bottom quarks of definite flavor and left handed chirality.  To complete the SM we also include the strong interaction by adding in an $SU_c(3)$ gauge coupling, whose gauge fields comprise 8 gluons, $g^a_\mu$. The covariant derivative now takes the form
\beq
D_\mu = \p_\mu -i g Y B_\mu- i g_2 \tau^a W^a_\mu - i g_3 \frac{\lambda^a}{2} g^a_\mu,
\label{eq:Twentysix}
\eeq
where $\lambda^a$ are the Gell-Mann matrices of the fundamental representation of $SU(3)$.   It should be understood that $u_{L,R}, \ d_{L,R}$ are $SU_c(3)$ color triplets of the form 
\beq
u_L = \lpa \bea{c} u^r_L \\ u^b_L \\ u^g_L \eea \rpa,
\label{eq:Twentyseven}
\eeq
with the $(r,b,g)$ indicating the color charge of the particle.  Even after adding in quarks and the strong interaction to the model, this cannot be the final picture, as we are left with massless, gauge bosons of $SU_L(2) \times U_Y(1)$ which experimentally is not correct. The standard solution to this problem is use of the Higgs mechanism.

\subsection{\label{sec:NuOsc}The Higgs Mechanism}

The Higgs mechanism allows us to generate masses for particles in the GWS model by adding a scalar field, the so called Higgs field.  The simplest model is generated by specifying the Higgs field as a complex $SU_L(2)$ doublet
\beq
H = \lpa \bea{c} \phi^+ \\ \phi^0 \eea \rpa.
\label{eq:Twentyeight}
\eeq
Using the $SU_L(2)\times U_Y(1)$ gauge invariance, we can choose a gauge where the Higgs field takes the convenient form
\beq
H\rightarrow e^{i \alpha} e^{i \theta^a \tau^a} H = \lpa \bea{c} 0 \\ \phi \eea \rpa,
\label{eq:Twentynine}
\eeq 
where $\phi$ is a real scalar function. The Higgs mechanism requires that we choose some potential for the Higgs field $V(\phi)$ such that its minimum value is non-zero.  The SM employes the ``Mexican hat" potential; which along with the kinetic gauge invariant terms, the scalar part of the SM Lagrangian takes the form
\beq
\CL_{\rm scalar} = \lpa D_\mu H \rpa^\dagger \lpa D^\mu H \rpa -\mu^2 \lvert H \rvert^2 - \lambda \lvert H \rvert^4.
\label{eq:Thirty}
\eeq 
If we were to proceed from this without further revision we would have difficulty doing any calculations as QFT calculations make wide use of the LSZ formula (for a review of the LSZ formula see~\cite{Srednicki:2007qs}) for calculating cross sections and decay rates.  The LSZ formula requires that the fields involved in the calculation have a zero vacuum expectation value (vev), that is we need $\langle 0 \lvert \phi  \rvert 0 \rangle = 0$.  It is possible to show through calculating the quantum action ~\cite{Ryder} for scalar fields that $\langle 0 \lvert \phi  \rvert 0 \rangle = \phi_c$ where $\frac{\p V(\phi)}{\p \phi}\lvert_{\phi_c}=0$ ,  that is $\phi_c$ is at the minimum of the effective potential which at leading order is the classical potential.  By calculating the Higgs field minimum below	
\beq
\bea{c}
\frac{\p V}{\p H^\dagger} = \lpa \mu^2 + 2 \lambda \lvert H \rvert \rpa H =0 \\
\phi_c = 0, \ \pm \sqrt{\frac{-\mu^2}{2 \lambda}}, \\
\eea
\label{eq:Thirtyone}
\eeq
we deduce the vev of the scalar field. However, we must not choose $\phi_c =0$ as the vev as this corresponds to a false vacuum as it is an unstable point of the potential.  Instead, we purpose that $-\mu^2 > 0$ and $\lambda > 0$, giving us a real minimum.  This allows us to  expand the Higgs field around the vacuum expectation value $\langle 0 \lvert \phi \rvert 0 \rangle = \frac{v}{\sqrt{2}} = \sqrt{\frac{-\mu^2}{2\lambda}}$, thereby putting the Higgs field in the form
\beq
H = \lpa \bea{c} 0 \\ \frac{1}{\sqrt{2}}\lpa v + h \rpa \eea \rpa,
\label{eq:Thirtytwo}
\eeq
where $h$ is a real scalar field that satisfies $\langle 0 \lvert h \rvert 0 \rangle = 0$ and thus is appropriate for use in the LSZ formula.  This expansion around the vev will give us mass terms in our Lagrangian, where we also assume that the Higgs field is electrically neutral, $Q H = 0$      
\beq
\bea{c}
\CL_{\rm scalar} = \frac{1}{2} \p_\mu h \p^\mu h + \frac{g_2^2 v^2}{4}W^-_\mu W^\mu_+ + \frac{g_2^2 v^2}{8 \cos^2(\theta_w)}Z^0_\mu Z^\mu_0 \\
+ \frac{g_2^2 v}{2} h W^-_\mu W^\mu_+ + \frac{g_2^2 v}{4} h Z^0_\mu Z^\mu_0 \\
+ \frac{g_2^2}{4} h^2 W^-_\mu W_+^\mu + \frac{g_2^2}{8 \cos^2(\theta_w)} h^2 Z^0_\mu Z^\mu_0 \\
- (-\mu^2) h^2 - \sqrt{-\mu^2 \lambda} h^3 - \frac{\lambda}{4} h^4.
\eea 
\label{eq:Thritythree}
\eeq
This allows us to identify the masses of the particles in the theory as well as relations between them
\beq
\bea{c}
M_W = \frac{g_2 v}{2}, \\
M_Z = \frac{g_2 v}{2 \cos(\theta_w)} \rightarrow M_W = M_Z \cos(\theta_w), \\
m_H^2 = -2 \mu^2.
\eea
\label{eq:Thirtyfour}
\eeq
The Higgs mechanism can also be used to give masses to the leptons by including Yukawa interactions of the form
\beq
\bea{c}
Y_e \lpa \bar{E}_L H \rpa e_R + h.c. = \frac{Y_e v}{\sqrt{2}}\bar{e}_Le_R + \frac{Y_e}{\sqrt{2}}h \bar{e}_Le_R + h.c. \\
= m_e \bar{e}e + \frac{m_e}{v} h \bar{e}e,
\eea
\label{eq:Thirtyfive}
\eeq
where $h.c.$ stands for the hermitian conjugate of the entirety of the preceding equation.  However this only generates a mass term for the lower spinor of the $SU_L(2)$ doublet.  It is easy to add a gauge invariant term that will generate a mass for the upper spinor of the $SU_L(2)$ doublet by adding a term like
\beq
Y \lpa \bar{E}_L i \sigma^2 H \rpa \nu_R + h.c. \ .
\label{eq:Thirtysix}
\eeq
This term is absent in the SM for leptons, and therefore leaves neutrinos massless.

This is where we encounter our first error in the SM. From 1968 the Homestake experiment performed by Davis, Jr. and collaborators~\cite{Davis} suggested that the flux of solar neutrinos was not what the SM would predict when coupled to the standard solar model of the time.  The generally accepted explanation for this is called neutrino oscillations. It takes the idea that the neutrino flavor states $\lpa \nu_e, \nu_\mu, \nu_\tau \rpa$ are not the same as the mass eigenstates, and can be expressed as the linear combination of the mass eigenstates
\beq
\lvert \nu_f \rangle = \sum_{m} U_{fm} \lvert \nu_m \rangle,
\label{eq:Thirtyseven}
\eeq
where $\nu_f$ are neutrino states of definite flavor and $\nu_m$ are neutrino states of definite mass. If we appeal to quantum mechanics we can calculate the probability that a neutrino of a definite flavor changes into a different flavor state. States of definite mass have a wave function of the form
\beq 
\lvert \nu_m(t) \rangle = e^{-i \lpa E t - p z \rpa} \lvert \nu_m \rangle \approx e^{-i \lpa \frac{m^2}{2 E} z \rpa} \lvert \nu_m \rangle.
\label{eq:Thirtyeight}
\eeq
Which follows from the approximation that the neutrino is highly relativistic.  This allows us to calculate the transition probability 
\beqa
P(f_2 \rightarrow f_1) &=& \lvert \langle \nu_{f_1} \mid \nu_{f_2}(t) \rangle \rvert^2 = \lvert \sum_{a,b} U^\dagger_{a f_1} U_{f_2 b} e^{-i \frac{m_b^2}{2 E} z } \langle \nu_a \mid \nu_{b} \rangle \rvert^2 , \nonumber
\\
&=& \lvert \sum_{b} U^\dagger_{b f_1} U_{f_2 b} e^{-i \frac{m_b^2}{2 E} z }\rvert^2, \nonumber
\\
&=& \delta_{f_1 f_2} - 4 \sum_{a>b} \hbox{Re}\lpa U^\dagger_{a f_1} U_{f_2 a} U_{f_1 b} U^\dagger_{b f_2}\rpa \sin^2\lpa \frac{\Delta m_{ab}^2 z}{4 E}\rpa  \nonumber
\\ 
&\ &+ 2 \sum_{a>b} \hbox{Im}\lpa U^\dagger_{a f_1} U_{f_2 a} U_{f_1 b} U^\dagger_{b f_2}\rpa \sin^2\lpa \frac{\Delta m_{ab}^2 z}{2 E}\rpa, \nonumber
\\
&\ & \Delta m_{ab}^2 = m_a^2 - m_b^2.
\label{eq:Thirtynine}
\eeqa
Based on the above, we see that if there are neutrinos with different but very small masses compared to their energies, then the flavor states oscillate between each other as they propagate through space.  Thus, it is possible that the deficit measurement of $\nu_e$ from solar neutrinos are from fluctuations into $\nu_\mu, \hbox{ and } \nu_\tau$.  Testing this idea with reactors and accelerator beams is still on going but has been confirmed multiple times such as in Kamiokande and Super-Kamiokande Cherenkov detectors, MiniBooNE and MINOS at Fermilab and the Soudan Mine, as well as others.

Adding the Yukawa interaction term with right chiral neutrinos requires an extra level of fine tuning to the SM formulation.  Fine tuning is considered theoretically un-natural,\ie to have various Yukawa couplings of strength $<\mathcal{O}(10^{-11})$ so that the neutrino masses come out to be $< 2 \hbox{ eV}$~\cite{Beringer:1900zz}, while the electron has a coupling of order $\mathcal{O}(10^{-6})$.  The SM is not without its own fine tuning as the top coupling is of order $\mathcal{O}(0.1)$, however it would only serve to increase the level of fine tuning by introducing right chiral neutrinos.  One mechanism to solve this naturalness problem is to introduce massive Majorana neutrinos as well and evoke the see-saw mechanism, which will allow for very light neutrinos and very massive additional neutrinos.  For a review of the see-saw mechanism, see~\cite{GonzalezGarcia:2007ib}. 

\subsection{Problems with the Higgs: Naturalness}
 
\begin{figure}[h]
\begin{center}
\postscript{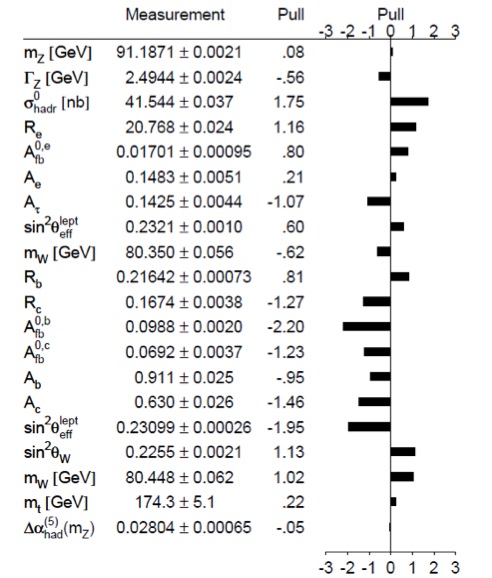}{0.6}
\caption[SM Predictions vs. Experiment]{\sglspc \small SM predictions versus experimental observation, where  $\lpa \hbox{Pull} \rpa \sigma_{\rm measured} =  M - P$ where $M$ is the measured observable and $P$ the predicted value  From Ref.~\cite{Novaes}. \dblspc}
\label{fig:SMT}
\end{center}
\end{figure}

Since the SM inception, it has been experimentally verified time and time again.  Shown in Fig.~\ref{fig:SMT} is a comparison of measured values versus their SM predicted values ~\cite{Novaes}.  The pull on the figure is an indication of how many standard deviations the observed value and the predicted value are separated by, $\lpa \hbox{Pull} \rpa \sigma_{\rm measured} =  M - P$ where $M$ is the measured observable and $P$ the predicted value. The average of the pulls gives $\langle \hbox{Pull} \rangle = -0.12$ showing how good of a theory the SM is.
%----------------Figure------------------
%-----------------------------------------
Of course we know the SM cannot be the final theory.  Many of the problems associated with the SM arise because of our use of the Higgs field.  To highlight this issue, we look at 1-loop corrections to the mass of the particles.  

\begin{figure}[h]
\begin{center}
\postscript{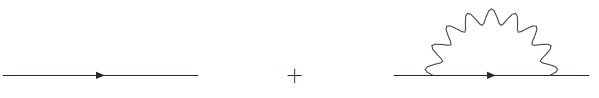}{0.8}
\caption[One Loop QED Fermion Propagator Correction]{\sglspc \small The fermion experiences self interactions with the photon of QED, this leads to a correction of the bare mass parameter of the Lagrangian for QED.}
\label{fig:OLQEDm}
\end{center}
\end{figure}

	The topic of renormalization is a complicated subject; however, it comes down to one issue: we do not know the relevant physics at high energies, and in loop calculations physics at all scales appear. It is known that if one could calculate all the observables in a theory via the bare parameter values then you would not have an issue with renormalization. However, since we cannot turn off and on interactions, we cannot measure the values of the bare parameters, even with this constraint, bare values makes a great conceptual tool.  The physical parameters measured will be related to their bare values through relations that can be calculated via Feynman diagrams.  For example, the lowest order corrections to QED fermion bare masses ($m_0$) is given by the diagram in Fig.~\ref{fig:OLQEDm}.
%----------------Figure------------------	
%-----------------------------------------
These diagrams result in the expression for the two point Green's function, in momentum space
\beq
\langle 0 | T\{\bar \psi(p) \psi(0)\} | 0 \rangle = \frac{i (p \cdot \gamma + m_0)}{p^2 -m_0^2} +  \frac{i (p \cdot \gamma + m_0)}{p^2 -m_0^2} i A  \frac{i (p \cdot \gamma + m_0)}{p^2 -m_0^2} \ .  
\eeq
This is understood as a modification of the $m_0$ parameter to the particle's physical mass ($m_p$) given by
\beqa
m_p &=& m_0 + i A, \nonumber \\
A &=& -e_0^2 \int \frac{d^4 k}{\lpa 2 \pi \rpa^4} \frac{\gamma^\mu \lpa \gamma^\nu k_\nu + m_0 \rpa \gamma_\mu}{\lpa k^2-m_0^2 \rpa k^2} \ , \nonumber \\
&=& -i m_0 \frac{e_0^2}{2 \pi^2} \int_0^\Lambda dk_E \frac{k_E}{\lpa k_E^2+m_0^2 \rpa}, \nonumber
\\
&\approx& -i m_0  \frac{2 \alpha_0}{\pi} \log \lpa \frac{\Lambda}{m_0}\rpa + \mathcal{O}\lpa \alpha_0^2 \rpa.
\label{eq:Forty}
\eeqa
Since quantum physics allows the particles in a loop to have any energy/momentum, then as we go up in energy scale through the integration, we should expect the relevant physics to change, so much so that we don't know if the current theory is correct through out the integration region.   To quantify this, we impose a cut-off of the momentum integral at $\Lambda$ in calculating the loop value, representing our confidence in the theory at scales below $\Lambda$. The physically observed mass of the particle is $m_p$ and we can relate the physical mass to the bare mass $m_0$ via
\beq
m_0 = m_p \lpa 1 - \frac{2 \alpha}{\pi} \log \lpa \frac{\Lambda}{m_p}\rpa + \mathcal{O}\lpa \alpha^2 \rpa \rpa.
\label{eq:Fortyone}
\eeq
We have replaced $\alpha_0$ by $\alpha$ because corrections to $\alpha_0$ are $\mathcal{O}\lpa \alpha_0^2 \rpa$ and thus do not appear at the $\alpha$ order.  From this we can deduce that even if $\Lambda = 10^{16} \hbox{ GeV}$ then $m_0 - m_p = -0.10$ amazingly even if we use the physical mass $m_p$ in our calculations when we should have been using the bare mass $m_0$, we are only making an error of $10\%$.  It is said to be natural in QED for fermions to have a low mass, as contributions from very high energies do not significantly change the bare mass of the particle.

\begin{figure}[h]
\begin{center}
\postscript{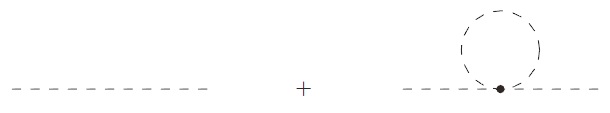}{0.8}
\caption[One Loop Scalar Field Propagator Correction]{\sglspc \small The scalar field experiences self interactions, which leads to a correction of the bare mass parameter of the Lagrangian for scalar fields that is much different than that of QED.}
\label{fig:OLSm}
\end{center}
\end{figure}

If we repeat this calculation for a scalar field we get a much different result. Corrections of a $\phi^4$ scalar field to the mass of the particle are given by the diagram in Fig.~\ref{fig:OLSm}
%----------------Figure------------------
%-----------------------------------------
This diagram results in the modification of the bare mass of the scalar field particle given as
\beqa
m_p^2 &=& m_0^2 -i A \ , \nonumber \\
A &=& \frac{\lambda_0}{2} \int \frac{d^4 k}{\lpa 2 \pi \rpa^4} \frac{1}{k^2-m_0^2} \approx -i \frac{\lambda_0}{32 \pi^2}\lpa \Lambda^2 -2 m_0^2 \log \left[ \frac{\Lambda}{m_0} \right] \rpa .
\label{eq:Fortytwo}
\eeqa 
From this we see that $m_p^2-m_0^2 \approx - \frac{\lambda_0}{32 \pi^2} \Lambda^2$.  Given the recent discovery of a Higgs-like particle with a mass of $m_H = 125 \hbox{ GeV}$, it seems that if $\Lambda = 10^{16} \hbox{ GeV}$ we are making a very large mistake using the tree level mass $m_p$ when we should be using the bare mass of the scalar field $m_0$, as it is natural for the bare mass to be very large.  This is odd as perturbation theory requires $\lambda_0$ to be small but it would seem to agree with experiment that $\lambda_0$ must be very, very small on the order of $10^{-32}$!  While the bare mass term would be on order of $10^2$ it would be very odd (un-natural) to have a theory where the parameters vary in strength so wildly!  This is also known as a fine-tuning problem, and is a main motivator for beyond the standard model physics, especially supersymmetry which solves this issue by removing exclusively scalar field quantum corrections.

\subsection{Problems with the Higgs: Stability}

\begin{figure}[ht]
\begin{center}
\postscript{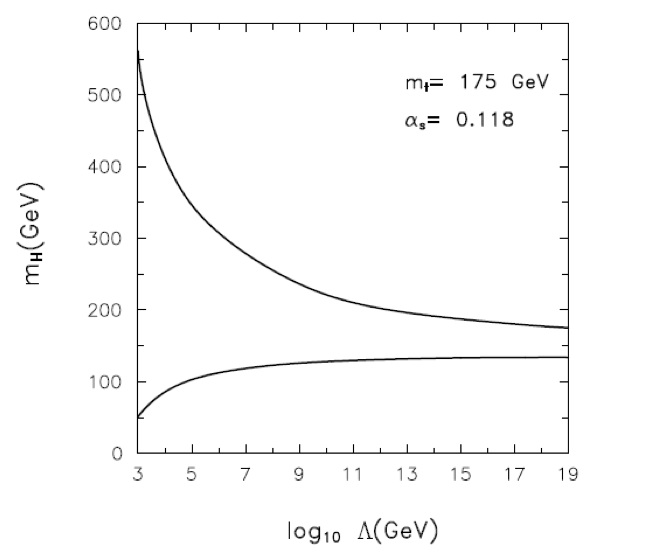}{0.6}
\caption[SM Higgs Stability Curves]{ \sglspc \small Stability of the Higgs vacuum under the SM requires the mass of the Higgs to fall with the curves at the given energy scale $\Lambda$ which you expect the SM to still be correct . Taken from Ref.~\cite{Novaes}.}
\label{fig:HiggsStab}
\end{center}
\end{figure}

The Higgs mechanism depends on a non-zero, real vacuum expectation value.  It must be that the observed non-zero vacuum expectation value measured at all energy interactions is non-zero and real.  So it must be that the physical couplings $\lambda(Q), \mu^2(Q)$ at the scale $Q$, must have $-\mu^2(Q) > 0 \rightarrow m_H^2(Q) >0 $, and $\lambda(Q)>0$.  If we consider the self coupling the dominate term in the Higgs interactions, then the renormalization for $\lambda(Q)$ at first order gives
\beq
\lambda(Q) = \frac{\lambda(q)}{1-\lambda(q)\frac{3}{4\pi^2} \log\lpa \frac{Q^2}{q^2} \rpa},
\label{eq:Fortythree}
\eeq
while $-\mu(Q)^2$ stays essentially fixed over a large range of $Q$.  If we rewrite the renormalization equation in terms of the Higgs mass we get a lower bound on what the Higgs mass can be,
\beq
Q < m_H \exp\lpa \frac{4 \pi^2 v^2}{3 m_H^2} \rpa,
\label{eq:Fortyfour}
\eeq
where $Q$ is the scale of new physics.  There is also another limit in which this calculation will no longer be correct, if $\lambda(Q) >1$.  The calculation is based on perturbation theory so it should be that $\lambda(Q) \ll1$, and using this as a bound gives
\beq
Q < M_H \exp \lpa \frac{4\pi^2}{3}\frac{v^2}{m_H^2}\lpa 1- \frac{m_H^2}{2 v^2}\rpa \rpa.
\label{eq:Fortyfive}
\eeq
If we include all of the couplings of the SM we get a bound on the Higgs mass and $\Lambda$ shown in Fig. \ref{fig:HiggsStab}.
%----------------Figure------------------
%-----------------------------------------
For a Higgs mass $m_H = 125 \hbox{ GeV}$, this indicates to us the SM vacuum is no longer stable at around $Q = 10^9 - 10^{10} \hbox{ GeV}$, and should be taken as an indication of some new physics around this scale.

\subsection{Accidental Symmetries of the SM}

Besides the gauge symmetries and space-time symmetries of the SM we also have accidental symmetries that occur.  Noether's theorem states that any symmetries of the action correspond to conserved currents.  These accidental symmetries are global $U(1)$ symmetries that are associated with Baryon number $B$ and Lepton number $L$ conservation, and are enacted by the transformations of the form
\beq
E_L \rightarrow e^{i \beta} E_L, \ e_R \rightarrow e^{i \beta} e_R;
\label{eq:Fortysix}
\eeq
with a similar $U(1)$ transformation for quarks that corresponds to Baryon number conservation.  In some grand unified theories (GUT) the decay of a proton is possible, though experiment suggests that the half life of the proton is at least $10^{33}$ years~\cite{SuperK}.  If proton decay is possible then the Baryon and Lepton number will be violated, however $B-L$ will not be.

\subsection{Anomalies}

\begin{figure}[ht]
\begin{center}
\postscript{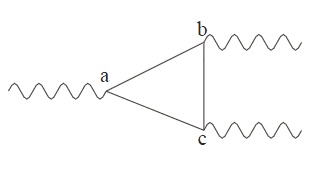}{0.45}
\caption[Triangle Anomaly]{\sglspc \small Triangle diagrams like this one can show the delicate balance required on gauge theories with chiral matter fields to remain renormalizable. \dblspc}
\label{fig:TriDia}
\end{center}
\end{figure}

Chiral theories coupled with gauge theories typically are not consistent.  The SM has interaction currents that have an axial term of the form
\beq
J^\mu_A = \bar{\nu} \gamma^\nu \gamma^5 e.
\label{eq:Fortyseven}
\eeq
This is a problem for QFT because renormalization of divergent gauge field theories such as the SM depend on interaction currents to respect their current conservation equations~\cite{Peskin}.  However, axial currents do not remain conserved at the quantum level.  Triangle diagrams of the type depicted in Fig.~\ref{fig:TriDia} have the potential to break the conservation of currents.
%----------------Figure------------------
%-----------------------------------------
In the SM we have both axial and vector currents and this problem persists.  The anomalies must cancel if the SM is to remain renormalizable, which is only possible by modifying the particle spectrum of the theory. 

The anomaly will no longer be present if $\hbox{Tr}[ T^a \{T^b,T^c\}] =0$ where $T^a,T^b,T^c$ are the generators of the gauge transformations of the SM at the interaction vertices of the triangle diagram. If we consider a triangle diagram containing only $SU_L(2)$ gauge bosons then
\beq
\hbox{Tr}[\sigma^a \{\sigma^b,\sigma^c\}] = 2 \delta^{bc} \ \hbox{Tr}[\sigma^a] = 0.
\label{eq:Fortyeight}
\eeq
We have no inherent anomalies with $SU_L(2)$, so there is no need to modify the particle spectrum. However, considering two vertices having $SU_L(2)$ gauge bosons $W^3_\mu$ and one vertex with a $U_Y(1)$ boson (the generator of the $U_Y(1)$ gauge symmetry is the hypercharge $Y$), the trace then evaluates to
\beq
\bea{c}
\hbox{Tr}[Y (\tau^3)^2] = \sum_{L} Y_i (\tau^3_i)^2 + \sum_{R} Y_i (\tau^3_i) \\
= 2(-1) + 3\cdot 2 \cdot \frac{1}{3}.
\eea
\label{eq:Fortynine}
\eeq
The use of the trace here is generalized to include a sum over all possible particles circulating in the loop of the triangle.  Let us examine this result, $2(-1)$ comes from the $2$ left chiral parts of $E_L$ while $3\cdot 2 \cdot \frac{1}{3}$ comes from the left chiral part of the quark content $Q_L$ (the $3$ because of the $3$ different colors the quark can posses).  If you repeat the process you are lead to the conclusion that anomalies can only cancel if the SM has complete $(e_R, E_L, Q_L, u_R, d_R)$ particle spectrums for each generation.  In fact this result was used to predict the top quark's existence ~\cite{Maskawa} in 1973 by Maskawa, and Kobayashi.

\subsection{Beyond the SM}

The symmetries allowed in the SM make up a great deal of how it works, but massive neutrinos and accidental symmetries have yet to be explained.   There are other problems plaguing the SM the most obvious of which is its absence of gravity.  Gravity presents a problem for the SM, as it is very very weak compared to the rest of the gauge couplings. Consider that that electroweak interactions occur at the scale of $M_Z = 91 \hbox{ GeV}$ while the relevant scale of gravitational interaction is $M_{pl} \approx 10^{19} \hbox{ GeV}$.  How can any unified theory overcome this enormous scale difference?  One popular theory that can solve such a problem is the inclusion of extra dimensions.
 If you consider Guass's equation to be fundamental\footnote{ \sglspc \small This derivation follows that of~\cite{Zwiebach}.} for any number of dimensions $D$ with $M^{(D)}_{pl}$ the fundamental gravitational scale for $D$ dimensions, then Guass's law is
\beq
\nabla \cdot \textbf{G}^{(D)} = \frac{4 \pi}{\lpa M^{(D)}_{pl} \rpa^2} \rho^{(D)}.
\label{eq:Fifty}
\eeq
If we then consider a ring of mass distributed around a compactified dimension that takes the shape of a cylinder of radius $R$ then we can then write $\rho^{(5)} = m \ \delta^{(3)}(\textbf{x})$ (the $5$ indicates the one extra spatial direction added from the usual $3+1$ dimensions we are used to).  If we integrate out the compactified dimension we would find a point mass in $3$ spatial dimensions with a mass density of the form $\rho^{(4)} = M \delta^{(3)}(\textbf{x})$ the question is what is the relation between the two mass densities?  Upon integration we can find the result:
\beqa
M &=& \int_{-\infty}^{\infty} dx_1 dx_2 dx_3 \int_0^{2\pi R} dx_4  m \ \delta^{(3)}(\textbf{x}), \nonumber
\\
M &=& m \ 2 \pi R \rightarrow \rho^{(5)} = \frac{1}{2\pi R} \rho^{(4)}.
\label{eq:Fiftyone}
\eeqa 
Adding this to our Guass's equation gives
\beq
\nabla \cdot \textbf{G}^{(4)} = \frac{4 \pi}{(2\pi R) \lpa M^{(5)}_{pl} \rpa^2} \rho^{(4)} = \frac{4 \pi}{\lpa M^{(4)}_{pl} \rpa^2} \rho^{(4)}.
\label{eq:Fiftytwo}
\eeq
This tells us the fundamental scale of gravity $M^{(4)}_{pl} = \sqrt{2\pi R} M^{(5)}_{pl}$, if $ 2\pi R > 1$  then  $M^{(4)}_{pl}$ becomes larger than the fundamental scale of gravitational physics.  This procedure generalizes to many compactified dimensions so that the final form is
\beq
M^{(4)}_{pl} = \sqrt{V_c} M^{(4+d)}_{pl},
\label{eq:Fiftythree}
\eeq
where $V_c$ is the volume of the $d$ additional compactified dimensions. With extra dimensions it becomes easier to unify gravity with the other three known forces and extra dimensions are a general attribute of string theory, though these extra spatial dimensions have yet to have been detected~\cite{Cullen:1999hc,Hoyle:2000cv,Hannestad:2001jv,Hannestad:2001xi,Anchordoqui:2001cg,Anchordoqui:2002vb,Chatrchyan:2011fq,Chatrchyan:2012tea,Aad:2012ic,Chatrchyan:2012me}.

Along with the absence of gravity in the SM, we must also ask why does the SM repeat the same rules for 3 generations.  This should be looked upon like a periodic table, the same behavior of certain elements in chemical reactions ends up being explained by a simpler unifying model of electrons and atomic shells.  We should take the 3 generations repeating in higher masses but same interactions as a hint of a substructure to the SM.  The SM is viewed as an effective field theory at the present moment and experiments conducted at the Large Hadron Collider (LHC) at CERN may offer hints at what is beyond the SM. Many theories are currently exploring possible extensions, one of which will be discussed in part I of this dissertation.

%---------------------------------Ch 1: Section 3-----------------------------------------
\section{Review of $\Lambda$CDM Cosmology}

Cosmology also can provide us with a window into BSM physics.  Using cosmology to explore new physics requires knowledge of the current standard model of cosmology known as the $\Lambda$CDM model.  We will review some of the main results of $\Lambda$CDM as it will be used for calculations of gravitational wave stochastic background which will be explored in Part II of this dissertation. As well, $\Lambda$CDM is used for searches for extra relativistic degrees of freedom during the early universe that the SM does not account for.  As this is not a comprehensive review, it is assumed the reader has knowledge of the theory of general relativity and has experience with these results.

\subsection{\label{sec:FRW} FRW metric}

The universe on the large scale (larger than $10^2$ Mpc) seems to be isotropic and homogenous.  If one asks what is the most generally isotropic and homogenous metric one can formulate, we are lead to the Friedmann-Robertson-Walker (FRW) metric~\cite{AFried,AFried1,HPRobert,HPRobert1,AGWalk}
\beq
ds^2 = dt^2 - a(t)^2\lpa \frac{dr^2}{1-k r^2} + r^2 d\Omega^2\rpa,
\label{eq:Fiftyfour}
\eeq
where $a(t)$ is known as the scale factor, and $k$ indicates the curvature of the universe, with $k = 1$ corresponding to closed, $k=0$ to flat, and $k=-1$ to open.  In equation  (\ref{eq:Fiftyfour}) the $\Omega$ is the solid angle familiar to spherical coordinates.  The use of this metric alone allows us to make an interesting observation, that the all objects a proper distance $d_p$ away (and considered at coordinate distance $R$ away from the origin) are moving with a velocity proportional to their proper distance as can be seen below, 
\beq
d_p = a(t) \int^R_0 \frac{dr}{ \sqrt{1-k r^2} } \rightarrow v_p =  \dot{a}(t) \int^R_0 \frac{dr}{ \sqrt{1-k r^2} } = \frac{\dot{a}(t)}{a(t)} d_p = H(t) d_p.
\label{eq:Fiftyfive}
\eeq
In  (\ref{eq:Fiftyfive}) the value $H(t)$ is the well known Hubble's constant~\cite{EHub}. Strictly, $H(t)$ actually isn't a constant in this equation, however if it varies slowly compared to observation time then it can be approximated to be constant.  This embodies the concept of the expanding universe, with everything at a radial distance $d_p$ moving away with the same velocity.  This immediately leads to the idea of the Big Bang, as all the objects move closer together as you reverse time. Solving for the age of the universe requires you to know the form of the scale factor $a(t)$ .  You can solve for $t$ in terms of $H_0$, the Hubble constant measured today, thus calculating the age of the universe. Even from this concept of a finite age of the universe many things can be deduced about what we should observe today, most importantly the CMB which we shall touch on in the next section.

If we apply the FRW metric to the Einstein field equations with the modification of a cosmological constant $\Lambda$, which does not break the general coordinate symmetry of GR, we arrive at the Friedmann equations
\beq
\bea{c}
\lpa\frac{\dot{a}}{a}\rpa^2 + \frac{k}{a^2} = \frac{8 \pi}{3}\frac{1}{M_{pl}^2} \rho + \frac{\Lambda}{3},
\\
\frac{\ddot{a}}{a}=-\frac{4 \pi}{3}\frac{1}{M_{pl}^2} \lpa \rho + 3 P \rpa + \frac{\Lambda}{3}.
\eea
\label{eq:Fiftysix}
\eeq
We have assumed the matter content is a homogenous perfect fluid with mass-energy density $\rho$ and pressure $P$. If we re-express this equation in terms of the Hubble constant, we get
\beq
\bea{c}
1 + \frac{k}{a^2 H^2} = \frac{8 \pi}{3}\frac{1}{M_{pl}^2 H^2} \lpa \rho + \frac{\Lambda M_{pl}^2}{8\pi} \rpa \rightarrow \frac{k}{a^2 H^2} = 1- \Omega,
 \\
\Omega = \frac{8 \pi}{3}\frac{1}{M_{pl}^2 H^2} \lpa \rho + \frac{\Lambda M_{pl}^2}{8\pi} \rpa = \frac{\sum_i \rho_i}{\rho_c},
\eea
\label{eq:Fiftyseven}
\eeq
where $\rho_i$ is the energy density content for different types of matter, \textit{e.g.} baryonic matter, radiation, dark energy, \textit{etc}.  We can see if $\Omega = 1$ then the universe is flat, $k = 0$.  This only occurs if the universe has an energy density equal to the critical density $\rho_c = 3 M_{pl}^2 H^2/8 \pi$.  Given that the universe is free to have any content, there are infinitely many more configurations of the universe not having this exact critical density, so it would be a surprise to find that the matter content is such that $\Omega = 1$.  Combining the two Friedmann equations gives
\beq
\dot{\rho} + 3 \frac{\dot{a}}{a} \lpa \rho + P \rpa = 0,
\label{eq:Fiftyeight}
\eeq
which is a different statement of the 1st law of thermodynamics for a universe with 3 spatial dimensions.  To demonstrate equation (\ref{eq:Fiftyeight}) is the 1st law of thermodynamics, consider a sphere of proper volume $V(t) = \frac{4\pi}{3} a(t)^3 r_0^3$, then $dQ = dU + P dV$ gives 
\beqa 
dQ &=& d(\rho V) + P dV,  \nonumber
\\
dQ &=& (d\rho) V  + \lpa \rho + P \rpa dV, \nonumber
\\
dQ &=& \dot{\rho} + \frac{\dot{V}}{V} \lpa \rho + P \rpa, \nonumber
\\
dQ &=& \dot{\rho} + 3  \frac{\dot{a}}{a} \lpa \rho + P \rpa = 0.
\label{eq:Fiftynine}
\eeqa
As a consequence of a homogenous universe, there can be no heat flow $dQ = 0$.  This forces the entropy to be constant, $dS = dQ/T = 0$. The conservation of entropy allows us to derive a useful result\footnote{derivation follows that of~\cite{Kolb:1990vq}.}
\beqa
0 &=& dQ = T dS = d(\rho V) + P dV = V d\rho + (\rho + P) dV, \nonumber
\\
T dS &=&T \lpa \lpa \frac{\p S}{\p T} \rpa dT + \lpa \frac{\p S}{\p V} \rpa dV   \rpa \rightarrow T \lpa \frac{\p S}{\p T} \rpa = V \frac{\p \rho}{\p T} , \ \lpa \frac{\p S}{\p V} \rpa = \frac{\rho + P}{T},  \nonumber
\\
\frac{\p^2 S}{\p V \p T} &=& \frac{\p^2 S}{\p T \p V} \rightarrow \frac{\p^2 S}{\p T \p V} = \frac{1}{T} \rho' =  -\frac{\rho + P}{T^2} + \frac{\rho' + P'}{T}, \nonumber 
\\
dP &=& \frac{\rho + P}{T} dT,
\label{eq:Sixty}
\eeqa
where $\rho' = d \rho/d T, \hbox{ and } P' = d P/d T$. This allows an expression for the entropy per unit volume
\beqa
dS &=& \frac{1}{T}d((\rho+P)V) - \frac{V}{T} dP =  \frac{1}{T}d((\rho+P)V) - \frac{V}{T^2} (\rho +P) = d\lpa \frac{(\rho+P)V}{T} \rpa, \nonumber 
\\
s &=& \frac{S}{V} = \frac{\rho + P}{T} + \hbox{const}.
\label{eq:Sixtyone}
\eeqa
We take as the definition of $s$ to be the case where $\hbox{const} = 0$.  For calculations we typically only count relativistic species as contributing towards the entropy per unit volume. Why is this?  We can make use of statistical mechanics to derive the pressure and energy density for fermions and bosons,
\beq
\rho =  \int \frac{d^3 p}{ (2 \pi)^3} \frac{g \ E}{e^{\beta E} \pm 1}, \ P =  \int \frac{d^3 p}{ (2 \pi)^3} \frac{p}{3 E} \frac{g}{e^{\beta E} \pm 1}, 
\label{eq:Sixtytwo}
\eeq
where $E = \sqrt{p^2 + m^2}, \ \beta = 1/T$, and $g$ is the degeneracy for the respective particle for the energy level $E$.  With $e^{\beta E} \pm 1$ we use the $+$ for fermions and $-$ for bosons. In the highly relativistic limit  ($m \beta \ll 1$) the particles behave as massless, and the results of equation  (\ref{eq:Sixtytwo}) give for bosons and fermions respectively,
\beq
P = \frac{1}{3} \rho = \lpa 1 , \   \frac{7}{8}  \rpa g \frac{\pi^2}{30} T^4 \rightarrow s =  \lpa 1 , \   \frac{7}{8}  \rpa  g  \frac{2 \pi^2}{45} T^3.
\label{eq:Sixtythree}
\eeq   
While in the non-relativistic limit $(m \beta \gg 1)$ the pressure and energy density are the same for bosons and fermions
\beq
\rho = g \ m \lpa \frac{ m T}{2\pi} \rpa^{3/2} e^{-\beta m}, \ P = \frac{\rho}{m} T \rightarrow s = \frac{g}{T} \lpa \frac{ m T}{2\pi} \rpa^{3/2} (1+ T/m)  e^{-\beta m}.
\label{eq:Sixtyfour}
\eeq
The suppressive nature of the $\exp(-\beta m)$ makes the non-relativistic particles contribute significantly less to the pressure, energy density, and entropy of the universe compared to the relativistic particles.  In most cases it will be appropriate to ignore the contributions from non-relativistic species when considering the thermodynamics of the universe. 

After our discussion of thermodynamics, we now can determine for the simplest cases, the equation of state of the perfect fluids dominating the stress energy tensor for the universe.  With this information we can determine the effect on the scale factor.  In general the equation of state can typically be related to the energy density $\rho$ through a temperature dependent constant $P = w \rho$. For example, if we consider radiation (highly relativistic particles), we see in equation  (\ref{eq:Sixtythree}) that $P = \rho/3$.  While dust (non-relativistic matter) would essentially have no internal kinetic energy ($T/m \approx 0$), we may then fix the pressure for dust as $P = 0$.  For these two cases the scale factor evolves like
\beq
\bea{c}
\dot{\rho} + 3 \frac{\dot{a}}{a} \lpa 1 + w \rpa \rho = 0, \\
\frac{d\rho}{\rho} = -3 \lpa 1 + w \rpa \frac{da}{a}, \\
\rho(a) a^{3(1+w)} = \rho_0 a_0^{3(1+w)},
\eea
\label{eq:Sixtyfive}
\eeq
which tells us that for matter dominated universes, we have $\rho_m \propto a(t)^{-3}$, and for radiation dominated, $\rho_r \propto a(t)^{-4}$.  The first Friedmann equation  (\ref{eq:Fiftyseven}) with $k=0$ (the $k=0$ result will be used later on) can give us more detail of the scale factor by inclusion of equation  (\ref{eq:Sixtyfive}) we have
\beq
\bea{c}
\frac{\dot{a}}{a} = \sqrt{\frac{8 \pi}{3}\frac{\rho_0}{M_{pl}^2}} a^{-\frac{3(1+w)}{2}} \rightarrow a^{\frac{3(1+w)}{2} -1} da =  \sqrt{\frac{8 \pi}{3}\frac{\rho_0}{M_{pl}^2}} dt,
\ \\
a(t) \propto t^{\frac{2}{3(1+w)}}.
\eea
\label{eq:Sixtysix}
\eeq
We have assumed the condition $a(t_0 = 0) = 0$ in the result.  With Eq.~(\ref{eq:Sixtysix}) the solutions for matter dominated universes is $a(t) \propto t^{\frac{2}{3}}$ and radiation dominated universes is $a(t) \propto t^{\frac{1}{2}}$.

\subsection{Cosmic Microwave Background}

If we look at the thermal history of the universe by rewinding time, we are also bringing all the the matter in the universe into a smaller volume, thus increasing the density.  When rewinding time the cosmic redshift of photons will reach a point at which all photons will have energies above $Q = 13.6 \hbox{ eV}$, the binding energy of hydrogen.  At this point the hydrogen disassociates into free protons and electrons. This matter will then be in a Baryon photon plasma state.  When the universe expands, there is a point where the photons ionizing the hydrogen will no longer be within the mean free path for ionization, so the photons start to freely stream in the universe.  Since the universe is assumed to be homogenous then we should expect this to occur everywhere in the universe.  We would expect to be able to observe these free streaming photons today redshifted to a longer wavelength.  We also expect that the spectrum of photons is that of a blackbody as they originate from thermal equilibrium of free protons and electrons.

To calculate the temperature at the time of recombination (the point at which stable hydrogen forms) we examine the process that produces thermal equilibrium in this era, namely the photo dissociation and recombination $H + \gamma \leftrightarrow p^+ + e^-$. It is useful to  define $X = n_p/(n_p+n_H)$, where $X$ is the fraction of ionized matter.  We should expect that the hydrogen (and other elements that are being produced) are non-relativistic and have a Boltzmann-like thermal distribution so that the number density for a type of particle $x$ with degeneracy $g_x$ is
\beq
n_x = g_x \lpa \frac{m_x}{2\pi \beta}\rpa^{3/2} e^{  -\beta m_x }.
\label{eq:Sixtyseven} 
\eeq
We also assume that the universe as a whole is electrically neutral which then requires $n_p = n_e$ (all electrically charged particles are in their most stable state).  This allows us to express the ratio of the number densities of hydrogen, protons, and electrons as,
\beqa
\frac{n_H}{n_p n_e} &=& \frac{g_H}{g_e g_p} \lpa \frac{m_H}{m_e m_p} \rpa^{3/2} (2\pi \beta)^{3/2} e^{ \beta (m_e + m_p - m_H) }, 
\\
&\approx& \lpa \frac{m_e}{2\pi \beta} \rpa^{-3/2} e^{\beta Q }.
\label{eq:Sixtyeight}
\eeqa
This is known as the Saha equation ~\cite{Weinberg_cosmo}.   In this expression $g_H = 4, \ g_p = 2, \ g_e = 2$ and to a good approximation $m_H = m_p$. In Eq. (\ref{eq:Sixtyeight}),$Q = 13.6 \hbox{ eV}$, the binding energy of the first orbital of the hydrogen atom.  Re-arranging our fractional ionization expression gives
\beq
n_H = \frac{1-X}{X}n_p,
\label{eq:Sixtynine}
\eeq
which can be related to the number density of photons $n_\gamma$ via an additional relation of $\eta = (n_p + n_H)/n_\gamma = n_p/(X n_\gamma)$ where $n_\gamma = (2 \zeta(3)/\pi^2) T^3$. This results in
\beq
\bea{c}
n_p = \eta X n_\gamma = \eta X  \frac{2 \zeta(3)}{\pi^2} \frac{1}{\beta^3},
 \\
\frac{1-X}{X^2} = \eta  \frac{2 \zeta(3)}{\pi^2} T^3 \lpa \frac{m_e T}{2\pi} \rpa^{-3/2} \exp [\beta (13.6 \hbox{ eV}) ],
\\
\frac{1-X}{X^2} = 3.84 \ \eta (m_e \beta)^{-3/2} e^{ \beta Q}. 
\eea
\label{eq:Seventy}
\eeq  
The time of recombination is defined as the time when $X = 1/2$ and we can find $\eta$ at the time of recombination by measuring $n_p+n_H$ and $n_\gamma$ today and taking into account that number densities scale like $a(t)^{-3}$ and thus $\eta$ remains fixed for all time.  Measurements of baryonic matter today show $\Omega_{b,0} \approx 0.05$, which is accomplished by measuring abundance of light elements in galaxies.  This allows us to find the number density of baryonic matter today. We can assume that the baryonic matter is non-relativistic so its energy is dominated by mass energy, which is mostly in the form of hydrogen or free protons which both have a mass of nearly $m_p = 938 \ \hbox{MeV}/c^2$.  We find
\beq
n_{b,0} = \frac{\rho_c \Omega_{b,0}}{m_p c^2} \approx 0.5 \ h^2 \ \hbox{m}^{-3},
\label{eq:Seventyone}
\eeq
where $H = 100 h \ \hbox{km/s/Mpc}$.  To make our calculation more precise we can use the measured CMB temperature of $T=2.7$ K.  An interesting note is without measuring a photon background you could use the fact that we cannot see a uniform background at night so it must be that $T_{\gamma,0} < 6 \ K$ so that the peak of the blackbody radiation today lies, at most, in the infrared, but the rest of this section takes the value of $T =2.7$ K. This gives $n_{\gamma,0} \approx 3.81 \times 10^{8} \ \hbox{m}^{-3} $. Using $h = 0.7$,  $\eta \approx 6.43 \times 10^{-10}$; this allows us to find the temperature at which recombination is half way complete to be approximately $T \approx 3,773 \ K$. 

We can also find the temperature at which the photon background decouples from interactions with electrons.  We can approximate the conditions for decoupling by using the equation for mean free path distance, is $d = 1/(n \sigma)$, where $\sigma$ is the interaction cross section and $n$ the number density of interaction points.  In an expanding universe a particle cannot exceed the particle horizon distance $d_p$, the furthest distance a particle moving at the speed of light can travel since the big bang. During the time of radiation dominance the horizon distance is $ d_p = 1/H$ and in the time of matter dominance is $d_p = 2/H$; in both cases the length scale is set by $1/H$ the Hubble horizon distance.  If the mean free path becomes larger than the particle horizon, then the particles can no longer interact and become free streaming.  This gives the relation, $d \geq d_p \rightarrow \Gamma = n \sigma \leq H$.  For photons scattering off bound electrons and non-relativistic free electrons, we approximate the interaction cross section by the dominant Thomson cross section $\sigma_e \approx 8\pi \alpha^2/(3 m_e^2) $.  This allows us to calculate the approximate temperature of the photons when they began free streaming by first calculating the interaction rate,
\beq
\Gamma(T) = n_e(T) \sigma_e = \frac{X(T) n_{b,0}}{a(T)^3} \sigma_e.
\label{eq:Seventytwo}
\eeq
The Friedmann equations allow us to replace the scale factor with a relation to the dominant matter term during decoupling (the universe is a matter dominated phase at this time)
\beq
H = \frac{H_0}{a^{3/2}} \Omega_{m,0}^{1/2}.
\label{eq:Seventythree}
\eeq
We use $\Omega_{m,0} \approx 0.3$, (a result determined by analysis of CMB anisotropies and the discrepancy between $\Omega_{m,0} > \Omega_{b,0}$ leading to more evidence of dark matter). With the decoupling condition $\Gamma(T) = H(T)$ we can determine when the photons decouple:
\beqa
\frac{X(T)}{a(T)^{3/2}} &=& \lpa \frac{H_0 \Omega_{m,0}^{1/2} }{n_{b,0} \sigma_e} \rpa,  \nonumber
 \\
X(T) T^{3/2} &=& T_0^{3/2} \lpa \frac{H_0 \Omega_{m,0}^{1/2} }{n_{b,0} \sigma_e} \rpa, \nonumber
\ \\
T^{\rm dec}_{\gamma} &\approx& 2945 \ K.
\label{eq:Seventyfour}
\eeqa
In the second line we used the fact that as the photons red-shift the spectrum retains its shape but moves to lower energies.  This allows us to relate the temperature of the photons with the scale factor as $T(t) = T_0/a(t)$, where $T_0$ is the temperature of the photons measured today and the convention $a(\rm today) = 1$ is used.  With a matter dominated universe from the point of free streaming to today, this corresponds to a time of $t \approx 539,066$ years after the big bang. 

Presumably there should also be stochastic backgrounds from all particle types that decouple from interactions at certain temperatures.  Consider a neutrino background which we can estimate would become free streaming at $\Gamma = n_e \langle \sigma \rangle \propto G_F^2 T^5$.\footnote{Valid when the reactions $e + \nu_e \leftrightarrow e + \nu_e$ and $e^+ + e^- \leftrightarrow \nu_e + \bar{\nu}_e$ falls out of equilibrium, which occurs when the electrons are still relativistic but have center of mass energies far from the $W^\pm$ resonance so Fermi's constant $G_F$ is appropriate to use.}  With $H \propto T^2/M_{pl}$, which gives us an order of magnitude estimate of (a similar argument is presented in \cite{Kolb:1990vq} p. 74)
\beq
\frac{G_F^2 T^5}{T^2/M_{pl}} \approx \lpa \frac{T}{1 \hbox{ MeV}} \rpa ^3.
\label{eq:Seventyfive}
\eeq
For temperatures around $T_\nu = 10^{10} \ K$, well before the photons decouple, the neutrinos decouple from the plasma and become free streaming.  The effect of theses additional relativistic particles will be seen when we examine CMB anisotropies, and will play a crucial role in searching for BSM, discussed in Sec. \ref{sec:DM}.  Furthermore, we expect a gravitational wave background at an even earlier time!  However this background will be overpowered by any other sources of stochastic gravitational waves produced at later times.  We will investigate such a process in Part II of this dissertation.

\subsection{Cosmic Parameter Measurement}

\begin{figure}[h]
\begin{center}
\postscript{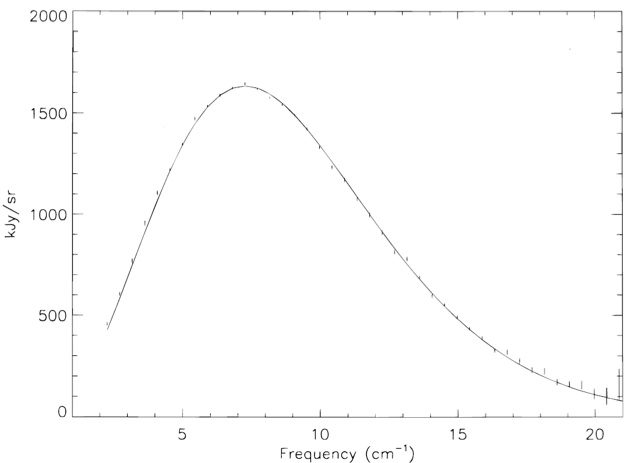}{0.8}
\caption[COBE Black Body Curve]{\sglspc \small Black-body curve corresponding to $T = 2.7$ K from the COBE, FIRAS detector ~\cite{COBE}. \dblspc}
\label{fig:COBEbbody}
\end{center}
\end{figure}

In 1965 Penzias and Wilson of Bell Labs detected a cosmic background of 3.5 K. However it was Dickem, Peebles and Wilkinson who would interpret the excess power at 3.5 K as the cosmic photon background~\cite{Penzias}. Penzias and Wilson would go on to receive the Nobel prize in physics in 1978 for this discovery.  The Cosmic Background Explorer (COBE) satellite would show that the sky is filled with a black-body radiation at a temperature of 2.7 K ~\cite{COBE} as depicted in Fig. \ref{fig:COBEbbody}.
%----------------Figure------------------
%-----------------------------------------

We expect to find that since the universe is homogenous on large scales today, it has always been homogenous, even during the period of photon decoupling.  It should not be surprising that each causal patch of the sky has a black-body spectrum with a temperature near 2.7 K because of this homogeneity.   Some differences from patch to patch are to be expected as the universe cannot be perfectly homogenous or else it is unclear how galaxies could form.  These small differences in temperatures in the CMB are called CMB anisotropies.  The CMB anisotropies were measured with a resolution of 10 arc-minutes by the Planck ESA-NASA satellite and whose results were released 20 March 2013 ~\cite{Ade:2013lta}. Previously, the Wilkinson Microwave Anisotropy Probe (WMAP) had measured the CMB anisotropies with the highest resolution and released its 9-year data in December 2012.  Surprisingly the differences in temperature from causal patch to causal patch are on the order of $10^{-5}$. How can it be that the early universe is so homogenous even for places out of causal contact?  This is known as the horizon problem and will be addressed in the next section.  

\begin{figure}[h]
\begin{center}
\postscript{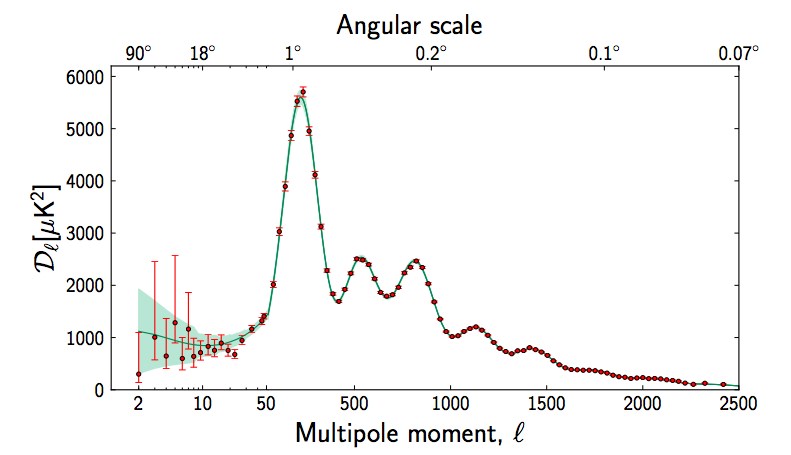}{0.8}
\ \\

WMAP 9 Year and Planck Cosmological Parameters Best Fit with $1 \sigma$ Variance. 
\begin{tabular}{| l | c | c |}
\hline
\textbf{Parameter} & \textbf{WMAP 9 yr} & \textbf{Planck 2013} \\ 
\hline
\hline
Age of the Universe  & $13.74\pm 0.11 \times 10^9$ yrs & $13.813 \pm 0.058 \times 10^9$ yrs \\ \hline
Hubble Constant $H_0$ & $70.0 \pm 2.2$  (km/s)/Mpc & $67.3 \pm 1.2$ (km/s)/Mpc \\ \hline
 $\Omega_b$ & $0.0463 \pm 0.0024$ & $0.0487 \pm 0.0023$ \\ \hline
$\Omega_{\rm CDM}$ & $0.233 \pm 0.023$ & $0.265 \pm 0.015$ \\ \hline
 $\Omega_{\Lambda}$ & $0.721 \pm 0.025$ & $0.686 \pm 0.020$\\ \hline
$1-\Omega$ & $-0.037 ^{+0.044}_{-0.042}$ & $0.000300 \pm 0.025$ \\
\hline
\end{tabular}
\caption[Planck Anisotropy Spectrum]{\sglspc \small ESA-NASA Planck mission anisotropy multipole data, with $\Lambda$CDM best fit curve that determines cosmological parameters. The anisotropy multipole map is decomposed into Legendre polynomials $\langle \frac{\delta T}{T}(\hat{n}) \frac{\delta T}{T}(\hat{n}') \rangle_{\hat{n}\cdot\hat{n}' =\cos(\theta)} =\frac{1}{4\pi} \sum_{l=0}^\infty (2l+1)C_l P_l(\cos(\theta))$, where $\CD_l = l(l+1) C_l/2\pi$, figure from Planck results ~\cite{PlanckFIG} and data for WMAP taken from reference~\cite{WMAP9yr}. \dblspc}
\label{fig:CSMTab}
\end{center}
\end{figure}

 Measurement of the anisotropies provides a method of measuring several cosmological parameters, such as the curvature, total matter content, and baryonic content, among other things.  A review of the analysis and interpretation of the CMB anisotropies can be found in reference ~\cite{Hu}.  The results of the measurement of the anisotropies forms the $\Lambda$CDM model of cosmology.  The main results of which are listed in Fig. \ref{fig:CSMTab}.
\sglspc
%----------------Figure------------------
%--------------------------------------------
 \dblspc
Measurements of these parameters suggests that we are living in a flat universe $(k = 0)$ in the FRW metric, which we saw can be accomplished if the energy content of the universe is exactly equal to the critical energy density.  However, we see that the baryonic matter can only account for $5 \%$ while the rest of the gravitationally interacting matter must be some other type of matter refereed to as dark matter, which accounts for $25 \%$ of the energy content of the universe.  The remainder of energy we assume is in some form of energy associated with free space known as dark energy which accounts for the remaining $70 \%$. Observations also suggest that we are now in an era of accelerating expansion, as discovered by Perlmutter, Schmidt, and Riess, who observed distant supernovae and from this deduced that the universe is currently transitioning from a matter dominated phase to a dark energy dominated phase~\cite{Riess:1998cb,Perlmutter:1998np, Perlmutter:2003kf}.  They received the 2011 Nobel prize in physics for this result.

It seems interesting that after 13.7 billion years of evolution that the energy content of the universe is still such that it suggests $k=0$. This is known as the flatness problem, as it seems odd that given all values possible for $\Omega$ it should be 1.  Furthering the mystery of this result, many theories that go beyond the standard model of particle physics predict magnetic monopoles which have never been discovered, so if these magnetic monopoles are created at very high energies where GUTs may be the correct physics, typically in the range of $E \approx 10^{16}$ GeV, then where are these magnetic monopoles today?  It would be in 1980 when Guth suggested a mechanism, known as inflation~\cite{Guth}, that would solve all problems associated with the CMB at once.

\subsection{\label{sec:INFLATE} Inflation}

 The scale factor can also have an exponential solution if we consider the cosmological constant $\Lambda$ to be the dominate form of matter in the universe.  If we look at the Einstein field equations with the cosmological term being the dominate term, we have
\beq
R_{\mu \nu} - \frac{1}{2} g_{\mu \nu} R \approx g_{\mu \nu} \Lambda,
\label{eq:Seventysix}
\eeq
Which allows us to associate the matter as a perfect fluid with $\rho = \Lambda$ and $P = - \rho$, so that the equation of state has $w = -1$.   The Friedmann equations then gives
\beq
\frac{ \ddot{a} }{a} = \frac{ 8 \pi }{3 M_{pl}^2} \Lambda > 0 \rightarrow a(t) \propto e^{H t} \ \hbox{with $H = \sqrt{\frac{8 \pi}{3 M_{pl}^2} \Lambda}$ }.
\label{eq:Seventyseven}
\eeq
The result is an era of exponential expansion, known as inflation.  An era of exponential expansion solves the flatness, horizon, and monopole problem all at once.  The Friedmann equation during this inflationary period is 
\beq
1-\Omega = \frac{k}{H^2 a(t)^2} \rightarrow 1-\Omega = \frac{k}{H^2} e^{-2 H t}.
\label{eq:Seventyeight}
\eeq
As a consequence of inflation we can see that $\Omega \rightarrow 1$ as time gets larger. With inflation, regardless of the initial value of $k$, the universe expands rapidly and forces $1-\Omega \rightarrow 0$.  The monopole problem can be solved by assuming that inflation takes place right around the time that GUTs are valid or when $k_B T \approx 10^{16}$ GeV which corresponds to a time $t \approx 10^{-36}$ s.  If any monopoles exist at the time of inflation they will be exponentially diluted as number density goes like $a(t)^{-3} \rightarrow e^{-3 H t}$.  The horizon problem can also be solved by inflation. Before the inflation of the universe we assume the universe is in a radiation dominated era. In this case the particle horizon is given by
\beq
d_p(t_b) = a(t_b) \int^{t_b}_0 \frac{dt'}{a(t')} = t_b^{1/2} \int^{t_b}_0 \frac{dt'}{t'^{1/2}} = 2 t_b \approx 6 \times 10^{-28} \ \rm m,
\label{eq:Seventynine}
\eeq
where the time before inflation is given as $t_b \approx 10^{-36}$ s.  Given that this is a very small distance, all of the universe should be in thermal equilibrium at this time.  At this point in the history of the universe inflation begins and ends at a time $t_e$.  We should also note that at the time of inflation the Hubble parameter is $H(t_b) = 5 \times 10^{35}$ Hz, which remains constant during inflation; the particle horizon then becomes
\beq
d_p(t_e)  =  2 t_b e^{N} - \frac{1}{H} + \frac{e^N}{H},
\label{eq:Eighty}
\eeq
where $N = H (t_e - t_b)$. If $N = 100$ (which corresponds to $t_e = 2 \times 10^{-34}$ s) then $d_p(t_e) \approx 3.2 \times 10^{16}$ m. The universe then proceeds along the standard thermal history where the large scale homogeneity is frozen in by the inflation. This allows the whole visible universe to be in causal contact at the time of recombination thus it is not surprising that the temperature variation of the CMB is so small.

The dynamics of inflation is typically described by a quasi-classical field called the inflaton.  Obviously during such small time scales before inflation, we should be using quantum field theory, but a classical field will capture the essence of inflation.  If we consider a scalar field with a generalized potential $V(\phi)$ with minimal coupling to the gravitational field, we can calculate the associated energy density and pressure that is coupled to gravity , with the the result
\beqa
S & = & \int d^4x \sqrt{-g} \lpa  \frac{1}{2} \p \phi^2 - V(\phi)\rpa  \rightarrow  T_{\mu \nu} = -\frac{2}{\sqrt{-g}} \frac{\delta ( \sqrt{-g} \CL )}{\delta g^{\mu \nu}}, \nonumber
\label{eq:Eightyone}
\eeqa
 Ignoring the spatial variations of the inflaton as we assume high levels of homogeneity, we have the terms of a perfect fluid
\beqa
\rho &=& \frac{1}{2}\dot{\phi}^2 + V(\phi), \nonumber
 \\
P &=& \frac{1}{2} \dot{\phi}^2 - V(\phi). 
\label{eq:Eightytwo}
\eeqa
If the potential initially is in a state of $\dot{\phi} \ll V$ then we have the conditions for inflation which are $\rho \approx V(\phi)$ and $P \approx -V(\phi)$, which as we saw gives exponential expansion when these conditions are met.  This initial state of the inflaton is the concept of slow roll inflation.  The least action principle for the inflaton gives
\beq
\ddot{\phi} + 3 \frac{\dot{a}}{a} \dot{\phi} = -\frac{dV}{d \phi}.
\label{eq:Eightythree}
\eeq
Equation (\ref{eq:Eightythree}) describes approximately the motion of an oscillator with a dampening term proportional to $H$.  If the inflaton starts in a false vacuum then it slowly rolls towards the minimum of $V$ where $dV/d\phi = 0$, then during this transition time we will have an inflationary era of the universe.  The inflaton is then dampened as it oscillates about the true vacuum and eventually will decay to the true vacuum value.  We will explore what this implies for gravitational wave observations in Part II.

\subsection{\label{sec:PreHeat} Reheating}

The inflationary scenario introduces a rapid expansion of space, and as a consequence the universe would be cold and have a low density.  This presents a problem for standard big bang cosmology where we need some sort of way to return to a state of high density and high temperature so that a radiation dominated phase occurs after inflation.  Work in the early 80's to the 90's (by Linde, Kofman, Starobinsky among others ~\cite{GarciaBellido:1997wm}), found a method of decaying inflaton field particles into standard model particles which would then interact and thus thermalize reheating the universe.  This thermalized state would be proceeded by a state of non-thermal equilibrium when the particles first come into existence, a pre-heating state.  During this state there can be large inhomogeneities  in the universe before thermalization smooths the inhomogeneities and thus this can lead to a stochastic gravitational wave background, similar to the CMB.

In the pre-heating phase the inflaton is oscillating about the minimum of its potential $V(\phi)$ after inflation has already occurred.  If another scalar field $\chi$ is coupled to the inflaton, then the equations of motion are
\beqa
\CL &=& \sqrt{-g}\lpa \frac{1}{2} \p \phi^2 + \frac{1}{2} \p \chi^2 - \frac{1}{2}g^2 \chi^2 \phi^2 - V(\phi)\rpa \ , \nonumber \\ 
&\downarrow& \nonumber \\
&\ddot{\phi}& -\frac{1}{a^2} \nabla^2 \phi + 3 \frac{\dot{a}}{a} \dot{\phi} + g^2 \chi^2 \phi   =  -\frac{\p V}{\p \phi} ,\nonumber \nonumber \\
&\ddot{\chi}& -\frac{1}{a^2} \nabla^2 \chi + 3 \frac{\dot{a}}{a} \dot{\chi} + g^2 \phi^2 \chi   =  0.
\label{eq:Eightyfour}
\eeqa
If the amplitude of $\phi \gg \chi$ then we can approximately ignore the term $g^2 \chi^2 \phi$ which implies the inflaton decays via the dampening term $3 \dot{a}/a$.  However for $\chi$ we can see its equation of motion resembles that of an oscillator with drag and a time dependent frequency term.  It is useful at this point to decompose $\chi$ into spatial Fourier modes which gives
\beq
\ddot{\chi}_k+ 3\frac{\dot{a}}{a} \dot{\chi}_k + \lpa \frac{k^2}{a(t)^2}+g^2 \phi(t)^2 \rpa \chi_k = 0.
\label{eq:Eightyfive}
\eeq 
This describes a field that has similar properties of a person on a swing.  Consider a person on a swing pumping their legs. What they are doing is changing the length of the pendulum they are attached to by extending and retracting their legs.  Done with the right frequency, they can increase their amplitude of the swing/pendulum~\cite{Price:2008hq}, a condition is known as parametric resonance.  Equation (\ref{eq:Eightyfive}) has a time dependent frequency, so we expect for some modes, the inflaton oscillating about its minimum will increase the amplitude of $\chi_k$ for particular modes $k$.

If we recall in flat space-time QFT, we can create a number operator for mode $k$, via the expansion of a scalar field as is done below,
\beq
\bea{c}
\phi(x) = \int \frac{d^3 k}{(2\pi)^3} \phi_k(t) e^{i \vec{k} \cdot \vec{x}}=\int \frac{d^3 k}{(2\pi)^3} \lpa a_k(t)  + a^\dagger_k(t) \rpa e^{i \vec{k} \cdot \vec{x}}, \\
\pi(x) = \p_0\phi(x) = \int \frac{d^3 k}{(2\pi)^3} \lpa \dot{a}_k(t)  + \dot{a}^\dagger_k(t) \rpa e^{i \vec{k} \cdot \vec{x}}, \\
\eea
\label{eq:Eightysix}
\eeq
where $a_k(t)$ are the solutions to spatially Fourier decomposed equations of motion. For example, in Minkowski space, we could have for a scalar field with potential $V = \frac{1}{2}m^2 \phi^2$, the equations of motion give
\beq
\ddot{\phi} - \nabla^2 \phi + m^2 \phi = 0 \rightarrow  \ddot{\phi}_k + \lpa k^2 + m^2 \rpa \phi_k = 0,
\label{eq:Eightyseven}
\eeq
which has solutions of the form $a_k(t) \propto e^{\pm i \sqrt{k^2+m^2} t}$.  We can construct a generalized number operator so long as $-2 i [a^\dagger_k(t=0), \dot{a}_{k'} (t=0)] = (2\pi)^3 \delta^{(3)}(k-k')$ (this is simply $[\phi(x),\pi(x')] = i \delta^{(3)}(x-x')$ in terms of mode expansion operators) is satisfied. The number operator for mode $k$ can then be expressed as
\beq
N_k(t) = a^\dagger_k(t) a_k(t).
\label{eq:Eightyseight}
\eeq
We can re-express this is in terms of the Fourier modes of the fields.  In our flat space example solution, this is
\beq
N_k(t) = \frac{\omega_k}{2} \lpa \phi_k^2 + \lpa \frac{\dot{\phi}_k}{\omega_k} \rpa^2 \rpa.
\label{eq:Eightynine}
\eeq
When a mode of $\chi_k$ increases, the generalization of this procedure can be interpreted at a quasi-classical level as the act of particle creation.  Thus, via parametric resonance with the inflaton, field particles can be generated, at which point standard interactions occur and thermalize the system.  The full description of this system is non-linear and numerical integration methods must be employed for detailed study.  In part II we will examine how we can calculate the effect of this particle creation on a stochastic background of gravitational waves, and how to do so in new inventive ways.

\section{The Dark Sector}

The $\Lambda$CDM model of cosmology predicts that we know almost nothing about approximately $94\%$ of the content of the universe, where $68\%$ is in the form of dark energy and $26\%$ is in the form of dark matter~\cite{Ade:2013lta}.  To explain these dark properties, it is proposed that one needs to modify general relativity,  possibly by modifying the Einstein field equation, adding some higher order curvature terms of the form
\beq
G_{\mu \nu} = R_{\mu \nu} - \frac{1}{2} g_{\mu \nu} R + \CO(R^2) = \frac{8 \pi}{3 M_{pl}^2} T_{\mu \nu},
\eeq 
or one can use models such as in modified newtonian dynamics theories (MoND)~\cite{Milgrom}.  Not modifying the Einstein field $G_{\mu \nu}$, one can explain dark properties by adding in new particles, so called dark matter.  

\subsection{Dark Matter Particles}
\begin{figure}[ht]
\begin{center}
\postscript{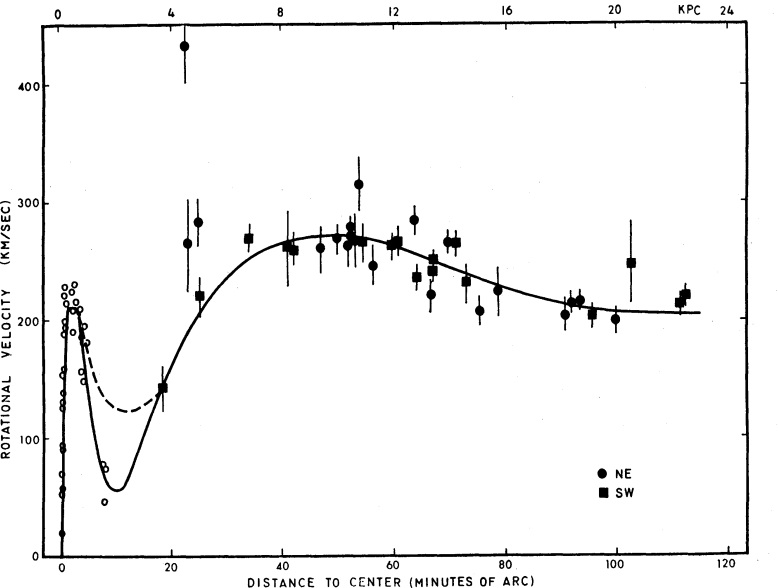}{0.6}
\caption[Galatic Rotation Curve]{ \sglspc \small  The rotation curve for the Andromeda galaxy M31, as observed in~\cite{VRubin} demonstrates that rather than the expected $v_r \propto r^{-1/2}$ fall off from Newtonian gravity, the rotational velocity of objects in M31 exhibit flattening suggesting additional non-electromagnetically interacting mass to keep the visible matter bound to the galaxy.}
\label{fig:GalRot}
\end{center}
\end{figure}

\begin{figure}[ht]
\begin{center}
\postscript{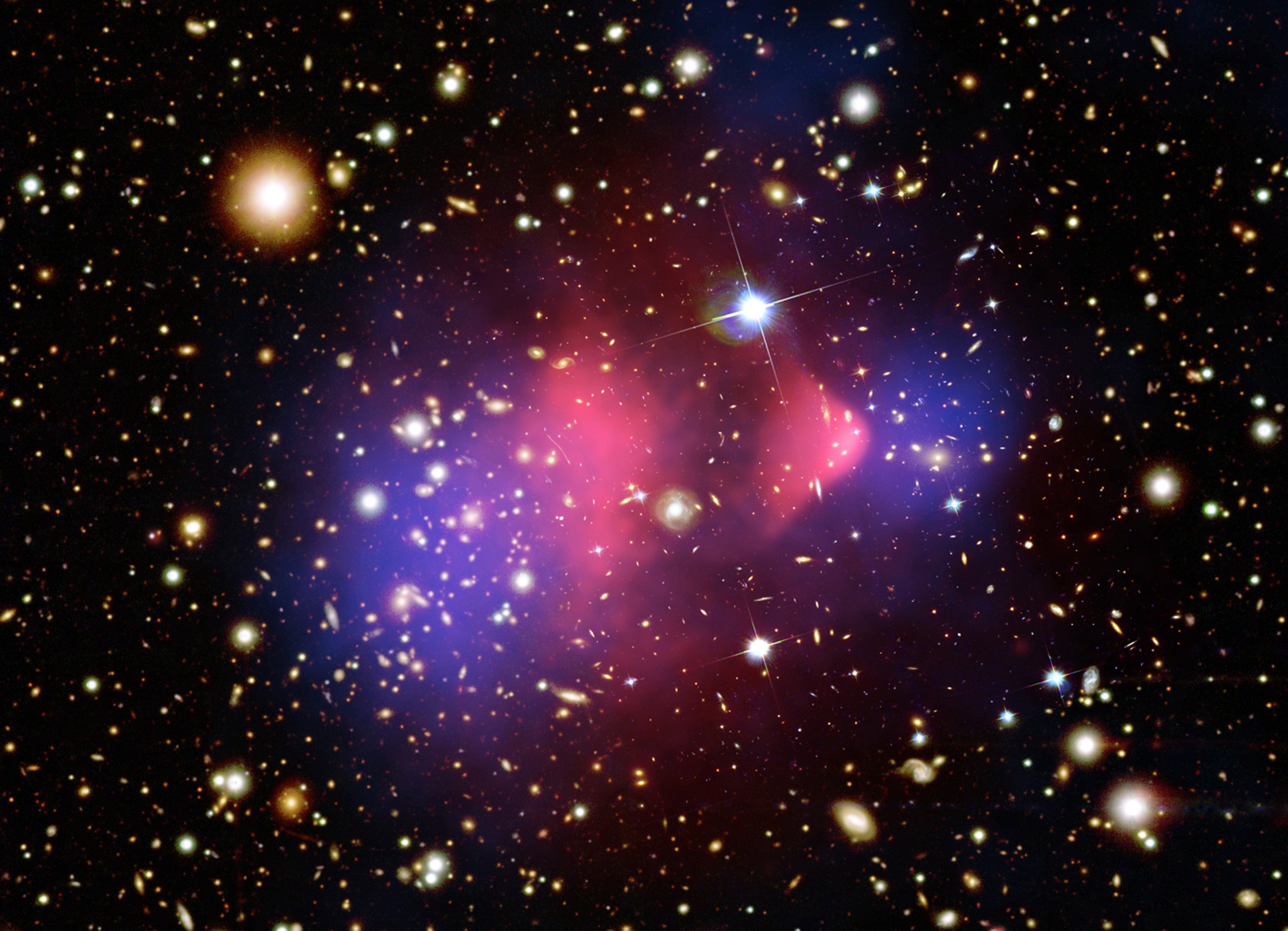}{0.7}
\caption[The Merging of the Bullet Cluster Dark Matter Lensing]{\sglspc \small Pictured is the the merging of the galaxy cluster in 1E0657-56.  The intracluster medium contains the majority of baryonic mass contained in galaxy clusters. This hot baryonic mass emits photons in the X-ray range, depicted in pink and imaged from NASA's Chandra space telescope. The shock wave on the right of the image indicates that the two clusters have passed through each other due to gravitational attraction.  However gravitational lensing maps the majority of the mass not with the X-ray emitting centers, but beyond this matter, mapped in blue/purple.  This indicates a form of matter has passed through the baryonic matter, under the influence of gravity and potentially a weakly interacting force. This image provides further evidence for dark matter particles as WIMPs~\cite{MarkevitchChand}.}
\label{fig:BulClust}
\end{center}
\end{figure}

\begin{figure}[h]
\begin{center}
\postscript{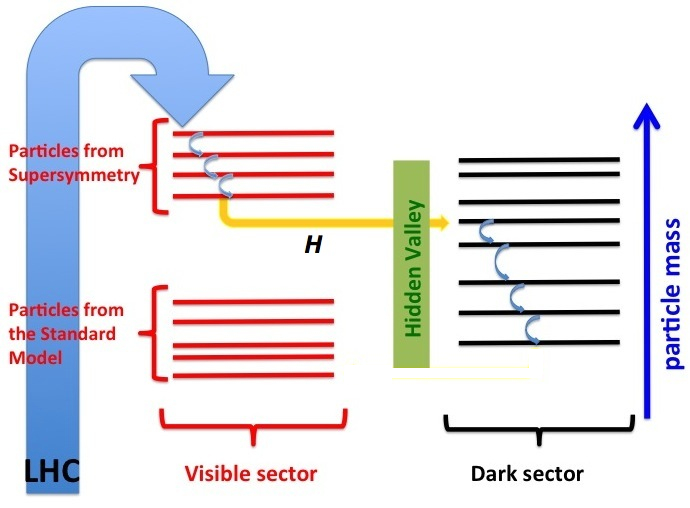}{0.7}
\caption[Higgs Portal to the Dark Sector]{\sglspc \small The SM is being probed by the LHC and may reveal SUSY particles which may make up some of the dark sector, while the Higgs particle (indicated by the H) may be the only mediator with the dark sector having its on dynamical theory of light particles, the lightest stable particle (LSP) making up the majority of the cold dark matter today.  Image modified from~\cite{Gagnon}}
\label{fig:HiggsPortal}
\end{center}
\end{figure}

 The theory of dark matter has gained evidence for existence through observations of galactic rotation curves, showing that objects far from the galactic core exhibit a higher velocity than that expected from Newtonian gravity, most famously demonstrated in~\cite{VRubin} and can be seen in Fig. \ref{fig:GalRot}. Gravitational lensing observations lend credence to the dark matter coming from massive particles, techniques of which are reviewed in~\cite{REllis1}.  One such dramatic example is that of the bullet cluster collision that demonstrates through lensing effects that dark matter interacts with visible matter weakly~\cite{Markevitch} as seen in Fig. \ref{fig:BulClust}.
%----------------Figure------------------
%-----------------------------------------
%----------------Figure------------------
%-----------------------------------------
It is popularly theorized that the explanation for dark matter comes in the form of weakly interacting massive particles (WIMPs)~\cite{TurnW}.  Weakly interacting in that they must not interact with ordinary matter very strongly or else we would have detected such matter.  The SM does not predict any weakly interacting matter other than neutrinos, however, neutrinos in the SM are massless and thus move at the speed of light and cannot form the cold dark matter needed to explain the evolution of the universe.  Models that extend the symmetry of the SM such as SUSY are popular theories for WIMPs because they predict as yet unseen particles that may explain the dark matter~\cite{Goldberg:1983nd}. It is then generally theorized that the full particle theory being incomplete consists of the visible sector, that which forms ordinary matter and is described by the SM, and that of the dark sector (DS), which forms the matter content of dark matter
\beq
\CL = \CL_{SM} + \CL_{DS} \ .
\eeq
To have any hope of detecting such a sector there must be some connection between the dark sector and the visible.   One of the simplest models of connecting the dark sector to the visible is through the least well probed particle of the SM\footnote{ \sglspc \small Least well probed at the time of this writing.}, the Higgs boson. Such models are deemed Higgs portal models~\cite{Englert:2011yb}, the structure of these models can be seen in Fig. \ref{fig:HiggsPortal}.
%----------------Figure------------------
%-----------------------------------------
  In Sec. \ref{sec:DM} we shall explore a Higgs portal model put forth by Weinberg~\cite{Weinberg:2013kea} and its validity to explain observed phenomena.
  One key observational requirement of any dark matter theory is its requirement to reproduce the correct thermal relic density of cold dark matter as observed today $\Omega_{\rm CDM} = 0.265 \pm 0.015$~\cite{Ade:2013lta}.  The connection to the relic density and the underlying particle physics model as formulated in~\cite{LeeW},  which applies the Boltzmann equation in an expanding universe, connects the annihilation cross section of dark matter particles calculated from a particle theory to that of the relic density. The Boltzmann equation in an expanding universe is given as\footnote{ \sglspc \small The following derivation follows closely that of~\cite{Kolb:1990vq}}
\beq
\dot{n}_w + 3 H n_w = - \langle \sigma v \rangle (n_w^2 - n_{\rm EQ}^2) \ ,
\label{eq:Boltz}
\eeq
where $n_w$ is the number density of WIMPS in a comoving frame, $\langle \sigma v \rangle$ the total thermally averaged annihilation cross section, M\"{o}ller velocity product, and $n_{\rm EQ}$ is the number density when the WIMPs are in thermal equilibrium. The number density in equilibrium is expressed by, for example taking the WIMPS to be fermions, as
\beq
n_{\rm EQ} =  \int \frac{d^3 k}{(2\pi)^3} \frac{g_{\rm f}}{e^{E/T} + 1} \ ,
\label{eq:NumbDens}
\eeq
where $g_{\rm f}$ is the degeneracies of the energy state $E$; typically the number of spin states the particle possesses and we will assume as such.  In the non-relativistic approximation $m_w \gg T$ ($m_w$ the mass of the WIMP) equation  (\ref{eq:NumbDens}) takes the form of equation  (\ref{eq:Sixtyseven}) repeated here for connivence
\beq
n_{\rm EQ} = g_{\rm f} \lpa  \frac{m_w T}{2\pi} \rpa^{3/2} e^{- m_w/T} .
\eeq
Equation  (\ref{eq:NumbDens}) also makes use of the total annihilation cross section for the WIMP with the M\"oller velocity a detailed calculation of which is model dependent.  By a change of variable $x = m_w/T$, $Y = n_w/s$, equation  (\ref{eq:NumbDens}) takes a much simpler form by virtue of the conservation of entropy $s a^3 = \rm const$,
\beq
\frac{d Y}{d x} = - \frac{x \langle \sigma v \rangle s}{H(m)} \lpa \frac{k}{1/2} \rpa \lpa Y^2 - Y_{\rm EQ}^2 \rpa \ , \ Y_{\rm EQ} = \sqrt{\frac{45}{32 \pi^7} } \frac{g_{\rm f}}{g_S(x)} x^{3/2} e^{-x} \ ,
\label{eq:BoltzY}
\eeq
where
\beq
s = g_S(x) \frac{2 \pi^2}{45} m_w^3 x^{-3}\ , \ H = H(m) x^{-2}= m_w^2 \frac{\sqrt{g_\rho(x)}}{M_{pl}} \sqrt{\frac{8\pi^3}{30}} \  x^{-2} \ , \ a(t) \propto t^k \ ,
\eeq
which results in
\beq
- \frac{x \langle \sigma v \rangle s}{H(m)} = - \sqrt{\frac{\pi}{45}} m_w M_{pl} \frac{g_S(x)}{\sqrt{g_\rho(x)}} x^{-2} \langle \sigma v (x) \rangle \ .
\eeq
The solutions of  (\ref{eq:BoltzY}) formally must be found by numerical methods, however approximate methods can be employed to give useful results.  To find approximate solutions we express  (\ref{eq:BoltzY}) in terms of $Y = Y_{\rm EQ} + \Delta$, where $\Delta$ expresses the deviation from equilibrium (we proceed with $k = 1/2$ for decoupling in the radiation dominated era)
\beq
\frac{d \Delta}{d x} = -\frac{d Y_{\rm EQ}}{dx} - \sqrt{\frac{\pi}{45}} m_w M_{pl} \frac{g_S(x)}{\sqrt{g_\rho(x)}} x^{-2} \langle \sigma v (x) \rangle \ \Delta ( \Delta + 2 Y_{\rm EQ}) \ . 
\label{eq:BoltzD}
\eeq
The thermal history of particle density can be understood by  (\ref{eq:Boltz}); if the averaged annihilation cross section term $\langle \sigma v \rangle$ is dominating the Hubble dampening term $3Hn_w$, then the particle distribution remains close to that of the equilibrium distribution any small deviation from equilibrium is resorted to the equilibrium distribution as can be seen in expanding  (\ref{eq:Boltz}) in first order deviations from $n_{\rm EQ}$ through $n_w = n_{\rm EQ} + \Delta_n$. Under the assumption of good thermal contact Eq. (\ref{eq:BoltzD}) becomes,
\beq
\frac{d \Delta_n}{dt} + 2 \Gamma \Delta_n \approx 0 \ ,  \Gamma = \langle \sigma v \rangle  n_{\rm EQ} \  ,
\eeq
which gives solutions $\Delta \propto \exp( - 2\Gamma t)$, killing any deviations on the scale $1/\Gamma$.
At some point the Hubble dispensation term starts to dominate and the Boltzmann equation admits solutions where the number density simply scales as $a^{-3}$ , 
\beq
\dot{n}_w \approx -3 H n_w \rightarrow n_w  = n_w(\tau) \lpa \frac{a(\tau)}{a(t)} \rpa^3 \ .
\eeq
At some cross over time when approximately $\Gamma \approx H$ the number density $n_w$ is frozen in and then scales as $a^{-3}$. We can exploit this fact to make approximate solutions to  (\ref{eq:BoltzD}).  Near the freeze out temperature, expressed as $x_f$, the deviation from equilibrium should be small and thus $\dot{\Delta} \approx 0$ and we consider only first order $\Delta$ deviations.  Under this restriction Eq. (\ref{eq:BoltzD}) takes the form
\beq
\frac{1}{Y_{\rm EQ}} \frac{d Y_{\rm EQ}}{dx} \approx - 2 \sqrt{\frac{\pi}{45}} m_w M_{pl} \frac{g_S(x)}{\sqrt{g_\rho(x)}} x^{-2} \langle \sigma v (x) \rangle \ \Delta \ .
\label{eq:BoltzFO}
\eeq
The left hand side can be simplified under the assumption that the freeze out temperature occurs at a non-relativistic temperature, typically given by the condition $x_f \gg 3$. So long as the freeze out occurs at a temperature where $g'_S(x)/ g_S(x) \approx 0$ then equation  (\ref{eq:BoltzFO}) admits solutions of the form
\beq
\Delta \approx \sqrt{\frac{45 \ g_\rho(x_f)}{\pi \ g_S^2(x_f)}} \frac{x_f^2}{m_w M_{pl}  \langle \sigma v(x_f) \rangle } \ .
\eeq
The convention is to consider a particle decoupled when $\Delta = c Y_{\rm EQ}$ with $c$ some number of order $1$, solving for this condition leads to
\beq
\sqrt{ \frac{45}{8 \pi^6}} c \frac{g_{\rm f}}{\sqrt{g_\rho(x_f)}} m_w M_{pl} x_f^{-1/2} \langle \sigma v(x_f) \rangle \approx e^{x_f} \ ,
\label{eq:Takec}
\eeq
which admits approximate solutions from self substitution of the form
\beqa
x_f &=& \ln \lsb \sqrt{\frac{45}{8 \pi^6}} c \frac{g_{\rm f}}{\sqrt{g_\rho(x_f)}} m_w M_{pl} x_f^{-1/2} \langle \sigma v(x_f) \rangle \rsb - \frac{1}{2} \ln \lsb x_f \rsb \ , \nonumber \\
x_f &\approx&   \ln \lsb \sqrt{\frac{45}{8 \pi^6}} c \frac{g_{\rm f}}{\sqrt{g_\rho(x_f)}} m_w M_{pl} x_f^{-1/2} \langle \sigma v(x_f) \rangle \rsb \nonumber \\
& &- \frac{1}{2} \ln \lsb \ln \lsb \sqrt{\frac{45}{8 \pi^6}} c \frac{g_{\rm f}}{\sqrt{g_\rho(x_f)}} m_w M_{pl} x_f^{-1/2} \langle \sigma v(x_f) \rangle \rsb \rsb + \dots \nonumber \\
\eeqa
After the freeze out occurs we know the distribution should only be slightly effected by the residual annihilations such that after freeze out $Y \gg Y_{\rm EQ}$ since $Y_{\rm EQ}$ has exponential decay, after freeze out $\Delta \approx Y$.  From equation  (\ref{eq:BoltzD}) we can find the value of $Y(x_0)$, where $x_0$ is the value taken today as
\beq
\frac{1}{Y(x_0)} \approx \frac{1}{Y_{\rm EQ}(x_f)} + \sqrt{\frac{\pi}{45}} m_w M_{pl} \int_{x_f}^{x_0} \frac{g_S(x)}{\sqrt{g_\rho(x)}} x^{-2} \langle \sigma v (x) \rangle \ .
\eeq
A key observation is that so long as $x^{-2} \langle \sigma v(x) \rangle$ dies off quickly as a function of $x$ then the term containing $\langle \sigma v \rangle$ contributes very little and $Y(x_0) \approx Y_{\rm EQ}(x_f)$, this will come back in Sec.~\ref{sec:BFit} as we consider a case where $\langle \sigma v \rangle$ grows for $ x \rightarrow \infty$. We will continue with the assumption that the term $\langle \sigma v \rangle$ doesn't contribute to the the yield $Y$ today.

We may now compute the thermal relic density by 
\beq
\Omega_{\rm CDM} = \frac{ \rho_w( {\rm today})}{\rho_c} = \frac{8 \pi \  m_w s_0 Y(x_0)}{3 H_0^2 M_{pl}^2} \ ,
\eeq
If we take $c = 1$ in equation  (\ref{eq:Takec}) and $T_0 = 2.7 \ \rm K$ the result is 
\beq
\Omega_{\rm CDM} h^2 \approx 4 \times 10^{-11} \frac{x_f^{3/2}}{\sqrt{g_\rho(x_f)} \langle \sigma v(x_f) \rangle} \ \rm GeV^{-2} \ .
\label{eq:RDEq}
\eeq
The Eq. (\ref{eq:RDEq}) gives for $\Omega_{\rm CDM} h^2 = 0.112$ ( computed with $h \approx 0.67$  from ~\cite{Ade:2013lta} ) and an example $x_f = 20$, $g_\rho(x_f) \approx 100 \rightarrow \ m_w = 200 \ \rm GeV$ gives the requirement
\beq
\langle \sigma v(x_f = 20) \rangle = 3.9 \times 10^{-26}  \rm cm^3 s^{-1} \ .
\eeq
though the flatness of $g_\rho$ in certain regions makes the result insensitive to the particle mass.

Besides a dark sector theory giving the correct relic density, dark matter also allows an explanation to recent observations of excess relativistic degrees of freedom from that of the SM in CMB analyses, the details of which are left to Sec.~\ref{sec:DM}.  
 
\subsection{Dark Matter Direct Detection}

\begin{figure}[ht]
\begin{center}
\postscript{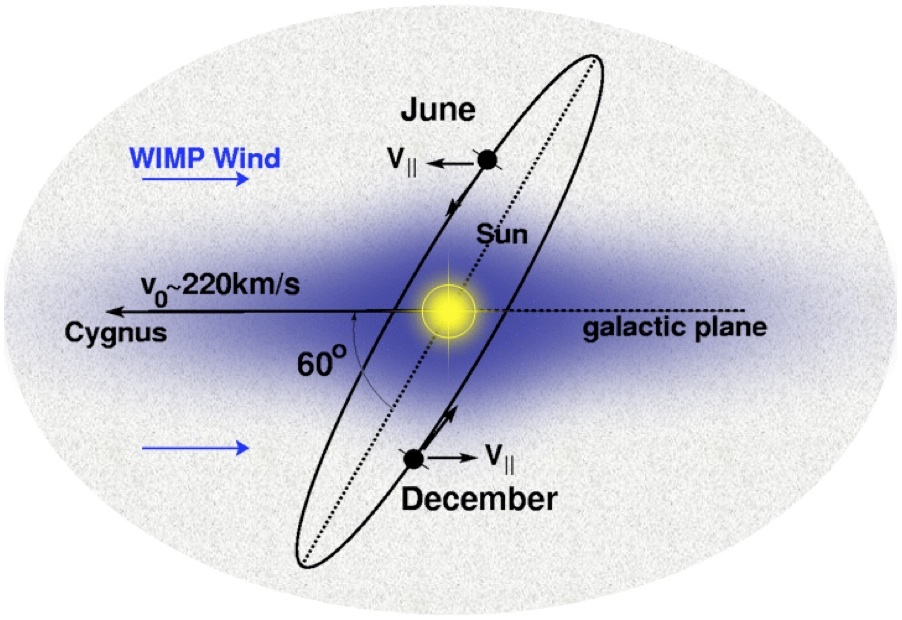}{0.5}
\caption[Earth Moves Through Dark Matter Halo]{ \sglspc \small As the Earth orbits the sun, the relative velocity of the Earth with the dark matter halo of the milky way has a period of 1 year.  This should have an effect on the number of incident collisions for dark matter direct detection searches (image taken from~\cite{Gagnon}).}
\label{fig:MovingThrou}
\end{center}
\end{figure}

On a more microscopic scale, if dark matter has interactions with the visible sector, then there should be some measurable effect of the particle's interactions.  Currently, direct detection measurements are being performed by several collaborations, which include DAMA/LIBRA~\cite{Bernabei:2010mq}, CoGeNT~\cite{Aalseth:2010vx,Aalseth:2011wp}, CRESST~\cite{Angloher:2011uu}, and CDMS~\cite{Agnese:2013rvf}.  The essential principle of direct detection searches is to measure the recoil and subsequent phonons (mK temperature changes) in various materials resulting from interactions of dark matter with stable nuclei. In this experimental setup, one might assume that as the Earth moves through the Galactic halo of dark matter particles as represented in Fig. \ref{fig:MovingThrou}; with this model, at various times of the year, one expects to observe different behavior in your measurements as the relative flux of dark matter particles changes as the Earth moves with and against the WIMP  wind.
%----------------Figure------------------
%-----------------------------------------
 %----------------Figure------------------
\begin{figure}[tpbh]
\begin{center}
\postscript{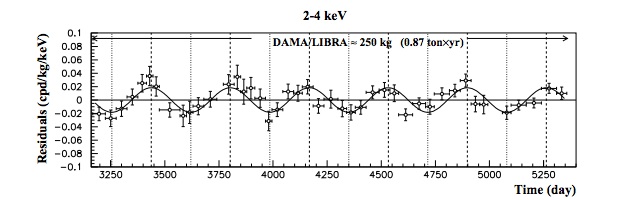}{0.9}
\caption[Annual Modulations In DAMA/LIBRA Direct Dark Matter Search]{ \sglspc \small Annual modulation data in dark matter direct detection searches observed from the DAMA/LIBRA Collaboration for a little over 5 years of data \cite{Bernabei:2010mq}.}
\label{fig:Annual}
\end{center}
\end{figure}
%-----------------------------------------
 %----------------Figure------------------
\begin{figure}[tpbh]
\begin{center}
\postscript{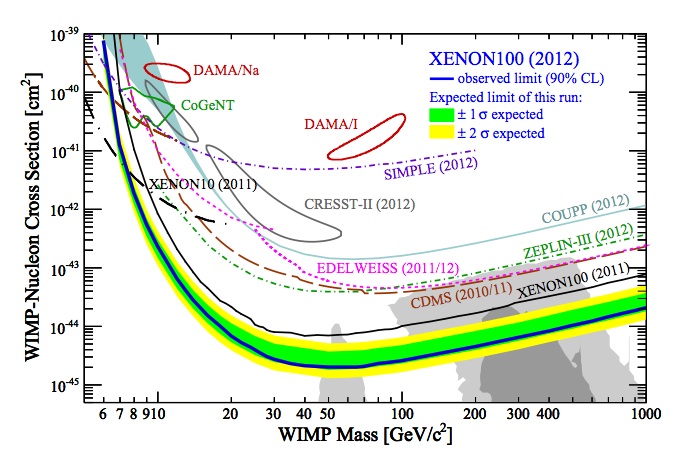}{0.8}
\caption[Upper Limits of Direct Detection Cross-Sections for Dark Matter Searches]{ \sglspc \small Limits on Nucleon WIMP cross sections by various collaborations. Cross sections are of the order $\CO(10^{-39}-10^{-45}) \ {\rm cm}^2$. Image adopted from ~\cite{Xenon100} }
\label{fig:IntXsec}
\end{center}
\end{figure}
%-----------------------------------------
 %----------------Figure------------------
\begin{figure}[tpbh]
\begin{center}
\postscript{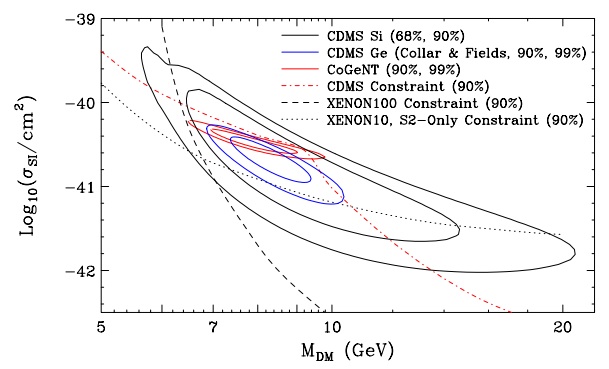}{0.8}
\caption[Direct Detection Hints at $m_w = 10$ GeV particle]{ \sglspc \small Several collaborations have started to converge on a direct detection limit that suggest a dark matter particle with mass around $10$ GeV~\cite{Hooper:2013cwa} though issues with the Xenon100 data exsist~\cite{Xenon100}. }
\label{fig:DDhoop}
\end{center}
\end{figure}
%----------------------------------------- 
As of 2010 the DAMA/LIBRA~\cite{Bernabei:2010mq} Collaboration's observations suggest such an annual modulation of dark matter interaction signals, in accordance to the Earth's relative motion to the galactic halo, the results are shown in Fig. \ref{fig:Annual}.
These direct detection searches also help to place upper limit constraints on dark sector physics. Interaction cross-sections limits can be seen in Fig. \ref{fig:IntXsec} from~\cite{Xenon100}.  
Interestingly, several collaborations, as of June 2013, have hinted at a possible signal in the $m_w \approx 10$ GeV range~\cite{Hooper:2013cwa} \ref{fig:DDhoop}.
In Sec.~\ref{sec:WWIMP} we will explore what the Higgs portal models can tell us in light of these new developments.

%--------------------------------------------Begin Document----------------------------
\newpage
\

\

\

\noindent\textbf{\Huge Part I:}

\

%---------------------------------Ch 2-----------------------------------------
\noindent\textbf{\LARGE $U_B(3) \times SU_L(2) \times U_L(1) \times U_{I_R}(1)\rightarrow {\rm SM}^{++}$}
\addcontentsline{toc}{chapter}{Part I - LHC Phenomenology on Effective Theory of BSM Physics}

\newpage

\thispagestyle{fancy}
\chapter{Extending the Standard Model}
\label{STaBSM}
\thispagestyle{fancy}
\pagestyle{fancy}

%---------------------------------Ch 2: Section 1-----------------------------------------
\section{\label{part:1} Strings, D-branes, and the SM}

  Exploring beyond the SM requires us to find new concepts that allow us to make modifications to the SM Lagrangian; one of the most highly regarded BSM theories is string theory.  The central idea of string theory is that all fundamental particles (leptons, quarks, and gauge bosons) are not point particles but rather one dimensional objects, strings. The use of strings as the fundamental objects of a QFT leads to a consistent quantum theory of gravity in its spectrum. String theory unites gravity and the other known forces into a consistent framework.  The use of quantum theory to the concepts of strings requires the theory to have more than the 3+1 dimensions of space-time, 26 in the bosonic string case and 10 in superstring theory. We do not observe a space-time dimension higher than 4 in current experiments, and do not find a departure from the inverse square law of gravity for distances greater than 56 $\mu$m, which sets a limit of one extra compact spatial dimension of $R \leq 44 \ \mu \rm m$~\cite{Kapner:2006ph}. To keep string theory as a viable theory, the idea of D-branes was introduced to connect string theory to experiment~\cite{Polchinski:1995mt,Polchinski:1996na}.  

   D$p$-branes are $p$ dimensional objects to which string end points can be attached to, (strings having Dirichlet boundary conditions, or considered fixed at their end points).  By attaching strings to D-branes, extra spatial dimensions are still a possibility as the particles of the SM can be forced to be bound to D-branes in the form of open strings,  while the mediators of gravity (gravitions), closed strings, are free to move into the extra-dimensional space.   As was shown earlier, the inclusion of extra spatial dimensions solves the hierarchy problem of the SM.  Within the spectrum of particles (string states) are vector gauge bosons, as well as the SM fermionic content \cite{Zwiebach}; with all the content of the SM + gravity it is natural to try to construct something similar to the SM.  Multiple D-branes provide a possible solution to construction of string-like SM. By having multiple D-branes, strings can attach one endpoint on one brane and its other endpoint on an entirely separate brane.  By attaching strings to multiple branes and having the branes at the same location (without the branes at the same location the gauge bosons may acquire mass, which we do not want) a $U(N)$ gauge symmetry is realized on the brane. 
	
	Stacking multiple branes together allows us to form something very close to the SM;  a brane world consisting of a 3-stack of branes, intersecting a 2-stack of branes and finally a single brane contains the SM ~\cite{Antoniadis:1998ig}. These 3 stacks of branes will contain the gauge group $U_B(3)\times U_L(2) \times U_Y(1)$,  which contains the SM gauge group as a subgroup.  To make the multiple brane approach to the SM applicable to experiment such as the LHC, the required string scale must be set at the TeV scale and  thus requires extra dimensions of the length 1 millimeter to a Fermi ($10^{-15}$ m )~\cite{Cullen:2000ef}. Regge recurrences (string excitations) most distinctly manifest in the $\gamma +$ jet~\cite{Anchordoqui:2007da,Anchordoqui:2008ac} and dijet~\cite{Lust:2008qc,Anchordoqui:2008di,Anchordoqui:2009mm} spectra resulting from their decay. The recent search for such narrow resonances in data collected during the LHC7 run now excludes a string scale below 4~TeV~\cite{Khachatryan:2010jd,Chatrchyan:2011ns, Lust:2013koa}.  Alternatively it is still possible that the string scale is actually closer to the Planck scale, $l_p \approx 1.6 \times 10^{-35}$ m and still get signals at the LHC as we will show below. 
\begin{figure}[ht]
\begin{center}
\postscript{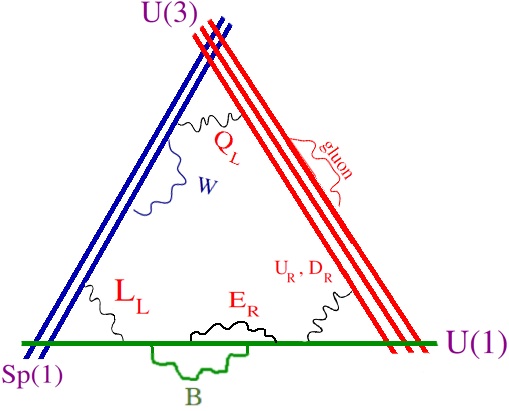}{0.6}
\caption[Three Brane Construction of SM]{ \sglspc \small The use of 3 brane stacks can be used to find the minimal representation of the SM within string theory, while still keeping the possibility of large extra spatial dimensions as a possibility. \label{fig:3BraneSM}}
\end{center}
\end{figure}
	
\begin{figure}[ht]
\begin{center}
\postscript{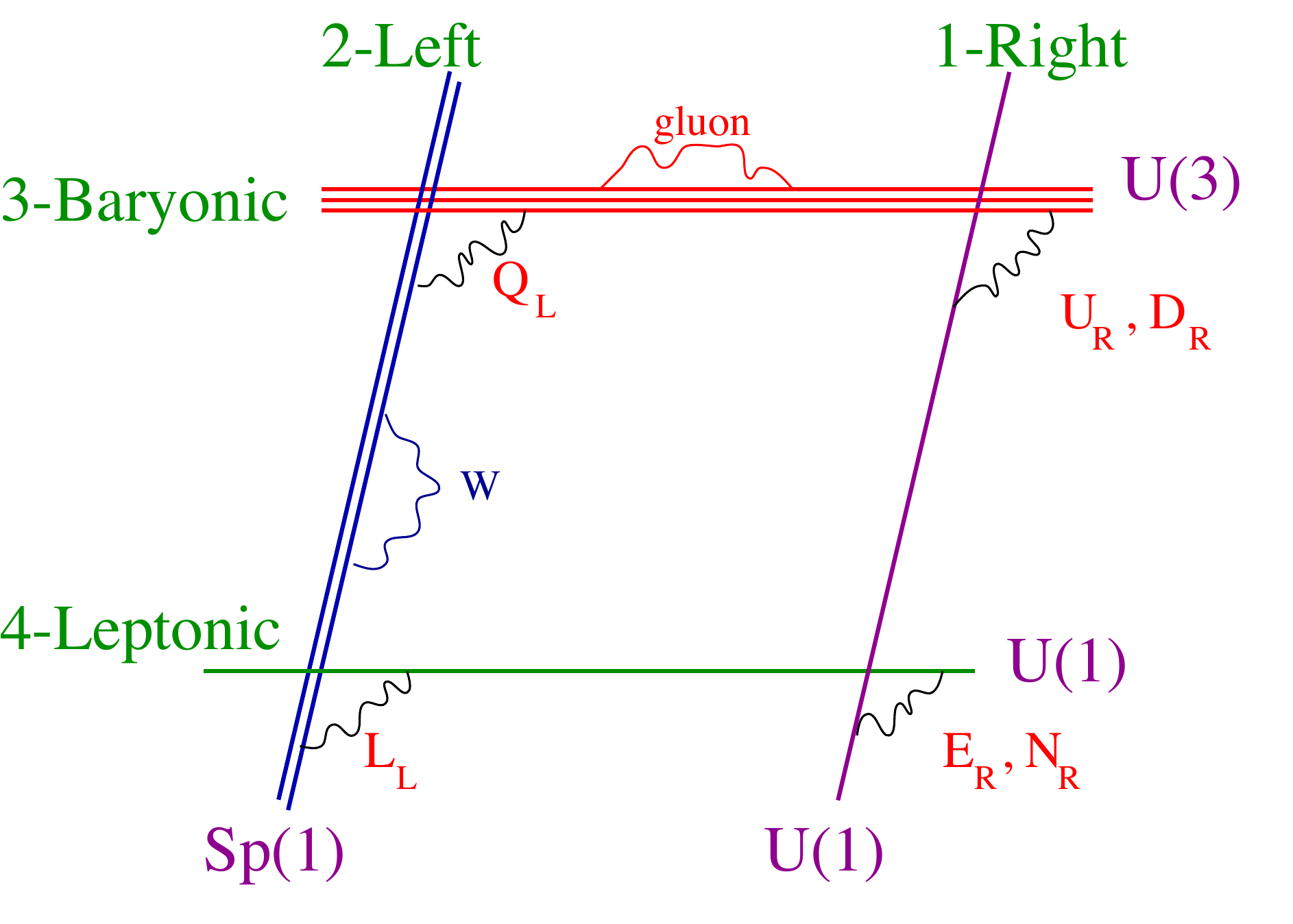}{0.8}
\caption[Four Brane Construction of SM++]{ \sglspc \small The addition of the $U_{I_R}(1)$ brane can alleviate the issues of a 1 TeV string scale of the 3 brane construction.  Above the gauge group of the 4 brane construction is  $U(3)_B \times Sp(1)_L \times U(1)_L \times U(1)_{I_R}$. \label{fig:4dbrane_SM}}
\end{center}
\end{figure}

	To increase the scale of string theory more towards the Planck scale, we include an extra 1-stack D-brane to the 3 stacks of D-branes that can represent the SM.  This additional 1-stack D-brane will supply us with an additional $U(1)$ gauge symmetry to that of the gauge group $U_B(3) \times U_L(2) \times U_Y(1) \rightarrow U_B(3)\times U_L(2) \times U_L(1) \times U_{I_R}(1)$,  where we have chosen to gauge lepton number via $U_L(1)$, this then naturally explains the accidental symmetry included with the SM of conservation of lepton number; we also include an additional charge $I_R$.  we have a special case for the $U_L(2)$ brane, since even powers of the $U(2N)$ group can be reduced to members of the symplectic group $Sp(N)$. With this in mind, we have $U_L(2) \rightarrow Sp_L(1) \cong SU_L(2)$, which alleviates the need for an extra, as yet unobserved gauge boson associated with $U_L(2)$, as depicted in Fig. \ref{fig:3BraneSM} for the 3 intersecting d-brane stacks, and Fig. \ref{fig:4dbrane_SM} for the 4 intersecting d-brane stacks.

	To understand the new consequences of this model, it helps to identify the SM from its construction.  This can be done by recognizing that any unitary group $U(N)$ can be decomposed to
\beq
U(N) \rightarrow U(1) \times SU(N).
\label{eq:P1One}
\eeq
By performing the decomposition to $U(1)$ and $SU(N)$ we must consider the coupling strengths of the decomposed $U(1)$ part and $SU(N)$.  It should be that at the string scale the underlying symmetry of $U(N)$ is restored, while at scales less than the string scale a symmetry breaking can make them appear to have different coupling strengths.  We ensure the $U(N)$ symmetry is released at the string scale by requiring that the gauge transformations have a common normalization. We write our $U(N)$ gauge transformation as a multiplication of a $U(1)$ transformation expressed as $\exp(i X)$ with $X$ proportional to a $N\times N$ identity matrix, and a $SU(N)$ transformation written as $\exp(i T^a \theta^a)$, where the $T^a$ are generators of the $SU(N)$ symmetry.  We can use the trace operator normalization of the $SU(N)$ group as a guide to the proper normalization of the $X$ operator via~\cite{Anchordoqui:2011eg} 
\beqa
\hbox{Tr} \lpa T^a T^b \rpa &=& \frac{1}{2} \delta^{ab} \rightarrow \hbox{Tr} \lpa X^2 \rpa = \frac{1}{2}, \nonumber
\\
\hbox{Tr}\lpa X^2 \rpa &=& \hbox{Tr} \lpa c^2 I \rpa =  c^2 N \rightarrow c = \frac{1}{\sqrt{2N}}.
\label{eq:P1Two}
\eeqa
We can then identify $X$ as the identity as long as we transfer the $1/ \sqrt{2N}$ to the coupling associated with $X$; this ensures that we will recover the $U(N)$ symmetry at the string scale. With the decomposition of $U_B(3)$, the gauge group of the 4 brane model becomes  $U_B(1) \times SU_c(3) \times SU_L(2) \times U_L(1) \times U_{I_R}(1)$, where now we have gauge field associated with baryon number via $U_B(1)$. Again one of the accidental symmetries of the SM naturally becomes gauged in this model; this then can be used to ensure proton stability.  With this gauge group as the underlying symmetry, a covariant derivative can be constructed as
\beq
D_\mu = \p_\mu - i g_3 T^a G^a_\mu - i g'_3 Q_B C_\mu - i g_2 \tau^a W^a_\mu - i g'_1 Q_{I_R} B_\mu - i g'_4 Q_{L} X_\mu,
\label{eq:P1Three}
\eeq 
where it must be enforced at the string scale that $g'_3(M_s) = g_3(M_s)/\sqrt{3\cdot 2}$ and we identify $C_\mu$ as the the $U(1)$ gauge boson associated with conservation of baryon number.
  
  In addition to extra gauge bosons, we also include in the model right chiral neutrino fields $n_R$, so that we can agree with the observation of neutrino oscillations. As long as the right chiral neutrino fields remain neutral with respect to hypercharge, it will not break the SM.  The matter content of the 4 brane model is summarized in Table \ref{tab:P1ChiFerSpect}.
 %-----------------------TABLE------------------------- 
\begin{table}
\caption[Chiral Fermion Spectrum for SM$^{++}$]{ \sglspc \small Chiral fermion spectrum of the $U(3)_B \times Sp(1)_L \times U(1)_L \times U(1)_{I_R}$ D-brane model with charges $Q_B$ Baryon number, $Q_L$ Lepton number, $Q_{I_R}$ additional U(1) charge, $Q_Y$ hypercharge of the SM.  The label for the field will be used to compactify expression in other sections.  The fields are labeled such that $\CF_i$ is field $i$ , for example $\CF_1 = u_R, \ \CF_2 = d_R, \dotso$}
\begin{center}
\begin{tabular}{ccccccccc}
\hline
\hline
~~~Label~~~&~~~Fields~~~ & ~~~Representation~~~ & ~~~$Q_B$~~~ & ~~~$Q_L$~~~ & ~~~$Q_{I_R}$~~~ & ~~~$Q_Y$~~~ \\
\hline
1 & $u_R$ &  $( 3,1)$ & $1$ & $\phantom{-}0$ & $\phantom{-} 1$ & $\phantom{-}\frac{2}{3}$  \\[1mm]
2 &  $d_R$ &  $( 3,1)$&    $1$ & $\phantom{-}0$ & $- 1$ & $-\frac{1}{3}$   \\[1mm]
3&  $E_L$ & $(1,2)$&    $0$ &  $\phantom{-}1$ & $\phantom{-}0$ & $-\frac{1}{2}$ \\[1mm]
4&  $e_R$ & $(1,1)$&   $0$ & $\phantom{-}1$ &  $- 1$ & $- 1$ \\[1mm]
5& $Q_L$ & $(3,2)$& $1$ & $\phantom{-}0 $ & $\phantom{-} 0$ & $\phantom{-} \frac{1}{6}$    \\[1mm]
6&  $n_R$& $(1,1)$&  $0$ & $\phantom{-}1$ &  $\phantom{-} 1$ & $\phantom{-} 0$ \\
7 & $H$ & $(1,2)$ & $0$ & $\phantom{-}0$ & $\phantom{-}1 $ & $ \phantom{-} \frac{1}{2}$  \\ [1mm]
8 & $H''$ & $(1,1)$ & $0$ & $-1$ & $-1$  & $ \phantom{-} 0$ \\ 
\hline
\hline
\label{tab:P1ChiFerSpect}
\end{tabular}
\end{center}
\end{table}
%------------------------------------------------------
The SM makes use of a Higgs doublet field to give mass to three gauge bosons.  Allowing a Higgs doublet field and a single scalar field to the 4 brane model, we can give masses to 3 gauge bosons using the Higgs doublet and give mass to 1 of the additional gauge bosons.   This leaves 8 massless fields, which we will take to be the gluons of the SM; 1 massless gauge boson we can eventually associate to the photon field.  This still leaves an additional $U(1)$ gauge boson massless.   We will give a mass to this additional $U(1)$ via a St\"{u}ckelberg mechanism. More will be said on the St\"{u}ckelberg mechanism in the following sections.

\subsection{\label{sec:SM++} The Effective Lagrangian}

	The particle content in the theory has been specified, and at this point we construct an effective Lagrangian within which we can identify the SM.  We identify $SU_B(3)$ with $SU_c(3)$, the color gauge group from the SM which is represented by 8 gluons with field strength
\beq
G^a_{\mu \nu} = \lpa \p_\mu G^a_\nu -G \p_\nu G^a_\mu + g_3 f^{abc} G^b_\mu G^c_\nu \rpa, \ i f^{abc} T^a=[T^b,T^a], \ T^a \in SU(3).
\label{eq:P1Four}
\eeq
Associated with the $U_B(1), U_L(1),$ and $U_{I_R}(1)$ we have the gauge fields $C_\mu, X_\mu,$ and $B_\mu$, respectively. The field strengths are given by the expressions
\beqa
F^{(1)}_{\mu \nu} &=& \p_\mu B_\nu - \p_\nu B_\mu, \nonumber \\
F^{(3)}_{\mu \nu} &=& \p_\mu C_\nu - \p_\nu C_\mu, \nonumber \\
F^{(4)}_{\mu \nu} &=& \p_\mu X_\nu - \p_\nu X_\mu.
\label{eq:P1Five}
\eeqa
We can construct the field strength for the $SU_L(2)$ gauge field as
\beq
W^a_{\mu \nu} = \lpa \p_\mu W^a_\nu - \p_\nu W^a_\mu + g_2 \epsilon^{abc} W^b_\mu W^c_\nu \rpa, \ i \epsilon^{abc} \tau^a=[\tau^b,\tau^a], \ \tau^a \in SU(2).
\label{eq:P1Six}
\eeq
The kinetic expression for the gauge fields, also known as the Yang-Mills part of the Lagrangian, is the gauge invariant expression
\beq
\CL_{YM} = -\frac{1}{4}\lpa G^a_{\mu \nu}G^{\mu \nu}_a + W^a_{\mu \nu}W^{\mu \nu}_a + F^{(1)}_{\mu \nu}F^{\mu \nu}_{(1)} + F^{(3)}_{\mu \nu}F^{\mu \nu}_{(3)} + F^{(4)}_{\mu \nu}F^{\mu \nu}_{(4)}\rpa.
\label{eq:P1Seven}
\eeq
We include the fermion fields via our standard gauge invariant expression as
\beqa
\CL_{\rm fermion} &=& i \bar{Q}_L \gamma_\mu D^\mu Q_L + i \bar{u}_R \gamma_\mu D^\mu u_R + i \bar{d}_R \gamma_\mu D^\mu d_R + i \bar{E}_L \gamma_\mu D^\mu E_L  \nonumber \\  
&\ &+ \ i \bar{e}_R \gamma_\mu D^\mu e_R + i \bar{n}_R \gamma_\mu D^\mu n_R,
\label{eq:P1Eight}
\eeqa
where it is understood that we repeat these interactions for each generation of particles.  To give the particles mass, we include the Yukawa interactions as they were in the SM, however we include a Yukawa term for the right chiral neutrino field
\beq
\CL_Y = -Y_d \lpa \bar{Q}_L H \rpa d_R - Y_u \lpa \bar{Q}_L i \sigma^2 H^* \rpa u_R - Y_e \lpa \bar{E}_L H \rpa e_R - Y_N \lpa \bar{E}_L i \sigma^2 H^* \rpa n_R + h.c. \ ,
\label{eq:P1Nine}
\eeq
where the Yukawa couplings $Y_i$ are matrices in flavor space (\textit{i.e.} the CKM matrix is absorbed into the definitions of $Y_i$). 

	Since this is an effective Lagrangian of the underlying string theory, it is possible to have terms that have mass dimension greater than 4, which would be considered non-renormalizable and thus non-fundamental, but knowing that this is an effective theory, terms with mass dimension greater than 4 are considered acceptable. However, we avoid this complication in the Lagrangian by setting the string scale well outside the scope of LHC physics, where our phenomenological analysis takes place in the following sections.  By choosing the string scale  $M_s \geq 10^{14}$ GeV, we ensure that operators with mass dimension greater than 4 in the Lagrangian are suppressed in our phenomenological analysis.   We must also state that there cannot be any terms involving $H''$ and have mass dimension 4 ensuring conservation of lepton number. 
	 	
To include the Higgs fields, we form the scalar sector of the effective Lagrangian as
\beq
\CL_{\rm scalar} = \lpa D^\mu H \rpa^\dagger D_\mu H + \lpa D^\mu H'' \rpa^\dagger D_\mu H'' - V(H,H''),
\label{eq:P1Ten}
\eeq
with the potential given as
\beq
V(H,H'') = \mu_1^2 |H|^2  + \mu_2^2 |H''|^2 + \lambda_1 |H|^4 + \lambda_2 |H''|^4 + \lambda_3 |H|^2 |H''|^2.
\label{eq:P1Eleven}
\eeq
Minimization of the potential results in two separate vevs for each Higgs field; those vevs are represented as
\beq
\langle\,  H\,\rangle =
 \frac{1}{\sqrt{2}} \begin{pmatrix}
  0 \\
  v  \\
 \end{pmatrix}  
\quad {\rm and} \quad
\langle H''\rangle  = \frac{1}{\sqrt{2}}  v''  \, .
\label{eq:P1Twelve} 
\eeq  
The doublet vev denoted $v$ is taken to be that of the SM while the singlet vev denoted $v''$ will be used to give mass to one of the gauge bosons not part of the SM. 

The effective theory in this form makes it difficult to identify the SM.  We require a change of basis of the 3 additional $U(1)$ gauge bosons, such that a linear combination of $C_\mu, B_\mu, X_\mu$ form the hypercharge gauge boson of the SM $Y_\mu$, along with two other $U(1)$ gauge bosons, $Y'_\mu,$ and $Y''_\mu$.  We accomplish this via an $SO(3)$ matrix, or a rotation, parameterized by three Euler angles as is shown below $\lpa \theta, \psi, \phi \rpa$~\cite{Anchordoqui:2011ag} 
\begin{equation}
\mathbb{R}=
\left(
\begin{array}{ccc}
 C_\theta C_\psi  & -C_\phi S_\psi + S_\phi S_\theta C_\psi  & S_\phi
S_\psi +  C_\phi S_\theta C_\psi  \\
 C_\theta S_\psi  & C_\phi C_\psi +  S_\phi S_\theta S_\psi  & - S_\phi
C_\psi + C_\phi S_\theta S_\psi  \\
 - S_\theta  & S_\phi C_\theta  & C_\phi C_\theta
\end{array}
\right) \,,
\label{eq:P1Thirteen}
\end{equation}
where we have $C_\theta = \cos(\theta)$ and $S_\psi = \sin(\psi)$ and similar notation for the other angles. After performing this rotation on the $U(1)$ gauge bosons with the relation
\beq
\left(
\begin{array}{c}
C_\mu   \\
X_\mu  \\
B_\mu
\end{array}
\right) = 
\mathbb{R} \left(
\begin{array}{c}
Y_\mu   \\
Y'_\mu  \\
Y''_\mu
\end{array}
\right) \ ,
\label{eq:P1Thirteen-1} 
\eeq
our covariant derivative is brought into the form
\begin{eqnarray}
D_\mu & = & \partial_\mu -i Y_\mu \left(-S_\xt g'_1 Q_{I_R} + C_\theta S_\psi  g'_4  Q_{L} +  C_\theta C_\psi g'_3 Q_B \right) \nonumber \\
 & - & i Y'_\mu \left[ C_\theta S_\phi  g'_1 Q_{I_R} +\left( C_\phi C_\psi + S_\theta S_\phi S_\psi \right)  g'_4 Q_{L} +  (C_\psi S_\theta S_\phi - C_\phi S_\psi) g'_3 Q_B \right] \label{linda}  \nonumber \\
& - & i Y''_\mu \left[ C_\theta C_\phi g'_1 Q_{I_R} +  \left(-C_\psi S_\phi + C_\phi S_\theta S_\psi \right)  g'_4  Q_{L} + \left( C_\phi C_\psi S_\theta + S_\phi S_\psi\right) g'_3 Q_B \right]  \nonumber \\ 
&+& \dots
\label{eq:P1Fourteen}
\end{eqnarray}
where the $\dots$ include gauge bosons of the non-abelian groups. By using an orthogonal matrix $\mathbb{R}$, we can find relations for the charges for the particles in the new basis. 

To illustrate this point, consider the situation of a collection of $U(1)$ gauge bosons $X_n$ and then perform a rotation on them.   We can write them in a new basis $Y_m$
\begin{eqnarray}
D_\mu &=& \p_\mu - i \sum_n g_n Q_n X_n + \dotso ,  \nonumber \\
&=& \p_\mu - i \sum_{nm} g_n Q_n \mathbb{R}_{mn} Y_m +\dotso = \p_\mu - i \sum_m g'_m Q'_m Y_m,
\label{eq:P1Fifteen}
\end{eqnarray}
where we take $X_n = \sum_m \mathbb{R}_{mn} Y_m$.
This allows us to write a relation among the charges and couplings of each basis as
\beq
g'_m Q'_m = \sum_n g_n Q_n \mathbb{R}_{nm}.
\label{eq:P1Sixteen}
\eeq
We are motivated to enforce that $Q'_1 = \sum_n c_n Q_n$ ($Q'_1$ is chosen as an example and has no significance as opposed to $Q'_2$ or any other charge) so that the rotated basis couples to some specifically chosen charge and where $c_n$ are real numbers.  Then, for example. the coupling to the 1$^{\rm st}$ gauge boson becomes
\beq
g'_1 Q'_1 = g'_1 \sum_n c_n Q_n =  \sum_n g_n Q_n \mathbb{R}_{n1}.
\label{eq:P1Seventeen}
\eeq
We now promote $Q_n$ to vectors $\textbf{Q}_n$ whose components $(\textbf{Q}_n)_p$ denote the charge $Q_n$ for the p$^{\rm th}$ particle.  In standard form this vector is
\beq
\textbf{Q}_n = \lpa \bea{c} Q_{n,1} \\ Q_{n,2} \\ \vdots \\ Q_{n,p} \eea \rpa,
\label{eq:P1Eightteen}
\eeq
where again, for clarification, $Q_{n,1}$ is the $Q_n$ charge of particle 1 and so on.  If we ask that the $\textbf{Q}_n$ vectors are orthogonal, then we can deduce that
\beq
\bea{c}
g'_1 \sum_n c_n \textbf{Q}_n \cdot \textbf{Q}_m =  \sum_n g_n \textbf{Q}_n \cdot \textbf{Q}_m \mathbb{R}_{n1} \rightarrow g'_1 c_m = g_m \mathbb{R}_{m1}, \\
\mathbb{R}_{m1} = g'_1 c_m/g_m.
\label{eq:P1Nineteen}
\eea
\eeq
With this result, and using the orthogonality of $\mathbb{R}$ we arrive at our final result, 
\beq
\sum_m \lpa \mathbb{R}_{m1} \rpa^2 = 1 \rightarrow \frac{1}{g_1^2} - \sum_m \lpa \frac{c_m}{g_m} \rpa^2 = 0.
\label{eq:P1Twenty}
\eeq
This orthogonality relation of equation  (\ref{eq:P1Twenty}) was shown to be true to the one loop order in ref~\cite{Anchordoqui:2011eg}.  When we apply this result to the 4 brane model we have
\beq
\frac{1}{g_Y^2} - \sum_m \lpa \frac{c_m}{g'_m} \rpa^2 = 0.
\label{eq:P1Twentyone}
\eeq
We then make the identification of the hypercharge of the SM as
\beq
Q_Y = c_1 Q_{I_R} + c_3 Q_B + c_4 Q_L,
\label{eq:P1Twentytwo}
\eeq
where $c_1 = 1/2, \ c_3 = 1/6, c_4 = -1/2$ and identify $B = Q_B/3$ and $L = Q_L$ (baryon number and lepton number respectively). The $c_i$ are chosen such that the charges are, baryon number, lepton number, and a combination of hypercharge and baryon number minus lepton number.  The specific relations for the charges are give as
\beq
Q_B = 3B ; \ Q_L = L ; \ Q_{I_R} = 2 Q_Y - \lpa B - L \rpa.
\label{eq:P1Twentythree}
\eeq
Thus in this basis we have gauged baryon number $B$ and lepton number $L$.  This, however, presents a problem as gauged baryon number and lepton number are anomalous charges in gauge theories.  Conversely the last charge, $Q_{I_R}$, is a combination of $Q_Y$ and $B-L$ which is non-anomalous.  The anomalies of $B$ and $L$ are solved by associating the bosons with a St\"{u}ckleberg~\cite{Ruegg:2003ps} mass, which will be discussed in the next subsection. 

Before addressing the anomalies, we consider that the charge coupling to the $Y_\mu$ gauge boson is the hypercharge,  which then fixes two Euler angles $\theta$ and $\psi$ of the rotation via
\beq
c_1 g_Y = - g'_1 \sin(\theta), \ c_4 g_Y = g'_4 \cos(\theta)\sin(\psi).
\label{eq:P1Twentyfour}
\eeq
Next we demand that $Y''_\mu$ couples to a linear combination of anomaly-free $I_R$ and $B-L$, that is we require the terms proportional to $B$ and $L$ must be equal so that they form a charge proportional to $B-L$.  This can be accomplished by fixing the Euler angle $\phi$ as
\beq
\tan \phi = -\sin \theta \frac{3 g'_3 \cos \psi + g'_4 \sin \psi}{3 g'_3 \sin \psi - g'_4 \cos \psi}.
\label{eq:P1Twentyfive}
\eeq
After fixing the angles, we have two non-anomalous gauge bosons $Y_\mu$ and $Y''_\mu$, along with one anomalous $Y'_\mu$ gauge boson. We must rid ourselves of this anomaly in order to consider the theory self consistent.

\subsection{Getting Rid of the Anomaly}

  We can rid ourselves of the anomaly associated with $Y'_\mu$ by canceling it using the 4D version~\cite{Witten:1984dg,Dine:1987xk,Atick:1987gy,Lerche:1987qk,Ibanez:1999it} of the Green-Schwarz mechanism~\cite{Green:1984sg}.  By using the Green-Schwarz mechanism, $Y'_\mu$ will acquire a mass on the order of the string scale $M_s$.  

We use the underlying string theory to help us because in the spectrum of the closed strings there is an anti-symmetric rank 2 tensor field, the Kalb-Ramond field $B_{\mu \nu}$, from which we can construct a gauge invariant interaction with an anomalous $U(1)$ gauge field that we will call $C_\mu$ in the example below. The gauge invariant Lagrangian with the Kalb-Ramond field is
\beq
\CL = -\frac{1}{12} H^{\mu \nu \rho}H_{\mu \nu \rho} - \frac{1}{4} F_{\mu \nu}F^{\mu \nu} + \frac{c}{4} \epsilon^{\mu \nu \rho \sigma} B_{\mu \nu} F_{\rho \sigma},
\label{eq:P1Twentysix}
\eeq
where
\beq
H_{\mu \nu \rho} = \p_\mu B_{\nu \rho} + \p_\rho B_{\mu \nu} + \p_\nu B_{\rho \mu}
\label{eq:P1Twentyseven}
\eeq
is the field strength of $B_{\mu \nu}$ and $c$ is some arbitrary constant. The anomalies associated with $C_\mu$ can be canceled by the proper choice of $c$, but this ends up giving mass to the gauge boson; this is the Green-Schwarz mechanism. We will re-write the Lagrangian so that the mass of the gauge boson is apparent by using
\beq
\frac{c}{4} \epsilon^{\mu \nu \rho \sigma} B_{\mu \nu} F_{\rho \sigma} = \frac{c}{2} \epsilon^{\mu \nu \rho \sigma} B_{\mu \nu} \p_\rho C_\sigma
\label{eq:P1Twentyeight}
\eeq
and integration by parts (remember this is part of an action and thus integrated over) of the last term of  (\ref{eq:P1Twentysix}), which leads us to
\beq
\frac{c}{4} \epsilon^{\mu \nu \rho \sigma} B_{\mu \nu} F_{\rho \sigma} = -\frac{c}{2} \epsilon^{\mu \nu \rho \sigma} \lpa \p_\rho B_{\mu \nu} \rpa  C_\sigma = -\frac{c}{6} \epsilon^{\mu \nu \rho \sigma} H_{\rho \mu \nu}  C_\sigma.
\label{eq:P1Twentynine}
\eeq
We also make use of the Bianchi identity, $ \epsilon^{\mu \nu \rho \sigma}\p_\mu H_{\nu \rho \sigma} = 0$. Adding this term to the Lagrangian doesn't change anything because it must be zero; we enforce this by adding a Lagrange multiplier field $\eta$ to the Lagrangian
\beq
\CL  = -\frac{1}{12} H^{\mu \nu \rho}H_{\mu \nu \rho} - \frac{1}{4} F_{\mu \nu}F^{\mu \nu} -\frac{c}{6} \epsilon^{\mu \nu \rho \sigma} H_{\rho \mu \nu}  C_\sigma - \frac{c}{6} \eta  \epsilon^{\mu \nu \rho \sigma}\p_\mu H_{\nu \rho \sigma}.\
\label{eq:P1Thrity}
\eeq
Again integration by parts on the final term of (\ref{eq:P1Thrity}) allows us to solve the equations of motion for $H_{\mu \nu \rho}$  in the form
\beq
H^{\mu \nu \rho} = - c \epsilon^{\mu \nu \rho \sigma} \lpa C_\sigma + \p_\sigma \eta \rpa.
\label{eq:P1Thirtyone}
\eeq
Inserting this solution back into the Lagrangian we arrive at the effective theory for the gauge field $C_\mu$, which is
\beq
\CL = -\frac{1}{4} F_{\mu \nu} F^{\mu \nu} - \frac{c^2}{2} \lpa C_\sigma + \p_\sigma \eta \rpa^2.
\label{eq:P1Thirtytwo}
\eeq
An appropriate choice of gauge for $C_\sigma$ can ``eat" the additional $\eta$ field and thus $C_\mu$ acquires a mass.  Through this mechanism we give mass to the $Y'_\mu$ gauge boson of the 4 brane model, of the order of $M_s$, and we will eliminate the anomalies associated with it. When the Higgs fields acquire non-zero vevs will generate additional mass terms for $Y'_\mu$, which will formally introduce mixing with the other gauge bosons, but these will be of order $(\hbox{TeV}/M_s)^2$.  We neglect such effects and identify $Y'_\mu \approx Z'_\mu$.

\subsection{Identifying the SM and Extension}

  With the anomaly of $Y'_\mu$ removed, we return to the covariant derivative in the basis of $Y_\mu, \  Y'_\mu, \ Y''_\mu$ and identify the photon $A_\mu$ and weak force mediators $W^+_{\mu},W^-_{\mu}, Z^0_{\mu}$ by performing a Weinberg transformation on this basis of the form
\beq
\lpa
\bea{c}
  A_\mu \\
  Z^0_\mu \\
  W^+_\mu \\
  W^-_\mu \\
\eea
\rpa
 = 
\lpa
\bea{cccc}
  \phantom{-}C_{\theta_W} & S_{\theta_W} & 0 & \phantom{-}0  \\
  -S_{\theta_W} & C_{\theta_W} & 0 & \phantom{-}0 \\
  \phantom{-}0 & 0 & 1/\sqrt{2} & \phantom{-}1/\sqrt{2} \\
 \phantom{-} 0 & 0 & 1/\sqrt{2} & -i/\sqrt{2} \\
\eea
\rpa
\lpa
\bea{c} 
  Y_\mu \\
  W^3_\mu \\
  W^1_\mu \\
  W^2_\mu \\ 
 \eea
\rpa \, ;
\label{eq:P1Thirtythree}
\eeq
this transformation then puts the covariant derivative in the form of
\begin{eqnarray}
D_\mu & = & \partial_\mu  -  \frac{i}{\sqrt{2}} \, g_2 \, \sigma^- W^+_\mu - \frac{i}{\sqrt{2}} \, g_2 \, \sigma^+ W^-_\mu   \nonumber \\
&-& i g_2 \, \cos \theta_W \, \left(\sigma^3/2 - Q_Y \tan^2 \theta_W \right) Z^0_\mu \nonumber \\
 &-&   i g_2 \sin \theta_W \left(\sigma^3/2 + Q_Y \right) A_\mu  -  i g_{Y'} Q_{Y'} Z'_\mu - i g_{Y''} Q_{Y''} Y''_\mu \,,
\label{eq:P1Thirtyfour}
\end{eqnarray}
with $\sigma^{\pm} = \left(\sigma^1 \pm i \sigma^2\right)/2$ , and $g_Y/g_2 = \tan\theta_W$. From  (\ref{eq:P1Fourteen}) and  (\ref{eq:P1Thirtyfour}) we find
\begin{eqnarray}
Q_Y H & = & H/2 \,, \nonumber \\ 
g_{Y'} Q_{Y'} H & = & (g'_1 C_\theta S_\phi ) H \,, \nonumber \\ 
g_{Y''} Q_{Y''} H
 & = & (g'_1 C_\theta C_\phi) H \, , \nonumber \\
 Q_Y H'' & = & 0 \,, \nonumber \\
g_{Y'} Q_{Y'} H'' & = &
-[g'_1 C_\theta S_\phi + g'_4(C_\phi C_\psi + S_\theta S_\phi
S_\psi)]H'' \,, \nonumber \\
g_{Y''} Q_{Y''} H'' & = & -[g'_1C_\theta C_\phi +
g'_4 (C_\phi S_\theta S_\psi - C_\psi S_\phi)]H'' \, .
\label{eq:P1Thirtyfive}
\end{eqnarray}

We turn now to exploring the masses of the gauge bosons. Expanding the Higgs fields about their respective vevs, the Higgs kinetic terms of equation  (\ref{eq:P1Ten}) together with the Green-Schwarz mass term, which is $\frac{1}{2} {M'}^2 Z'_\mu Z'^\mu$, gives 
\beq
\mathscr{B} = \frac{1}{2} [D^\dagger_\mu \left( 0 \ v\right)] \left[
    D^\mu \left(\begin{matrix} 0 \\ v \end{matrix}\right) \right] +
 \frac{1}{2}  (D_\mu v'')^\dagger(D^\mu v'') + \frac{1}{2} {M'}^2
Z'_\mu Z'^\mu  \, .
\label{eq:P1Thirtysix}
\eeq
Expanding equation  (\ref{eq:P1Thirtysix}) gives
\begin{eqnarray}
&\mathscr{B}& =  \frac{(g_2\, v)^2}{4} W^+_\mu W^{-\mu} + \frac{ (g_2 v)^2}{8 \cos^2(\theta_W) } \, Z^0_\mu Z_0^{\mu} \nonumber 
\\
&-& \frac{v^2}{2 \cos(\theta_W)} g'_1 g_2 \ C_\theta \left(S_\phi Z'_\mu + C_\phi Y''_\mu\right)   Z_0^{\mu} \nonumber 
\\
&+&   \frac{1}{2} (g'_1 v \ C_\theta)^2 \left(S_\phi Z'_\mu + C_\phi Y''_\mu\right)\left(S_\phi Z'^\mu + C_\phi Y''^\mu\right) + \frac{1}{2} {M'}^2 Z'_\mu Z'^\mu  \nonumber 
\\
&+&  \frac{1}{2} {v''}^2 \left\{ g'_1 C_\theta (S_\phi \, Z'_\mu + C_\phi \, Y''_\mu) + g'_4 \left[ (C_\phi C_\psi + S_\theta S_\phi S_\psi) \, Z'_\mu \right.\ \right.\  \nonumber 
\\
& &\left.\  \left.\ + \lpa S_\psi S_\theta C_\phi - C_\psi S_\phi \rpa \, Y''_\mu \right] \right\}^2 \ .  \nonumber \\
& \simeq&  \frac{(g_2\, v)^2}{4} W^+_\mu W^{-\mu}  + \frac{ (g_2 v)^2}{8 \cos^2(\theta_W) } \, Z^0_\mu Z_0^{\mu}  - \frac{v^2}{2 \cos(\theta_W)} g'_1 g_2 \ C_\theta  C_\phi Y''_\mu Z_0^{\mu} \nonumber
\\
& +&  \frac{1}{2} (g'_1 v \ C_\theta C_\phi)^2 Y''_\mu Y''^\mu  + \frac{1}{2} {v''}^2\left( g'_1 C_\theta  C_\phi + g'_4 \lpa S_\psi S_\theta C_\phi - C_\psi S_\phi \rpa \right)^2 Y''_\mu Y''^\mu \nonumber \\
& +& \dots
\label{eq:P1Thirtyseven}
\end{eqnarray}
where the omitted terms only contain the $Z'$ couplings, which if you recall we ignore since they will be significantly smaller than the mass term provided from the Green-Schwarz mechanism. By inspection of  (\ref{eq:P1Thirtyseven}) we can see the $W^\pm$ masses are given by usual tree level formula, and the mass of the $Z^0$ particle associated with the SM is given by $M_{Z_0} = g_2 v/(2 \cos(\theta_W))$ before mixing. 

	Now we use the relation $g'_1 S_\theta = g'_4 C_\theta S_\psi$ from equation  (\ref{eq:P1Twentyfour}) to conveniently rewrite  (\ref{eq:P1Thirtyseven}) as
\begin{eqnarray}
\mathscr{B} & \simeq &  \frac{(g_2\, v)^2}{4} W^+_\mu W^{-\mu}  + \frac{ (g_2 v)^2}{8 \cos^2(\theta_W) } \, Z^0_\mu Z_0^{\mu}  - \frac{v^2}{2 \cos(\theta_W)} g'_1 g_2 \ C_\theta  C_\phi Y''_\mu Z_0^{\mu} \nonumber
\\
& + & \frac{(g'_1 v'')^2}{2} \lpa \lpa \frac{C_\phi}{C_\theta} - \frac{C_\psi S_\phi S_\theta}{C_\theta S_\psi} \rpa^2 + \lpa \frac{v}{v''} \rpa^2 C^2_\theta C^2_\psi \rpa Y''_\mu Y''^\mu  + \dots 
\\
& \simeq & \frac{1}{4} \, (g_2 v)^2 \, W^+_\mu W^{-\mu} +\lpa \bea{cr} Z^0_\mu & Y''_\mu \eea \rpa  \cdot \textbf{M} \cdot \lpa \bea{c}Z^\mu_0 \\ Y''^\mu \eea \rpa  + \dots \ ,
\label{eq:P1Thirtyeight}
\end{eqnarray}
where $\textbf{M}$ is a non-diagonal mass matrix.  To identify the masses of the system we need to make another change of basis such that $\textbf{M}$ becomes diagonal.  In doing so linear combinations of $Y''_\mu$ and $Z^0_\mu$ become the massive bosons we observe in experiment.  However, if we make the assumption that $v'' \gg v$, then the mass matrix is automatically diagonalized into the SM plus an additional massive gauge boson of the scale $v''$, the equation  (\ref{eq:P1Thirtyeight}) becomes
\beqa
\mathscr{B}& \simeq & \frac{(g_2\, v)^2}{4} W^+_\mu W^{-\mu} + \frac{(g_2 v)^2}{8 \cos^2(\theta_W)} Z^0_\mu Z_0^\mu \nonumber
\\
& +& \frac{(g'_1 v'')^2}{2} \lpa C_\phi C^{-1}_\theta - C_\psi S^{-1}_\psi S_\phi T_\theta \rpa^2 Y''_\mu Y''^\mu +\ \CO \lpa  \lpa \frac{v}{v''} \rpa^2 \rpa \ .
\label{eq:P1Thirtynine}
\eeqa
We can see we that we preserve the SM results of the ratio of $M_{Z_0}$ to the mass of $M_W$.  Furthermore we make the identification that $Z'' \simeq Y'' + {\rm small \ corrections}$.

  Finally, we must check our assumption of orthogonality of charges. Table~\ref{tab:P1ChiFerSpect} shows the charges $\mathbf{Q_B}$, $\mathbf{Q_L}$, and $\mathbf{Q_{I_R}}$ are mutually orthogonal in the fermion space, {\em i.e. } \beq \sum_f (\mathbf{Q}_i)_f (\textbf{Q}_j)_f=0 \ , \eeq for $i\neq j$. The orthogonality relation will be satisfied to one loop~\cite{Anchordoqui:2011eg} for the fermions. However, the charges assigned to $H''$ will violate the orthogonality condition. The non-orthogonality of $H''$ is only a minor problem, as contributions from $H''$ to the running of $g'_1$ are at the 0.9\% level from the string scale to the TeV scale, and at the level of 0.3\% for the running of $g'_4$. These are of the same order as the two loop contributions from the fermion sector, so we may ignore the nonorthogonality introduced by $H''$ in the context of one loop considerations.

In summary, we have constructed a model that extends the SM by using concepts introduced from string theory; by considering the symmetries of the underlying string theory, we constructed an effective field theory that has all the features of the SM plus two additional $U(1)$ gauge bosons whose masses are at the string scale and at the scale of an additional Higgs field vev.  With these additional bosons, Lepton and Baryon number become gauged conservation laws.  Because of the addition of two $U(1)$ gauge bosons while still retaining the SM we name this model the SM$^{++}$, and in the next section we look for potential signals from the SM$^{++}$ model in LHC data.
	
\subsection{The LHC Era}
 At the time of writing the LHC is the largest circular proton-proton collider ever constructed. With a circumference of $27$ km, and with a design collision energy of $14$ TeV, it will be the highest energy collider to date once full collision energy is achieved. Science runs at the LHC, which is located in the Geneva on the Swiss-French boarder, commenced during the years 2011 and 2012.  To compare the model of the SM$^{++}$, we must first understand what is observed at the LHC experiments.  One concept key to new physics discoveries is the concept of beam luminosity.  The beam luminosity indicates that the number of particles crossing a unit of area per unit time may or (more likely) may not collide with the intersecting beam at the interaction points (centered around the various detectors that form the LHC).  The number of a certain type of events the detectors have are capable of detecting is given by the formula
\beq
N_{\rm events} = \sigma \int \CL_{\rm I}(t') dt',
\label{eq:P1Forty}
\eeq
where $\sigma$ is the cross section of the process of interest and $\int \CL_{\rm I} (t') dt'$ is the integrated luminosity, which is typically given in units of inverse barns, where $1 \ \rm bn = 10^{-28} \ \rm m^2 = 10^2 \ \rm fm^2$.  The design instantaneous luminosity of the LHC is $\CL_{\rm I}(t) = 10^{34} \ \rm cm^{-2} s^{-1} = 10^{-5} \ \rm fb^{-1} s^{-1}$~\cite{Vankov} which would give an annual integrated luminosity of $315 \ \rm fb^{-1}$ if the LHC was to run 24 hours a day, 365 days a year however, it does not and cannot. 

 Detection of new physics is (like all scientific measurement) a question of statistics; the more events we observe of a rare process will make smaller the standard deviation of the null hypothesis (recall from the mean value theorem the standard deviation goes like $1/\sqrt{N}$, with $N$ as the number of observations) and thus any deviation above $5$ standard deviations of the null hypotheses is the generally accepted criteria for discovery.  So a larger luminosity means better possible detection of new physics.  The data we will use to compare the SM$^{++}$ was taken during the spring of 2012 with integrated luminosities varying from $3.6 \pm 0.2 \ \rm fb^{-1}$ to $4.1 \pm 0.2 \ \rm fb^{-1}$~\cite{cms_dijet8tev}~\cite{atlas_dijet8tev}.  During this time the LHC was running at collision energies of $\sqrt{s} = 8 \ \rm TeV$, the highest collision energies ever produced in a lab.  The $s$ in the last equation is the common Lorentz invariant term known as a Mandelstam variable.  Its value is given in the center of mass frame as
 \beq
 s = (k+p)^2 = m_1^2 + m_2^2 + 2 \left( \sqrt{p^2 + m_1^2} \sqrt{k^2 + m_2^2} + p^2 \right) = (E_1 + E_2)^2 = E_{\rm c.m.}^2,
 \label{eq:P1Fortyone}
 \eeq
  with $k, p$ the incoming particle $4$-momentum. Expressing collision energies is typically represented by $\sqrt{s}= E_{\rm c.m.}$ with $E_{\rm c.m.}$ as the center of mass total energy.  
 
When searching for new particles, a common technique is looking for resonances or peaks in the cross section as a function of invariant mass.  The invariant mass method is a technique that takes advantage of the conservation of momentum/energy that is enforced by Lorentz invariance.  Searches can be done with multiple products in the decay of the new physics particle, however we will focus on only dilepton and dijet  processes.  We can see the usefulness of this method by imagining in the collision process a massive particle is produced; with a mass of $M$ this particle then decays and produces two new particles of mass $m'_1$ and $m'_2$. The conservation of energy requires
\beq
\sqrt{p^2 + M^2} = \sqrt{p_1^{'2} + m_1^{'2}} + \sqrt{p_2^{'2}+m_2^{'2}},
\label{eq:P1Fortytwo}
\eeq
where $p$ is the initial momentum of the particle of mass $M$.  These particles then decay, and may decay again, and so on, and the conservation of energy must then read
\beq
\sqrt{p^2 + M^2} = \sum_{\rm products} E_i,
\label{eq:P1Fortythree}
\eeq
where we are summing over all the final products. We also can use the conservation of momentum to enforce
\beq
\textbf{p} = \sum_{\rm products} \textbf{p}_i,
\label{eq:P1Fortyfour}
\eeq
where $\textbf{p}$ is the 3-momentum of the decaying particle of mass $M$ and $\textbf{p}_i$ the 3-momentum of the decay products.  Because of this equality, equation  (\ref{eq:P1Fortythree}) can be expressed as
\beq
M^2 = \left(\sum_{\rm products} E_i \right)^2 - \left(\sum_{\rm products} \textbf{p}_i \right)^2.
\label{eq:P1Fortyfive}
\eeq
The value of $M^2$ is called the invariant mass or rest mass of the particle that has undergone the decay process.  In the analysis we will use to set limits on the SM$^{++}$, we use dilepton (detected $e^+ e^-$, or $\mu^+\mu^-$ final states)  and dijet  (two nearly back-to-back jets (in transverse momentum) of multiple hadrons having nearly identical 3-momentums, can be resolved to two seed partons at the interaction point (IP)) analyses preformed by the CMS and ATLAS experiments~\cite{cms_dijet8tev}~\cite{atlas_dijet8tev}.  

\begin{figure}[tbp] 
\postscript{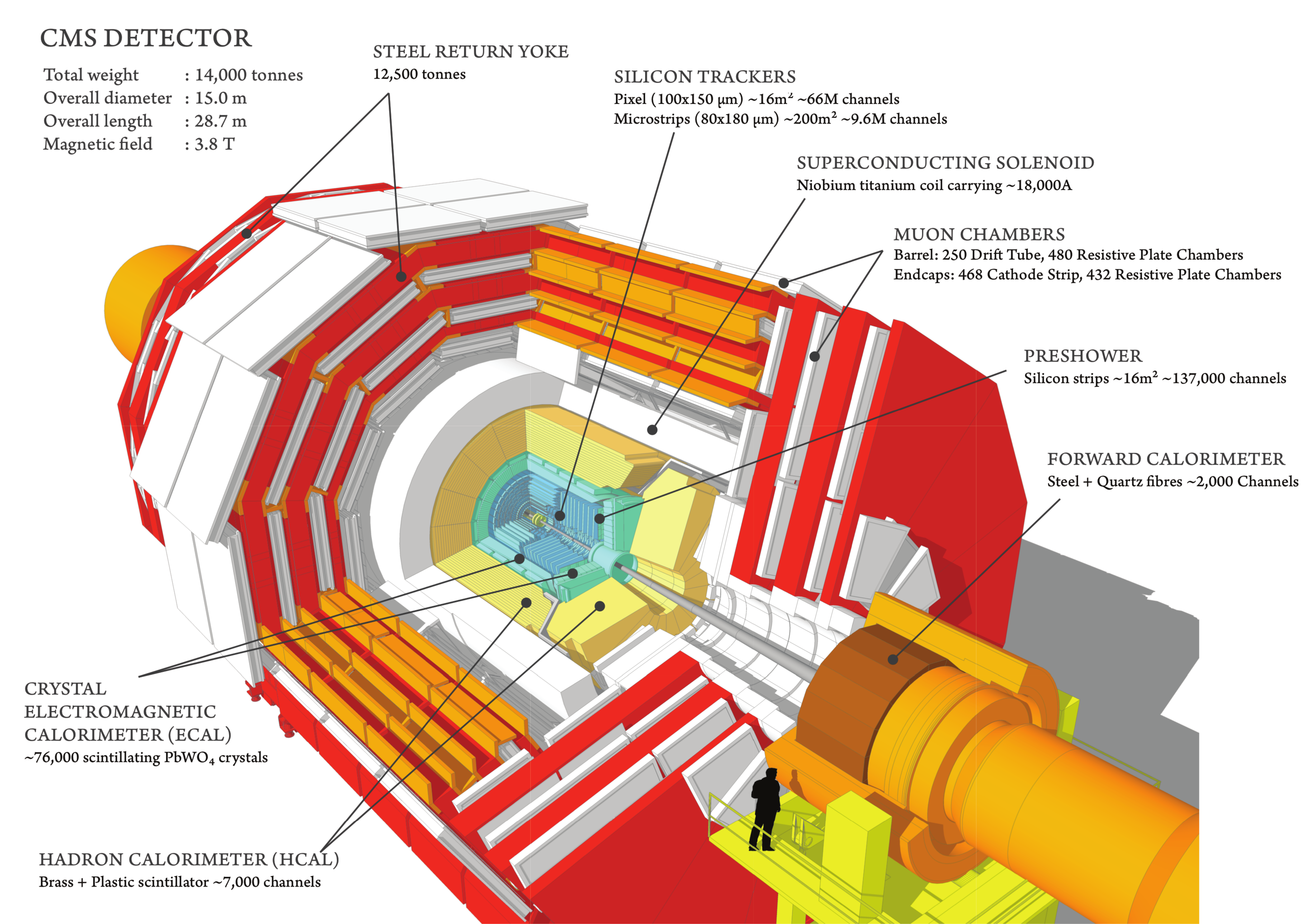}{0.9}
 \caption[CMS Detector Systems Cutaway]{ \sglspc \small The CMS detector at the LHC cut-away shows the layering of the detector systems in the barrel calorimeter geometry. Figure taken from~\cite{CMSCut}.}
\label{fig:P1CMScutaway} 
\end{figure}

Further understanding of the experimental apparatus to compare with theory is required to understand the analysis.  The two main detectors for generalized particle physics at the LHC are the Compact Muon Solenoid (CMS) and the A Toroidal LHC Apparatus (ATLAS) detectors.   Both have a similar design, as both are barrel calorimeter detectors.  As seen in Fig. \ref{fig:P1CMScutaway} the CMS consists of several layers of detector systems.  
The inner-most detectors are the silicon strip and pixel detectors, which are designed to track the charged debris from collision of two partons of the colliding protons.  These detectors help deduce the particles' identity and momentum, as the curvature of the track is a measurement of the particles' momentum in the magnetic field of the detector.  After the inner-most tracker, the electromagnetic calorimeter (ECAL) is  layered in a cylindrical fashion around the IP.  This consists of 76,000 lead-tungstate scintillating crystals.  The ECAL is designed to use the pair productions from near approaches of photons or of $e^+ e^-$ to the nucleus of the individual atoms of the crystal structure creating photonic showers, which are then read out by photo-multiplier tubes (PMTs)~\cite{TullyNutShell}.  Hadrons, however, being much more massive, move through this region hardly impeded and then interact with the next layer the brass scintillator hadronic calorimeter (HCAL).  This region does essentially the same as the ECAL but has the power to stop hardons.  Sandwiched between the superconducting magnet that produces an axial magnetic field of $3.8$ T for the CMS is the muon detection system. The muon system consists of several types of detectors, such as drift tubes and capacitive plates (these measure ionization of the internal gas like a series of Geiger counters); the muon system is the final system as the muon is so massive compared to other elementary particles (but not so massive that it decays before being detected) that it plows through the whole detector and is only finally detected when interacting in the iron yoke region of the detector (the iron yoke is the manifold the detector is built upon and used to direct the magnetic field in the CMS).

\begin{figure}[tbp] 
\postscript{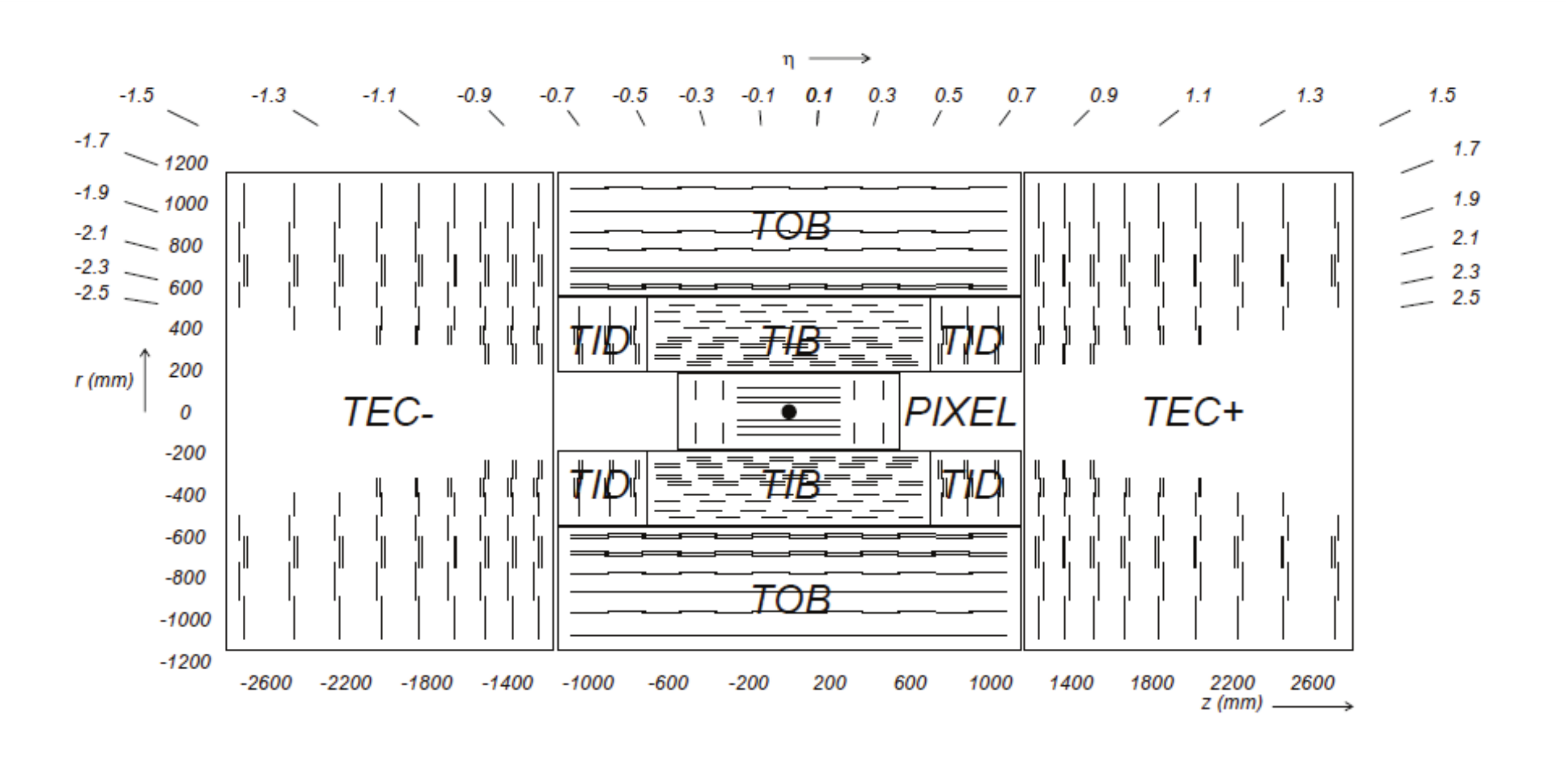}{1.0}
 \caption[CMS Detector Pseudo-Rapidity Coverage]{ \sglspc \small The CMS detector pseudo-rapidity coverage available from the detector layout. TEC (the end caps), TOB (the outer barrel), TIB/D (the inner barrel/detector). Figure taken from~\cite{KrammerM}.}
\label{fig:P1pseduorap} 
\end{figure}

For experimental analysis a common variable, pseudo-rapidity, is used rather than the scattering angle, as it makes algorithms for jet reconstructions easier.  Pseudo-rapidity is given by the equation
\beq
\eta = - \ln\left(\tan\left(\frac{\theta}{2} \right)\right),
\label{eq:P1Fortysix}
\eeq
with $\theta$ as the scattering angle against the beam direction (commonly taken as the z-direction).  The CMS covers a region of rapidity from 0 to 2.5 as seen in Fig. \ref{fig:P1pseduorap}.
To find new particles we must, as in all experiments, battle against noise of the detector.  Besides detector uncertainty, which ranges from trigger rates (recording proper events), to offline efficiencies (radiation can damage detectors, taking them offline), to detector acceptance (the efficiency of the detector to actually detect a particle when it is there).  We have an enormous amount of background events that come from well understood physics steaming from glancing interactions or other high energy interactions that have already been understood.  Many of these unwanted processes are discarded by the detector itself through using a two-layer trigger system. Fast electronics make determinations if certain events are worth recording, and make the data coming from the detector more manageable to record.  One way to cut down on noise events is to use selection cuts.  Selection cuts correspond to only analyzing dijet, dilepton (in our case) events that correspond to having a transverse momentum (transverse to the beam direction) $p_T$ greater than a certain value.  This helps cut down on noise, as particles with high transverse momentum necessarily underwent hard collision processes rather than glancing collisions.    Even with selection cuts, many processes of QCD interactions make it into the data. To understand this background noise a Monte Carlo (MC) simulation of QCD processes and the detection system is done in the invariant mass window of interest via the program {\sc PYTHIA} for a certain value of integrated luminosity, as was done by the CMS Collaboration in references~\cite{cms_dijet8tev} and~\cite{cms_dilepton8tev}. The MC program also determines the level of standard deviations given the integrated luminosity in order to determine if new physics is or is not found.   Comparison of the MC generated background, experiment, and theory allows us to set limits on the SM$^{++}$ model.      

\subsection{LHC Phenomenology of SM$^{++}$}

The LHC allows us the possibility to detect the extra gauge boson $Z''$ if it has a mass the scale of several TeV.  The SM$^{++}$ free parameters include three new coupling strengths $g'_1,\, g'_3,$ and $g'_{4}$, which are encoded into three Euler angles $\lpa \theta,\phi,\psi \rpa$ of the $U(1)$ gauge field rotation.

The baryon number coupling $g'_3$ is fixed to be $g_3(M_s)/\sqrt{6}$ where $g_3(M_s)$ is the $SU_c(3)$ coupling from the SM at the scale of $U_B(3)$ unification which restores the underlying string theoretical symmetry at the string scale.  Therefore $g'_3$ is determined at all energies through the renormalization group (RG) running of the $U_B(1)$ gauge field and is not a free parameter.  We take the string scale to be $M_s = 10^{14}~{\rm GeV}$ for running down the $g'_3$ coupling to the TeV region that is accessible to the LHC; we take special note that we are {\em ignoring mass threshold effects of stringy states}, which yields $g'_3 (M_s)= 0.231$. Varying the string scale does not significantly affect the running of the $g'_3$ within the LHC range. 

 To ensure perturbativity of $g'_4$ between the TeV scale and the string scale so that the renormalization group one loop equations are valid requires that $g'_4(M_s) \leq 1$ (in fact we would like it to be much less than 1, but this is less restrictive).  Enforcing this limit in equation  (\ref{eq:P1Twentyone}) requires that $g'_1(M_s) > 0.4845$ by knowing that the electromagnetic coupling at the mass of the $Z^0$ scale is $\alpha_{\rm EM}(M_{Z^0}) = 1/127.9$~\cite{Beringer:1900zz}, we can determine $g_Y(M_{Z^0})$ via
 \beqa
 \alpha_{\rm EM} &=& \frac{ e^2}{4\pi} = \frac{g_2^2}{4 \pi}  \sin^2(\theta_W), \nonumber
 \\
 g_Y(M_{Z^0}) &=& g_2(M_{Z^0}) \tan(\theta_W) = \frac{\sqrt{4 \pi \  \alpha_{\rm EM}( M_{Z^0} ) } }{\cos(\theta_W) }, \nonumber
 \\
 g_Y(M_{Z^0} = 91 \ {\rm GeV} ) &\approx& 0.357, \nonumber
 \\
 g_Y(M_s) &\approx& 0.429.
 \eeqa
 
   In the last line, the one loop RG equations for $U_Y(1)$ are used to determine $g_Y(M_s)$, which then allows us the limit previously mentioned.  Similarly we take $g'_1(M_s) \leq 1$ in order to ensure perturbativity at the string scale and all scales below (this is ensured because it is a $U(1)$ coupling).  Because of the constraint of  (\ref{eq:P1Twentyone}) and the requirement of the string scale restored symmetry of $U_B(3)$, only two free parameters are allowed the string scale $M_s$, which we choose to be $M_s = 10^{14} \ \rm GeV$ and one coupling; we take $g'_1$ to be free and it must lie in the range $0.4845 < g'_1(M_s) < 1.0$.

\subsection{\label{sec:ic}Branching Ratios of $Z'$ and $Z''$}
	
Here we calculate the branching ratios ($\mathcal{B}$) of various decays types for the $U(1)$ gauge bosons of the SM$^{++}$.  Branching ratios are defined by
\beq
\mathcal{B} (Z \rightarrow l) = \frac{\Gamma_{Z \rightarrow l} }{\sum_k \Gamma_{Z \rightarrow k} }, 
\eeq
where $\Gamma_{Z \rightarrow l}$ is the decay width of particle $Z$ into type $l$.  The first case we consider is setting $g'_1 (M_s) \approx 1$. This leads to $\psi (M_s) = -1.245$, $\theta (M_s) = -0.217$, $\phi (M_s) =-0.0006$, $g'_4 (M_s) = 0.232$, and $g'_3(M_s) = 0.231$. Substituting these values into (\ref{linda}), we find the vector bosons $(Y, Z'_\mu,Z''_\mu)$ couple to currents
\begin{eqnarray}
J_Y& = & \lpa 2.1 \times 10^{-1}\rpa ~Q_{I_R} +\lpa 2.1 \times 10^{-1} \rpa ~(B-L) \nonumber 
\\
J_{Z'} &=& \lpa 5.8 \times 10^{-4} \rpa ~ Q_{I_R} + \lpa 6.6\times 10^{-1} \rpa ~B + \lpa 7.4 \times 10^{-2} \rpa ~L\nonumber
\\     
J_{Z''} & = & \lpa 9.8 \times 10^{-1}\rpa ~Q_{I_R}  - \lpa 4.7 \times 10^{-2} \rpa ~(B-L)  \, ,
\end{eqnarray}
at the string scale. To find the couplings down at the TeV region we must use $U(1)$ running equations given by
\begin{equation} 
\frac{1}{\alpha_Y(Q)} = \frac{1}{\alpha_Y (M_s)} - \frac{b_Y}{2\pi} \, \ln(Q/M_s) \,, 
\end{equation} 
\begin{equation}
 \frac{1}{\alpha_i(Q)} = \frac{1}{\alpha_i (M_s)} - \frac{b_i}{2\pi}\, \ln(Q/M_s) \,, 
\label{RGbi} 
\end{equation} 
where 
\begin{equation}
b_i = \frac{2}{3} \, {\sum_f} \, Q_{i,f}^2 \, + \frac{1}{3} \, {\sum_s} \, Q_{i,s}^2, 
\end{equation} 
with $f$ and $s$ indicating contribution from fermion and scalar loops, respectively~\cite{Anchordoqui:2011eg}.  This result can be found in many standard texts, such as Peskin and Schroeder~\cite{Peskin}. By setting the exchange boson energy scale appropriate for current (at the time of writing) LHC data to $Q = 4~{\rm TeV}$, we obtain from (\ref{RGbi}) the couplings: $g'_1 = 0.406$, $g'_3 = 0.196$, $g'_4 = 0.218$, $\theta = -0.466$, $\psi = -1.215$, and $\phi = -0.0003$.  In terms of currents, this is
\begin{eqnarray}
J_{Y}& = & \lpa 1.8 \times 10^{-1} \rpa~Q_{I_R} + \lpa 1.8 \times 10^{-1} \rpa~(B - L) \nonumber
 \\
 J_{Z'} &=& \lpa 1.1 \times 10^{-4} \rpa Q_{I_R} +\lpa 5.5 \times 10^{-1} \rpa B + \lpa 7.6 \times 10^{-2} \rpa L \nonumber
 \\
J_{Z''} & = & \lpa 3.6 \times 10^{-1} \rpa ~Q_{I_R}  - \lpa 9.2 \times 10^{-2}\rpa ~(B-L)  \, ,
\end{eqnarray}
where we have assumed that $H''$ has developed its vev $v''$ at this energy.  Decay rates for a particle $Z''$ are calculated as
\beqa
d \Gamma_{Z''\rightarrow k' p'} &=& \frac{| \CM |^2}{2 M_{Z''}} \frac{d^3 k'}{(2 \pi)^3 2  E_{k'}}  \frac{d^3 p'}{(2 \pi)^3 2  E_{p'}}   (2\pi)^4 \delta^{(4)}(M_{Z''} - k' - p'), \nonumber
\\
|\CM |^2 &\simeq& | \langle k', p'| T\{J^\mu_{Z''}  Z''_\mu \}| M_{Z''} \rangle |^2 \propto J_{Z''}^2.
\eeqa
The semi equals is due to the fact that the conservation of 4 momentum is already enforced in the differential decay width formula. Due to the vector nature of these $U(1)$ couplings, each decay channel has a common kinematic term (assuming each decay product is highly relativistic so that we can ignore their individual masses), and the end results only depend on the coupling strengths as indicated in the currents. Summing over fermionic decay channels done in a generalized trace operator. We have ${\rm Tr}~[Q_{I_R} \, B] = {\rm Tr}~[Q_{I_R} L] = {\rm Tr}~[B L] =  0$, which give the $Z'$ and $Z''$ total decay widths as
\beqa
\Gamma_{Z'} &=& \Gamma_{Z' \rightarrow Q_{IR}} + \Gamma_{Z' \rightarrow B} + \Gamma_{Z' \rightarrow L}, \nonumber
\\
&\propto& \lpa 1.1 \times 10^{-4} \rpa^2 \hbox{Tr} [Q_{I_R}^2] +\lpa 5.5 \times 10^{-1} \rpa^2 \hbox{Tr} [B^2] + \lpa 7.6 \times 10^{-2} \rpa^2 \hbox{Tr}[L^2] \nonumber
\\
&\propto& 9.7 \times 10^{-8} + 4.0 \times 10^{-1} + 2.3 \times 10^{-2},
\\
\Gamma_{Z''} & = & \Gamma_{Z'' \to Q_{I_R}}  + \Gamma_{Z'' \to B-L} , \nonumber \\
& \propto & ( 3.6 \times 10^{-1})^2 \, {\rm Tr}[Q_{I_R}^2]  + (9.2\times 10^{-2})^2 {\rm Tr}\left[(B-L)^2 \right] \nonumber \\
& \propto & 1.0 + 4.5 \times 10^{-2} \, .
\eeqa
We can determine generalized branching ratios with this information.  We generalize the branching ratios into a sum over particles having a non-zero charge specified below, such as summing over all decay products with non-zero baryon number $B$ would result in a branching ratio for, say, the $Z'$ particle into these particles as  $Z' \rightarrow B : \ 0.946$.  We present the results for the remaining branching ratios as
\begin{equation}
\begin{tabular}{ccccccccccc} 
$\mathcal{B}$ & $Z' \to B$ & : & $\mathcal{B}$ &$Z' \to L$ & : & $\mathcal{B}$ &$Z'' \to Q_{1R}$  &:& $\mathcal{B}$ &$Z'' \to B-L$   \\
  & $0.946$ & : & & $0.054$ & : & & $0.959$ &  : & & $0.041$ \, .
\end{tabular}
\end{equation}
Though not relevant for LHC phenomenology due to the string scale mass, we see that $Z'$ is very nearly all in $B$, with $\mathcal{B}$: $Z' \to B = 0.946$ and $\mathcal{B}$: $Z' \to L = 0.054$. Of course, there can be variation in decay channels particle by particle, as can be seen by the different individual charges in Table~\ref{tab:P1ChiFerSpect}.  The physical couplings of the $Z''$ to fermionic fields given in Table~\ref{tab:P1ChiFerSpect} are consistent with the bounds presented in~\cite{Williams:2011qb} from a variety of experimental constraints.

\begin{figure}[tbp] 
\begin{minipage}[t]{0.49\textwidth}
\postscript{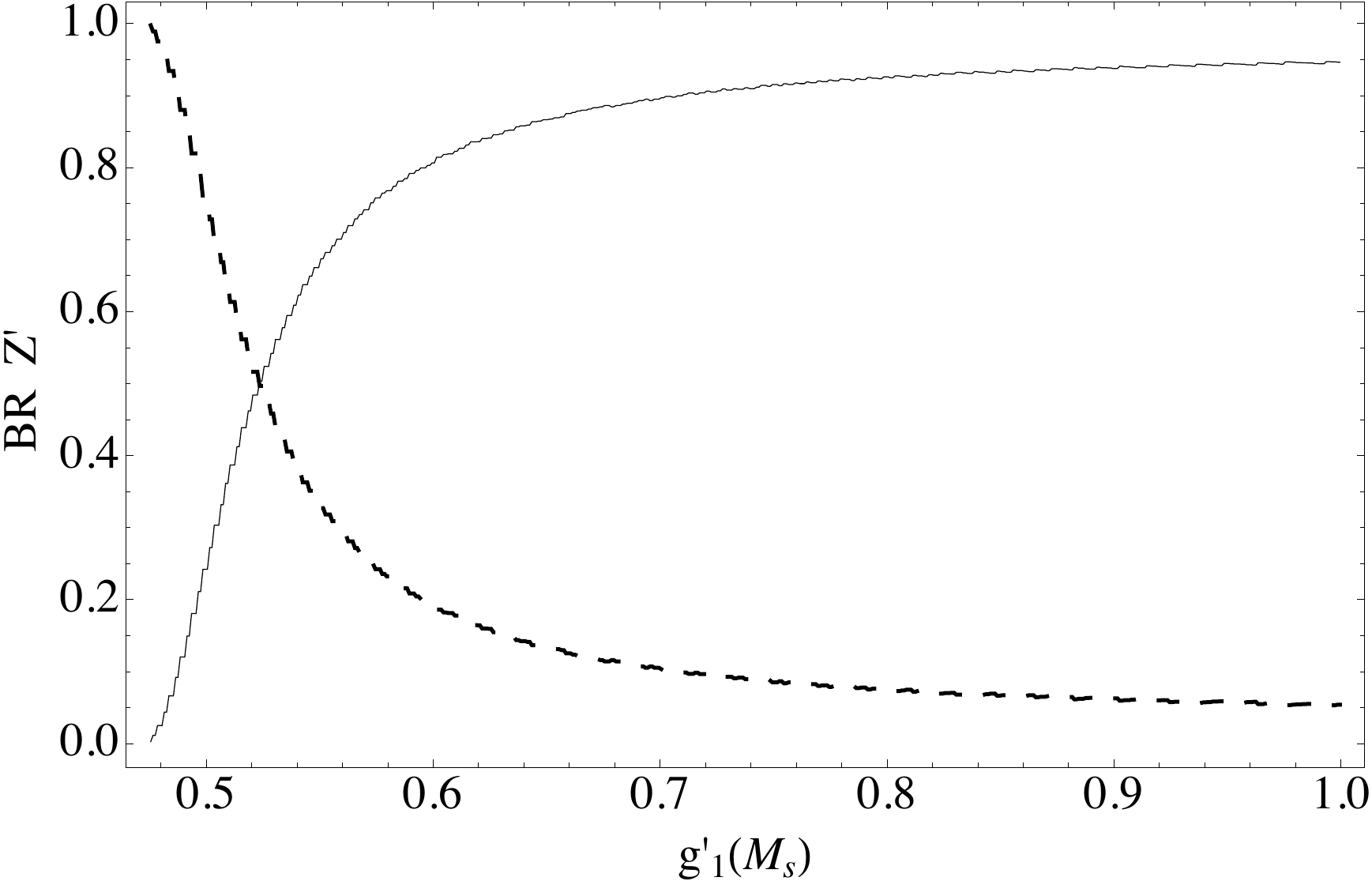}{0.99} \end{minipage}
\hfill \begin{minipage}[t]{0.49\textwidth}
\postscript{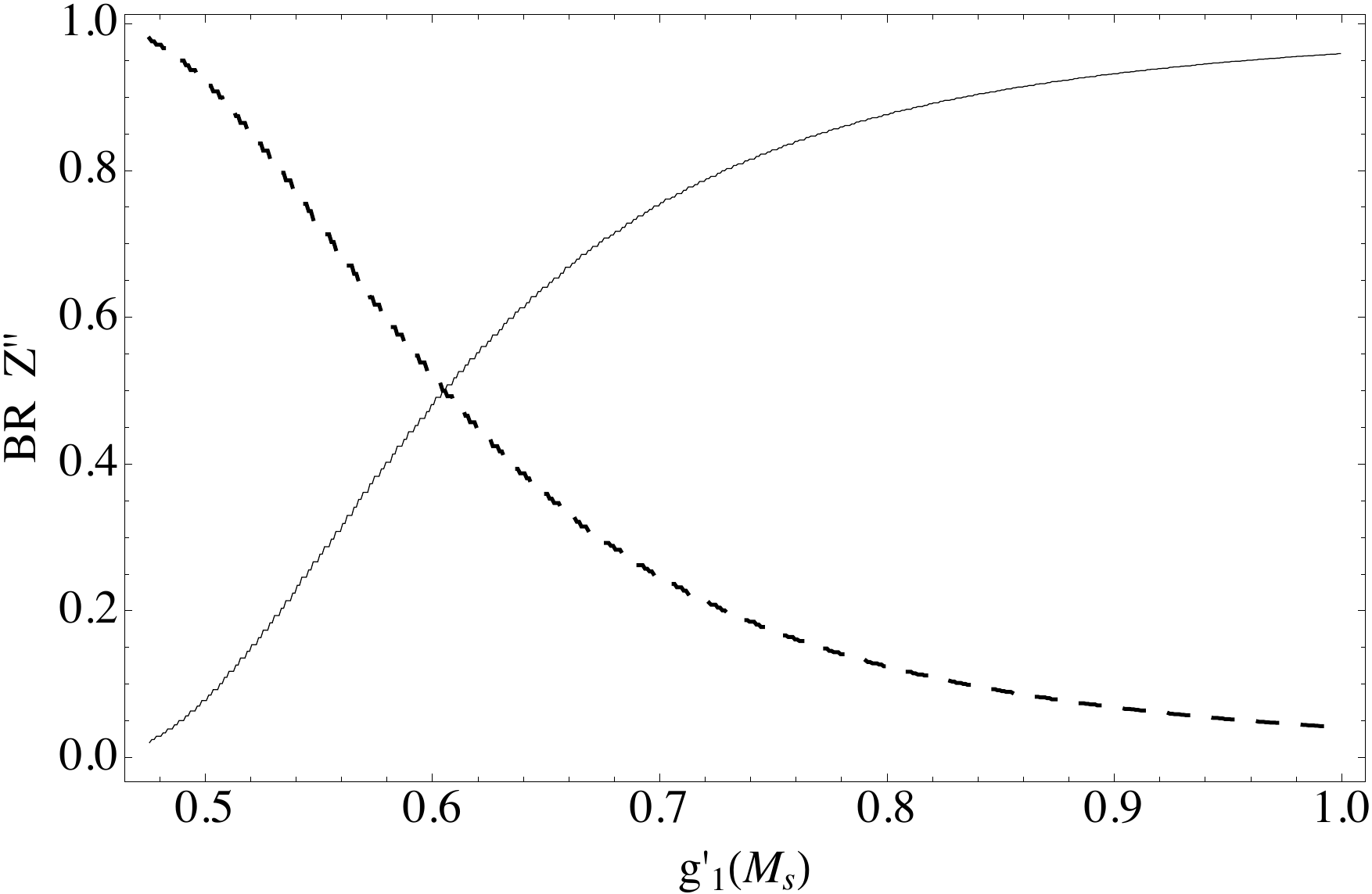}{0.99} \end{minipage} 
\caption[Branching Ratios of SM$^{++}$ for $Z'$ and $Z''$ Bosons]{ \sglspc \small Branching ratios of $Z'$ (left) and $Z''$ (right) as a function of $g'_1(M_s)$. The solid lines denote the branching into $B$ (left) and $I_R$ (right). The dashed lines denote the branching into $L$ (left) and B-L (right).}
\label{sm++_BR} 
\end{figure}

Now, we duplicate the procedure for $g'_1(M_s) = 0.4845$, for which we obtain
\begin{equation}
\begin{tabular}{ccccccccccc} 
$\mathcal{B}$ & $Z' \to B$ & : & $\mathcal{B}$ &$Z' \to L$ & : & $\mathcal{B}$ &$Z'' \to Q_{1R}$  &:&
 $\mathcal{B}$ &$Z'' \to B-L$   \\
  & $0.066$ & : & & $0.934$ & : & & $0.039$ &  : & & $0.961$ \, .
\end{tabular}
\end{equation}
The chiral couplings of $Z'$ and $Z''$ gauge bosons decay mostly into $L$ and $B-L$, respectively. The individual charges in this case are given in Table~\ref{newtable}. Figure \ref{sm++_BR} displays the branching ratios for differing values of  $g'_1(M_s)$ that are allowed by perturbativity constraints.
%-------FIGURE------------------
%----------------------------------
\subsection{Dijet and Dilepton LHC $Z''$ Phenomenology}

Using a data set of $pp$ collisions at $\sqrt{s} = 8~{\rm TeV}$, with
an integrated luminosity of $4.0~{\rm fb}^{-1}$, the CMS Collaboration
has searched for narrow resonances in the dijet  invariant mass
spectrum~\cite{cms_dijet8tev}. Each event in the search is
required to have its two highest $p_T$ jets with (pseudo-rapidity)
$|\eta_j| < 2.5$. The acceptance of selection requirements
is reported to be $ \CA \approx 0.6$.  

\begin{figure}[h]
\begin{minipage}[t]{0.49\textwidth}
\vskip -1.5 cm
\postscript{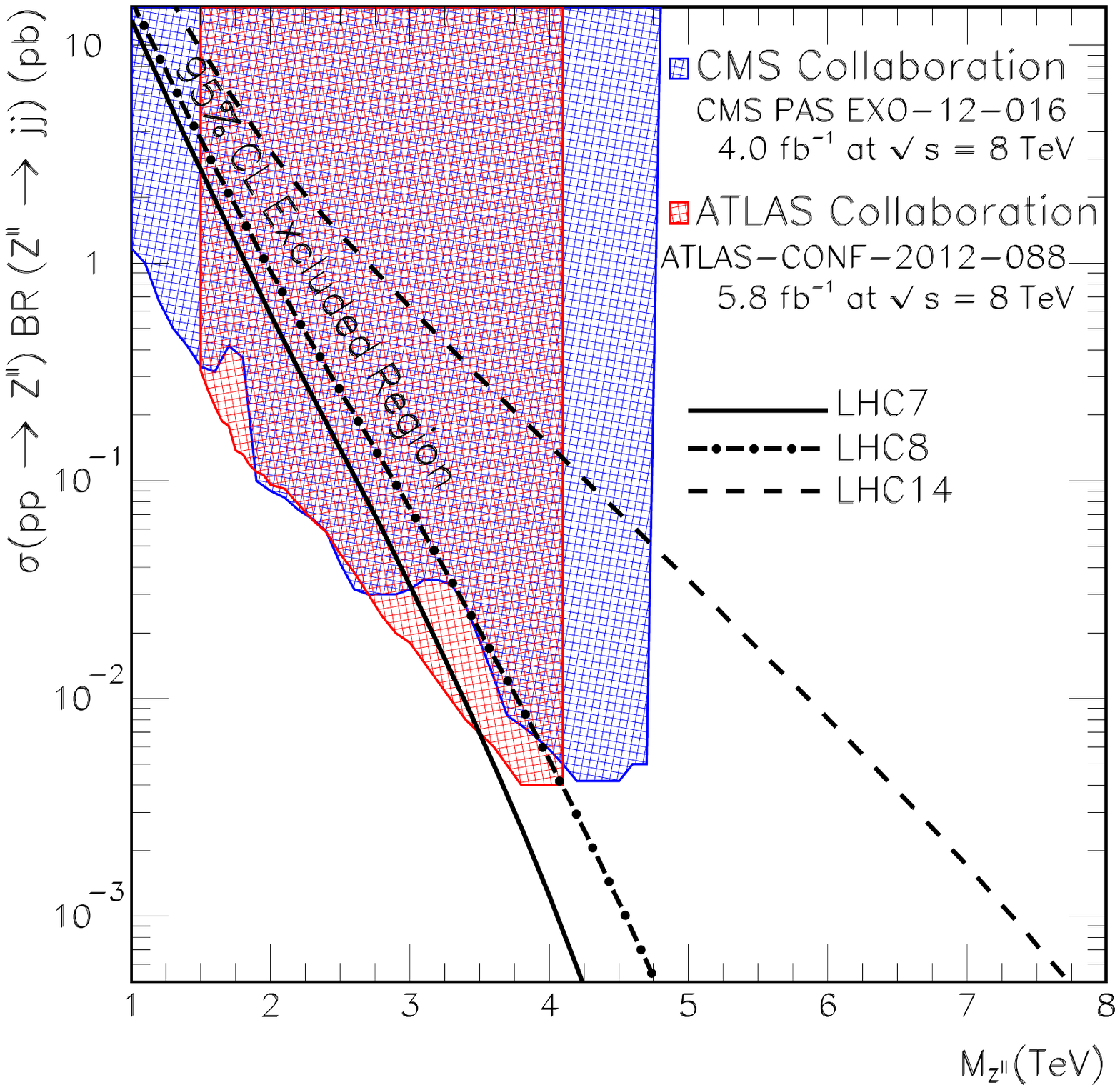}{0.99} \end{minipage}
\hfill 
\begin{minipage}[t]{0.49\textwidth}
\vskip -1.5 cm
\postscript{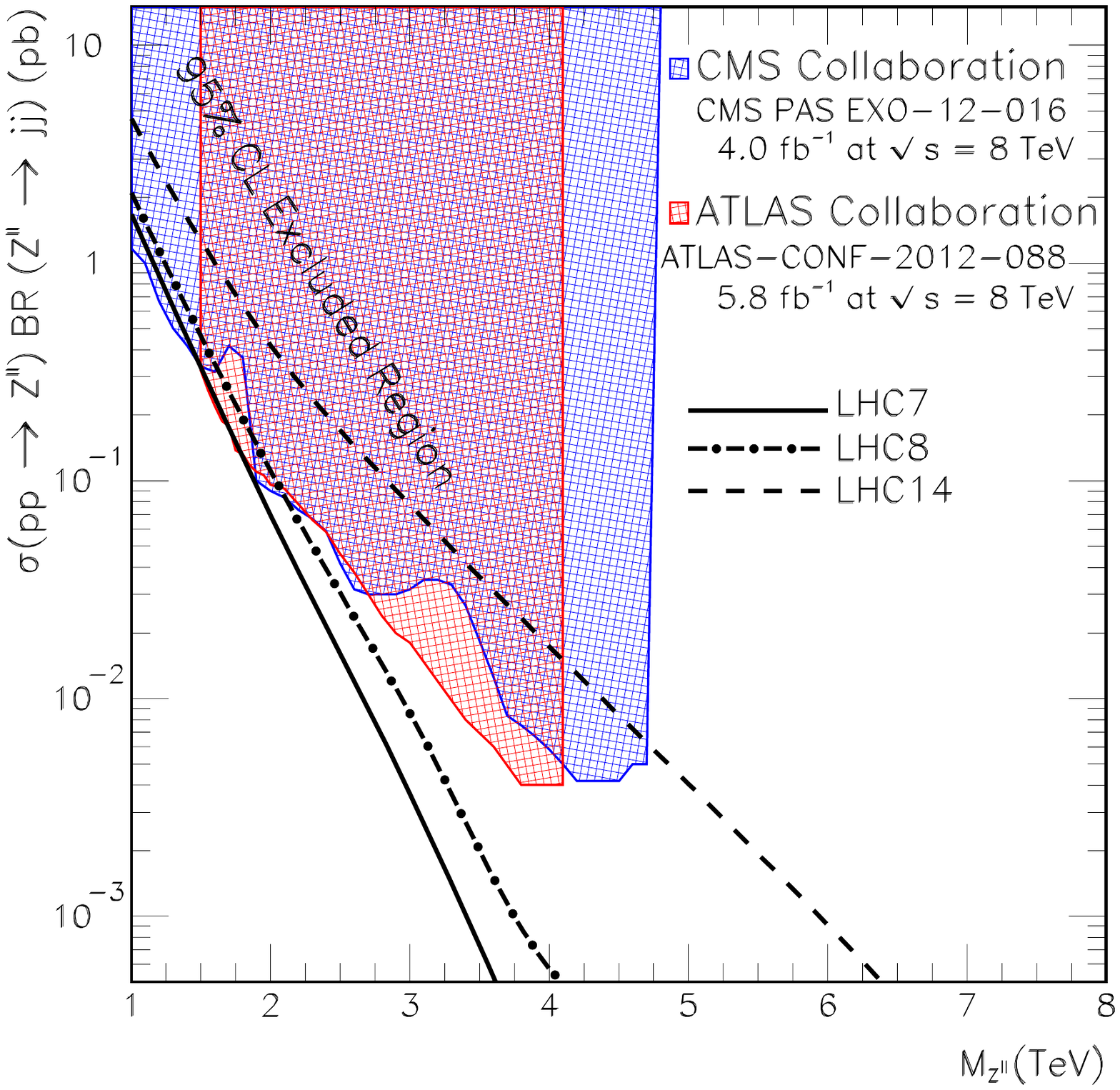}{0.99} \end{minipage} 
\vskip -1.75cm
\caption[Dijet Invariant Mass Spectrum From CMS and ATLAS]{ \sglspc \small Comparison of the (pre-selection cut) total cross section for the production of $p p \to Z'' \to jj$ with the 95\% CL upper limits on the production of a gauge boson decaying into two jets as reported by the CMS and ATLAS Collaborations (corrected by acceptance).  The case in which $Z''$ is mostly diagonal in $I_R$ is shown in the left panel and the case in which it is mostly $B-L$ in the right panel.}
\label{fig:CMS} 
\end{figure}

The invariant mass spectra fit the SM expectations and thus lower mass limits can be inferred from the cross section times branching ratio for $Z''$ into two jets. Similar lower mass limits have been obtained by the ATLAS Collaboration using $5.8~{\rm fb}^{-1}$ of data collected at $\sqrt{s} = 8~{\rm TeV}$~\cite{atlas_dijet8tev}. These results, which are displayed in Fig.~\ref{fig:CMS}, extend previous exclusion limits from runs at $\sqrt{s} = 7 \ {\rm TeV}$ done in LHC7~\cite{Khachatryan:2010jd,Chatrchyan:2011ns,Aad:2011fq,cms_dijet7tev,atlas_dijet7tev}.

\begin{figure}[h]
\begin{minipage}[t]{0.49\textwidth}
\vskip -1.5 cm
\postscript{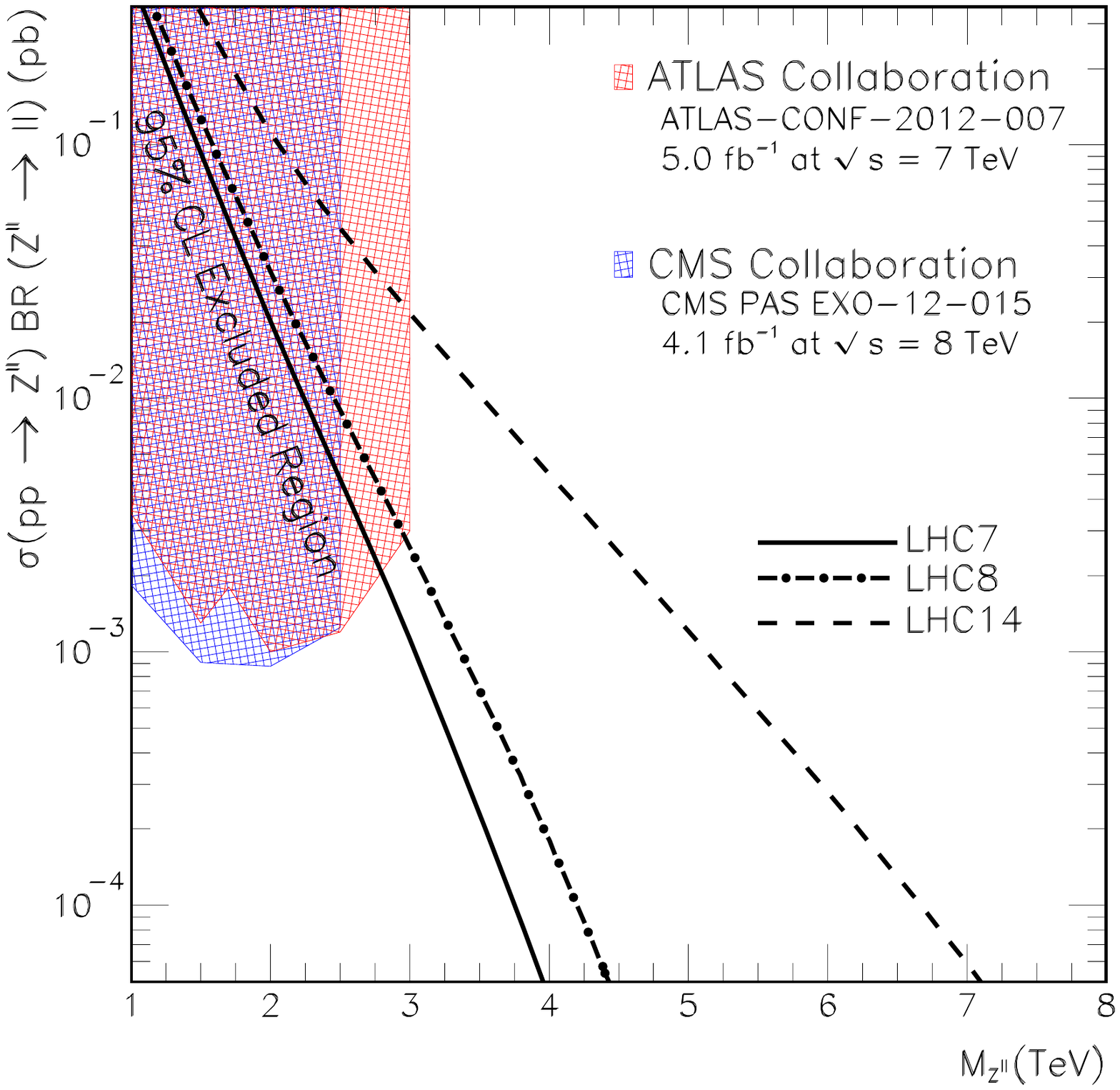}{0.99} \end{minipage}
\hfill 
\begin{minipage}[t]{0.49\textwidth}
\vskip -1.5 cm
\postscript{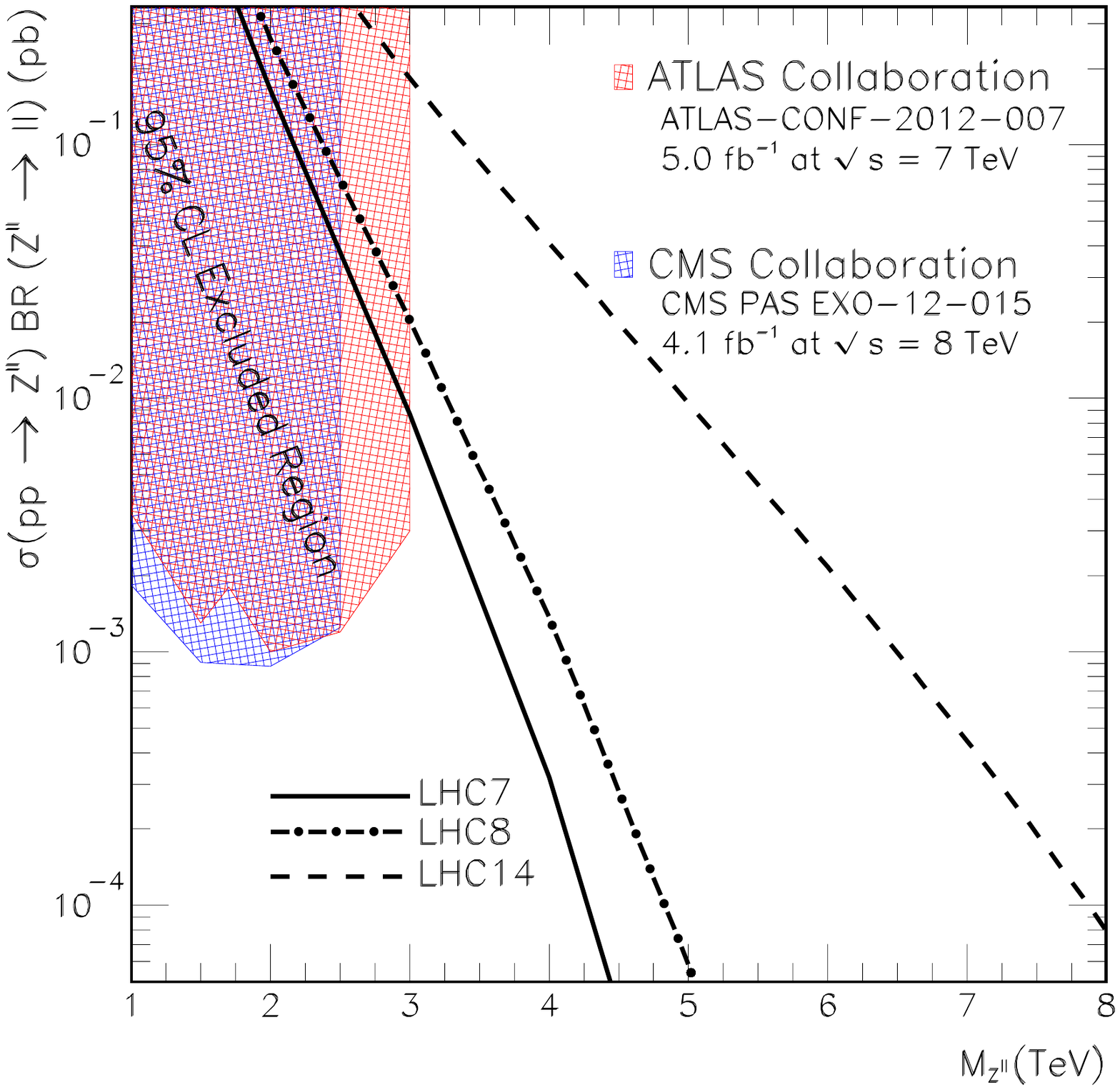}{0.99} \end{minipage}  
\vskip -1.75cm
\caption[Dilepton Invariant Mass Spectrum From CMS and ATLAS]{ \sglspc \small Comparison of the (pre-selection cut) total cross section for the production of $p p \to Z'' \to \ell \ell$ with the 95\% CL upper limits on the production of a gauge boson decaying into two leptons, as reported by the ATLAS and CMS Collaborations. The case in which $Z''$ is mostly diagonal in $I_R$ is shown in the left panel and the case in which it is mostly $B-L$ in the right panel.}
\label{fig:ATLAS} 
\end{figure}

The ATLAS Collaboration has also searched for narrow resonances in the
invariant mass spectrum of dimuon and dielectron final states at $\sqrt{s} = 7~{\rm TeV}$ with an integrated luminosity of $4.9~{\rm fb}^{-1}$ and $5.0~{\rm fb}^{-1}$, respectively~\cite{atlas_dilepton7tev}. The spectra fit
with SM expectations and thus upper limits on the cross section times branching ratio for $Z''$ into lepton pairs can be set. More recently, the CMS Collaboration updated the LHC7 results using $4.1~{\rm fb}^{-1}$ of data collected at $\sqrt{s} = 8~{\rm TeV}$~\cite{cms_dilepton8tev}. The combined upper limits from LHC7 and LHC8 are shown in Fig.~\ref{fig:ATLAS}.  Previous dilepton searches by the LHC experiments have been reported in~\cite{Collaboration:2011dca,Chatrchyan:2012it}.

To set upper limits on the SM$^{++}$ model we need to compute the dijet  and dilepton cross sections along with the relevant branching ratios.  The Lagrangian term for $\bar{f} \gamma^\mu f Z''_\mu$ coupling can be expressed in the traditional form of electro-weak interactions as
\begin{eqnarray}
\mathscr {L} &=& \sum_f \bigg((g_{Y''}Q_{Y''})_{f_L^i} \, \bar f_L^i \gamma^\mu
f_L^i +  (g_{Y''}Q_{Y''})_{f_R^i} \, \bar f_R^i \gamma^\mu f_R^i \bigg) \, Z''_\mu, \nonumber
\\
& = & \frac{1}{2}   \sqrt{g_Y^2 + g_2^2} \ \sum_f
\bigg(\epsilon_{f^i_L} \, \bar f_L^i \gamma^\mu f_L^i +
\epsilon_{f_R^i} \, \bar f_R^i \gamma^\mu f_R^i \bigg) \, Z''_\mu, \nonumber
\\
& = &  2^{1/4} \sqrt{G_F} M_{Z^0} \ \sum_f
\bigg(\epsilon_{f^i_L} \, \bar f_L^i \gamma^\mu f_L^i +
\epsilon_{f_R^i} \, \bar f_R^i \gamma^\mu f_R^i \bigg) \, Z''_\mu,
\label{lagrangian}
\end{eqnarray}
where $f_{L \, (R)}^i$ are fermion chiral fields and $\epsilon_{f_L^i,f_R^i} = v_q \pm a_q$, with $v_q$ and $a_q$ the vector and axial couplings, respectively, and $G_F$ is the Fermi coupling constant taken at the $4$ TeV scale via
\beq
G_F = \frac{1}{4\sqrt{2}} \frac{g_2^2}{M_{Z^0}^2 \cos^2 \theta_W} =  \frac{1}{4\sqrt{2}} \frac{g_Y^2+g_2^2}{M_{Z^0}^2}. 
\eeq
In order to compare LHC experimental searches in dilepton and dijet events we need to consider the production cross section in the narrow $Z''$ width approximation of the Breit-Wigner distribution,
\beq
\sigma (s) \propto \frac{\Gamma^2}{(\sqrt{s}-M_{Z''})^2 + \Gamma^2/4} \rightarrow 2 \pi \ \Gamma \delta(\sqrt{s} -M_{Z''}), 
\eeq
in the limit that $\Gamma \ll M_{Z''}$.  The cross-section for two quarks to $Z''$ is given by 
\begin{equation}
\hat \sigma (q \bar q \to Z'')  =   2 \pi \frac{K}{3} \, \frac{G_F \, M_Z^2}{\sqrt{2}}  \left[v_q^2 (\phi, g'_1)+ a_q^2 (\phi, g'_1) \right] \, \delta \left(\hat s - M_{Z''}^2 \right) \,,
\end{equation}
where the $K$-factor represents enhancements from higher order QCD processes estimated to be $K \simeq 1.3$~\cite{Barger}. We include hats $(\hat s, \ \hat \sigma)$ to indicate that these are the values of the partons. To understand what the detector observes we must integrate this result against the possible internal momentum configurations of the proton known as parton distribution functions (PDF).  These cannot be solved by perturbation theory as they are described by QCD in the non-perturbative regime.  After folding (integration over) $\hat \sigma$ with the CTEQ6 parton distribution functions~\cite{Pumplin:2002vw}, we can determine the resonant production cross section for $\sigma(pp \rightarrow Z'')$. In Figs.~\ref{fig:CMS} and \ref{fig:ATLAS} we compare the predicted $\sigma (p p \to Z'') \times \mathcal{B} (Z'' \to jj)$ and $\sigma (p p \to Z'') \times \mathcal{B} (Z'' \to \ell \ell)$ production rates with 95\% CL upper limits as reported by the CMS and ATLAS Collaborations. We conclude that if $Z''$ is mostly $I_R$, then the predicted production rates for $M_{Z''} \approx 4~{\rm TeV}$ at $\sqrt{s} = 8~{\rm TeV}$ lie at the current dijet limits.  On the other hand, if $Z''$ is mostly $B-L$ then the lower limit on the boson mass, $M_{Z''} \geq 3~{\rm TeV}$, is determined primarily from dilepton searches.
\newpage
%-----FIG-----------------
%-------------FIG------------------
%---------------------------------------
\newpage
\begin{table}
\begin{center}
\caption[Chiral Couplings of $U(1)$ Bosons With $Z'$ Mostly $B$ ]{Chiral couplings of $Y$, $Z'$, and $Z''$ gauge bosons. All fields in a given set have common $g_{Y'} Q_{Y'},\ g_{Y''} Q_{Y''}$ couplings. We have taken $Z'$ to be mostly $B$ and and $Z''$ to be mostly $I_R$. }
\ \\
\begin{tabular}{cccc}
\hline
\hline
~~~~~Fields~~~~~ & ~~~~~$g_Y Q_Y$~~~~~  & ~~~~~$g_{Y'} Q_{Y' }$~~~~~  &  ~~~~~$g_{Y''} Q_{Y''}$~~~~~  \\ \hline
$u_R$ &  $\phantom{-}0.2434$ & $\phantom{-}0.1836$ & $\phantom{-}0.3321$ \\
$d_R$ &  $-0.1214$ & $\phantom{-}0.1838$ & $-0.3933$ \\ 
$E_L$ &  $-0.1826$ & $\phantom{-}0.0759$ & $\phantom{-}0.0918$ \\ 
$e_R$ &   $-0.3650$ & $\phantom{-}0.0760$ & $-0.2709$ \\ 
$Q_L$ &  $\phantom{-}0.0610$ & $\phantom{-}0.1837$ & $-0.0306$ \\
$n_R$ &  $\phantom{-}0.0000$ & $\phantom{-}0.0758$ & $\phantom{-}0.4545$ \\
$H\phantom{''}$ &  $\phantom{-}0.1824$ & $\phantom{-}0.0000$ & $\phantom{-}0.3627$ \\
$H''$ &  $\phantom{-}0.0000$ & $-0.0758$ & $-0.4545$ \\
\hline
\hline
\label{tab:2}
\end{tabular}
\end{center}
\end{table}
\ \\
\begin{table}
\vskip -3.2cm
\caption[Chiral Couplings of $U(1)$ Bosons With $Z'$ Mostly $L$]{ Chiral couplings of $Y$, $Z'$, and $Z''$ gauge bosons. All fields in a given set have common $g_{Y'} Q_{Y'},\ , g_{Y''} Q_{Y''}$ couplings.  We have taken $Z'$ to be mostly $L$ and and $Z''$ to be mostly $B-L$.}
\begin{center}
\begin{tabular}{cccc}
\hline
\hline
~~~~~Fields~~~~~ & ~~~~~$g_Y Q_Y$~~~~~  & ~~~~~$g_{Y'} Q_{Y' }$~~~~~  &  ~~~~~$g_{Y''} Q_{Y''}$~~~~~  \\ \hline
$u_R$ & $\phantom{-} 0.2435$ & $\phantom{-}0.1101$ & $-0.0763$ \\
$d_R$ &  $-0.1217$ & $\phantom{-}0.1101$ & $-0.2242$ \\ 
$E_L$ &  $-0.1825$ & $\phantom{-}0.7165$ & $\phantom{-}0.4509$ \\ 
$e_R$ & $-0.3651$ & $\phantom{-}0.7165$ & $\phantom{-}0.3769$ \\ 
$Q_L$ &  $\phantom{-}0.0609$ & $\phantom{-}0.1101$ & $-0.1503$ \\
$n_R$ &  $\phantom{-}0.0000$ & $\phantom{-}0.7165$ & $\phantom{-}0.5248$ \\
$H$ &  $\phantom{-} 0.1826$ & $-0.0000$ & $\phantom{-}0.0739$ \\
$H''$ &  $-0.0000$ & $-0.7165$ & $-0.5248$ \\
\hline
\hline
\end{tabular} 
\end{center}
\label{newtable}
\end{table}
\newpage

\
\subsection{Future Detection Possibilities}

For the discovery potential in the high mass region, the dijet channel is statistically a better discriminator than lepton pairs. Therefore, we investigate at the parton level the LHC14 sensitivity for a $Z''$ resonance (which we take as mostly $I_R$) in the dijet  invariant mass $M$. After setting selection cuts on the different jet rapidities, $|y_1|, \,|y_2| \le 1$ and transverse momenta $p_{\rm T}^{1,2}>50$~GeV, we calculate the differential cross section as
\beq
\frac{d \hat \sigma_{ij \to kl} }{d M^2} = \frac{| \CM |_{ij \to kl}^2}{2 \hat s} \frac{d^3 k}{(2\pi)^3 2 k^0}\frac{d^3 l}{(2\pi)^3 2 l^0} (2 \pi)^4 \delta^{(4)}(i+j-k-l) \delta(\hat{s} - M^2).
\label{eq:P1partonScat}
\eeq
Following~\cite{Anchordoqui:2008ac} we use the definitions $l^0 = l_\perp \cosh y_1$, $k^0 = k_\perp \cosh y_2$, and $\tau = \hat{s}/s$, where $k_\perp,\ l_\perp$ are the transverse momentum of the partons in the reaction, and we make a change of variable into rapidities via the relation
\beq
\frac{d^3 k}{2 k^0} = \frac{1}{2} d^2 k_\perp d y_2 = \pi k_\perp d k_\perp d y_2.
\eeq
Furthermore we use $y \equiv \frac{1}{2} (y_1-y_2)$ and a common transverse momentum $p_\perp$ defined via
$p^2 = \hat s = (k + l)^2 \approx 2 k\cdot l  = 4 p_\perp^2 \cosh 2 y $.  This allows the expansion of the delta function $\delta(\hat s - M^2)$ into
\beqa
\delta(\hat s - M^2) = \delta(4 p_\perp^2 \cosh^2 y - M^2) = \frac{1}{4 \cosh^2 y} \delta \left(p_\perp^2 - \frac{M^2}{4 \cosh^2 y} \right).
\eeqa
After changing the variable yet again to $k_\perp dk_\perp l_\perp dl_\perp \rightarrow k_\perp dk_\perp p_\perp dp_\perp$, we integrate out dependence on the transverse momentum extracting from the over energy conservation delta function the perpendicular part
\beq
\int d^2 k_\perp d^2 p_\perp \delta^{(2)}(k_\perp + p_\perp) \delta(p_\perp^2 - M^2/4 \cosh^2 y) = \pi. 
\eeq
The parallel components of the momentum of the partons can be converted into limits on the rapidities.
We must now fold the result with the PDFs for $pp$ collisions.  In terms of these variables, equation  (\ref{eq:P1partonScat}) folded with PDFs becomes
\beq
\frac{d \sigma(pp \to jj)}{d M} = \frac{1}{2} M \tau \int dy_1 dy_2 \frac{1}{\cosh^2 y} \sum_{ijkl} f_i(\sqrt{\tau} e^Y,M) f_j(\sqrt{\tau} e^{-Y},M) \frac{d \hat{\sigma}_{ij \to kl} }{d \hat t},
\eeq    
where we used 
\beq
|\CM |_{ij \to kl}^2 = 16 \pi \hat{s}^2 \ \frac{ d \hat{\sigma}_{ij \to lk} }{d\hat t}.
\eeq
We can express the cross section per interval of $M$ as
\begin{eqnarray}
\frac{d\sigma}{dM} & = & M\tau\ \sum_{ijkl}\left[
\int_{-Y_{\rm max}}^{0} dY \ f_i (x_a,\, M)  \right. \ f_j (x_b, \,M ) \
\int_{-(y_{\rm max} + Y)}^{y_{\rm max} + Y} dy \frac{d\hat{\sigma}_{ij\to lk}}{d\hat t} \ \frac{1}{\cosh^2
y} \nonumber \\
& + &\int_{0}^{Y_{\rm max}} dY \ f_i (x_a, \, M) \
f_j (x_b, M) \ \int_{-(y_{\rm max} - Y)}^{y_{\rm max} - Y} dy
\left.  \frac{d\hat{\sigma}_{ij \to kl}}{d\hat t} \
\frac{1}{\cosh^2 y} \right] \,,
\label{longBH}
\end{eqnarray}
where $Y\equiv \frac{1}{2} (y_1 + y_2)$ . We use for $f(x,M)$ the parton distribution functions of CTEQ6~\cite{Pumplin:2002vw}; we also have $\tau = M^2/s$, $x_a = \sqrt{\tau} e^{Y}$, $x_b = \sqrt{\tau} e^{-Y}$. The $Y$ integration range in Eq.~(\ref{longBH}) is $Y_{\rm max} = {\rm min} \{ \ln(1/\sqrt{\tau}),\ \ y_{\rm max}\}$, which comes from requiring the fraction of the total momentum of the parton to be less than one, $x_a, \, x_b < 1$,  and the rapidity cuts $y_{\rm min} < |y_1|, \, |y_2| < y_{\rm max}$. The kinematics of the scattering also provides the relation $M = 2p_T \cosh y$, which, when combined with $p_T = M/2 \ \sin \theta_s = M/2 \sqrt{1-\cos^2 \theta_s},$ yields $\cosh y = (1 - \cos^2 \theta_s)^{-1/2},$ where $\theta_s$ is the center-of-mass scattering angle.  Additionally, the Mandelstam invariants are given by $\hat s = M^2,$ $\hat t = - (M^2/2)\ e^{-y}/ \cosh y,$ and $\hat u = -(M^2/2) \ e^{+y}/ \cosh y.$

The spin/color averaged square amplitude (for incoming quark/anti-quark pair $q \bar{q}$ and outgoing quark/anti-quark pair $q' \bar{q}'$) is given by 
\beqa 
|{\cal M} (q\bar q \stackrel{Z''} {\to} q'\bar q {}')|^2  &= &\frac 1 4 \lsb g_{Y''}^2 Q_{Y''}^2(q_L)  + g_{Y''}^2 Q_{Y''}^2(q_R) \rsb \lsb g_{Y''}^2 Q_{Y''}^2(q_L{}')  +  g_{Y''}^2 Q_{Y''}^2(q_R{}') \rsb \nn
& \times & \left [\frac{2(  u^2+   t^2)}{( s-M_{Z''}^2)^2 + (\Gamma_{Z''}\ M_{Z''})^2} \right],
\label{Zdprimecross}
\eeqa 
where $g_{Y''}Q_{Y''}(q_L)$ and $g_{Y''} Q_{Y''}(q_R)$ are the couplings of $Z''$ to quarks (note that we have not summed over the flavors).

The decay width of $Z'' \to f\bar f$ is given by
\begin{equation}
\Gamma( Z'' \to f \bar f) = \frac{G_F M_Z^2}{6 \pi \sqrt{2}}  N_c
 \, M_{Z''} \sqrt{1 -4x} \left[v_f^2 (1+2x) + a_f^2 (1-4x)
\right] \, \left(1 + \frac{\alpha_s}{\pi}\right) \, ,
\end{equation}
where  $\alpha_s = \alpha_s(M_{Z''})$ is the strong coupling constant at the scale $M_{Z''}$, $x = m_f^2/M_{Z''}^2$; $v_f$, $a_f$ are the vector and axial couplings, and $N_c =3$ or 1 if $f$ is a quark or a lepton, respectively~\cite{Feldman:2006wb}. For our values of $g'_1$ where $Z''$ is mostly $I_R$, we obtain $v_u^2 + a_u^2 = 0.396$ and $v_d^2 + a_d^2 = 0.554$.

In Table~\ref{table-v} we calculate prospective signal-to-noise ratios for different possible $M_{Z''}$ masses and integrated luminosity values for LHC14 data.  The signal rate $\CS$  is estimated in the invariant mass window $[M_{Z''} - 2 \Gamma, \, M_{Z''} + 2 \Gamma]$ bin size, and is given as
\beq
\CS =  \sigma_{Z''}  \int \CL_I(t') dt', 
\eeq
where $\CS$ is the expected number of events to be observed in the invariant mass window based on the SM$^{++}$ cross section.  To accommodate the minimal acceptance cuts on dijets from the CMS and ATLAS proposals~\cite{Bhatti:2008hz}, an additional kinematic cut, $|y_{\rm max}|<1.0$, is included in the calculation. The noise (${\cal N}$) is the square root of the expected number of QCD background events ($N_{\rm bg}$) in the same dijet  mass interval for the same integrated luminosity, coming from SM processes. This gives the expected signal-to-noise ratio as
\beq
\frac{\CS}{\CN} = \frac{ \sigma_{Z''}}{\sqrt{\sigma_{\rm bg}}}  \lpa \int \CL_I(t') dt' \rpa^{1/2},
\eeq
where $\sigma_{\rm bg}$ is the total background cross section into dijets in the invariant mass window being probed from SM processes. We conclude that the LHC provides discovery potential for $Z''$ which is mostly $I_R$ for $M_{Z''} \leq 5$ TeV. The discovery potential of a $Z''$ that is mostly $B-L$ is controlled by the sensitivity of LHC14  to dilepton searches. For $300~{\rm fb}^{-1}$, the projected LHC sensitivity is again for masses $M_{Z''} \leq 5~{\rm TeV}$~\cite{Chiang:2011kq}.
\sglspc
\begin{table}[htb]
\begin{center}
\caption[Signal-To-Noise Ratio for LHC14 Luminosities]{\centering Signal-to-Noise Ratio at LHC14 for Different Integrated Luminosities.}
\label{table-v}
\begin{tabular}{c|@{}ccc|@{}ccc|@{}ccc}
  \hline
  \hline
  & \multicolumn{3}{@{}c|}{10~fb$^{-1}$} & \multicolumn{3}{@{}c|}{100~fb$^{-1}$} & \multicolumn{3}{@{}c}{1000~fb$^{-1}$} \\
  \cline{2-4} \cline{5-7} \cline{8-10}
  ~$M_{Z''}~({\rm TeV})$~ & ~~${\cal S}$~~ & ~~${\cal B}$~~ & ~~${\cal S}/{\cal N}$~~ & ~~${\cal S}$~~ & ~~${\cal B}$~~ & ~~${\cal S}/{\cal N}$~~ & ~~${\cal S}$~~ & ~~${\cal B}$~~ & ~~${\cal S}/{\cal N}$~~ \\
  \hline
  4 & ~39 & ~579	& 1.62	  & 391   & 5789   & ~5.14   & ~3910  & ~57895     & 16.25 \\
  5 & ~~7& ~176	& 0.50	  & ~67   & 1759   & ~1.60   & ~~670  & ~17590      & ~5.05 \\
  6 & ~~1 & ~~66	& 0.14	  & ~11   & ~664   & ~0.44   & ~~113 & ~~6646      & ~1.39 \\
  \hline
  \hline
\end{tabular}
\end{center}
\end{table}
\dblspc
	
\section{Stability of Extended Higgs Mechanism}	

\quad Now we turn to the stability of the SM$^{++}$ model.  For the symmetry-breaking Higgs mechanism to work, we require that the potential of the SM$^{++}$, 
\beq
V^{++} \left(H, H'' \right) = \mu_1^2 \left| H \right|^2 +{ \mu_2}^2 \left| H''
\right|^2 + \lambda_1 \left| H \right|^4 + \lambda_2 \left| H''
\right|^4 + \lambda_3 \left| H \right|^2 \left| H'' \right|^2 \ ,
\label{higgsV}
\eeq 
has a non-zero real minimum for all values of the interaction energies.  The values of the parameters in quantum field theories change depending on the energies that the fields interact with.  Luckily, renormalizable quantum field theories allow one to compute the values of the parameters at any interaction energy (or at least tell you when the quantum field theory is no longer an accurate description of the system), or else the theory couldn't make any predictions. We say that the parameters of the theory \emph{run} with the interaction energy; the computation of the parameters is addressed in the renormalization group equations.  

\subsection{The Renormalization Group}

\begin{figure}[ht]
\begin{center}
\postscript{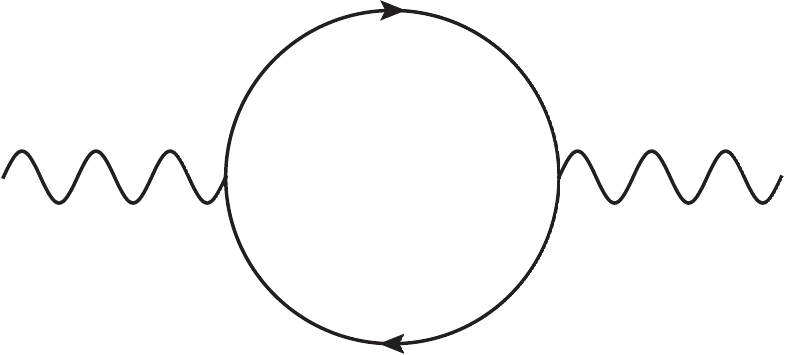}{0.4}
\caption[One Loop QED Photon Propagator Diagram]{ \sglspc \small Higher order terms involving photons include loops of virtual pair production of fermions particles circulating in the loop.  This leads to a correction of the bare charge parameter of the Lagrangian for QED.}
\label{fig:OLQEDscat}
\end{center}
\end{figure}

The renormalziation group (RG), in my opinion, can best be explained by example. Here we will use quantum electro-dynamics (QED) as our toy model.  In quantum electrodynamics we have the $U(1)$ gauge invariant Lagrangian
\beq
\CL = \bar \psi \lpa i D \cdot \gamma - m \rpa \psi - \frac{1}{4} F^2  = \bar \psi \lpa i \gamma \cdot \p  + e A \cdot \gamma - m\rpa \psi - \frac{1}{4} F^2 \ .  
\label{eq:QEDLag}
\eeq    
From (\ref{eq:QEDLag}) we can compute scattering cross sections of fermions undergoing interactions with the photon.  At lowest order (known as tree level) we assume that the parameters in the Lagrangian are those we measure at low energies, such as for the fermion being an electron, where $m = 0.511 \ \rm MeV$ is the mass of the electron, $e = 1.6 \times 10^{-19} \ \rm C$ is the elementary charge for the electron, $\psi, A$ are the fields for the electron and photon, respectively.  However, if we want to compute observables to a higher order, then we require loops in our Feynman diagrams.  One such example for QED is the correction to the photon propagator shown in Fig.~\ref{fig:OLQEDscat}.
These loops present a problem as they are formally infinite.  As stated in the introduction, we can understand the infinities, by assuming the infinities arise because we extend the theory beyond its region of validity to energy scales larger than when the theory is correct.  We can prevent extending the theory beyond its reach, by introducing a cut-off scale denoted $\Lambda$.  An example of this can be done by computing the value of the diagram in Fig. \ref{fig:OLQEDscat}, which gives
\beq
-e^2 \int \frac{d^4 l}{(2\pi)^4} {\rm Tr}\left[ \frac{\gamma^\mu((k+l)\cdot \gamma + m) \gamma^\nu (l \cdot \gamma + m)}{((k+l)^2-m^2)(l^2-m^2)}\right] \ .
\label{eq:PhotonOL}
\eeq
We add the cut-off such that $|l|\leq \Lambda$.  However, the value of $\Lambda$ cannot show up in experiment and the result of the any calculation must be independent of the choice of $\Lambda$.  This can only be possible if the parameters appearing the Lagrangian are themselves functions of $\Lambda$.  The Eq. (\ref{eq:QEDLag}) can be put into the form
\beq
\CL = \bar \psi(\Lambda) \lpa i \gamma \cdot \p  + e(\Lambda) A(\Lambda) \cdot \gamma - m(\Lambda) \rpa \psi(\Lambda) - \frac{1}{4} F(\Lambda)^2 \ .  
\label{eq:QEDLagEff}
\eeq  
Equation (\ref{eq:QEDLagEff}) is said to be the effective theory for energy scales $< \Lambda$.  It is important to note here that in practice, we will not literally take a cut-off in the computation of the integral (\ref{eq:PhotonOL}) as this creates issues for gauge invariance. We will, however, use a regularization scheme known as dimensional regularization that re-parameterized the cut-off scale $\Lambda$ into a variable $\epsilon$, which is related to the dimension in which we will calculate (\ref{eq:PhotonOL}). It is still useful to discuss the RG in terms of a cut-off as it is physically more meaningful. With the understanding of  (\ref{eq:QEDLagEff}) as an effective theory, we want to believe there is an overall field theory that describes QED from an energy scale $\Lambda = \infty$ and downward.  To this end we assume there is a QED field theory for $\Lambda = \infty$ that takes the same form as the effective field theory and no extra terms (this defines a renormalizable theory as the number of divergences at the effective theory level can be absorbed into the already existing number of parameters or less; at that level, this ensures that the number of terms of the Lagrangian at the $\Lambda = \infty$ theory does not proliferate):
\beq
\CL = \bar \psi_0 \lpa i \gamma \cdot \p  + e_0 A_0 \cdot \gamma - m_0 \rpa \psi_0 - \frac{1}{4} F_0^2 \ .  
\label{eq:QEDLagBare}
\eeq 
The values of the fields and parameters are now written as their \emph{bare} values, the values valid at the highest energy scale that do not depend on $\Lambda$.  We can find the values at the scale $\Lambda \neq \infty$ from the bare values by expressing the fields and parameters as 
\beqa
\psi_0 &=& Z^{1/2}_\psi(\Lambda) \psi(\Lambda) \ , \nonumber
\\
A^\mu_0 &=& Z^{1/2}_A (\Lambda) A^\mu(\Lambda) \ , \nonumber
\\
e_0 &=& Z_e(\Lambda) Z_A^{-1/2}(\Lambda) Z^{-1}_\psi (\Lambda) \Lambda^\epsilon e(\Lambda) \ , \nonumber
\\
m_0 &=&  Z_m(\Lambda) Z^{-1}_\psi (\Lambda) m(\Lambda) \ .
\label{eq:BareVals} 
\eeqa
The $\Lambda^\epsilon$ factor will be explained during explicit calculation of the running of the parameters. It is related to the dimension in which we calculate the divergent integrals, such that as $\Lambda \rightarrow \infty, \ \epsilon \rightarrow 0$.  With this we can write the true or bare Lagrangian as
\beq
\CL = i Z_\psi \bar \psi \gamma \cdot \p \psi  + Z_e \Lambda^\epsilon e A \cdot (\bar \psi \gamma \psi) - Z_m m \bar \psi \psi - \frac{Z_A}{4} F^2 \ ,  
\label{eq:QEDLagCT}
\eeq  
where the explicit dependence on $\Lambda$ is no longer displayed on $Z_i, \ e,$ or $m$.  It is easy to see, but nontrivial to prove that for gauge symmetry to be valid at all scales, that $Z_\psi = Z_e$. This is known as the Ward Identity and the proof can be found in many quantum field theory texts; we will simply assume gauge invariance is true for all scales.  One other interesting note is that if there was no interaction term $ e A \cdot \bar \psi \gamma \psi$, we would have two separate and free field theories for which $m(\Lambda) = m_0$ where the bare mass would take the physical mass of the particle. Therefore since we are calculating using perturbation theory, we can assume that the divergences of the integrals such as (\ref{eq:PhotonOL}) can be expressed as a series of the coupling
\beq
Z_i = 1 + \sum_{n=1}^\infty \tilde{a}_n(e) \Lambda^n  = 1 + \sum_{n=1}^\infty \frac{a_n(e)}{\epsilon^n} \ .
\eeq 
We will find it advantageous to take the logarithms of the bare values; for example, the bare charge can be written as
\beqa
\log e_0 &=& \log( Z^{-1/2}_A \Lambda^\epsilon e ) = \log(Z^{-1/2}_A) + \epsilon \log(\Lambda) + \log(e) \nonumber
\\
&=&\sum_{n=1}^\infty \frac{G_n(e)}{\epsilon^n}  + \epsilon \log(\Lambda) +\log(e),
\label{eq:RG2}
\eeqa   
with
\beq
\sum_{n=1}^\infty \frac{G_n(e)}{\epsilon^n} = \log(Z_A^{-1/2}).
\eeq
We see in Eq. (\ref{eq:BareVals}) that the bare values do not depend on $\Lambda$, we take advantage of this fact by taking derivatives with respect to $\log (\Lambda)$.  It is advantageous to take the derivatives of the logarithms of the bare values, as this separates the dependencies of the $Z_i$, the result of which is
\beqa
0 = \frac{d \log e_0}{d \log(\Lambda)} &=& \epsilon + \frac{1}{e}\frac{d e}{d \log(\Lambda)} + \sum_{n=1}^\infty \lpa \frac{\p G_n}{\p e} \frac{d e}{d \log(\Lambda)} \frac{1}{\epsilon^n} \rpa \nonumber
\\
0 &=&  \lpa 1+ e \sum_{n=1}^\infty \frac{\p G_n}{\p e}\frac{1}{\epsilon^n} \rpa \frac{d e}{d \log(\Lambda)}+\epsilon e \ . \nonumber
\\   
\label{eq:RG1}
\eeqa 
We now assume that if the theory is renormalizable, then $d e / d \log(\Lambda)$ must be finite in the limit that $\epsilon \rightarrow 0$ and we express this as
\beq
\frac{d e}{d \log(\Lambda)} = -\epsilon e + \beta(e) \ .
\label{eq:BetaRG}
\eeq
Using (\ref{eq:BetaRG}) allows us to write Eq. (\ref{eq:RG1}) order by order in $\epsilon$ and enforce that the coefficient of $\epsilon^{-n}$ is zero so that Eq. (\ref{eq:RG1}) remains true when $\epsilon \rightarrow 0$.  This procedure gives the conditions 
\beqa
\beta(e) &=& e^2 \frac{\p G_1}{\p e} \nonumber
\\
\beta \frac{\p G_n}{\p e} &=& e G_{n+1} \ . 
\eeqa 
From this we can finally express 
\beq
\frac{d e}{d \log(\Lambda)} = \beta(e) = e^2 \frac{\p G_1}{\p e}.
\eeq
Similar expressions can be found for the other parameters of the theory as well.  In the end, to know the values of the parameters at different scales $\Lambda$, one must calculate the beta function from the $Z_i$ functions.  As a note, when it was said that $\Lambda \propto \frac{1}{\epsilon}$, it was in effect choosing a regularization method of the divergent integrals known as dimensional regularization, which is the basis of the $\overline{\rm MS}$ (MS-bar) scheme~\cite{Srednicki:2007qs}.  Other schemes exist but all give the same experimental values, as they must. Now we will see the RG equations in action.

\subsection{Example Renormalization Group Use} 

\begin{figure}[ht]
\begin{flushleft}
\postscript{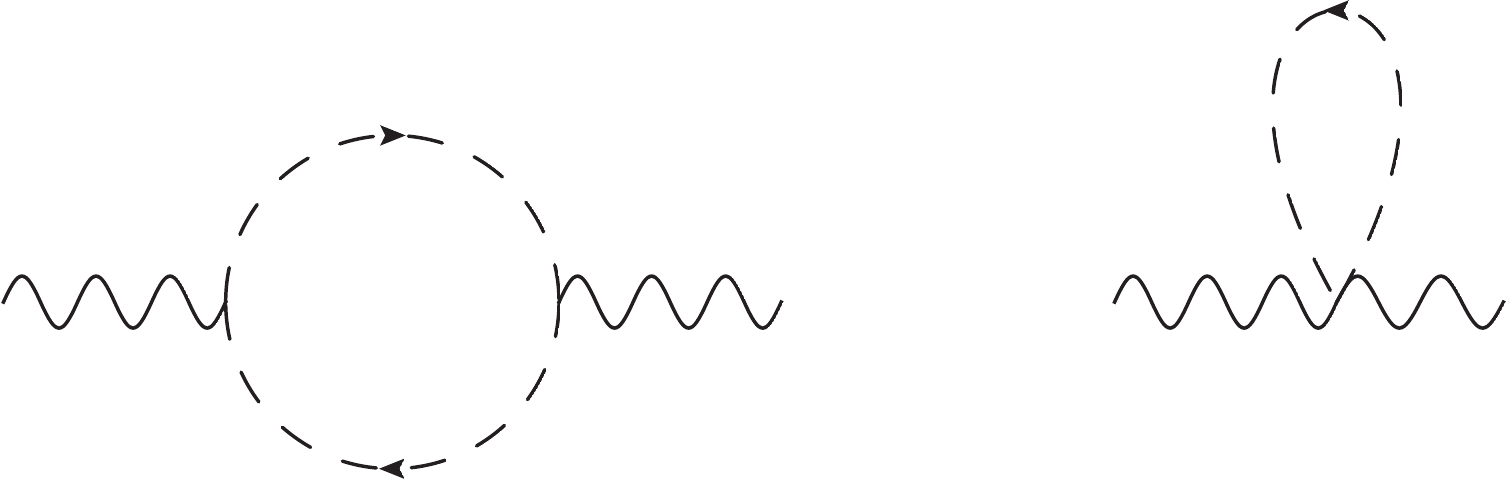}{0.8}
\caption[One Loop QED Photon Propagator Diagram with Scalars]{ \sglspc \small Higher order terms involving photons in a theory with scalar fields includes two diagrams.  This leads to a correction of the bare charge parameter of the Lagrangian for QED with scalar fields.}
\label{fig:QEDscalarFig}
\end{flushleft}
\end{figure}

We return to QED as our toy model, and we see from (\ref{eq:RG2}) that the requirement that gauge invariance at all scales makes solving the $\beta$ function for QED relativity simple to calculate.  We must calculate
\beq
\log(Z_A^{-1/2}) \approx \frac{G_1(e)}{\epsilon}. 
\eeq
We accomplish this again by rewriting the Lagrangian (\ref{eq:QEDLagCT}), this time in terms of canonically normalized fields as
\beqa
\CL = & i& \bar \psi \gamma \cdot \p \psi -  m \bar \psi \psi - \frac{1}{4} F^2 \nonumber 
\\  
&+& i \delta_\psi \bar \psi \gamma \cdot \p \psi - \delta_m m \bar \psi \psi - \frac{\delta_A}{4} F^2 \nonumber
\\
&+& Z_e \Lambda^\epsilon e A \cdot (\bar \psi \gamma \psi) \ ,
\eeqa
where $\delta_i = Z_i -1$.  The terms containing $\delta_i$ are known as counter terms, and can be viewed as additional interactions.  Let us now understand the origin of the $\Lambda^\epsilon$ term.  In 4 dimensional QED, the elementary charge is unit-less in natural units. Generally renormalizable nontrivial theories only contain dimensionless couplings~\cite{ZeeA}.  Rather than enforcing a hard cut-off $\Lambda$, we regularize the divergent integrals by using dimensional regularization.  That is, we compute the loop terms in a general $D = 4-2\epsilon$ dimensional space-time and in the end take the result in the limit that $\epsilon \rightarrow 0$.  Because we are using dimensional regularization, we like to keep the charge coupling dimensionless, so to keep that true in all dimensions, we leave $e$ dimensionless and add in a massive multiplier that holds the units of $e$ such that $e \rightarrow e \Lambda^{\epsilon}$ in $D$ dimensions.  
This can be accomplished via the Feynman diagram in Fig. \ref{fig:OLQEDscat} and the additional interactions from the counter terms is given as
\beq
-(Z_e e)^2 \Lambda^{2\epsilon} \int \frac{d^D l}{(2\pi)^D} {\rm Tr}\left[ \frac{\gamma^\mu((k+l)\cdot \gamma + m) \gamma^\nu (l \cdot \gamma + m)}{((k+l)^2-m^2)(l^2-m^2)}\right] + i \delta_A \lpa k^2 g^{\mu \nu} - k^\mu k^\nu \rpa \ .
\label{eq:PhotonOLCT}
\eeq
Evaluation of (\ref{eq:PhotonOLCT}) can be done through Feynman squaring method and Wick transformations and generalized trace theorems~\cite{Peskin}, which are done below.  Also we set $Z_e = 1$ since $Z_e = 1 + \CO(e^2)$, so at lowest order for $Z_A$ we do not include corrections to $Z_e$.
\beq
\bea{l}
=- e^2 \Lambda^{2\epsilon} \int \frac{d^D l}{(2\pi)^D} {\rm Tr}\left[ \frac{\gamma^\mu((k+l)\cdot \gamma + m) \gamma^\nu (l \cdot \gamma + m)}{((k+l)^2-m^2)(l^2-m^2)}\right] + i \delta_A \lpa k^2 g^{\mu \nu} - k^\mu k^\nu \rpa \nonumber
\\
=- D e^2 \Lambda^{2\epsilon} \int_0^1 dx \int \frac{d^D l}{(2\pi)^D} \frac{2 l^\mu l^\nu + 2 x(x-1) k^\mu k^\nu -(l^2 -m^2 + x(x-1) k^2)g^{\mu \nu} }{(l^2 -m^2 -x (x-1)k^2)^2} \nonumber
\\
+ i \delta_A \lpa k^2 g^{\mu \nu} - k^\mu k^\nu \rpa \nonumber
\\
=- D e^2 \Lambda^{2\epsilon} \int_0^1 dx \int \frac{d^D l}{(2\pi)^D} \frac{(2/D-1)l^2 g^{\mu \nu}+ 2 x(x-1) k^\mu k^\nu -(x(x-1) k^2 - m^2)g^{\mu \nu} }{(l^2 -m^2 -x (x-1)k^2)^2} \nonumber
\\
+ i \delta_A \lpa k^2 g^{\mu \nu} - k^\mu k^\nu \rpa \nonumber
\\
=- i D e^2 \Lambda^{2\epsilon} \int_0^1 dx \int \frac{d^D l_E}{(2\pi)^D} \frac{2 x(x-1)}{(l_E^2 +m^2 +x (x-1)k^2)^2}(k^2 g^{\mu \nu} - k^\mu k^\nu) \nonumber
\\
+ i \delta_A \lpa k^2 g^{\mu \nu} - k^\mu k^\nu \rpa \nonumber
\\
=- i \lpa D e^2 \Lambda^{2\epsilon} \int_0^1 dx \int \frac{d^D l_E}{(2\pi)^D} \frac{2 x(x-1)}{(l_E^2 +m^2 +x (x-1)k^2)^2} - \delta_A \rpa \lpa k^2 g^{\mu \nu} - k^\mu k^\nu \rpa \nonumber
\\
=- i \lpa  -\frac{e^2}{12\pi^2 \epsilon} + \int_0^1 dx \frac{2x(1-x)}{4\pi^2} \log \lpa \frac{\overline{\Lambda}^2}{m^2 +  x(x-1)k^2} \rpa - \delta_A \rpa \lpa k^2 g^{\mu \nu} - k^\mu k^\nu \rpa 
\eea
\eeq
with $\overline{\Lambda}^2 =4 \pi e^{-1/2-\gamma}\Lambda^2$. The rescaling of $\Lambda$ is known as the ``bar" part of the $\overline{ \rm MS}$ scheme.  The minimum we can do to make this finite is absorb the infinite part $e^2/12 \pi^2 \epsilon$ into the term $\delta_A$; this is the minimal subtraction (MS) in the $\overline{\rm MS}$ scheme
\beq
\delta_A = -\frac{e^2}{12\pi^2 \epsilon} \ .
\eeq
From this expression we can find the $\beta$ function for QED as $Z_A = 1 + \delta_A$. Therefore,
\beqa
\log(Z_A^{-1/2}) &\approx&  \frac{e^2}{24 \pi^2 \epsilon} \rightarrow G'_1(e) =  \frac{e}{12 \pi^2} \ , \nonumber
\\
\frac{d e}{d \log \Lambda} &=& \frac{d e}{d \log \overline{\Lambda}} =   \frac{e^3}{12 \pi^2}.  
\label{eq:BetaQED}
\eeqa
The result of (\ref{eq:BetaQED}) is a result that can be generalized for multiple copies of the fermionic sector of the QED Lagrangian, such that each particle has a charge $Q_f e$; we can account for each particles' contribution by summing the charges so that (\ref{eq:BetaQED}) becomes
\beq
\frac{d e}{d \log \Lambda} =  \frac{e^3}{12 \pi^2}\sum_{\rm fermions} Q_f^2.
\label{eq:BetaMultF}
\eeq
Inclusion of scalar fields that interact with the $U(1)$ field (photon in this case), can also be done, however an additional diagram must be included, and in scalar abelian gauge theories, there is also a term $e^2 \phi^\dagger \phi A^\mu A_\mu$.  The relevant diagrams are pictured in Fig. \ref{fig:QEDscalarFig}.
The result can be generalized to multiple scalar fields having couplings $Q_s e$, in which Eq. (\ref{eq:BetaMultF}) becomes
\beq
\frac{d e}{d \log \Lambda} =  \frac{e^3}{12 \pi^2}\lpa \sum_{\rm fermions} Q_f^2 + \frac{1}{4} \sum_{\rm scalars} Q_s^2 \rpa.
\label{eq:RGQEDFNL}
\eeq

\subsection{The SM Higgs Stability}

On July 4th, 2012, the CERN LHC experiment reported from Higgs searches a result that indicates a
Higgs boson candidate with a mass $m_H = 125$ GeV.  The
ATLAS~\cite{Gianotti} and CMS~\cite{Incandela} Collaborations
independently combined about $5~{\rm fb}^{-1}$ of data collected at $\sqrt{s} = 7~{\rm TeV}$
and more than $5~{\rm fb}^{-1}$ at $\sqrt{s} = 8~{\rm TeV}$. The
excess events at 125~GeV that was evident already in data from the 7~TeV
run~\cite{ATLAS:2012ae,Chatrchyan:2012tx} was consistently
observed by both experiments in the $\gamma \gamma$ invariant mass
spectrum with a local significance of $4.5\sigma$ and $4.1\sigma$,
respectively. In addition to the photon data, an excess of events in 
4 lepton final states (with $m_{4\ell} \simeq 125~{\rm GeV}$) can be interpreted as a signal
of the $H \to ZZ^* \to 4 \ell$ decay, and is observed by both experiments
with a significance of $3.4\sigma$ and $3.2\sigma$, respectively. The
CMS experiment also presented updated Higgs boson searches in $W^+W^-$
(a broad excess in the invariant mass distribution of $1.5\sigma$ is
observed), $b\bar b$ (no excess is observed), and $\tau \bar \tau$ (no
excess is observed) channels.  More recently, the ATLAS Collaboration
reported a $2.8\sigma$ deviation in the $H \to W^+ W^- \to 2 \ell \nu$
decay channel~\cite{Arnaez}.  When combining the data from the 7~TeV
and 8~TeV runs, both experiments separately reached the
sensitivity to the new boson with a local significance of
$5\sigma$~\cite{ATLASnew,CMSnew}.  Additionally, the CDF and
D0 Collaborations at Fermi-lab published an update on searches for the Higgs boson
decaying into $b \bar b$ pairs using $9.7~{\rm fb}^{-1}$ of data
collected at $\sqrt{s} = 1.96~{\rm TeV}$~\cite{Aaltonen:2012qt}. They
reported a $3.3\sigma$ deviation with respect to the background-only
hypothesis in the mass range between $120 - 135~{\rm GeV}$.

The data seems to indicate the existence of the long sought Higgs boson. 
The question we address is the stability of the Higgs mechanism.  That is, does Higgs
field obtain a non-zero real vacuum expectation value for all energy scales.
Next-to-leading order (NLO) constraints on SM vacuum stability based
on two-loop renormalization group (RG) equations, one-loop threshold
corrections at the electroweak scale (possibly improved with two-loop
terms in the case of pure QCD corrections), and one-loop effective
potential seem to indicate $m_H \approx 125 - 126~{\rm GeV}$ saturates
the minimum value that ensures a vanishing Higgs quartic coupling
around the Planck scale ($M_{\rm Pl}$) (the scale we assume that quantum gravitational effects cannot be ignored, and we assume some change to the SM would be included), see {\it
  e.g.}~\cite{Lindner:1988ww,Sher:1988mj,Diaz:1994bv,Casas:1994qy,Diaz:1995yv,Casas:1996aq,Isidori:2001bm,Isidori:2007vm,Hall:2009nd,Ellis:2009tp,EliasMiro:2011aa,Xing:2011aa}. However, a more recent NNLO
analysis~\cite{Bezrukov:2012sa,Degrassi:2012ry} yields a very
restrictive condition of absolute stability up to the Planck scale given as
\begin{equation}
m_H> \left[129.4 + 1.4 \left( \frac{m_t/{\rm GeV} -173.1}{0.7}
\right) - 0.5 \left(\frac{\alpha_s(m_Z) - 0.1184}{0.0007} \right) \pm
1.0_{\rm th} \right]~{\rm GeV}  \, .
\label{1}
\end{equation}
When combining in quadrature the theoretical uncertainty with
experimental errors on the mass of the top ($m_t$) and the strong
coupling constant ($\alpha_s$), one obtains $m_H > 129 \pm 1.8~{\rm
  GeV}$. The vacuum stability of the SM up to the Planck scale is
excluded at 2$\sigma$ (98\% C.L. one sided) for $m_H < 126~{\rm
  GeV}$~\cite{Bezrukov:2012sa,Degrassi:2012ry}. It seems achieving
stability will require some new BSM physics.

\subsection{RG Equations of SM$^{++}$}
\label{II}

In order for the Higgs mechanism to be valid for scales up to the underlying string theory of the SM$^{++}$, $M_s$,
we impose the positivity conditions~\cite{Barger:2008jx} on the parameters of SM$^{++}$ scalar potential
\begin{equation}
\lambda_1 > 0, \quad \quad \lambda_2 > 0, \quad \quad  \lambda_1
\lambda_2 > \frac{1}{4} \lambda_3^2 \, .
\label{VernonGabePaul}
\end{equation}
If the conditions (\ref{VernonGabePaul}) are satisfied, we can minimize 
$V^{++} (H, H'')$ and find two real, non-zero, VEVs for
the two Higgs fields of the SM$^{++}$. 
In the unitary gauge, the fields can be written as
\begin{equation}\label{min}
H  \equiv \frac{1}{\sqrt{2}} 
\left( \begin{array}{c} 0 \\ v_1 + h_1(x)\end{array} \right)
\quad {\rm and} \quad  H''  \equiv \frac{1}{\sqrt{2}} \left(
v_2 + h_2(x) \right)\, ,
\end{equation} 
with $v_1$ and $v_2$ the real and non-negative VEVs. The non-zero, real solutions to the minimization of (\ref{higgsV}) are
obtained for $v_1$ and $v_2$ and are given by
\begin{equation}
v_1^2 = \frac{-\lambda _2 \mu_1^2 + \frac{1}{2} \lambda _3 {\mu_2}
  ^2}{\lambda _1 \lambda _2 - \frac{1}{4} \lambda _3^2}
\quad {\rm and} \quad
v_2^2 = \frac{-\lambda _1 \mu_2^2 + \frac{1}{2} \lambda _3 \mu_1
  ^2}{\lambda _1 \lambda _2 - \frac{1}{4} \lambda _3^2} \, .
\label{minima}
\end{equation}
To compute the scalar masses, we must expand the potential
(\ref{higgsV}) around the minima (\ref{minima}). We denote by $h$ and
$h''$ the scalar fields of definite masses (mass matrix eigenstates), $m_{h}$ and $m_{h''}$,
respectively. After a bit of algebra, the explicit
expressions for the scalar mass eigenvalues and eigenvectors are given
by
\begin{eqnarray}
m^2_{h} &=& \lambda _1 v_1^2 + \lambda _2 v_2^2 - \sqrt{(\lambda _1
  v_1^2 - \lambda _2 v_2^2)^2 + (\lambda _3 v_1 v_2)^2} \, , \label{mh1}\\
m^2_{h''} &=& \lambda _1 v_1^2 + \lambda _2 v_2^2 + \sqrt{(\lambda _1 v_1^2 - \lambda _2 v_2^2)^2 + (\lambda _3 v_1v_2)^2} \, ,
 \label{mh2}
\end{eqnarray}
\begin{equation}
\left( \begin{array}{c} h\\h''\end{array}\right) = \left( \begin{array}{cc} \cos{\alpha}&-\sin{\alpha}\\ \sin{\alpha}&\cos{\alpha}
	\end{array}\right) \left( \begin{array}{c} h_1\\h_2\end{array}\right) \, ,
\end{equation}
where $\alpha \in [-\pi/2, \pi/2]$ also  fullfils
\begin{eqnarray}\label{sin2a}
\sin{2\alpha} &=& \frac{\lambda _3 v_1v_2}{\sqrt{(\lambda _1 v_1^2 - \lambda _2 v_2^2)^2 + (\lambda _3 v_1v_2)^2}} \, ,\\ 
\cos{2\alpha} &=& \frac{\lambda _1 v_1^2 - \lambda _2 v_2^2}{\sqrt{(\lambda _1 v_1^2 - \lambda _2 v_2^2)^2 + (\lambda _3 v_1v_2)^2}}\, .
\end{eqnarray}
Now, it is convenient to invert (\ref{mh1}), (\ref{mh2}) and (\ref{sin2a}), 
to extract the parameters in the Lagrangian in terms of the physical quantities $m_{h}$, $m_{h''}$ and $\sin{2\alpha}$
\begin{eqnarray}
\label{12}
\lambda _1 &=& \frac{m_{h''}^2}{4v_1^2}(1-\cos{2\alpha}) +
\frac{m_{h}^2}{4v_1^2}(1+\cos{2\alpha}), \nonumber \\ 
\lambda _2 &=& \frac{m_{h}^2}{4v_2^2}(1-\cos{2\alpha}) +
\frac{m_{h''}^2}{4 v_2^2}(1+\cos{2\alpha}),\\ 
\lambda _3 &=& \sin{2\alpha} \left( \frac{m_{h''}^2-m_{h}^2}{2v_1v_2}
\right). \nonumber
\end{eqnarray}

One-loop corrections to (\ref{higgsV}) can be implemented by
  making $\lambda_1$, $\lambda_2$, and $\lambda_3$ field dependent
  quantities. Equation (\ref{VernonGabePaul}) then needs to be imposed
  in the regions where this is the case. When we talk about the
  stability of (\ref{higgsV}) at some energy $\Lambda = Q$ (with the use of the
  couplings at that scale), we are thinking that the field values are
  at the scale $Q$. For $\lambda_3>0$, the third condition in
  (\ref{VernonGabePaul}) is only invalidated for field values $v_1$
  around $m_{h''}$, regardless of the renormalization scale
  $Q$~\cite{EliasMiro:2012ay}.   We can find the instability regions by expressing the scalar potential as
  \beqa
V^{++} = & \lambda_1(Q)& \lpa |H|^2 - \frac{v_1^2}{2} \rpa^2 + \lambda_2(Q) \lpa |H''| - \frac{v_2^2}{2}\rpa^2 \nonumber \\
 & + & \lambda_3(Q) \lpa |H|^2 - \frac{v_1^2}{2} \rpa \lpa |H''|^2 - \frac{v_2^2}{2} \rpa .
\eeqa
The instability region ($V^{++} < 0$) is given by, $|H''| \approx 0$ and
\begin{equation}
v_2 < \frac{m_{h''}}{\sqrt{2\lambda_2}}, \quad   Q_- < v_1 < Q_+,\quad
    \left. Q^2_\pm = \frac{m_{h''}^2 \lambda_3}{8 \lambda_1 \lambda_2} \left(1
  \pm \sqrt{1 - \frac{4\lambda_1 \lambda_2}{\lambda_3^2}} \right) \right|_{Q_*} \,,
\label{rolfi}
\end{equation}
where $Q_*$ is some energy scale where the third condition of (\ref{VernonGabePaul})
is violated~\cite{EliasMiro:2012ay}. Thus, $Q_\pm \sim m_{h''}$ when the third
condition is saturated, i.e. $\lambda_1 \lambda_2 =
\lambda_3^2/4$. From (\ref{rolfi}) we see that $Q_\pm \sim m_{h''}$
when all the $\lambda_i$ are roughly at the same scale. If one of the
$\lambda_{1,2}$ is close to zero, then $Q_+$ can be $\gg m_{h''}$, but
this region of the parameter space is taken care of by the condition
$\lambda_{1,2}>0$. The stability for field values at $m_{h''}$ is then
determined by the potential with coupling at scale $m_{h''}$ (instead
of $Q$). Therefore, for $\lambda_3>0$, we impose the third
condition of (\ref{VernonGabePaul}) in the vicinity of $m_{h''}$ only.  Even though the potential
appears to be unstable at $Q \gg m_{h''}$, it is actually stable when
all the field values are at the scale $Q$. Note that the potential
with $\lambda_i(Q)$ can only be used when the physical quantities
(field values $v_1$, $v_2$) are at the scale $Q$. 

On the other hand, the instability region for $\lambda_3<0$ occurs for $|H''| > v_2/\sqrt{2}$. Because of this,
we can neglect the mass parameters and approximate the potential as
\beq
V^{++} \approx \lambda_1(Q) |H|^4 + \lambda_2(Q) |H''|^4 + \lambda_3(Q) |H|^2 |H''|^2.
\eeq
The instability region is then given as
\begin{equation}
v_2 > \frac{m_{h''}}{\sqrt{2\lambda_2}}, \quad c_- < \frac{v_1}{v_2}
  <c_+, 
    \quad \left. c_{\pm}^2 = - \frac{\lambda_3}{2\lambda_1} \left( 1 \pm \sqrt{1
        - \frac{4\lambda_1 \lambda_2}{\lambda_3^2} }\right) \right|_{Q_*}\,.
\end{equation}
We can see in this case the ratio of $v_1$ and $v_2$ determines the instability region, which can be
reached even with both $v_1$ and $v_2$ being $\gg
m_{h''}$~\cite{EliasMiro:2012ay}. Therefore, for $\lambda_3< 0$, we
impose the third condition at all energy scales.  Note that
the asymmetry in conditions on $\lambda_3$ will carry over into an asymmetry in
$\alpha$.

Calculations similar to the ones that lead to Eq. (\ref{eq:RGQEDFNL}) can be done for all the parameters
of the SM$^{++}$ model.  The RG equations for the five
parameters in the scalar potential~\cite{Basso:2010jm} are
\begin{eqnarray} \label{RG}
 \frac{d \mu_1^2}{dt} &=&
\frac{\mu_1^2}{16\pi ^2}\left( 12\lambda _1 +6Y_t^2+2\frac{\mu_2
    ^2}{\mu_1^2}\lambda _3
  -\frac{9}{2}g_2^2-\frac{3}{2}g_Y^2-\frac{3}{2} g_{\cal Y}^2\right)\,
, \nonumber \\ 
\frac{d \mu_2 ^2}{dt} &=& \frac{\mu_2^2}{16\pi
   ^2}\left( 8\lambda _2 +4\frac{\mu_1^2}{\mu_2^2}\lambda _3 - 24
   g_{B-L}^{2}\right)\, , \nonumber \\  \frac{d\lambda_1}{dt} & = &
 \frac{1}{16\pi ^2}\left( 24\lambda _1^2+\lambda _3^2
-6Y_t^4 +\frac{9}{8}g_2^4 +\frac{3}{8}g_Y^4 +\frac{3}{4}g_2^2g_Y^2
+\frac{3}{4}g_2^2
g_{\cal Y}^2  +\frac{3}{4}g_Y^2 g_{\cal Y}^2+\frac{3}{8} g_{\cal Y}^4 \right.
\nonumber \\ & + &
\left.  12\lambda _1 Y_t^2 
 -   9\lambda _1 g_2^2-3\lambda _1 g_Y^2-3\lambda _1
  g_{\cal Y}^2
	\right)\, ,\\ 
 \frac{d \lambda _2}{dt} &=&  \frac{1}{8\pi ^2}\left( 10\lambda
   _2^2+\lambda _3^2 +48 g_{B-L}^{4} -24\lambda _2g_{B-L}^{2}
 \right)\, , \nonumber \\ 
\frac{d \lambda _3}{dt} &=&  \frac{\lambda _3}{8\pi
  ^2}\left( 6\lambda _1+4\lambda _2+2\lambda
  _3+3Y_t^2-\frac{9}{4}g_2^2-\frac{3}{4}g_Y^2-\frac{3}{4} g_{\cal Y}^2
  - 12 g_{B-L}^{2} \right) \nonumber \\
  &+& \frac{3}{4 \pi^2} \, g_{\cal Y}^2 \, g_{B-L}^{2}  \nonumber ,
\end{eqnarray}
where $t = \ln Q$ and $Y_t$ is the top Yukawa coupling, with
\begin{equation}\label{RGE_yuk_top}
\frac{dY_t}{dt} = \frac{Y_t}{16\pi ^2}\left(
  \frac{9}{2}Y_t^2-8g_3^2-\frac{9}{4}g_2^2-\frac{17}{12}g_Y^2-\frac{17}{12}
  g_{\cal Y}^2 -\frac{2}{3}g_{B-L}^{2}-\frac{5}{3} g_{\cal Y} g_{B-L} \right)
\end{equation}
and $Y_t^{(0)} = \sqrt{2} \, m_t/v$.
The RG running of the gauge couplings follow the standard form
\begin{eqnarray}
\frac{dg_3}{dt} &=& \frac{g_3^3}{16\pi ^2}\left[ -11+\frac{4}{3}n_g
\right] = - \frac{7}{16} \, \frac{ g_3^3}{\pi ^2} \, , \nonumber \\
\frac{dg_2}{dt} &=& \frac{g_2^3}{16\pi ^2}\left[
  -\frac{22}{3}+\frac{4}{3}n_g+\frac{1}{6}\right] = -\frac{19}{96}
\,\frac{g_2^3}{\pi ^2} \,,  \nonumber \\
\frac{dg_Y}{dt}  &=& \frac{1}{16\pi ^2}\left[A^{YY}g_Y^3 \right]\, , \\ 
\frac{dg_{B-L}}{dt} &=& \frac{1}{16\pi ^2}\left[A^{(B-L)
    (B-L)}g_{B-L}^3+2A^{(B-L) Y}g_{B-L}^2 g_{\cal Y}+A^{YY}g_{B-L}
  g_{\cal Y}^2 \right] \, , \nonumber \\ 
\frac{dg_{\cal Y}}{dt} &=& \frac{1}{16\pi ^2}\left[A^{YY} g_{\cal
    Y}\,(g_{\cal Y}^2+2g_{Y}^2)+2A^{(B-L)Y}g_{B-L}(g_{\cal
    Y}^2+g_{Y}^2) \right.\ \nonumber \\
    &+& \left.\ A^{(B-L) (B-L)}g_{B-L}^2 g_{\cal Y} \right] ,
\nonumber 
\end{eqnarray}
where $n_g =3$ is the number of generations and
\begin{equation}\label{A_charge}
A^{ab} = A^{ba} = \frac{2}{3} \sum _f Q_{a,f} Q_{b,f} + \frac{1}{3}\sum _s
Q_{a,s} Q_{b,s}\, , \qquad (a,b=Y,\ B-L) \,,
\end{equation}
with $f$ and $s$ indicating
contribution from fermion and scalar loops, respectively.

For energies below the mass of the heavier Higgs $H''$, the
  effective theory is the SM. Thus the effective scalar Lagrangian in the 
low energy regime must take on the form appropriate for the SM 
\begin{equation}
\mathscr{L}_s =\left( {\cal D}^{\mu} H\right) ^{\dagger} {\cal
  D}_{\mu} H - \mu^2 \left| H \right|^2 - \lambda \left| H \right|^4 \ .
\label{higgsSM}
\end{equation}
The RG equations in this regime must simplify to those of SM. To obtain the matching
conditions connecting the two theories so that they reflect a consistent theory, we can 
follow ~\cite{EliasMiro:2012ay} and integrate out the field $H''$ to
obtain the effective Lagrangian of the form (\ref{higgsSM}).  To find the effective Lagrangian is to perform
the Feynman path integral over the $H''$ field
\beq
\int \CD H'' \ e^{i S(H,H'')} = e^{i S_{\rm eff}(H)},
\eeq
from which you can read the potential $V_{\rm eff}$ of the form (\ref{higgsSM}).  Rather than carrying out this rather complicated task, we can approximate this integral by expanding the action $S(H,H'')$ around the field configuration that gives $\delta S/\delta H''  = 0$, which is the condition for the classical equations of motion. This gives
\beq
\int \CD H'' \ e^{i S(H,H'')}  \approx   e^{i S(H,H_{cl})} \int \CD H'' e^{i \int \frac{1}{2} \frac{\delta^2 S}{\delta H''^2}|_{H''_{cl}}(H'' - H''_{cl})^2 + \dots} \ ,
\eeq
where above $H''_{cl}$ is the classical solution where $\delta S/\delta H''|_{H''_{cl}} = 0$ is satisfied.  From this expression we can see that $S_{\rm eff}$ can be approximated by $S(H,H_{cl})$ so long as $(H''-H_{cl})$ is small for all cases considered. Since we are in the phase where the energies of interaction are below the scale of exciting $H''$, we can assume the quantum corrections are small.  By solving the equations of motion for $H''$ and neglecting the derivative terms, as we are assuming that there is little excitation of the $H''$ field, we have the result
\beq
\pd{H''} V(H,H'') = 0 \rightarrow | H''|^2 = -\frac{\mu_2^2 + \lambda_3 |H|^2}{2 \lambda_2}. 
\label{eq:EffLagriangian}
\eeq
Replacing (\ref{eq:EffLagriangian}) back into $V(H,H'')$ allows you to
identify the quadratic
and quartic terms in the potential, which yields 
\begin{equation}
\mu^2 = \mu_1^2 - \mu_2^2 \ \frac{\lambda_3}{2 \lambda_2}  
\end{equation}
and
\begin{equation}
\lambda = \lambda_1 \ \left(1 - \frac{\lambda_3^2}{4 \lambda_1 \lambda_2} \right) \, ,
\end{equation}
respectively.
The matching conditions are consistent with the continuity of $v \leftrightharpoons 
v_1$; namely
\begin{equation}
v^2 = - \left. \frac{\mu^2}{\lambda} \right|_{Q = m_{h''}} = -
\left. \frac{\mu_1^2 - \mu_2^2 \ \lambda_3 /(2 \lambda_2)}{ \lambda_1
    \ 
\left[1 - \lambda_3^2/(4 \lambda_1 \lambda_2) \right]} \right|_{Q = m_{h''}} \, ,
\end{equation}
or equivalently
\begin{equation}
\left. v^2 \right|_{Q = m_{h''}} = \left. v_1^2 \right|_{Q = m_{h''}} \,,
\end{equation}
with $v_1$ given by (\ref{minima}). The quartic interaction between
the heavy scalar singlet and the Higgs doublet provides an essential
contribution for the stabilization the scalar field
potential~\cite{EliasMiro:2012ay}.

\subsection{Running the Couplings}
\label{III}

Now that we have the equations that determine how the values of the parameters
of the scalar potential change with scale $Q$, we must solve these coupled differential equations.
In order to ensure perturbativity of $g'_4$ between the TeV scale and the
string scale, we find from  (\ref{eq:P1Twentyone}) that $g'_1 >
0.232$. We also take $g'_1(M_s) \simeq 1$ in order to ensure perturbativity
at the string scale. Let us first study the region of the parameter
space constrained by $g'_1 (M_s) \simeq 1$. The string-scale values
of the other abelian couplings are fixed by previous considerations
 (\ref{eq:P1Twentyone}) and  (\ref{eq:P1Two}). The Euler angles at $M_s$
are also fixed by  (\ref{eq:P1Twentyfour}), and  (\ref{eq:P1Twentyfive}).  All
the couplings and angles are therefore determined at all energies
through RG running. As an illustration, we set $M_s = 10^{14}~{\rm
  GeV}$; this leads to $g'_3 (M_s)= 0.231$, $g'_4 (M_s) = 0.232$,
$\psi (M_s) = -1.245$, $\theta (M_s) = -0.217$, and $\phi (M_s) =
-0.0006$. 

Now we take $Q_{\rm min} = 125~{\rm GeV}$ and normalize $t
= \ln(Q/125~{\rm GeV})$ and $t_{\rm max} = \ln(M_s/125~{\rm
  GeV})$. From the $M_s$ scale we run the couplings and angles down to the TeV
region giving $g'_1 = 0.406$, $g'_3 = 0.196$, $g'_4 = 0.218$, $\theta =
-0.466$, $\psi = -1.215$, and $\phi = -0.0003$.

We are now in a position to randomly choose $v''$ and $m_{h''}$ at the scale $Q = m_{h''}$.  We can then use the SM relation $m_H^2 = - 2 \mu^2$, where $m_H \simeq 125~{\rm
  GeV}$, and $v^2 = 246~{\rm GeV}$, both taken at the same energy scale $Q
= 125~{\rm GeV}$ to find the initial conditions for the parameters $\mu$
and $\lambda$ at the $Q_{\rm min}$ scale.  
It should be noted that we take the top Yukawa coupling
evaluated at the scale $m_t$. This introduces a small but unnoticeable error.
On the other hand, $m_t$ is taken to be the physical top mass; if we
used the running mass instead, as is done in~\cite{Casas:1994us}, the running
of the quartic coupling $\lambda$ would be much slower, with the
instability scale pushed to almost $10^9~{\rm GeV}$.  

Then we run the SM couplings $\lambda$, and $\mu$ from 125~GeV up to the mass scale $Q = m_{h''}$.  After having done this, we then use the matching conditions to determine $v$, which in turns allows one to
solve algebraically for $m_h$ at the scale $Q = m_{h''}$.  This process ensures that we match the SM results when $Q < m_{h''}$. 

After completing this task, there is one free
parameter left to be fixed at the TeV-scale: $\alpha$.  The initial values of $g_Y$, $g_{\cal Y}$ and $g_{B-L}$
are then fixed by previous considerations as in Sec.~\ref{sec:ic}. 

Rather than using the parameters $v'', \ \alpha, \ {\rm and } \ m_{h''}'$ as the free ones, we can use
the relation $M_{Z''} = g'_1 \, C_\phi \, v_2/C_\theta$~\cite{luisnMe1}, so that we can take $(M_{Z''},\,
\alpha,\, m_{h''})$ as the free parameters of the model.\footnote{ \sglspc \small For
  $M_s=10^{14}~{\rm GeV}$, the $v_2 \leftrightharpoons M_{Z''}$
  relation implies that if $7~{\rm TeV} < v_2 < 13~{\rm TeV}$, then
  $3.2~{\rm TeV} < M_{Z''} < 6.0~{\rm TeV}$. For a different $M_s$
  the range of $M_{Z''}$ is altered because of changes in $g'_1$,
  $\theta$, and $\phi$; {\it e.g.}  for $M_s = 10^{19}~{\rm GeV}$, the
  range becomes $2.8~{\rm TeV} < M_{Z''} < 5.8~{\rm TeV}$.} 

\begin{figure}[tbp] 
    \postscript{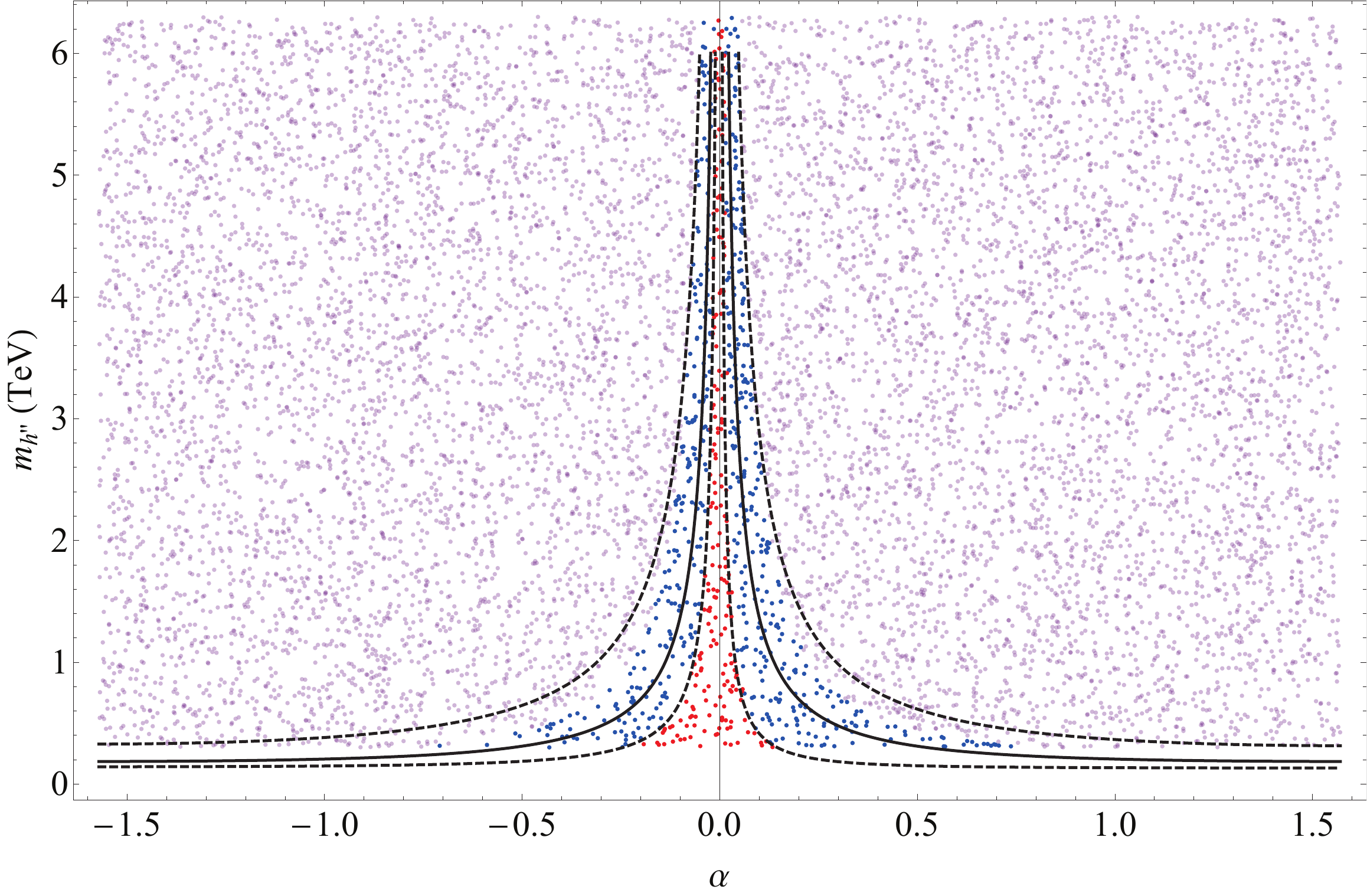}{1.0} 
  \caption[SM$^{++}$ Monte-Carlo Method of Stable Solution Finding]{ \sglspc \small The SM$^{++}$ vacuum stability patterns in
    the $m_{h''}$ vs $\alpha$ plane, for $M_{Z''} = 4.5~{\rm
      TeV}$. The analysis is based on a scan of $10^4$ random
    points with $M_s = 10^{14}~{\rm GeV}$. The points yielding a
    stable vacuum solution up to $M_s$ are blue-printed, those leading
    to unstable vacuum solutions are red-printed, and points giving
    runaway solutions ({\em i.e.}, those in which the Higgs doublet
    self-coupling blows up) are purple-printed.
Fits to the  boundaries defining the region with stable
    vacuum solutions (dashed lines) and to the average value of the scatter
    points contained in that region (solid lines) are also shown~\cite{luisnMe2}.}
\label{sm++_f2} \end{figure}

\begin{figure}[tbp] 
\begin{minipage}[t]{0.49\textwidth}
    \postscript{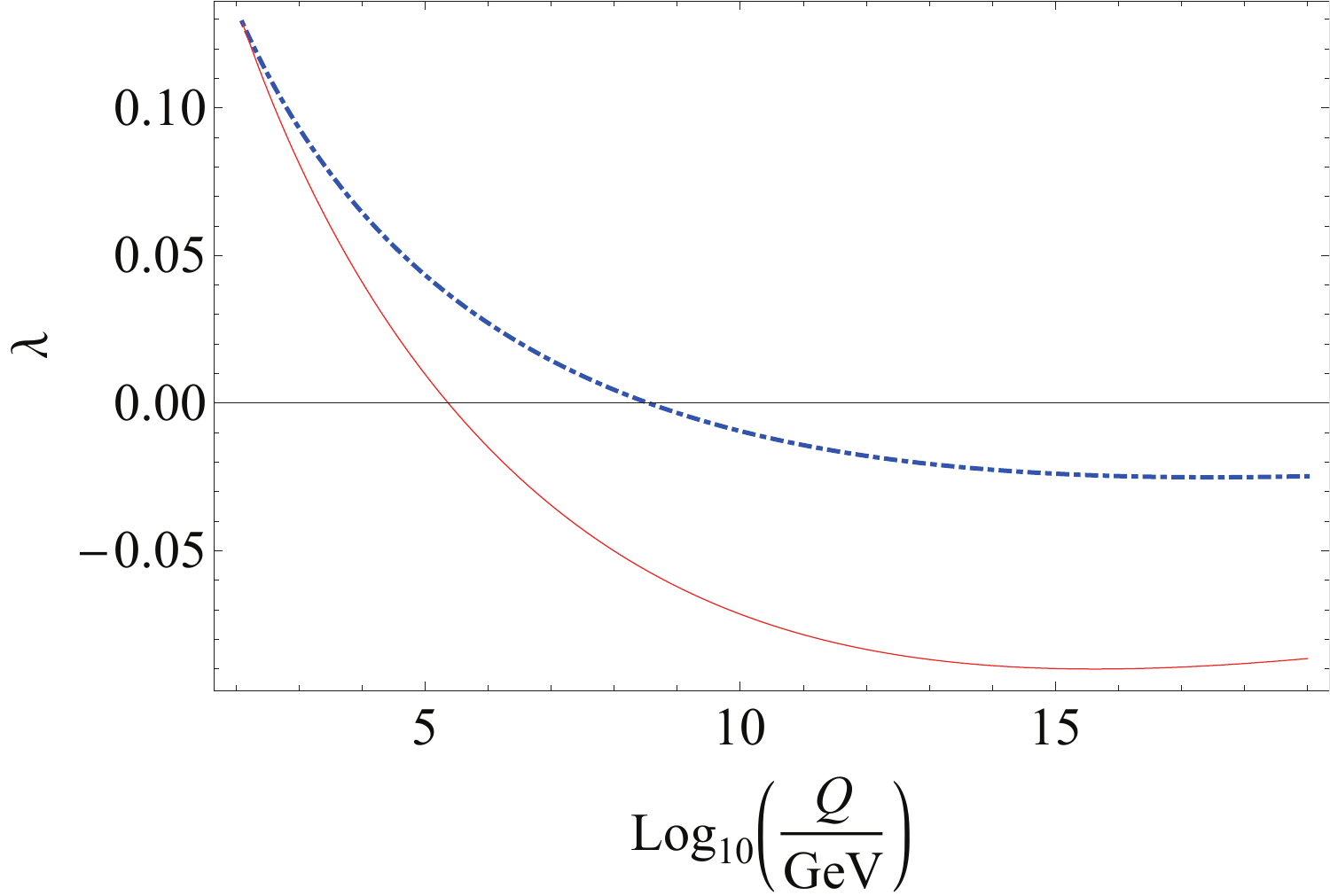}{1.0} \end{minipage}
  \begin{minipage}[t]{0.49\textwidth}
    \postscript{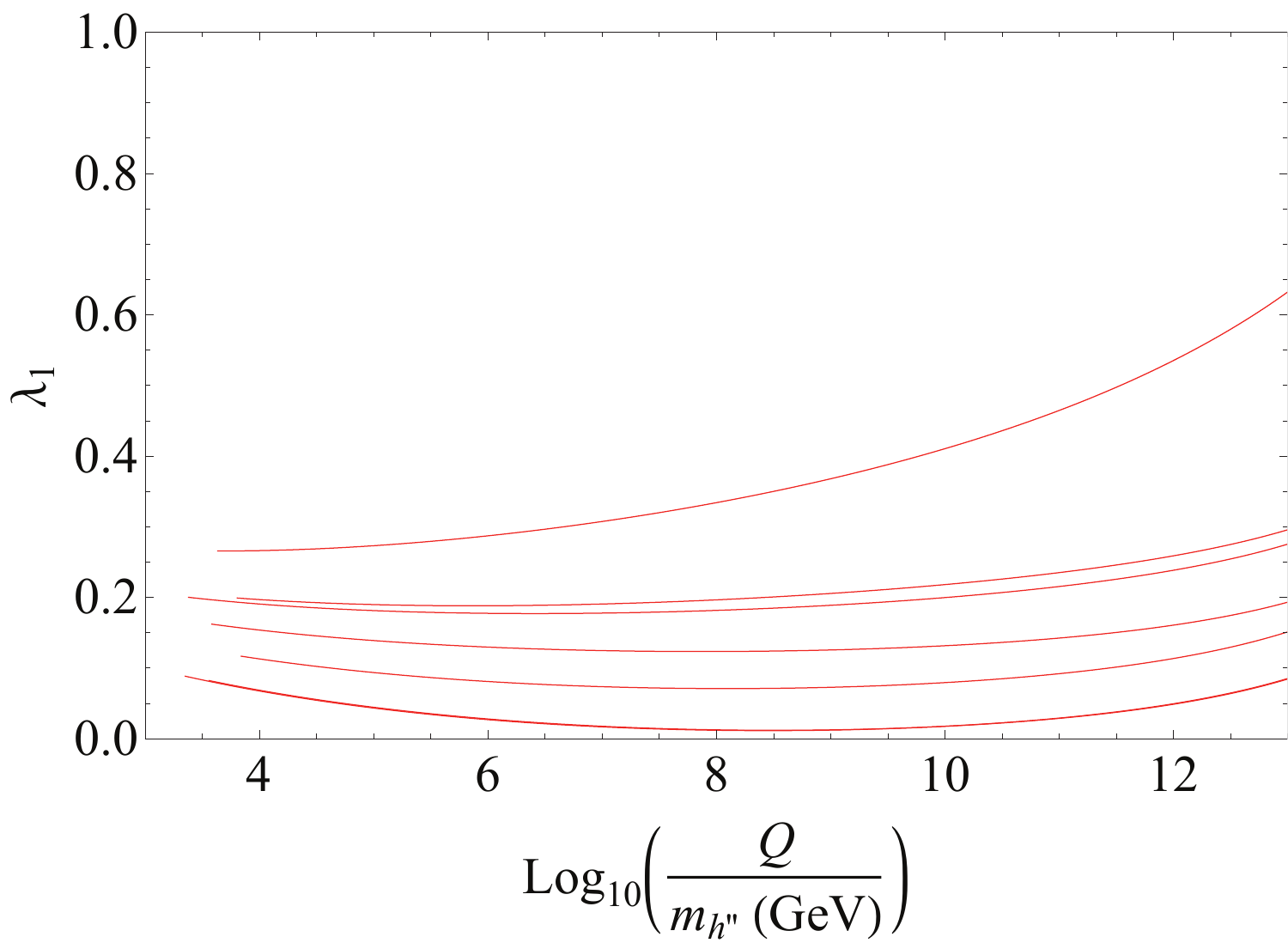}{1.0} \end{minipage} 
\begin{minipage}[t]{0.49\textwidth}
    \postscript{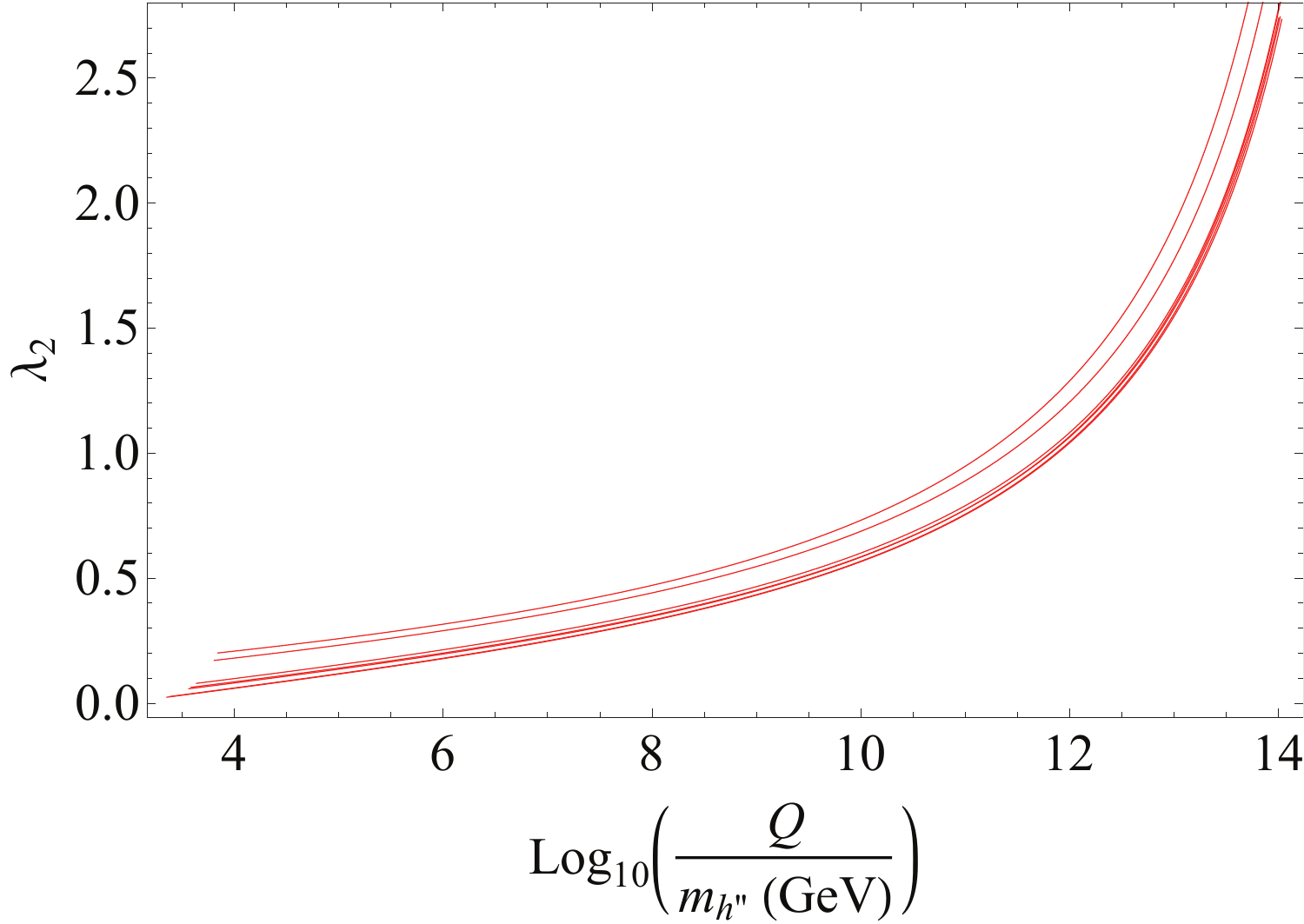}{1.0} \end{minipage} 
  \begin{minipage}[t]{0.49\textwidth}
    \postscript{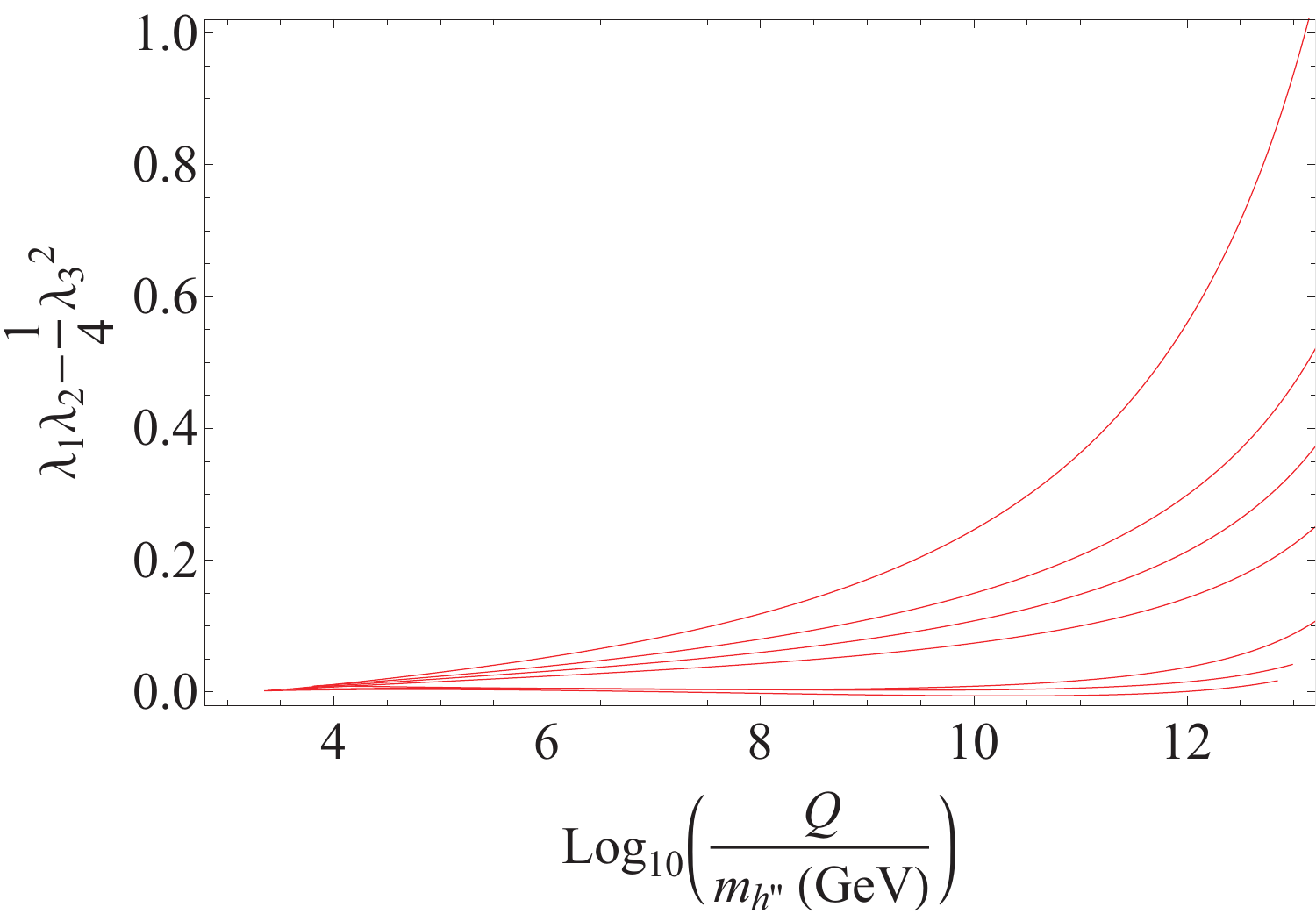}{1.0} \end{minipage} 
  \caption[Running of the Couplings for SM$^{++}$]{ \sglspc \small From left to right downwards: the first panel shows the
    running of $\lambda$ from its value at $125~{\rm GeV}$ (red solid
    line $m_t = 172.9~{\rm GeV}$ and blue dot-dashed line $m_t =
    164~{\rm GeV}$); the second and third panels show the typical
    behavior of the running couplings $\lambda_1(t)$ and $\lambda_2
    (t)$ for the average value of the initial condition, $\langle
    \lambda_1^{(0)} \rangle = 0.28$ in the integration of (\ref{RG});
    the fourth panel shows the behavior of the extra positivity
    condition for $\alpha <0$. In the running of $\lambda_i$ we have
    taken $M_s = 10^{14}~{\rm GeV}$.}
\label{sm++_f3} 
\end{figure}

\begin{figure}[tbp] 
    \postscript{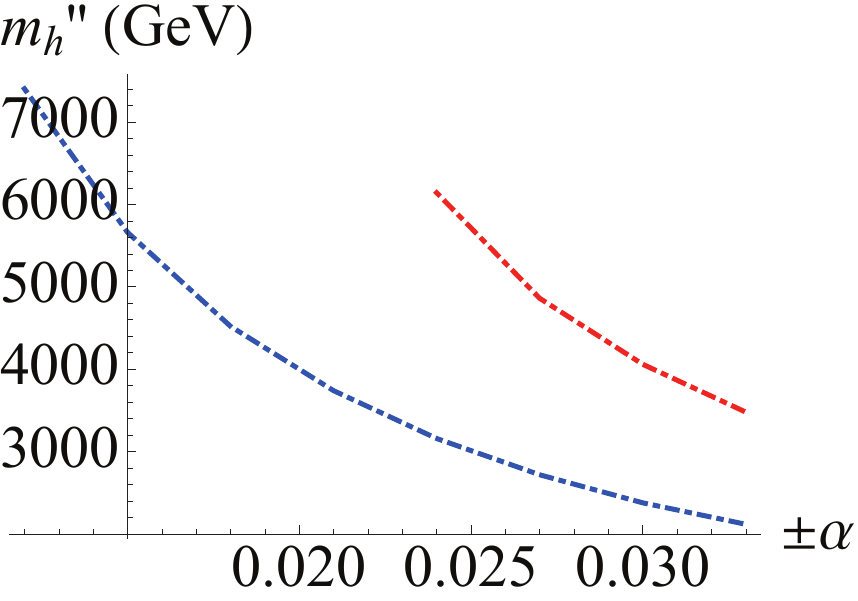}{0.8} 
    \caption[Asymmetry of Stability Requirements Close Up]{ \sglspc \small The lower boundary of the allowed parameter space in the
      $m_{h''}-\alpha$ plane under the vacuum stability constraint of
      Eq.~(\ref{VernonGabePaul}), for the positive alpha (blue) and negative alpha
      (red). We have taken $M_{Z''} = 4.5~{\rm TeV}$ and $M_s = 10^{14}~{\rm GeV}$.}
\label{sm++_f4} 
\end{figure}

\begin{figure}[tbp] 
    \postscript{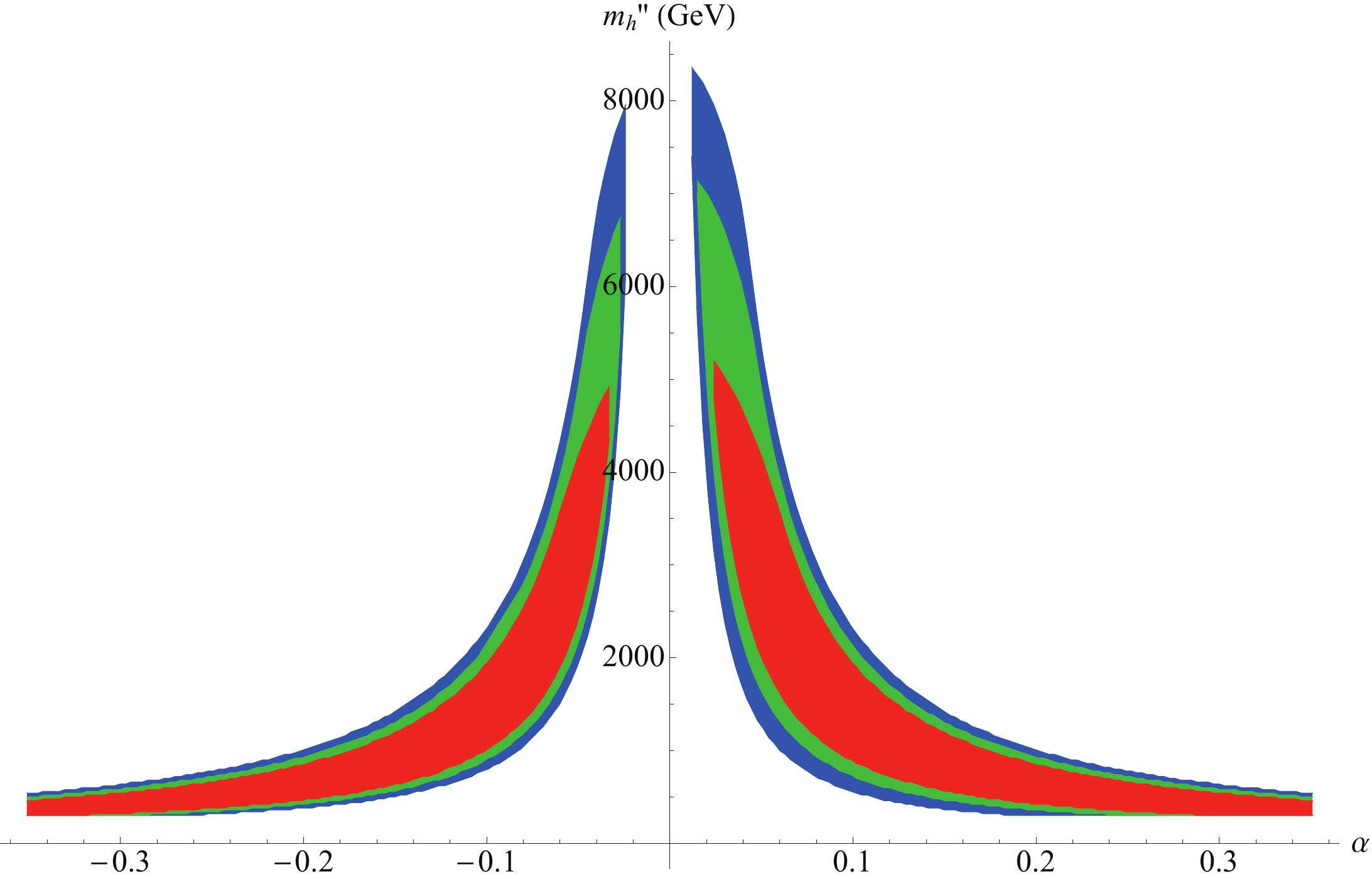}{0.8}
    \caption[Allowed Parameter Space for SM$^{++}$ Model Stability]{ \sglspc \small The allowed SM$^{++}$ parameter space in the $m_{h''}$ vs
      $\alpha$ plane under the vacuum stability constraint of
      Eq.~(\ref{VernonGabePaul}), for the case $M_{Z''} = 4.5~{\rm
        TeV}$, with $M_s = 10^{14}~{\rm GeV}$ (blue), $M_s =
      10^{16}~{\rm GeV}$ (green), and $M_s = 10^{19}~{\rm GeV}$
      (red). The perturbative upper bound is defined by $\lambda_i <
      2 \pi$.}
\label{sm++_f5} 
\end{figure}

\begin{figure}[tbp] 
    \postscript{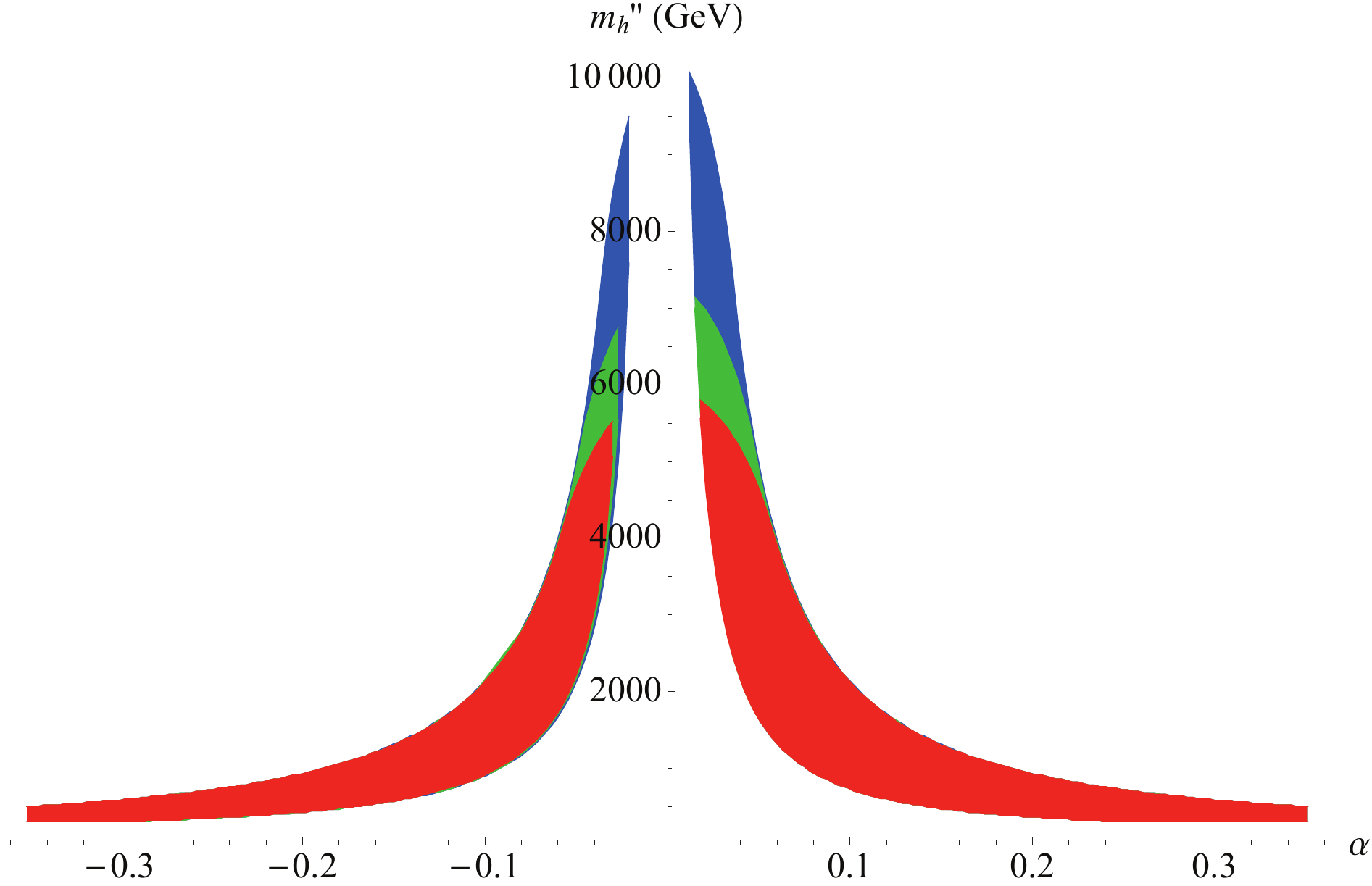}{0.8}
    \caption[Variation of SM$^{++}$ Stability Regions with Respect to $M_{Z''}$]{ \sglspc \small Variation of SM$^{++}$ vacuum stability regions with
      $M_{Z''}$. We have taken $M_s = 10^{16}~{\rm GeV}$, $M_{Z''} =
      3.5~{\rm TeV}$ (red), $M_{Z''} = 4.5~{\rm TeV}$ (green), and
      $M_{Z''} = 6.0~{\rm TeV}$ (blue). The perturbative upper bound
      is defined by $\lambda_i < 2 \pi$.}
  \label{sm++_f6}
\end{figure}

\begin{figure}[tbp] 
    \postscript{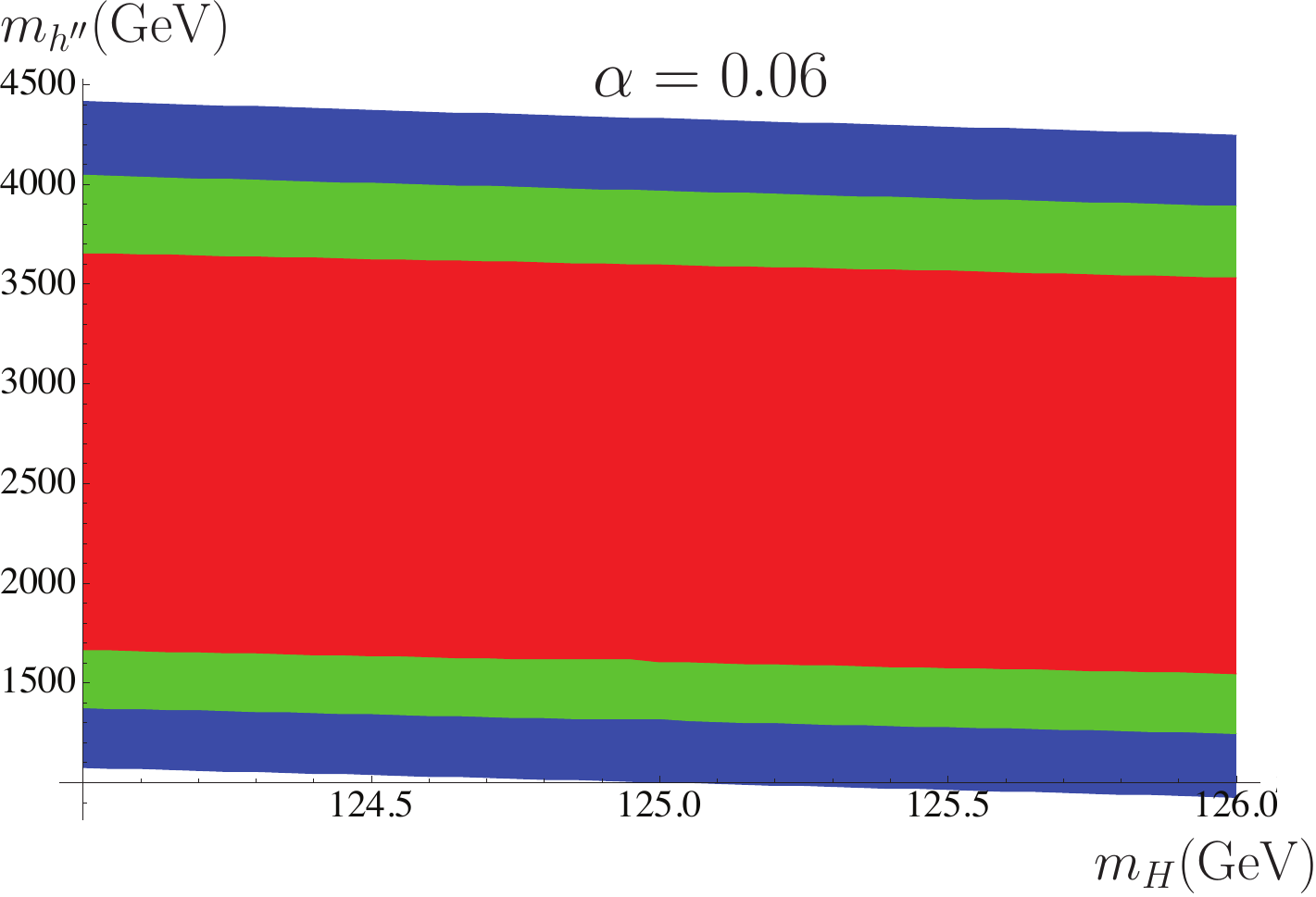}{0.8} 
\caption[Variation of SM$^{++}$ Stability Regions with Respect to $M_s$]{ \sglspc \small Variation of SM$^{++}$ vacuum stability regions with
    $m_H$. We have taken $\alpha = 0.06$, $M_{Z''} = 4.5~{\rm TeV}$,
    $M_s = 10^{14}~{\rm GeV}$ (blue), $M_s = 10^{16}~{\rm GeV}$
    (green), $M_s = 10^{19}~{\rm GeV}$ (red).  The perturbative upper bound is defined by $\lambda_i < 2 \pi$.}
  \label{sm++_f7}
\end{figure}

\begin{figure}[tbp] 
    \postscript{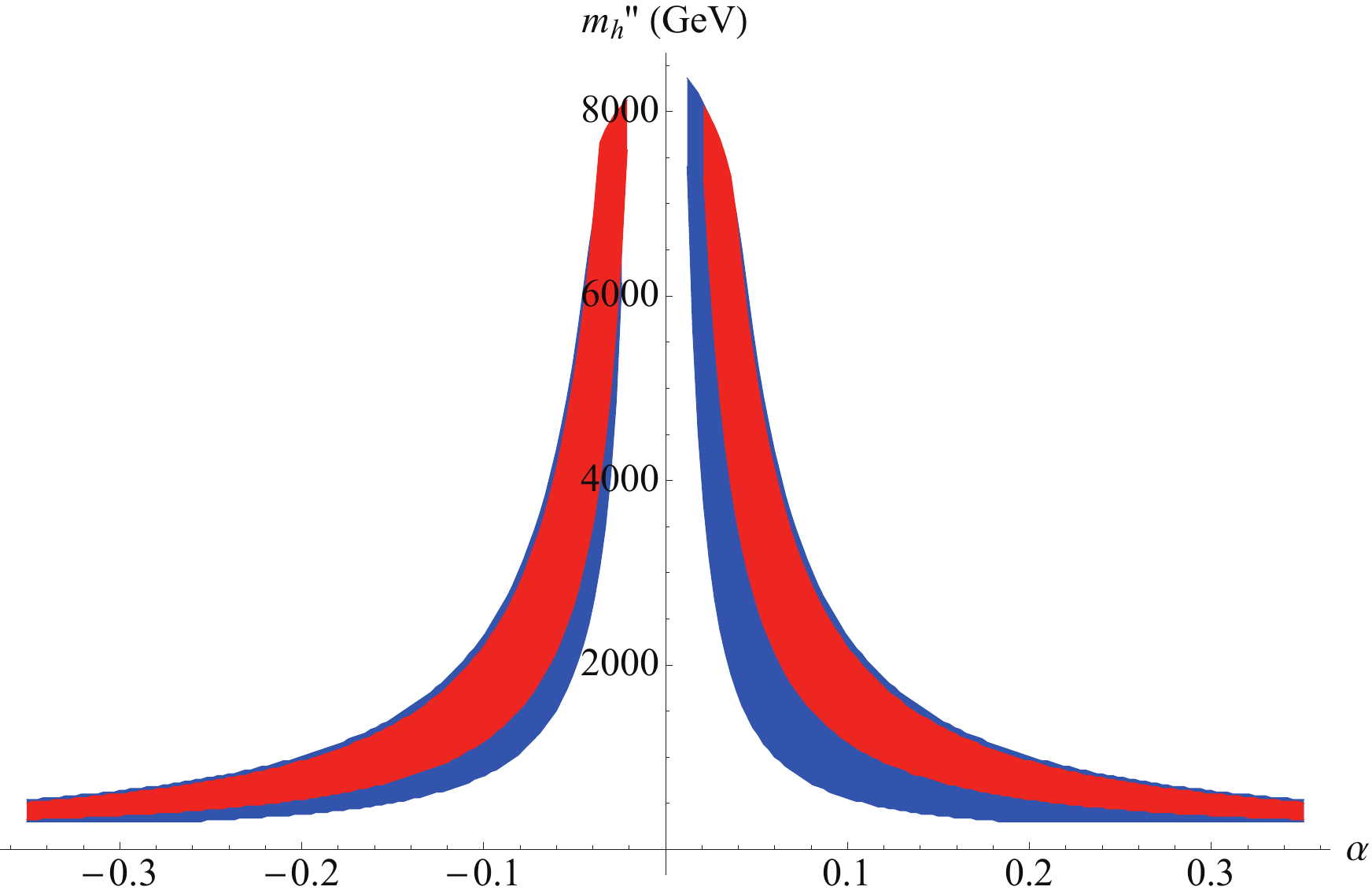}{0.8} 
    \caption[Variation of SM$^{++}$ Stability Regions with Respect to $g_1'(M_s)$]{ \sglspc \small Variation of SM$^{++}$ vacuum stability regions with
      $g'_1(M_s)$. The stable regions correspond to $g_1'(M_s) = 1.000$
      (blue), $g_1'(M_s) =  0.232$ (red). We have taken $M_s =
      10^{14}~{\rm GeV}$, $M_{Z''} = 4.5~{\rm TeV}$, $m_H = 125~{\rm
        GeV}$. The perturbative upper bound is defined by $\lambda_i
      < 2 \pi$.}
  \label{sm++_f8}
\end{figure}

For $M_s = 10^{14}$~GeV, we perform a scan of $10^4$ random values
of $(M_{Z''},\, \alpha,\, m_{h''})$ points, and using (\ref{12}) we
obtain the initial conditions ($\lambda_1^{(0)}$, $\lambda_2^{(0)}$,
$\lambda_3^{(0)}$) at the $Q = m_{h''}$ value,  after which we integrate the RG equations (\ref{RG}). 
For each set of random points, we verify that the positivity condition (\ref{VernonGabePaul}) 
is fulfilled all the way to the $M_s$ scale. The $10^4$ trials are
duplicated for $M_s= 10^{16}$ and $M_s = 10^{19}$~GeV.  The results
are encapsulated in Figs.~\ref{sm++_f2} to \ref{sm++_f8}. 
%-----FIG----------------------
%-----FIG----------------------
Figure~\ref{sm++_f2} shows the entire scan for $M_s =
10^{14}~{\rm GeV}$ and $M_{Z''} = 4.5~{\rm TeV}$. The points yielding a
stable vacuum solution up to $M_s$ are blue-printed, those leading
to unstable vacuum solutions are red-printed, and points giving
runaway solutions are purple-printed.  A stable vacuum solution is
one in which the positivity condition (\ref{VernonGabePaul}) is
fulfilled all the way to $\Lambda = M_s$. An unstable solution is
one in which the stability conditions of the vacuum ($\lambda_1 > 0$,
$\lambda_2 > 0$, $\lambda_1 \lambda_2 > \lambda_3^2/4$) are
violated (recall that for the case $\lambda_3 >0$ there is no need
to impose the third condition in (\ref{VernonGabePaul}) at all
scales, but only in the vicinity of $m_{h''}$). A runaway solution
is one in which the RG equations drive the Higgs doublet
self-coupling to non-perturbative values, thus invalidating the p-theory results. 
The perturbative upper bound (sometimes referred to as `triviality' bound) is
given by $\lambda_1 < 2 \pi$, so that at any point in the RG evolution of the $\lambda_1(t)$ parameter, the triviality bound is violated we take $\lambda_1$ as runaway at that point~\cite{Ellis:2009tp}.  
The vacuum stability condition is driven by the behavior of $\lambda_1$,
and actually is largely dominated by the initial condition
$\lambda_1^{(0)}$. Indeed, if the extra gauge boson $Z''$ gets its
mass through a non-Higgs mechanism and the scalar potential
(\ref{higgsV}) is that of SM ({\em i.e.} when  $v_2= \lambda_2 =
\lambda_3 = 0$), the RG evolution becomes that of SM and there
are no stable solutions.\footnote{ \sglspc \small Of course, even if $v_2 =
\lambda_2 = \lambda_3 = 0$, with an extra gauge boson the RG
evolution of $\lambda_1$ is not exactly that of SM, see
(\ref{RG}).} 

To determine the range of initial conditions on $\lambda_1^{(0)}$
yielding stable vacuum solutions, we fit the boundaries of the blue
band in the scatter plot. The resulting curves, which are shown as
  dashed lines in Fig.~\ref{sm++_f2}, correspond to $0.16 <
  \lambda_1^{(0)} < 0.96$ when $\alpha <0$, and $0.15 < \lambda_1^{(0)}
  < 0.96$ when $\alpha >0$. The lower limit of $\lambda_1^{(0)}$, which
  defines the boundary between stable and unstable solutions, is close
  to the value required for vacuum stability of the SM potential, as
  shown in (\ref{1}). Specifically, by substituting $m_h = 130~{\rm GeV}$ and
  $\alpha = 0$ in (\ref{12}) we obtain $\lambda_1^{(0)} = 0.14$.  The
  similarities between the minimum value of $m_H$ that allows absolute
  stability up to the Planck scale within SM and the minimum value of
  $m_h$ in the decoupling limit of (\ref{12}) reinforces the previous
  statement concerning the strong dependence of the RG evolution with
  the initial condition $\lambda_1^{(0)}$.  The average value of the initial condition $\lambda_1^{(0)}$ can be performed 
	through a fit to the blue points in the scattered plot. The result, which is
  shown as solid lines in Fig.~\ref{sm++_f2}, corresponds to
  $\langle \lambda_1^{(0)} \rangle = 0.28$. 
%-----FIG----------------------
%-----FIG----------------------	
	
	The behavior of $\lambda$, together with the typical behavior of $\lambda_1$ and $\lambda_2$
  for the average value of the initial condition $\langle
  \lambda_1^{(0)} \rangle$, are shown in Fig.~\ref{sm++_f3}. Note that
  $\lambda_1$ heads towards the instability and reaches a minimum
  greater than zero; thereafter it rises towards the Landau point (divergence).  This
  behavior is characteristic of models with scalar
  singlets~\cite{Kadastik:2011aa}. Also shown in Fig.~\ref{sm++_f3}
  is the typical behavior of $\lambda_1 \lambda_2 -\lambda_3^2/4$ for
  $\alpha <0$ and $\langle \lambda_1^{(0)} \rangle = 0.28$.
%-----FIG----------------------
%-----FIG----------------------
 The asymmetry between $\pm \alpha$ appears to be
 small on Fig.~\ref{sm++_f2}, but it is actually not insignificant. For a given
 $\alpha$, the lower boundary sometimes changes by a factor of two. For example, at $\alpha = 0.24$, the lower boundary
 changes from $6,140~{\rm GeV}$ to $3,160~{\rm GeV}$ as seen in Fig. \ref{sm++_f4}. However, the effect on the area is less
    noticeable. The reason is that we can only change the lower
    boundaries of the accepted parameter space. The upper boundary is
    determined by the constraint that $\lambda_i$ (usually $\lambda_2$)
    remains perturbative. This constraint is symmetric with respect to
    $\alpha$. So the area cannot be enlarged indefinitely. Even if
    somehow we can send the lower boundary to zero, the area would only
    increase by another 20\%  to 30\%.
%-----FIG----------------------
%-----FIG----------------------
To determine the sensitivity of the RG evolution with respect to
  the choice of the string scale, the analysis is duplicated for $M_s =
  10^{16}~{\rm GeV}$ and $M_s = 10^{19}~{\rm GeV}$. The contours displayed
  in Fig.~\ref{sm++_f5} (for $M_{Z''} = 4.5~{\rm TeV}$) show that the
  region of stable vacuum solutions shrinks as $M_s$ increases.  The
  allowed range of initial conditions with stable vacuum solutions
  therefore depends on the value of the string scale; {\it e.g.}  for
  $M_s = 10^{16}~{\rm GeV}$, the stability region is $0.17 < \lambda_1^{(0)} < 0.83$
  when $\alpha < 0$, and $0.16 < \lambda_1^{(0)} < 0.83$ when $\alpha
  >0$. For $M_s = 10^{19}~{\rm GeV}$, the stability region is $0.18 <
  \lambda_1^{(0)} < 0.69$ when $\alpha<0$, and $0.17 < \lambda_1^{(0)} <
  0.69$ when $\alpha >0$. The corresponding average value for $M_s =
  10^{16}~{\rm GeV}$ is $\langle \lambda_1^{(0)} \rangle = 0.31$, and
  for $M_s = 10^{19}~{\rm GeV}$ is $\langle \lambda_1^{(0)} \rangle =
  0.32$.
%-----FIG----------------------
%-----FIG----------------------
In Fig.~\ref{sm++_f6} the sensitivity of the RG
evolution with respect to $M_{Z''}$ is displayed. For large values of $|\alpha|$ there is no
  variation in the contour regions. For $\alpha > -0.05$ and
  $\alpha < 0.06$ there are some small variances.  These small
  differences show the effect of the initial conditions of
  $\lambda_2^{(0)}$ and $\lambda_3^{(0)}$ on the evolution of the system.
%-----FIG----------------------
%-----FIG----------------------
Figure~\ref{sm++_f7} verifies that there is no significant variation of the
SM$^{++}$ vacuum stability regions within the $m_H$ uncertainty.  An example for $\alpha = 0.06$ and $M_{Z''} = 4.5~{\rm TeV}$
is displayed in Fig.~\ref{sm++_f7}.
%-----FIG----------------------
%-----FIG----------------------
Figure~\ref{sm++_f8} displays the variation of the
  results of the analysis with respect to varying $g'_1(M_s)$. It is clearly seen that for $0.232 < g'_1(M_s)
  < 1.000$ the dependence on $g_1'$ seems to be fairly
  weak.  The stability of SM$^{++}$ vacuum is then nearly
  independent of the $Z''$ branching fractions~\cite{luisnMe1}.

While it is true that the low energy effective theory discussed in this dissertation requires
  a high level of fine tuning, this can be explained by
  applying the anthropic landscape of string
  theory~\cite{Bousso:2000xa,Susskind:2003kw,Douglas:2003um}. Alternatively, 
  the fine tuning can be circumvented with a more complete broken SUSY
  framework.  Since in pure SUSY the vacuum is automatically stable,
  the stability analysis perforce involves the soft SUSY-breaking
  sector. Hence rather than simply searching for the Higgs
  self-coupling going negative in the ultraviolet, the stability
  analysis would involve finding the local and global minima of the
  effective potential in the multi-dimensional space of the
  soft-breaking sector~\cite{Claudson:1983et}. However, the Higgs mass
  range favored by recent LHC data may be indicative of high-scale
  SUSY breaking~\cite{Hall:2009nd}; perhaps near the high energy
  cut-off of the field theory, beyond which a string description
  becomes a necessity~\cite{Hebecker:2012qp}.

\section{Summary of the Results and Conclusions}

We have discussed the phenomenology of a Standard-like Model inspired by string theory, in which the gauge fields are localized on D-branes, whose intersection can give rise to chiral fermions. The energy scale associated with string physics is assumed to be near the Planck mass. To develop our program in the simplest way, we worked within the construct of a minimal model with gauge-extended sector $U (3)_B \times Sp (1)_L \times U (1)_{I_R} \times U (1)_L$. The resulting $U (1)$ content gauges the baryon number $B$, the lepton number $L$, and a third additional abelian charge $I_R$. All mixing angles and gauge couplings are fixed by rotation of the $U(1)$ gauge fields to a basis diagonal in hypercharge $Y$ and in an anomaly free linear combination of $I_R$ and $B-L$. The anomalous $Z'$ gauge boson obtains a string scale St\"uckelberg mass via a 4D version of the Green-Schwarz mechanism. To keep the realization of the Higgs mechanism minimal, we added an extra $SU(2)$ singlet complex scalar, which acquires a VEV and gives a TeV-scale mass to the non-anomalous gauge boson $Z"$. The model is fully predictive and can be confronted with dijet and dilepton data from LHC8 and, eventually, LHC14. We have  shown that $M_{Z"} \approx 3 - 4$~TeV saturates current limits from the CMS and ATLAS Collaborations. We have also shown that for $M_{Z"} \leq 5$~TeV, LHC14 will reach discovery sensitivity $\approx 5\sigma$. After that, we  derived the complete set of renormalization group equations at one loop order and we pursue a numerical study of the system to determine the triviality and vacuum stability bounds, using a scan of $10^4$ random set of points to fix the initial conditions. We have shown that, if there is no mixing in the scalar sector, the top Yukawa coupling drives the quartic Higgs coupling to negative values in the ultraviolet and, as for the SM, the effective potential develops an instability below the Planck scale. However, for a mixing angle $-0.35 \leq \alpha \leq -0.02$ or $0.01 \leq \alpha \leq 0.35$, with the new scalar mass in the range $500~{\rm GeV} \leq m_{h"} \leq 8~{\rm TeV}$, the SM$^{++}$ ground state can be absolutely stable up to the string scale. Our results are largely independent of TeV-scale free parameters in the model: the mass of the non-anomalous $U(1)$ gauge boson and its branching fractions.

%---------------------------------------------

\newcommand{\bx}{\bar{x}}

% End of macros

%--------------------------------------------Begin Document----------------------------

\newpage

\

\

\

\noindent\textbf{\Huge Part II:}

\

%---------------------------------Ch 3-----------------------------------------
\noindent\textbf{\huge Gravitational Waves From \\ Post-Inflationary Sources}
\addcontentsline{toc}{chapter}{Part II - Gravitational Waves From Post-Inflationary Sources}

\

\noindent\textbf{\huge $\ddot{h}^{\rm TT}_{ij} + 3 \frac{\dot{a}}{a}\dot{h}^{\rm TT}_{ij} -\frac{1}{a^2} \nabla^2 h^{\rm TT}_{ij} = \frac{16\pi}{M_{pl}^2} T^{\rm TT}_{ij}$}

\newpage

\thispagestyle{fancy}
\chapter{Gravitational Waves from the Early Universe}
\thispagestyle{fancy}
\pagestyle{fancy}

%---------------------------------Ch 3: Section 1-----------------------------------------

\section{Gravitational Waves Generated in the Early Universe}
\label{sec:GWStart}

  	In section~\ref{sec:FRW} we discussed how the universe on scales larger than $10^2$ Mpc best described by the flat, isotropic and homogenous, represented by the metric
\beq
ds^2 = -dt^2 + a(t)^2 d {\bf x}^2.
\eeq
In this section of the dissertation, we investigate what potential new discoveries can be made by observing perturbations of the FRW metric, and their sources.  

Specifically we want to investigate radiative (tensor) perturbations to the FRW metric known as gravitational waves.   Gravitational waves are generically predicted by GR. An excellent laboratory of GR is in the extreme locations in the universe where there are high concentrations of matter, and relativistic speeds, one such example being pulsar binaries.  The 1993 Nobel Prize awarded to Hulse and Taylor was for the study of a binary pulsar system, PSR B1913+16 and it's unique properties.  This particular binary system's orbital radius was found to be decaying at a rate in agreement with predictions of the decay via energy carried away by gravitational radiation with a discrepancy between observation and theory of $0.13 \pm 0.21 \%$~\cite{Hulse}.  The concept of gravitational waves is a unique resource for astronomers and cosmologists as all known data from the cosmos comes in the form of electromagnetic radiation propagating through the universe.  The disadvantage of the use of electromagnetic radiation is that it is strongly interacting compared to gravity so the information we receive has undergone secondary scatterings, effectively blurring or adding noise to what would be the most energetic and interesting sources.  Gravity being a comparatively weak with interactions suppressed to by the scale $M_{pl}^{-2}$, gravitons\footnote{The analogous particle form of gravitational waves as photons are to the particles of electromagnetic waves.} undergo significantly less secondary scattering events.  This leaves the radiated information from energetic sources to essentially propagate un-affected, allowing a probe of some of the most interesting astrophysical sources with unprecedented clarity.  Unfortunately this advantage is also its downfall as such a weak coupling to matter means it will be extremely difficult to detect.  
  The task of direct detection is at the forefront of current physics research projects such as Advanced LIGO, VIRGO, GEO 600, and TAMA 300.  These detectors are based on the concept of time of travel of light in a laser interferometer apparatus; LIGO for example, uses arm lengths of 4 km.  The large length is necessary as the effect of a passing gravitational waves is proportional to the arm length of the interferometer~\cite{Maggiore}. Many excellent texts have been written regarding gravitational waves and interferometer detection systems such as~\cite{Creighton},~\cite{Maggiore}.  We will be concerned with the form of theoretical signals from stochastic sources generated in the early universe, and not on the detectors themselves.
  
  As was stated earlier, the weak coupling of gravity allows one to detect radiation from deep within areas from which electromagnetic radiation cannot escape.  One such case of interest is the opaque photon matter plasma of the early universe.  Before the time of photon decoupling and the creation of the CMB ($t \approx \CO(10^6)$ years~\cite{Kolb:1990vq}), the photons are confined by the multiple scatterings of the various constituents of the plasma.  Because of this we cannot see earlier then the time of the CMB electromagnetically, similarly to why you cannot see far through a thick fog, photons originating from a further distance undergoes multiple scattering events thus not making it to your eyes.  Effectively this blinds us from any direct observations of the physics before the CMB era via photons.  To directly probe the physics before the CMB era we can turn to gravitational waves. We are motivated to search for clues for BSM physics at higher energies than we can produce in a lab. Specifically we saw that inflation occurs at times when the relevant energies are at the GUT scale $10^{16}$ GeV, therefore the time immediately following inflation presents a unique unexplored region of high energy physics that can in principle be probed by gravitational waves.
  
  The act of inflation itself is theorized to produce a stochastic background of gravitational waves
  ~\cite{Starobinsky:1979ty,Allen:1987bk}, unfortunately it is too weak to be directly detected with existing
instruments. However, the possibility remains that these primordial gravitational waves can be observed via B-modes in the cosmic  microwave background
(CMB)~\cite{Spergel:2006hy,Peiris:2006sj}  but doing so requires precise measurement of the polarization of the CMB.

The period directly following inflation does present several mechanisms for the creation of a stochastic gravitational wave background such as the reheating and preheating phase discussed in~\cite{Kofman:1994rk,Khlebnikov:1997di,Greene:1997fu,
  Kofman:1997yn,GarciaBellido:1997wm,GarciaBellido:1998gm,
  Easther:2006gt,Easther:2006vd,GarciaBellido:2007af,
  GarciaBellido:2007dg,Dufaux:2007pt,Easther:2007vj,Easther:2008sx,
Caprini:2007xq,JonesSmith:2007ne}, and
 cosmological phase transitions~\cite{Kosowsky:1991ua,Kosowsky:1992rz,Kosowsky:1992vn,Krauss:1991qu,Kamionkowski:1993fg,Felder:2000hj,Kosowsky:2001xp}.

\subsection{Perturbing the FRW Metric}

  In order to make predictions of the form and scale of the stochastic background spectrum coming from these sources, we require the Einstein field equations in the linear perturbation to the FRW metric approximation, specifically for tensor modes. The general analysis can be found in ref~\cite{Weinberg_cosmo}. The qualitative answer can be described in the following manner.  We make a small perturbation to the FRW metric of the form
\beq
g_{\mu \nu} = \bar g_{\mu \nu} + h_{\mu \nu}, \ {\rm Min}\{ \bar g_{\mu \nu} \} \gg h_{\mu \nu}, 
\eeq
where $\bar g_{\mu \nu}$ is the flat FRW background metric for which the over bar indicates quantities with respect to this background metric. In the above ${\rm Min}\{ \bar g_{\mu \nu} \}$ is the minimum value of the non-zero components of $\bar g_{\mu \nu}$.  The last condition ensures that $h_{\mu \nu}$ is a small perturbation to the FRW background.  The Einstein field equations,
\beq
R_{\mu \nu} - \frac{1}{2} g_{\mu \nu} R= \frac{8 \pi}{M_{pl}^2} T_{\mu \nu},
\eeq can be expanded to linear order in $h_{\mu \nu}$
\beqa
\bar R_{\mu \nu} + R^{(1)}_{\mu \nu}   - \frac{1}{2} \bar g_{\mu \nu} \bar R -\frac{1}{2} \lpa h_{\mu \nu}  \bar R +  \bar g_{\mu \nu} R^{(1)} \rpa &\approx& \frac{8 \pi}{M_{pl}^2} \lpa \bar T_{\mu \nu} + T^{(1)}_{\mu \nu} \rpa \ , \nonumber
\\
R^{(1)}_{\mu \nu}-\frac{1}{2} \lpa h_{\mu \nu}  \bar R +  \bar g_{\mu \nu} R^{(1)} \rpa &\approx& \frac{8 \pi}{M_{pl}^2} T^{(1)}_{\mu \nu},
\eeqa
with 
\beqa
R^{(1)}_{\mu \nu} &=& \frac{1}{2} \lpa \bar D^\alpha \bar D_\mu h_{\nu \alpha} + \bar D^\alpha \bar D_\nu h_{\mu \alpha} - \bar D^2 h_{\mu \nu} - \bar D_\nu \bar D_\mu h \rpa, \nonumber
\\
R^{(1)} &=& \bar g^{\mu \nu} R^{(1)}_{\mu \nu} - h^{\mu \nu} \bar R_{\mu \nu}, \nonumber \\
\eeqa
where $\bar D_\alpha$ is the covariant derivative operator with respect to the background $\bar g_{\mu\nu}$. All terms with $(1)$ superscript indicate the linear order perturbation of the quantity. Note that common notation for $T^{(1)}_{\mu \nu}$ is  $\pi_{\mu \nu}$, and corresponds to the anisotropic energy-stress tensor perturbation to the homogenous and isotropic background matter content that governs the Friedmann equations.  Following~\cite{Weinberg_cosmo} the perturbations to the FRW metric are decomposed into 
\beqa
h_{00} &=& - E, \nonumber \\
h_{i 0} &=& a \lpa \p_i F + G_i \rpa ,  \nonumber \\
h_{i j} &=& a^2 \lpa A \delta_{ij} + \p^2_{ij} B + \p_j C_i + \p_i C_j + D_{ij} \rpa, \ (i,j = 1,2,3) \ ,
\label{eq:DecompFRW}
\eeqa
along with the conditions
\beq
\p_i C_i = \p_i G_i = 0, \ \p_i D_{ij} =0, \ D_{ii} = 0
\eeq
This decomposition allows three separable cases of the perturbed Einstein field equations, the most important to us is the tensor perturbations $h_{i j}$, taking the form
\beq
\ddot{D}_{ij} + 3 \frac{\dot{a}}{a} \dot{D}_{ij} - \frac{1}{a^2} \nabla^2 D_{ij} = \frac{16 \pi}{M_{pl}^2} \pi^T_{ij},
\label{eq:TensPert}
\eeq
where $\pi^T_{ij}$ is the traceless part of the anisotropic stress. 
We see this is the wave equation in the FRW background of a spatially transverse ($\p_i D_{ij} = 0$) and spatially traceless ($D_{ii} = 0$) perturbation.  The question remains what are the values of the vectorial components, $h_{i 0}$, and scalar component, $h_{00}$ of the metric perturbation.  

To answer this question we note that the Einstein field equations posses coordinate invariance, one may say gravity possesses general gauge invariance to changes in the metric.  Interesting comparisons between the gauge symmetries of the SM and GR are found in~\cite{Ryder}.  We can use this gauge invariance to our advantage to find a gauge in which the scalar and vector components of the metric perturbation are zero, the so called transverse-traceless gauge.  The details of gauge transformation of the field equations can be found in ref~\cite{Weinberg_cosmo}.  The results are that by choosing an appropriate coordinate system $x_\mu \rightarrow x_\mu + \epsilon_\mu$ the metric perturbation is modified (gauge transformed) to
\beq
h_{\mu \nu} \rightarrow h_{\mu \nu} + \Delta h_{\mu \nu},
\eeq
where
\beqa
\Delta h_{ij} &=& -\p_j \epsilon_i - \p_i \epsilon_j + 2 a \dot{a} \delta_{ij} \epsilon_0, \nonumber \\
\Delta h_{i 0} &=& -\p_0 \epsilon_i - \p_i \epsilon_0 + 2 \frac{\dot{a}}{a} \epsilon_i, \nonumber \\
\Delta h_{00} &=& - 2 \p_0 \epsilon_0.
\label{eq:GaugeTran}
\eeqa
Above the system of equations (\ref{eq:GaugeTran}) is written in the coordinates for which the background is expressed as \beq \bar g_{\mu \nu} = {\rm diag}\{-1, a(t)^2,a(t)^2,a(t)^2 \}. \eeq  It should be noted that a change of gauge also changes the form of $\pi^T_{ij}$ as it must now be expressed in the same coordinates/gauge.  Given the gauge freedom of the field equations, we choose the transverse-traceless (TT) gauge such that
\beqa
h_{\mu 0} &=& 0, \nonumber \\
\partial_i D_{ij} &=&0,\nonumber \\
D_{ii}&=&0.
\eeqa
From this another change of coordinates is used that does not effect the transverse-traceless relations to change from cosmological/comoving time to that of conformal time via
\beq
d \eta = \frac{dt}{a},
\eeq
allowing us to write the fully perturbed metric in the transverse-traceless gauge by use of (\ref{eq:DecompFRW}) as
\beq
ds^2 = a(\eta)^2 \lpa -d\eta^2 + (\delta_{ij} + h^{\rm TT}_{ij})d x^i dx^j \rpa,
\eeq
where the perturbation is governed by Eq. (\ref{eq:TensPert}) expressed in conformal time
\beq
\ddot{h}^{\rm TT}_{ij} + 2\frac{\dot{a}(\eta)}{a(\eta)}
\dot{h}^{\rm TT}_{ij} -\nabla^2h^{\rm TT}_{ij} = \frac{16\pi}{M_{pl}^2} \pi_{ij}^{\rm TT},
\label{eq:DifEQGrav}
\eeq
where over dots indicate derivatives with respect to conformal time $\eta$.

The process of finding the transverse-traceless part of any tensor is accomplished by the use of a projection operator that gives the TT part of a tensor.  The TT projection operator is constructed first by forming the spatially transverse projection operator given as 
\beq
P_{ij}({\bf k}) = \delta_{ij} - \frac{k_i k_j}{k^2}, \ k_i P_{ij}({\bf k}) = 0.
\eeq
The transverse operator ensures that the projected tensor is transverse to the vector $k_i$.  Applying the projection operator to a rank 2 tensor creates a transverse to the vector $k_i$ version 
\beq
P_{im}({\bf k})P_{pj}({\bf k})A_{jm} = A^T_{ip} , \  k_i A^T_{ip} = k_p A^T_{ip} =  0. 
\eeq
Subtracting out the trace of the transverse tensor allows the formation of the TT projector of the form
\beqa
A^{TT}_{ij} &=& \lpa P_{i m}({\bf k}) P_{j n}({\bf k}) - \frac{1}{2} P_{ij}({\bf k}) P_{mn}({\bf k})\rpa A_{mn}, \nonumber \\
A^{TT}_{ij} &=& P_{i m, j n}({\bf k})  A_{mn}.
\label{eq:Projector}
\eeqa
Applied to the perturbed stress-energy tensor gives
\beqa
T^{TT}_{mn} &=& P_{i m, j n}({\bf k}) T_{mn} = P_{i m, j n}({\bf k}) (-a(t)^2 \delta_{mn} P + \pi_{mn}), \nonumber \\
&=& -a(t)^2 P_{i n, j n}({\bf k}) \ P + \pi^{TT}_{mn} , \nonumber \\
&=&  \pi^{TT}_{mn} .
\eeqa
Thus we may replace $\pi^{\rm T T}_{ij}$ in (\ref{eq:DifEQGrav}) by $T_{ij}^{\rm TT}$ as the only surviving term after the TT gauge transformation is $\pi^{\rm TT}_{ij}$.

\subsection{Energy From Gravitational Waves}
  When calculating stochastic backgrounds, it is useful to compute the energy density from the gravitational radiation in a given frequency range $k, k+dk$.  In order to extract this information the covariant expression for the energy tensor in gravitational waves as given in~\cite{MTW} is used
\beq
T^{gw}_{\mu \nu} = \frac{M_{pl}^2}{32 \pi} \langle \bar D_\mu H^{\rm TT}_{\alpha \beta} \bar D_\nu H^{\alpha \beta, {\rm TT}} \rangle \ ,
\label{eq:StressGW}
\eeq
where $H^{\rm TT}_{\mu \nu}$ is related to the full metric via $g_{\mu \nu} = \bar g_{\mu \nu}  + H^{\rm TT}_{\mu \nu}$.  For the case at hand $H^{\rm TT}_{i j} = a(\eta)^2 h^{\rm TT}_{ij}$.  In Eq. (\ref{eq:StressGW}) the brackets indicate a spatial average over several wavelengths of the gravitational waves.  The energy density seen by a comoving observer is given by via
\beq
\rho_{gw} = u^\mu u^\nu T^{gw}_{\mu \nu} \ ,
\eeq 
with $u^\mu$ the four-velocity of a comoving observer\footnote{An observer embedded in the FRW background metric, and without motion relative to the background}.  A comoving observer of the background FRW space-time is an observer with fixed spatial coordinates and 4-velocity given by $u^\mu = (1 , {\bf 0})$ in the comoving frame, and $u^\mu = (a^{-1} , {\bf 0})$ in the conformally flat frame.
From this we find the energy density to be
\beqa
\rho_{gw} &=& \frac{M_{pl}^2}{32 \pi a(\eta)^2} \sum_{ij} \langle \bar D_0 \lpa a^2 h^{\rm TT}_{i j} \rpa \bar D_0 \lpa a^{-2} h^{\rm TT}_{ij} \rpa \rangle, \nonumber \\ 
&=& \frac{M_{pl}^2}{32 \pi a(\eta)^2} \sum_{ij} \langle  \dot{h}^{\rm TT}_{i j}  \dot{h}^{\rm TT}_{ij} \rangle.
\label{eq:GWEdens}
\eeqa
To find the energy density per frequency interval, we make a spatial Fourier transformation
\beq
h_{ij}^{\rm TT}(\eta, {\bf x}) = \int  \frac{d^3 {\bf k}}{(2\pi)^3} h_{ij}^{\rm TT} (\eta, {\bf k}) e^{i \bf k \cdot x} \ . 
\label{eq:GravFour}
\eeq
Using Parseval's theorem~\cite{Gradshteyn} we can express (\ref{eq:GWEdens}) as an integral over the wavevector $\bf k$ as follows
\beqa
\langle  \dot{h}^{\rm TT}_{i j}  \dot{h}^{\rm TT}_{ij} \rangle &=& \frac{1}{V} \int_V    \dot{h}^{\rm TT}_{i j}  \dot{h}^{\rm TT}_{ij} d^3 {\bf x} \nonumber \\
&=& \frac{1}{V} \int_V \int  \frac{d^3 {\bf k}}{(2\pi)^3} \frac{d^3 {\bf k'}}{(2\pi)^3} e^{i \bf( k' +k) \cdot x} \dot{h}_{ij}^{\rm TT} (\eta, {\bf k'}) \dot{h}_{ij}^{\rm TT} (\eta, {\bf k}) d^3 {\bf x}  \nonumber \\
&=& \frac{1}{L_x L_y L_z}  \int  \frac{d^3 {\bf k}}{(2\pi)^3} \frac{d^3 {\bf k'}}{(2\pi)^3} \prod_{i=1}^3 \frac{2 \sin((k'+k)_i L_i) }{(k'+k)_i}  \dot{h}_{ij}^{\rm TT} (\eta, {\bf k'}) \dot{h}_{ij}^{\rm TT} (\eta, {\bf k}) d^3 {\bf x} \nonumber \\
&\approx& \frac{1}{V}  \int  \frac{d^3 {\bf k}}{(2\pi)^3} \dot{h}_{ij}^{\rm TT} (\eta, {\bf k}) \dot{h}_{ij}^{\rm TT} (\eta, -{\bf k}),  \nonumber \\
&\approx& \frac{1}{V}  \int  \frac{d^3 {\bf k}}{(2\pi)^3} | \dot{h}_{ij}^{\rm TT} (\eta, {\bf k}) |^2 \ , 
\eeqa
In the last line we used the fact that $h^{\rm TT}_{ij}(\eta,{\bf x})$ is real.  The  approximation in the fourth line, is valid when $(k'+k)_i << L_i$, where $L_i$ is the length of the $i^{\rm th}$ side of the volume in question (assuming a rectangular prism region).  Technically this is not always true, as the range of $k, k'$ extends from $-\infty$ to $\infty$, however the use of a Fourier transform in a finite volume $V$ should be replaced by a discrete Fourier transformation for which the last result is is an exact equality, and is the result of Parseval's theorem. This allows the expressions
\beqa
\rho_{gw} &=& \frac{M_{pl}^2}{32\pi a(\eta)^2} \frac{1}{V} \sum_{ij} \int  \frac{d^3 {\bf k}}{(2\pi)^3} | \dot{h}_{ij}^{\rm TT} (\eta, {\bf k}) |^2 \ , \nonumber \\
k \frac{d \rho_{gw}}{d k} &=& \frac{d \rho_{gw}}{d \log k} = \frac{M_{pl}^2}{32\pi a(\eta)^2} \frac{1}{V} \sum_{ij}   \frac{k^3}{(2\pi)^3} \int | \dot{h}_{ij}^{\rm TT} (\eta, {\bf k}) |^2  d\Omega_k.
\label{eq:GWEpF}
\eeqa
The use of $\exp( i \bf k \cdot x )$ and natural units, tells us that $k = \omega$ the angular frequency for the gravitational waves in the region $k,k+dk$.
Above we should view $ d \log k = d k/k$ as unit-less though technically $\log k$ does not make any physical sense.  This equation will be very useful in computation of the energy spectrum.  However before making use of Eq.  (\ref{eq:GWEpF}) we need a method to solve Eq. (\ref{eq:DifEQGrav}) for $\dot h^{\rm TT}_{ij}$.

\subsection{Solutions to the Field Equations}
  In order to solve Eq. (\ref{eq:DifEQGrav}) we cast it into a simplified form, by use of Fourier transformation, Eq. (\ref{eq:GravFour})
\beq
\ddot{h}^{\rm TT}_{ij}(\eta, {\bf k}) + 2\frac{\dot{a}(\eta)}{a(\eta)}
\dot{h}^{\rm TT}_{ij}(\eta, {\bf k}) + k^2 h^{\rm TT}_{ij}(\eta, {\bf k}) = \frac{16\pi}{M_{pl}^2} T_{ij}^{\rm TT}(\eta, {\bf k}), \ k^2 = \bf k \cdot \bf k ,
\label{eq:GravF_Dif}
\eeq
Solving Eq. (\ref{eq:GravF_Dif}) can be done numerically~\cite{Giblin}, using methods such as Runge-Kutta, Leap-Frog, and Implicit Euler,  or other numerical differential equation solvers.  The advantage of using one of these methods, is that in a self consistent calculation the scale factor $a(\eta)$ is unknown prior to simulation.  The fields that make up the matter content typically have complicated solutions that must be solved numerically.  This causes the form of $a(\eta)$ to be unknown as it depends on the matter evolution. Therefore numerical methods where $a(\eta)$ is not a known quantity is useful.  The use of numerical methods pervades the subject of stochastic gravitational backgrounds and is worth the time of understanding the complications within it as in Sec. \ref{sec:AGFM} we will explore a solution to a stability issue that arrises in some numerical methods.

When using a numerical method to solve a set of differential equations, there is always a balance between accuracy and speed of a simulation. To clarify, the speed of an algorithm is not considered in actual seconds or minutes of a clock but rather how many operations are performed, when fewer operations are performed the algorithm is considered faster.  Usually algorithms that are fast suffer from inaccuracies, however increasing the accuracy usually requires the use of more memory and thus increasing the number of operations.  To illustrate this trade off we will briefly review common, simple, methods of solving differential equations via numerical simulation.

In all numerical methods the full solution to some differential equation $dy/dt = f'(t,y(t))$ can be approximated by a method where you iteratively make small steps towards the final evaluation parameter, for example the final time of an evolution of a system of equations. In the simplest method, the Euler method, you advance your solution by assuming your step size is small enough that the final result error is small.  You iteratively find the value of your function $y$ by use a Taylor expansion
\beq
y( t + \delta t) \approx y(t) + f'(t,y(t)) \delta t + \CO\lpa \frac{1}{2} f''(t, y(t)) \delta t^2 \rpa,
\label{eq:EulerMethod}
\eeq
Above we're now using big $\CO$ notation that indicates the order of the dominant error comes from.  Typically the terms $\frac{1}{2} f''(t,y(t))$ are dropped, and it is understood that some over all constant is involved when writing the order of the solution. 

The method in (\ref{eq:EulerMethod}) is the {\it explicit} Euler first order method.  This means that each progressive step depends on the step before it \eg you explicitly use the previous iteration in the next.  It is first order even though each step has error $\CO(\delta t^2)$, the accumulated error is of order $\CO(\delta t)$. This results from dividing the interval of interest, which we will call $t: (a, b)$, into $N$ discrete steps of size $\delta t$ the accumulated error is then estimated as
\beq
E = {\rm Max}\{\frac{1}{2} f''(t,y(t)) \} N \delta t^2 =  \frac{ (b-a) }{2}  {\rm Max}\{f''(t,y(t)) \}\delta t \simeq \CO\lpa \delta t\rpa,
\eeq
where $ {\rm Max}\{f''(t,y(t)) \}$ is the maximal value of $f''(t,y(t))$ in the interval of $(a,b)$.  The error estimate in first order methods scales with the step size $\delta t$.  Obviously to make the error smaller we should decrease the step size $\delta t$, however this requires us to increase $N$, requiring more computational time.   To increase the accuracy with the same number of steps $N$, or less, methods like RK second order algorithms are used. However, by increasing the accuracy we require more storage of additional variables.  Note that in terms of physical time (seconds, hours, etc.), if a computer can store additional variables more rapidly than it can run through serial operations (iterations), then there is little disadvantage to these types of methods.  
  To quantitatively understand how second order algorithms work, we will use the example of an RK method.  Runge-Kutta methods employ the concept that evaluating the term $f'(t,y(t))$ at some intermediate point(s) between $t$ and $t+\delta t$ we decrease the error of each step.  We can generically express this idea for a RK $s^{\rm th}$ order term by 
  \beqa
  y(t+\delta t) &=& y(t) + \sum_{m = 1}^s b_m k_m  + \CO \lpa \delta t^{s+1} \rpa\nonumber \\
  k_1 &=& \delta t f'(t,y(t)) \nonumber \\
  k_2 &=& \delta t f'(t + c_2 \delta t, y(t) + a_{21} k_1) \nonumber \\
 & \vdots& \nonumber \\
 k_s &=& \delta t f' \lpa t + c_s \delta t, y(t) + \sum_{m = 1}^s a_{s m} k_m \rpa
 \eeqa
 For the $2^{\rm nd}$ order RK algorithm we have
 \beq
   y_{n+1} = y_n + \delta t  \lpa b_1 f'(t,y(t))  + b_2  f'(t + c_2 \delta t, y(t) + a_{21} k_1) \rpa  + \CO \lpa \delta t^3 \rpa ,
   \label{eq:RK2nd}
 \eeq
where the short hand $y_n = y(t)$ and $y_{n+1} = y(t+\delta t)$ is used.  We solve for $b_i, c_2, a_{21}$ via Taylor expansion to the same order and comparing the results.  First expand (\ref{eq:RK2nd}) to give
\beqa
 y_{n+1} &=& y_n + \lpa  b_1 f'_n +  b_2  f'_n \rpa \delta t  \nonumber \\
 &\ &+\lpa a_{21} b_2 \frac{\p f'}{\p y}  f'_n  + c_2 b_2 \frac{\p f'}{\p t} \rpa \delta t^2   + \CO \lpa \delta t^3 \rpa.
\label{eq:RK2Expand}
\eeqa
Now we expand the general solution of the Taylor series
\beqa
y_{n+1} &\approx& y_n + f'_n \delta t +\frac{1}{2} \frac{d f'}{d t} \delta t^2 + \CO(\delta t^3), \nonumber \\
&\approx& y_n +  f'_n \delta t +\frac{1}{2} \lpa \frac{\p f'}{\p t}  + \frac{\p f'}{\p y} f'_n \rpa \delta t^2 + \CO(\delta t^3).
\label{eq:TaylorExpand}
\eeqa
Comparing (\ref{eq:TaylorExpand}) and (\ref{eq:RK2Expand}) allows one to solve $b_i, c_2, a_{21}$ via
\beqa
b_1 + b_2 & =& 1, \nonumber \\
a_{21} b_2 &=& \frac{1}{2}, \nonumber \\
c_2 b_2 &=& \frac{1}{2}.
\eeqa
One common solution is $b_1 = 0, \ b_2 = 1, \ a_{21} = c_2 = 1/2$.  This results in
\beq
y_{n+1} \approx y_n + \delta t  f' \lpa t + \frac{ \delta t}{2} , y_n + \frac{ \delta t}{2} f'_n \rpa + \CO(\delta t^3).
\label{eq:MidPt}
\eeq
Equation (\ref{eq:MidPt}) is one solution to the RK 2nd order solutions, this particular solution is known as the mid-point method.  Which has the explanation that the step term is evaluated at the mid-point between $t$ and $t+\delta t$.  From this exercise it is important to note that in (\ref{eq:MidPt}) we require two evaluations of $f(t,y(t))$ and thus increases the number of operations.  We will continue to explore further complications with numerical methods in the next sections.

An attractive alternative to using numerical differential equation solving algorithms is the use of Green's functions solutions to (\ref{eq:GravF_Dif}).  For known scale factors of the form
\beq
a(\eta) = \alpha \eta^n,
\eeq
with $\alpha$ and $n$ constant, it was realized in~\cite{Price:2008hq} that (\ref{eq:GravF_Dif}) has the solutions
\ \\
\\
$h^{\rm TT}_{ij}(\eta, {\bf k}) = $ \sglspc \\
\beqa 
\frac{16\pi}{M_{pl}^2}  \frac{k}{\eta^{n-1}} \int_{\eta_i}^{\eta_f} d \eta' \ (\eta')^{n+1} \lpa j_{n-1}(k \eta') y_{n-1}(k \eta) - j_{n-1}(k \eta) y_{n-1}(k \eta') \rpa \ T^{\rm TT}_{ij}(\eta', {\bf k}). \nonumber \\ 
\label{eq:GreensFuncs}
\eeqa \dblspc
where $j_n(x)$ and $y_n(x)$ are spherical Bessel functions of the first and second kind, respectively. Here $\eta_i$ is the starting time of the source, and $\eta_f$ is the final time at which $h_{ij}^{\rm TT}$ is evaluated, ordinarily the time after which the emission of gravitational waves is negligible. The advantage of (\ref{eq:GreensFuncs}) is that it is exact.  In practice we must evaluate the integral in (\ref{eq:GreensFuncs}) numerically, however integrals evaluated numerically do not inherently suffer from instability issues as numerical differential equation methods do.  A disadvantage of using the Green's function approach is that we require the form of $a(\eta)$ before evaluation of (\ref{eq:GreensFuncs}).  In section \ref{sec:AGFM} it will be shown that this may present an issue when attempting to simulate physics of pre-heating.

\subsection{Sources of Anisotropic Stress}
  With the equations to solve for the energy density per unit frequency, and a method for solving for $\dot h^{\rm TT}_{ij}$, we still need to calculate the source term $T^{\rm TT}_{ij}(\eta, {\bf k})$ and the scale factor $a(t)$.  The scale factor is governed by the Friedmann equations for a flat background
\beq
\lpa \frac{\dot{a}(t)}{a(t)} \rpa^2 = \frac{8\pi}{3 M_{pl}^2} \rho, \ \frac{\ddot{a}(t)}{a(t)} = -\frac{4 \pi}{3 M_{pl}^2} \lpa \rho + 3 P \rpa ,
\eeq
where $\rho, \ P$ are the isotropic energy density, and isotropic internal pressure of the comoving perfect fluid respectively.  However we do not know what the particle content of $\rho$ and $P$ is. In order to understand these terms in the era immediately after inflation we need to express the isotropic energy density and pressure in terms of the relevant fields. Immediately after inflation ends, we assume we can use classical analysis of the fields in a quasi-classical setting.  To find expressions for $\rho$ and $P$ we will use a Lagrangian for $N$ real scalar fields assumed to be the dominant fields at the time in question. The inflaton is identified as $\Phi_0 = \phi(x,t)$ and fields $\Phi_i = \chi_i(x,t)$ represents fields which may cause pre-heating, as was outlined in Sec.~\ref{sec:PreHeat}, or be involved in phase transitions.  The Lagrangian appropriate to describe gravity and these fields is given by the Hilbert action~\cite{DHilbert}, note that for fields with non-integer spin it it may be more appropriate to use the Cartan-Einstein formulation of GR~\cite{Hehl:1976kj}
\beq
S = \int d^4 x \ \sqrt{-\bar g} \lpa  \frac{M_{pl}^2}{16 \pi} \bar R + \CL(\Phi_i) \rpa , \ \CL(\Phi_i) = \frac{1}{2} \sum_{i = 1}^N \bar g^{\mu \nu} \bar D_\mu \Phi_i \bar D_\nu \Phi_i - V(\Phi_i).
\label{eq:GLa} 
\eeq
Variation of (\ref{eq:GLa}) with respect to $\bar g^{\mu \nu}$ and $\Phi_i$ gives
\beqa
\bar R_{\mu \nu} - \frac{1}{2} \bar g_{\mu \nu} \bar R &=& \frac{8 \pi}{M_{pl}^2}  \lpa -\frac{2}{\sqrt{-\bar g}} \frac{\delta}{\delta \bar g^{\mu \nu}} \sqrt{-\bar g} \CL(\Phi_i) \rpa, \nonumber \\
 \p_\mu \p^\mu \Phi_i + 3 \frac{\dot{a}(t)}{a(t)} \dot{\Phi}_i &=& -  \frac{\p V}{\p \Phi_i}
 \label{eq:EQofMo}
\eeqa
which then allows the identification
\beq
T_{\mu \nu} = -\frac{2}{\sqrt{-\bar g}} \frac{\delta}{\delta \bar g^{\mu \nu}} \sqrt{-\bar g} \CL(\Phi_i) = -2 \frac{\delta \CL}{\delta \bar g^{\mu \nu}} + \bar g_{\mu \nu} \CL  . 
\eeq
With the given form of (\ref{eq:GLa}) 
\beq
T_{\mu \nu} = - \sum_{i = 1}^N \p_\mu \Phi_i \p_\nu \Phi_i   + \frac{1}{2} \bar g_{\mu \nu} \sum_{i = 1}^N \p_\kappa \Phi_i \p^\kappa \Phi_i - \bar g_{\mu \nu} V. 
\label{eq:SEFldTens}
\eeq
Now we can find the expressions for energy density $\rho$ and internal pressure $P$ by comparison with the stress-energy tensor of a perfect fluid
\beq
T_{\mu \nu} = \lpa \rho+ P \rpa u_{\mu} u_{\nu} + \bar g_{\mu \nu} P \rightarrow T_{00} = \rho+ 2 P, \ T_{ii} = -a(t)^{2} P \ \hbox{(no sum on $i$)}
\label{eq:PFlu}
\eeq
 and our expression for stress-energy in terms of fields, Eq. (\ref{eq:SEFldTens}), giving
\beq
\rho = \sum_{i = 1}^N  \lpa \frac{1}{2} \dot{\Phi}_i ^2 + \frac{1}{2 a^2}( \nabla_x \Phi_i)^2+ V \rpa, \ P =  \sum_{i = 1}^N \lpa  \frac{1}{2} \dot{\Phi}_i ^2 + \frac{1}{2 a^2}( \nabla_x \Phi_i)^2 - V  \rpa \ ,
\eeq
where above $\nabla_x \Phi_i ^2$ is the spatial derivatives of the field squared, to clarify in cartesian coordinates it takes the form $ \nabla_x \Phi_i ^2= \p_x \Phi_i^2+ \p_y \Phi_i^2+\p_z \Phi_i^2$.
The expression for $\rho$ and $P$ is not in general isotropic.  To separate the isotropic parts of $\rho$, and $P$ and the perturbations to the perfect fluid, we define the isotropic part as $\rho_{\rm iso} = \langle \rho \rangle$, and $P_{\rm iso} = \langle P \rangle$ where here $\langle  \dots \rangle$ indicate a spatial average over a large enough volume such that perturbations resulting from the terms $\nabla_x \Phi_i$ average to zero, that is we take a course grained version of the energy density and pressure to determine the isotropic contributions.  We can find the anisotropic stress, the source of gravitational waves, by applying the TT operator (\ref{eq:Projector}) to the field stress energy tensor (\ref{eq:SEFldTens}).  After the TT operator is applied terms proportional to the background metric $\bar g_{\mu \nu}$ will be completely zero thus the only anisotropic parts of the fields is from the term $-\sum_{i=1}^N \p_\mu \phi_i \p_\nu \phi_i$ that is fields with spatial gradients will source gravitational waves. These spatial variances in-fact grow in the pre-heating process which results in a stochastic background of gravitational waves.
  
  It should be noted that different types of the matter fields can be included in $\CL(\Phi_i)$ here, however, we will keep strictly to scalar fields. The evolution of the stress-energy tensor $T^{\mu \nu}$ under (\ref{eq:EQofMo}) is complicated, as is the solution to the scale factor $a(t)$.  In order to make progress on these generally intractable problems we again resort to numerical methods, now for solving the field Eq. (\ref{eq:EQofMo}) and scale factor $a(t)$ evolutions.  
  
\subsection{Lattice Simulations}  
  Since the matter field equations are a system of $N$ partial differential equations one convenient method of solving these numerically is through finite differencing methods~\cite{NumRec}.  Finite differencing methods solve partial differential equations by discretizing the field on a lattice, for which in $3+1$ dimensions takes the form 
\beq
\Phi(t,{\bf x}) \rightarrow \Phi(t, {\bf x_0 }+ {\bf \hat{n}} \Delta x), \ {\bf \hat{n}} =  n \ {\bf \hat{e}_1}+ m \ {\bf \hat{e}_2} +  p \ {\bf \hat{e}_3},
\eeq
with $(n,m,p)$ Integers denoting the 3 dimensional location on the lattice along the directions ${\bf \hat{e}_i}$, which are unit vectors in the separate directions comprising the lattice, finally $\Delta x$ is the discrete lattice point spacing.
  We can solve for the field configurations at the lattice points for successive iterations of some fixed variable, typically time, in physical applications. The matter field equations we have in (\ref{eq:EQofMo}) fully expanded yield
\beq
\frac{\p^2 \Phi_i}{\p t^2} - \frac{1}{a(t)^2} \nabla_{\bf x}^2 \Phi_i + 3 \frac{\dot{a}(t)}{a(t)} \dot{\Phi}_i = - \frac{\p V}{\p \Phi_i}.
\label{eq:Dif2}
\eeq
We can approximate the spatial derivatives in the above as finite differences, which can be expressed in terms of a Taylor series expansion of the form
\beqa
\Phi_i(t,x_1+\Delta x,x_2,x_3) &\approx& \Phi_i(t,{\bf x}) + \p_1   \Phi_i(t,{\bf x}) \Delta x + \frac{1}{2} \p^2_1 \Phi_i (t,{\bf x}) \Delta x^2 , \nonumber \\
\Phi_i(t,x_1-\Delta x,x_2,x_3) &\approx& \Phi_i(t,{\bf x}) - \p_1   \Phi_i(t,{\bf x}) \Delta x + \frac{1}{2} \p^2_1 \Phi_i (t,{\bf x}) \Delta x^2 , \nonumber \\
\p_1 \Phi_i (t,{\bf x}) &\approx&  \frac{\Phi_i(t,x_1+\Delta,x_2,x_3)  - \Phi_i(t,x_1-\Delta,x_2,x_3)}{2 \Delta x} - \CO( \Delta x^2) , \nonumber \\
\eeqa 
and
\beqa
 \p^2_1 \Phi_i (t,{\bf x}) =  \Delta_1^2 \Phi_i &\approx& \frac{\Phi_i(t,x_1+\Delta,x_2,x_3)  - 2 \Phi_i(t,x) + \Phi_i(t,x_1-\Delta,x_2,x_3)}{\Delta x^2} \nonumber \\
 &-& \CO( \Delta x^2). 
\label{eq:FDscheme}
\eeqa
We then use a numerical method to iteratively step forward in the fixed variable $\Phi_i(t + \delta t, {\bf x})$ at each lattice point $\bf x$. This must be done for each lattice point which in a $3+1$ dimensional field theory we have $N^3$ terms to evaluate $\p^2 \Phi_i / \p t^2$ for a lattice consisting of $N$ points per orthogonal direction.  We must update each point, and we need the first order derivatives to compute the anisotropic stress.  For this reason we require storing both first and second order derivatives at each point thus doubling the number of stored points $2 N^3$.
  In the next section we address the issues associated with lattice simulations.

\subsection{\label{sec:CompStability} Computational Stability and Discrete Fourier Transforms}
 When doing lattice simulations, there are several effects that we must account for.  The first of these is we handle the finite nature of the lattice when simulating an infinite space, specifically, what do we do at the lattice boundaries? The boundaries of the lattice can be fixed, periodic, or free.  For the simulation considered periodic is a natural choice, because the lattice can be viewed as a representative volume of the total universe. The isometric and homogenous assumption of the universe, allows the interpretation that effects occurring on the boundaries re-entering the lattice on the opposite side can be understood as a field configuration entering the lattice from the next identical volume element in the universe.  The periodic condition can be expressed as
\beq
\Phi_i(t,{\bf x} + {\bf \hat N_j} \Delta x) = \Phi_i(t, {\bf x}) , \ {\bf \hat N_j} = N {\bf \hat e_j}.
\eeq

Using a discrete lattice also requires careful consideration of Fourier transformations, for which we require to calculate (\ref{eq:GWEpF}).  The Sampling Theorem does not allow one to find a Fourier amplitude of frequencies higher than the Nyquist critical frequency~\cite{NumRec} given as
\beq
f_c = \frac{1}{2 \Delta x} = \frac{N}{2} \frac{1}{L},
\eeq
with $L$ the size of one lattice side $L = N \Delta x$.  Discretizing the Fourier transform of the fields that form the stress-energy tensor perturbation gives
\beq
\Phi_i(t, {\bf k}) = \int d^3 x \Phi_i(t,  {\bf x}) e^{-i {\bf k \cdot x}} \rightarrow \Phi_i(t, {\bf \hat n_k}\Delta k) \approx \Delta x^3 \sum_{n,m,p=1}^N \Phi_i(t, {\bf \hat n} \Delta x) e^{-i {\bf \hat n_k \cdot \hat n} \Delta k \Delta x},
\label{eq:DFT}
\eeq 
where $\Delta k = 1/L$ and ${\bf \hat n_k} =  n_k \ {\bf \hat{e}_{1k}}+ m_k \ {\bf \hat{e}_{2k}} +  p_k \ {\bf \hat{e}_{3k}},$ with ${\bf \hat{e}_{ik}}$ is a unit vector in the k-space lattice where $n_k,m_k,p_k$ each range over $-N/2$ to $N/2$. 
  
  It is useful to see how computationally demanding Fourier transforms are. For each k-space lattice point, of which there are $N^3$, we must calculate (\ref{eq:DFT}) which itself is an order $N^3$ operation.  To obtain the entire discrete Fourier transformed (DFT) field it requires $N^3 \times N^3$ operations.  Taking Fourier transforms and simultaneously solving the differential equations as stated previously, each step in the system is an order $2 N^3 \times N^3 \times N^3$ operation, even further complicated by the fact that each Fourier mode is complex which requires us to store $2$ variables for each Fourier mode, compounding the entire process by $2$ leaving an overall operation of order $4 N^9$.  We can see even for small lattice sizes of $N = 32$ at each step we have approximately $1.4 \times 10^{14}$ operations to perform! Because of this reason it is rare that someone would use a DFT. An advanced algorithm known as Fast Fourier Transforms (FFT) for which a review of the most basic form can be found in~\cite{NumRec}, allows faster execution of Fourier transformations.  These algorithms change DFT from being a $\lpa N \times N \rpa^3$ operation to a $\lpa N \log_2 N  \rpa^3$ operation saving us many operations.  We also cut down on the number of operations with the use of symmetry relations, such as with the stress-energy tensor $T^{\rm TT}_{ij} = T^{\rm TT}_{ji}$.
  
  The final issue we will address associated with use of simulation is stability.  If a method to solve differential equations is insensitive to changes in the discrete stepping size $\Delta x$,  it is said to be stable.  To illustrate, consider the differential equation
\beq
\frac{dy}{dx} = - \alpha y,
\label{eq:StabExamp}
\eeq
where $\alpha$ is a positive real constant.  The solution to (\ref{eq:StabExamp}) is $y(x) = A e^{- \alpha x}$ with $A$ some arbitrary constant to be fit with initial conditions. Any good numerical method should return a result near the exact solution, however this type of equation suffers from instability, for example if we use the explicit $1^{\rm st}$ order Euler method, one iteration gives
\beq
y(x + \delta x) = y(x)(1 - \alpha \delta x),
\eeq
This result will result in a perfectly acceptable numerical solution so long as $1/\alpha <  \delta x$. This condition implies that each progressive step satisfies $y(x+\delta x ) < y(x)$ as the full solution agrees with.  However if $1/\alpha  > \delta x$ then $y(x+\delta x) >y(x)$ and each step increases the value of $y(x)$. This results in an unbounded solution that will go to infinity if the final $x$ parameter value is sufficiently large.  These are known as run away solutions.  Obviously this behavior is not the intended result, we now have the surprising result that a choice of step-size can drastically alter the numerical results. 

In order to combat step size sensitivity implicit methods may be used. To demonstrate, the implicit $1^{\rm st}$ Euler method, applied to (\ref{eq:StabExamp}) results in the ability to use any step size to get the correct numerical behavior,  as seen below\beqa
y_{n+1} &=& y_n + f'(x+\delta x, y_{n+1}) \delta t, \nonumber \\
y_{n+1} &=& y_n - \alpha y_{n+1} \delta t, \nonumber \\
y_{n+1} &=& \frac{y_n}{1+\alpha \delta t}.
\label{eq:ImpEul}
\eeqa
Each step gives $y_{n+1} < y_n$ as expected, regardless of step size.  In implicit methods we must solve an algebraic equation for $y_{n+1}$ which may not in general be solvable by analytical methods, nor result in a stable method. 

  Finite differencing methods can also be unstable, the analysis of which goes under the name von Neumann stability analysis~\cite{NumRec}.  We deduce requirements on the lattice spacing by analyzing the discrete Fourier modes of the numerical solutions.  Expressing (\ref{eq:Dif2}) with finite differencing as well as choosing the finite difference described in (\ref{eq:FDscheme}) for both the time and spatial derivatives results in
 \beq
\Delta_t^2 \Phi_{n} = \frac{1}{a^2_n} \sum_{j=1}^3 \Delta^2_j \Phi_n - 3 \frac{\dot{a}_n}{a_n} \Phi_n - \frac{\p V}{\p \Phi_n},
\label{eq:FD}
\eeq
where above it is implicitly implied that $\Phi$ is $\Phi_i$ and we now call $\Phi$ any of the fields governed by the differential Eq. (\ref{eq:Dif2}).  Again we use the short hand $\Phi_n = \Phi_i(t,{\bf x})$ and $\Phi_{n+1} = \Phi_i (t + \delta t, {\bf x})$  as well we use the notation
\beq
F = 3 \frac{\dot{a}_n}{a_n} \Phi_n + \frac{\p V}{\p \Phi_n}.
\eeq
Expanding and rearranging (\ref{eq:FD}) gives 
\beq
\Phi_{n+1} = 2 \Phi_n  - \Phi_{n-1} + \frac{\delta t^2}{a^2_n} \sum_{j=1}^3 \Delta^2_j \Phi_n - \delta t^2 F_n. 
\label{eq:FSol}
\eeq
We can analyze the stability of this method by using the spatial Fourier transformation of the numerical solutions $\Phi_n$ and expressing the numerical solution as the exact solution plus some error, which in a finite volume $L^3$ takes the form
 \beq
 \Phi(t,{\bf x}) = \Phi_{\rm exact}(t, {\bf x}) + \frac{1}{V} \sum_{{\bf m} = -\infty}^\infty \xi_{\bf m}(t) e^{i \pi \bf{m}\cdot  {\bf x}/L^3} , \ {\bf m} = (n,m,p).
 \eeq 
 Applying this form of the fields on (\ref{eq:FSol}) gives
 \beq
\xi_{{\bf m},n+1}= 2 \xi_{{\bf m},n}  - \xi_{{\bf m},n-1} -\frac{4 \delta t^2}{a^2_n \delta x^2} \sum_{j=1}^3 \xi_{{\bf x},n} \sin^2 \lpa  \frac{\pi \Delta x_j}{2 V} \rpa - \delta t^2 \sum_{{\bf m} = -\infty}^\infty  e^{-i \pi \bf{m}\cdot  {\bf x}/L^3} F_n.
 \eeq
The exact part of the solution has dropped from the equations since it satisfies the differential equation to all orders.  For the sake of discussion we drop the possibly non-linear terms contained in $F_n$ as this complicates the analysis.  For stability we require that $|\xi_{{\bf m},n+1}/\xi_{{\bf m},n}| \leq 1$ the error does not grow unbounded step by step for any mode.  For the worst case scenario the error propagates from the previous step at the same level so $\xi_{{\bf x},n-1} = \xi_{{\bf x}, n}$ we then resolve
\beq
\left| 1-3 \frac{4 \delta t^2}{a^2_n \Delta x^2} \sin^2 \lpa  \frac{\pi \Delta x}{2 V} \rpa \right| \leq 1.
\eeq
To ensure absolute stability in this method requires
\beq
\sqrt{3} \frac{\delta t}{a_n \Delta x} \leq \frac{1}{2}.
\label{eq:Courant}
\eeq
This is the famous Courant condition for wave equations~\cite{NumRec}.  We can see however in (\ref{eq:Courant}) that this condition cannot be satisfied at all times since $a_n \leq 1$ because of the normalization $a({\rm today}) = 1$ so this form is completely unstable! We can avoid this issue by changing coordinates and scaling the fields, so that $1/a_n$ drops for this condition.  

  Felder and Tkachev, created a program {\sc LatticeEasy}~\cite{Felder:2000hq}, with all statements above were addressed.  The program {\sc LatticeEasy} performs self consistent calculations of the Friedmann equations and field evolutions while enforcing the Courant condition, and checking stability of the fields throughout the simulation. The fields and coordinates are scaled via
\beq
  x_{pr} = B x, \ \Phi_{pr} = A a^r \Phi, \  dt_{pr} = B a^s dt,
  \eeq
  where $B,\ A, \ r$, and $s$ are constants.
  These transformations change the field equations (\ref{eq:Dif2}) into
  \beqa
 & \Phi''_{pr} &+ (s - 2 r + 3) \frac{a'}{a} \Phi'_{pr} -  a^{-2s-2} \nabla^2_{pr} \Phi_{pr} - \lpa r(s-r+2)\lpa \frac{a'}{a} \rpa^2 + r \frac{a''}{a} \rpa \Phi_{pr} \nonumber \\
 & + &   \frac{\p}{\p \Phi_{pr}} \lpa \frac{A^2}{B^2} a^{-2 s + 2r} V \rpa =0,
  \eeqa
  where $\Phi', \ a'$ denote derivation with respect to program time $t_{pr}$.  In attempts to ensure stability of the field equations the constants $A, \ B, \ s, \ {\rm and}  \ r$ take specific values that depend on the dominant part of the potential which we assume takes the form 
  \beq
  V = \frac{\alpha}{\beta} \Phi^\beta.
  \eeq
First {\sc LatticeEasy} scales the fields to their initial values to prevent unnecessarily large or small field values possibly causing over or underflows
\beq
A = \frac{1}{\Phi_0}, 
\eeq 
Next {\sc LatticeEasy} removes the dampening term $a' \Phi'_{pr}/a$ by enforcing $ s- 2r + 3 = 0 \rightarrow s = 2r - 3$. This allows us to check the Courant condition for stability as the matter field equations now take on the form of wave equations with a source. To avoid large sources that may drive large fluctuations in the field configuration, the dominant potential term is scaled to a similar order as the rescaled fields, such that $B$ and $r$ satisfy
\beqa
B &=& \sqrt{\alpha} \Phi_0^{-1+\beta/2}, \nonumber \\
 \frac{A^2}{B^2} a^{-2 s + 2r} V  &=&  \frac{a^{6 - 2 r -\beta r}}{\beta} \Phi_{pr}^\beta, \nonumber \\
 6 - 2r - \beta r =0 \rightarrow r &=& \frac{6}{2+\beta}.  
\eeqa
With these scalings {\sc LatticeEasy} uses a staggered-leapfrog method of progressing the evolution of the fields.  Using {\sc LatticeEasy} we can calculate the anisotropic stress-energy tensor, use a FFT, and project the TT part for sourcing the solution of the metic perturbation, for which one can use some numerical differential equation solving method, or approximate the scale factor evolution by a fixed power law and use a numerical integration of an appropriate Green's function to solve for the metric perturbation, and therefore calculate the power spectrum.  

\subsection{Numerical Solid Angle Integration on a Lattice} 
 
\begin{figure}[h]
\begin{center}
\postscript{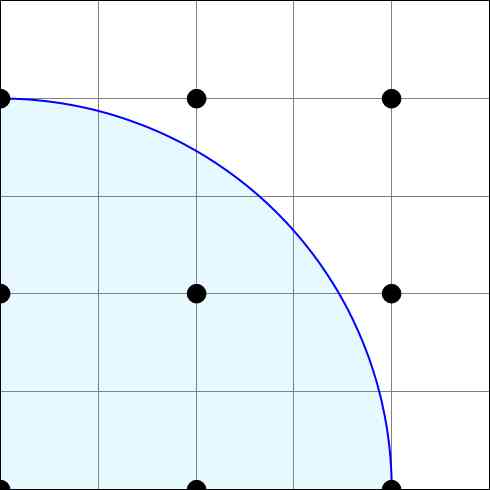}{0.5}
\caption[Lattice Points in a Solid Angle Integration]{ \sglspc \small Sweeping out a solid angle, as seen in blue, in a rectangular lattice requires us to use some interpolation method to find the intermediate values of the integrand function as the values of the integrand are only known at the lattice points depicted in black dots.}
\label{fig:LatticePts}
\end{center}
\end{figure}

 When we compute the power spectrum we require an integration over the solid angle in k-space of the form
 \beq
 \int \left| \dot h^{\rm TT}_{ij}(\eta,{\bf k}) \right|^2 d \Omega_k \ .
 \eeq
  To compute this on a three dimensional rectangular lattice we must choose some k- value of which only can be evaluated on the discrete rectangular k-space values.  For example if we choose some k -value along one axis as the radius to evaluate the solid angle integral.   Sweeping through the solid angle sphere at that radius we will not have the value of $\left| \dot h^{\rm TT}_{ij}(\eta,{\bf k}) \right|^2$ at each point as seen in Fig. \ref{fig:LatticePts}.
  %----------------Figure------------------
%-----------------------------------------
%----------------Figure------------------
\begin{figure}[th]
\begin{center}
\postscript{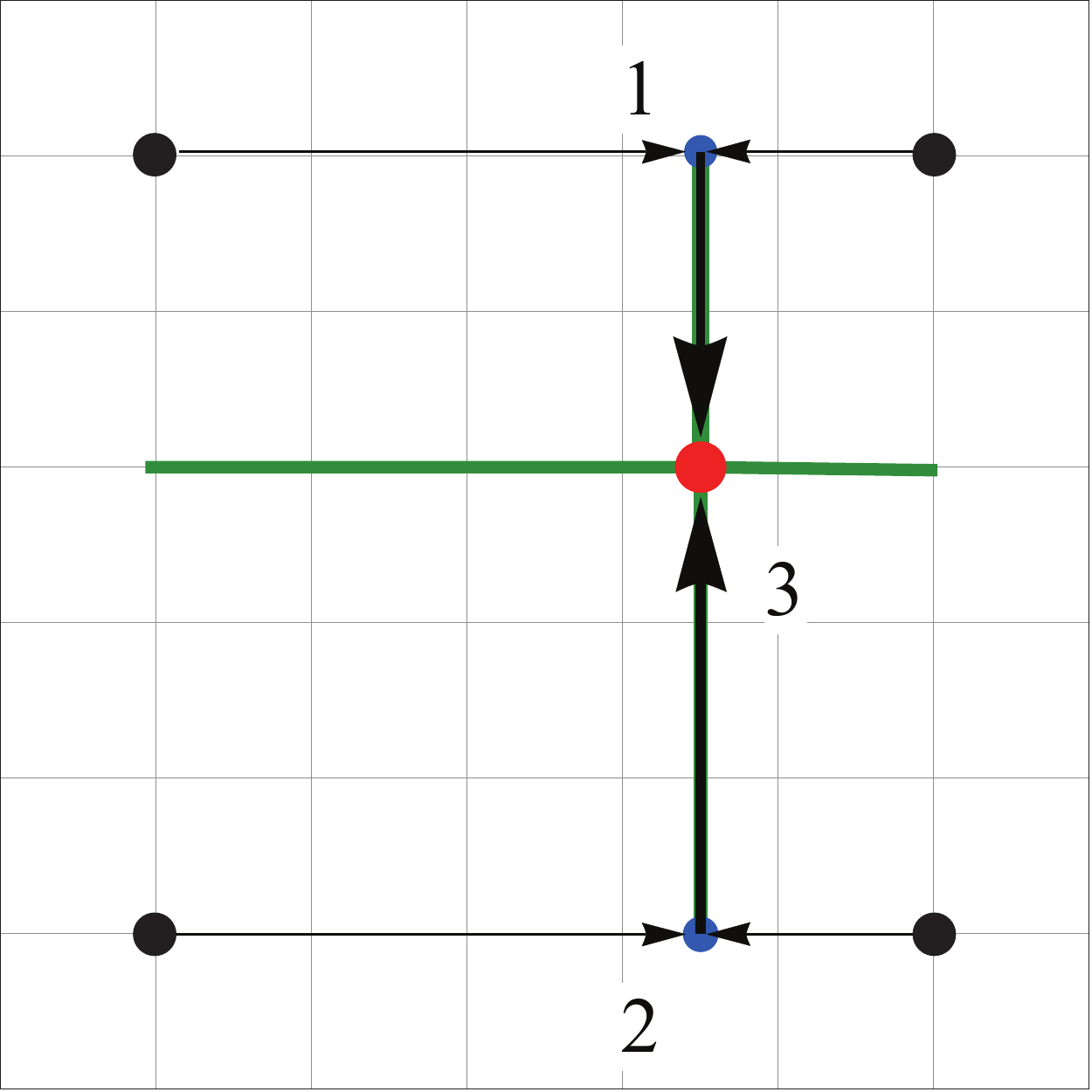}{0.5}
\caption[Interpolation Method on a Lattice]{ \sglspc \small First one determines the interpolated values of the function on each opposite side of the square, as seen by the arrows towards points 1 and 2.  From which the the interpolated value along the y direction is found as seen by following the arrows towards point 3.  This process is then repeated in the z direction to fully interpolate the value of the function in a 3 dimensional lattice.}
\label{fig:Interp}
\end{center}
\end{figure}
%-----------------------------------------
    To remedy this we will use a 3-dimensional interpolation method to obtain an approximate value of $\left| \dot h^{\rm TT}_{ij}(\eta,{\bf k}) \right|^2$  so long as the $\bf k$ value lies within the k-lattice values.  Interpolation of the scalar function $\left| \dot h^{\rm TT}_{ij}(\eta,{\bf k}) \right|^2$ is done by expanding the function in a Taylor series around each of the eight nearest lattice points that create a cube around the desired k-value. We will demonstrate this method with a scalar function $f({\bf x})$ where we interpolate the function at some point $\bf x$.  First we express the point ${\bf x}$ in component form for a rectangular coordinate system $(x,y,z)$.  To find the desired point we interpolate the value of the function on the opposite faces of the cube along the z-direction.  This is done by finding the interpolated values of the function on each face by first interpolating the value of the function along the x-direction on the opposite sides of the square making in the x-y plane as depicted in Fig. \ref{fig:Interp} by the blue points 1 and 2, then finding the interpolated value in the y-direction for each face.  Quantitatively the value of the function at point $1: (x,y_1,z_1)$ of Fig. \ref{fig:Interp} is found by expanding the function at the lattice points for which we do have the value (at the lattice points which the arrows pointing towards 1 originate from)
 \beqa
  f({\bf x}_1) &=& f_1\approx f( x,y_1,z_1)  + \p_x f(x,y_1,z_1) ( x_1 - x) \ , \nonumber \\
  f({\bf x}_2) &=& f_2\approx f( x,y_1,z_1)  + \p_x f(x,y_1,z_1) ( x_2 - x). 
  \eeqa
  We approximate $\p_x f ( {\bf x} )$ by
  \beq
  \p_x f (x,y_1,z_1 ) \approx \frac{ f_2 - f_1}{\Delta x} = \frac{ f_2 - f_1}{x_2-x_1} .
  \eeq
  Which then gives
 \beq
 f_1 \lpa \frac{x_2-x}{x_2-x_1} \rpa + f_2 \lpa \frac{x-x_1}{x_2-x_1} \rpa \approx  f( x,y_1,z_1) . 
 \eeq
 This process is then repeated for the y-value, which can be visualized from following the arrows from points 1 and 2 in Fig. \ref{fig:Interp}.  This is done for both opposite faces of the cube and then finally interpolated along the z-axis to give the fully interpolated value at the desired point $\bf x$.
 
\subsection{Transfer functions}
  Running the program we setup the initial conditions of the field such that we are immediately following inflation at some physical comoving time $t_0$, corresponding to a program time $t_{pr,0}$, such that the inflaton is oscillating with small amplitude about the minimum of its self interaction potential. This drives the amplification of various modes of the coupled scalar fields.  Eventually the expansion of the universe causes oscillations of the inflaton and coupled fields to dampen, so that fluctuations are no longer large enough to source energy into gravitational waves generated by this mechanism.  At this time the program stops corresponding to a final program time $t_{pr, f}$, this corresponds to some physical time in our early universe which can be solved for via
 \beq
  t_{f} = t_0 + \int_{t_{pr,0}}^{t_{pr,f}} a^{-s}(t_{pr}) dt_{pr}.
\eeq
Once we know the result of the power spectrum at this time $t_f$ we need to transfer these results to values we would observe today. 
  We take advantage of the fact that gravitational waves propagate at the speed of light and red-shifts in the same way that electromagnetic radiation does.  The redshift of objects moving at the speed of light can be understood because for speed of light travel $ds^2 = 0$.  This along with assuming the radiation moves along a straight path (radially) from the emitted time $t_e$ to the point of detection $t_d$ along a coordinate distance $R$ gives the relation
  \beq
  ds^2 = 0 = dt^2 - a(t)^2 dr^2 \rightarrow R = \int_{t_e}^{t_d} \frac{dt'}{a(t')}. 
  \eeq 
Assuming the first wave front is emitted at a comoving time $t_e$ and detected at a time $t_d$ and the following wave front is emitted at a time $t_e + \lambda_e$, where $\lambda_e$ is the emitted wavelength in natural units, when the second wave front is detected, one would associate the time between the first wavefront and second wavefront as the wave length in natural units.  The arrival of the second wave front occurs at $t_d + \lambda_d$, where $\lambda_d$ is the observed wavelength at the detection point in natural units.  Since both wave fronts travel along the same coordinate distance $R$ we find
\beqa
\int_{t_e}^{t_d} \frac{dt'}{a(t')}  = \int_{t_e  + \lambda_e}^{t_d + \lambda_d} \frac{dt'}{a(t')} &\rightarrow&  \int_{t_e}^{t_e + \lambda_e} \frac{dt'}{a(t')}   -  \int_{t_d}^{t_d + \lambda_d} \frac{dt'}{a(t')}  = 0 , \nonumber \\
\frac{\lambda_e}{a(t_e)} &=& \frac{\lambda_d}{a(t_d)},
\eeqa
where the second equality assumes that the variation of $a(t)$ time scale is much smaller than the times $\lambda_d, \ \lambda_e$ allowing us to assume $a(t)$ remains constant in the integration regions.  Using this relation on the frequencies for which we calculate the power spectrum, transfers the discrete frequencies calculated at the end of the simulation to the values detected today via
\beq
k({\rm today}) =  k(t_f) \frac{a(t_f)}{a(\rm today)}.
\eeq
This is simply red shifting the power in each frequency to the corresponding red-shifted frequency today.  The effect of the expansion of the energy density in the gravitational waves has the same diluting effect as electromagnetic radiation such that
\beq
\rho_{gw} ({\rm today}) = \rho_{gw} (t_f) \lpa \frac{a(t_f)}{a({\rm today})} \rpa^4.
\label{eq:TranFunRho}
\eeq
We can see that the energy contained in the bin $k, k+ dk$ gets redshifted and diluted.  

To display these results it is common practice to display the results not in energy density per frequency but rather as $h^2 d \Omega_{gw} /d \log k$ where $\Omega_{gw} = \rho_{gw}/\rho_c$, with $\rho_c$ the critical energy density to close the universe.  Representing the data this way ensures the exact value of the Hubble constant $H = h \ 100 \ \rm km/s/Mpc$ does not effect the final result.  Take note that it is common practice to express $h^2 d \Omega_{gw} /d \log k$ as simply $h^2 \ \Omega_{gw}(k)$ with the $d/d \log k$ understood, the context should make it clear.

In order to complete the transfer functions we need a method to determine the ratio $a(t_f)/a(\rm today)$. We appeal to the conservation of entropy in the universe in order to calculate this ratio.  The total entropy in SM matter can be expressed normalized to the photon entropy in the form
\beqa
S &\propto& a(t)^3 g_S(t) T_\gamma(t)^3 = \rm const, \nonumber \\
g_S(t) &=& \sum_{\rm boson} g_i \lpa \frac{T_i^{\rm F.O.} }{T_\gamma(t)}\rpa^3 + \frac{7}{8} \sum_{\rm fermion} g_i \lpa \frac{T_i^{\rm F.O.} }{T_\gamma(t)}\rpa^3,
\eeqa
where $g_i, \ T_i^{\rm F.O.}$ are the internal spin degeneracies of particle type $i$ and the freeze out temperature for species $i$ respectively.  Conservation of entropy gives
\beq
\frac{a(t_f)}{a({\rm today})} = \lpa \frac{g_S({\rm today})}{g_S(t_f)} \rpa^{1/3} \frac{T_\gamma({\rm today})}{T_\gamma(t_f)},
\eeq
where $g_S({\rm today}) = 2  + \frac{7}{8} \cdot 3 \cdot 2 \cdot \lpa \frac{4}{11} \rpa = 3.91$ consisting of photons and three neutrino species.  We can write $T_\gamma$ in terms of the total energy density from SM matter.  The total energy density from SM matter is dominated by particles that are relativistic at a temperature $T_\gamma$. Because of this the total energy density in SM matter is normalized to photons, and referred to as energy density in radiation (as relativistic particles behave as such) expressed as
\beq
\rho_{\rm rad} = \frac{\pi^2}{30} g_\rho(T_\gamma) T_\gamma^4 \rightarrow \lpa  \frac{30}{\pi^2} \frac{\rho_{\rm rad}}{g_\rho} \rpa^{1/4} = T_\gamma.
\eeq
where the degrees of energy freedom is given as
\beq
  g_\rho (t) = \sum_{\rm boson} g_i \lpa \frac{T_i^{\rm F.O.} }{T_\gamma(t)}\rpa^4 + \frac{7}{8} \sum_{\rm fermion} g_i \lpa \frac{T_i^{\rm F.O.} }{T_\gamma(t)}\rpa^4.
  \eeq
From this it follows that 
\beq
\frac{a(t_f)}{a({\rm today})} = \lpa \frac{g_S({\rm today})}{g_S(t_f)} \rpa^{1/3} \lpa \frac{g_\rho(t_f) \rho_{\rm rad}({\rm today}) }{g_\rho({\rm today}) \rho_{\rm rad}(t_f) }\rpa^{1/4}.
\eeq  
For times in the early universe $g_S \approx g_\rho$, we will extend this approximation to today, so that the final expression takes the form
\beq
\frac{a(t_f)}{a({\rm today})} =  \lpa \frac{\rho_{\rm rad}({\rm today})}{\rho_{\rm rad}(t_f)} \rpa^{1/4}  \lpa \frac{g_S({\rm today})}{g_S(t_f)} \rpa^{1/12},
\label{eq:TranFunScaleFact}
\eeq
where $\rho_{\rm rad}(t_f)$ is calculated in terms of the fields at final simulation time.

\section{\label{sec:GWSPT} Gravitational Waves from Second Order Global Phase Transitions}
  
\quad  We now apply the methods developed in the previous section to simulate the gravitational wave spectrum to be produced today from a second order global phase transition sourced from different local, non-zero, vacuum expectation values of a broken $SO(N)$ field symmetry of a scalar field $\phi_i$.  We are motivated to study this model because particle physicists are motivated to increase the symmetry group of the SM and thus unite the strong and electroweak forces into a GUT.  The larger symmetry must become broken as the universe cools in the case the transition is first-order, bubbles of the broken phase nucleate and coalesce.  In this model the phase transition happens very rapidly---the entirety of the universe can end up in a unique state in less than a Hubble time.  This process is likely to produce gravitational radiation 
\cite{Witten:1984rs,Kosowsky:1991ua,Kamionkowski:1993fg,Child:2010} as bubbles collide 
and coalesce. Another case could be the phase transition could be second-order where the fields
smoothly transition to the broken phase as the temperature of the universe drops. 
If the broken phase has degeneracies, the effects of the existence of this phase transition can lead to observational effects. At various initially causally disconnected regions of the universe, the $SO(N-1)$ degeneracy of the minimization of the potential associated with the $SO(N)$ symmetry allows domains of space-time where the field values have different vacuum expectation values.  As these domain walls collide, the differences in field value settle out, and self order, sourcing a stochastic gravitational wave background. It is worth noting that a scale-invariant spectrum of gravitational radiation is a key prediction of inflation \cite{Starobinsky:1979ty,Rubakov:1982df}. It has been noted by some authors that phase transitions of the type described above can however, mimic the scale-invariant inflationary signal \cite{Krauss:1991qu,JonesSmith:2007ne,Fenu:2009qf}.  In previous studies, authors have relied on large-$N$ approximations to calculate the gravitational wave signal, but here we  make no approximations other than those associated with numerics.  

We begin with the assumptions that the universe is radiation dominated
at the time when the phase transition occurs, and that the energy associated with the
fields undergoing the phase transition is a small fraction, $\alpha$, of the 
total energy density $\rho_T$ at the time of the transition.  The total energy
density, at any time, is given by
\begin{equation}
\rho = \rho_{\rm rad} + \rho_{\phi} , \ \rho_\phi = \sum_i \frac{1}{2}\left( \dot{\phi}^2 + \frac{1}{a^2} \nabla_x \phi_i^2 \right) + V(\phi_i,T)
\label{tdeppot}
\end{equation}
where $\rho_\phi$ is the energy density associated with the scalar fields $\phi_i$ and
\begin{equation}
\rho_{\rm rad} = (1-\alpha) \rho \ .
\end{equation}
Since the universe is dominated by the radiation energy-density in the relevant eras, we will only consider cases where $\alpha \ll 1$, so that the universe is well described by assuming $H^2 \propto a^{-4}$ because the potential in (\ref{tdeppot}) is temperature dependent, we may simulate the phase transition by tracking the temperature throughout the simulation. 

\subsection{\label{sec:Thermal} Statistical Mechanics of Quantum Fields}
 In order to arrive at the temperature dependent potential one must resort to the study of statistical mechanics of quantum fields.  We can compute the potential by first expressing the fundamental quantity in quantum statistical mechanics, the partition function~\cite{BailinL}, as
\beq
Z_p = {\rm Tr} \ e^{-H/T} ,
\eeq
with $H$ the Hamiltonian of the system and $T$ the temperature.  The trace can be represented in field configuration space via
\beq
Z_p = \int \CD \phi \ \langle \phi| e^{-H/T} | \phi \rangle = \int \CD \phi \ \langle \phi | e^{i H (i \beta)} | \phi \rangle \ .
\label{eq:ParFun}
\eeq
Equation (\ref{eq:ParFun}) is specifically written in a form that displays the quantum propagator with the substitution $t \rightarrow i \beta = i/T$. With this in mind we can use the Feynman path integral formulation of the propagator to arrive at
\beq
Z_p = \int \CD \phi \   \int_{\phi(0, {\bf x})}^{\phi(i\beta, {\bf x})}  \CD \chi \ e^{i \int^{i \beta}_0 \CL (\chi) d^4x } \ .
\eeq
We now make a change of variable, known as a Wick rotation, from $t \rightarrow i \tau$ such that the action for scalar fields takes the form
\beq
i S = - S_E =  - \int \int^\beta_0 \lpa \frac{1}{2} \p_\tau \phi( \bx ) ^2 + \frac{1}{2}  \nabla_x \phi( \bx)^2 + V(\phi(\bx)) \rpa d \tau d^3 x\ , 
\label{eq:Eucled}
\eeq 
where now the fields are functions of $\bx = (\tau, {\bf x})$, and we compact the notation by using the 4-dimensional Euclidean metric $\bar \p ^2 = \p_\tau^2 + \nabla _x ^2$ and writing $\phi(\bx)  =\bar \phi$.  At this point it is worth noting that Eq. (\ref{eq:Eucled}) takes on the form of the classical Hamiltonian with the replacement $t \rightarrow \tau$.  With this notation we may now write the partition function as
\beq
Z_p = \int \CD \bar \phi  \int^{\bar \phi(\beta, {\bf x})}_{\bar \phi(0,{\bf x})}  \CD \bar{\chi} \ e^{- S_E[ \bar \chi] } =  \int \CD \bar \phi  \int^{\bar \phi}_{\rm PBC} \CD \chi \ e^{- S_E[ \bar \chi] }  .
\label{eq:ParFunc2}
\eeq
In the second equality the integration limits can be taken over all field configurations with the condition that the fields $\bar \chi$ have periodic boundary conditions (PBC) where the period is $\beta$ {\it i.e.}
\beq
\bar \phi(\tau, {\bf x}) = \bar \chi(\tau,{\bf x}) = \bar \chi( \tau + \beta, {\bf x})= \bar \phi( \tau + \beta, {\bf x})  \ .
\eeq
Equation (\ref{eq:ParFunc2}) also allows identification of the probability distribution of field configurations which will be useful later on, specifically
\beq
Z_p = \int \CD \bar \phi \ P_R[\bar \phi] \ \rightarrow \  P_R[\bar \phi] \propto e^{\beta H_{\rm eff}[\bar \phi]} =\int^{\bar \phi}_{\rm PBC} \CD \chi \ e^{- S_E[ \bar \chi] },
\label{eq:Heff}
\eeq
where $P_R[\bar \phi]$ is the relative probability of the field configuration $\bar \phi$.  From Eq. (\ref{eq:Heff}) we can extract the thermal potential by finding the quantum, thermal effective Euclidean action $S^{(\rm eff)}_{E}$ from the path integral in (\ref{eq:Heff}). 

The effective action can be formulated by following~\cite{BailinL} and identifying the effective Hamiltonian via
\beq
P_R[\bar \phi] = \int^{\bar \phi}_{\rm PBC} \CD \chi \ e^{- S_E[ \bar \chi]  } = \CN(\beta) e^{\beta H_{\rm eff}[\bar \phi]} \ , 
\label{eq:EffAction}
\eeq
where $\CN(\beta)$ is some temperature dependent function. We can make an approximate evaluation of (\ref{eq:EffAction}) by expanding $S_E[\bar \chi]$ about the field configuration $\bar \phi_0$ that satisfies
\beq
\left. \frac{\delta S_E}{\delta \bar \chi} \right|_{\bar \chi = \bar \phi_0} = 0.
\label{eq:ClasFld}
\eeq
Now we make a change of variable and expand the action around the fixed field configuration $\bar \phi_0$
\beq
\bar \chi = \bar \phi_ 0 + \bar \phi, \ \CD \bar \chi = \CD \bar \phi \ .
\eeq
Under this change of variable, we retain the first term in the expansion of $S_E$ to give
\beq
P_R \approx e^{- S_E[ \bar \phi_0] } \int_{\rm PBC} \CD \bar \phi \ e^{- \frac{1}{2!} \bar \phi \cdot \frac{\delta^2 S_E}{\delta \bar \phi_0^2} \cdot \bar \phi } \ ,
\label{eq:PR896}
\eeq
where above we use the shorthand 
\beq
\bar \phi \cdot \frac{ \delta^2 S_E }{ \delta \bar \phi_0^2} \cdot \bar \phi  = \int d^4 \bx d^4 \bx' \ \bar \phi(\bx') \left. \frac{ \delta^2 S_E }{ \delta \bar \phi(x) \delta \bar \phi(x')} \right|_{\bar \phi = \bar \phi_0} \bar \phi(\bx) \ .
\label{eq:ShortHand}
\eeq
To continue with the analysis we must now choose a potential for the $N$ scalar fields to break.  We use a potential that supports a non-zero vev of the form
\beq
V(\phi) = \frac{\lambda}{8} \lpa \phi^2  -  \frac{v^2}{2} \rpa^2 = - \frac{1}{2} m_0^2 \phi^2 + \frac{\lambda}{8}  \phi^4 + \frac{v^4 \lambda}{32} ,
\eeq
where $\phi^2 = \sum_i \phi_i^2$ and $m_0^2 = \lambda v^2/4$.  Finding the functional derivatives in (\ref{eq:ShortHand}) gives
\beq
\bar \phi \cdot \frac{ \delta^2 S_E }{ \delta \bar \phi_0^2} \cdot \bar \phi= \bar \phi \cdot \lpa - \bar \p^2 +  m_0^2 + \frac{3\lambda}{2} \bar \phi_0^2  \rpa  \cdot \bar \phi \ .
\eeq
The integration in (\ref{eq:PR896}) can be shown to be proportional to
\beq
P_R \propto e^{- S_E[ \bar \phi_0] } \exp\left[-\frac{1}{2} \ {\rm Tr}  \ln \lpa-\bar \p_{\bx}^2 + m_0^2 + \frac{3\lambda}{2} \bar \phi_0^2 \rpa\right].
\eeq
We evaluate the trace in Fourier space, keeping in mind the periodic behavior of  $\bar \phi$.  The periodicity is enforced by the Fourier series
\beq
\bar \phi(\bx) = \frac{1}{\beta} \sum_{n = -\infty}^\infty \phi_n({\bf x}) e^{i k_n \tau}, \ k_n = \frac{2 \pi n}{\beta}.
\eeq 
Under this Fourier series the trace becomes 
\beq
\frac{1}{2} \ {\rm Tr} \ \ln \lpa  -\bar \p^2 + m_0^2 +  \frac{3\lambda}{2} \bar \phi_0^2  \rpa = \frac{1}{2} \int d^3 x \ \sum_{n = -\infty}^\infty  \int \frac{d^3 k}{(2\pi)^3} \ \ln \lpa \bar k^2 + \bar m^2 \rpa  ,
\eeq
where $\bar k^2 = k_n^2 + {\bf k}^2$ and $\bar m^2 = m_0^2 + 3 \lambda \bar \phi_0^2 /2 $.  The sum over frequencies is know as \emph{Matsubara frequency sums}.  A sum of this kind is equivalent, up to an overall constant to a contour integral~\cite{BailinL}, giving a sum over residues of the form 
\beq
\frac{1}{2}  \int \frac{d^3 k}{(2\pi)^3} \
 (\beta \sqrt{k^2 + \bar m^2} + 2 \ln (1- e^{-\beta \sqrt{k^2 + \bar m^2}}) + k \ {\rm  independent  \ constant} ),
\eeq
The $k$ independent constant and integration of $\sqrt{k^2 + \bar m^2}$ can be absorbed into the overall temperature dependent proportionality constant, so that we only retain the $\ln$ term. Evaluation of this last term results in 
\beq
\frac{1}{2 \pi^2 } \int_0^\infty \ k^2  \ln (1- e^{-\sqrt{ \beta ^2k^2 + \beta^2 \bar m^2}}) \approx - \frac{\pi^2}{90 \beta^3} + \frac{\bar m^2}{24 \beta} \ , \ \bar m \beta \ll 1 \ .
\eeq
We can now express the effective Hamiltonian as
\beq
\beta H_{\rm eff} \approx S_E + \int d^3 x  \frac{\bar m^2}{24 \beta} = \beta \int d^3 x \ \lpa \frac{1}{2} \p_x \bar \phi_0^2 + V(\bar \phi_0) +\frac{\bar m^2}{24 \beta} \rpa \ .
\eeq
In the second equality, we assumed that $\bar \phi_0$ has no $\tau$ dependence.  We can now find to leading order in temperature,  the effective potential 
\beq
V(\phi_i, T ) \approx \frac{\lambda v^2}{8} \lpa  \frac{ T}{2 v^2} - 1 \rpa\phi^2 + \frac{\lambda}{8} \lpa  \phi^4 + \frac{v^4}{4} \rpa = m^2_{\rm eff}(T)\phi^2 +
\frac{\lambda}{8}\left(\phi^4+\frac{v^4}{4}\right),
\label{tpot}
\eeq
where the temperature dependent effective mass is parameterized by
\begin{equation} 
m_{\rm eff}^2 = \frac{\lambda v^2}{8} \left(\frac{T}{T_c}-1\right).
\end{equation} 
At temperatures higher than the
critical temperature, $T_c$, the effective mass is positive, the potential has 
a unique minimum at $\phi_i = 0$, and this minimum has full $SO(N)$ 
symmetry. When the temperature drops to $T_c$ where the effective mass of the field vanishes, a phase transition occurs as the potential now allows a new stable vacuum state with an $SO(N-1)$ symmetry, a state for which
the fields take non-zero vevs. The simulation can keep track of the temperature with the relations between energy and the Hubble value at the start of the simulation
\beq
T(t) = \frac{T_0}{a(t)} , \  \rho_0 = \frac{\pi^2}{30} g_\rho(T_0) T_0^4  = \frac{3 M_{pl}^2}{8\pi} H_0^2  \ , 
\eeq
where $T_0$ the initial temperature to be solved for as a function of the initial Hubble value $H_0$.  We will take the number of relativistic degrees of freedom to initially be $g_\rho(T_0) = 1000$.

\subsection{Initial Field Conditions on the Lattice}
Initially the simulation will start at $T_0 = T_c$  where the field has a mean value, $\phi=0$; however, there is a variance associated with this temperature, 
\begin{equation}
\sigma^2 = \left\langle\phi^2\right\rangle -\left\langle\phi \right\rangle^2 =  \left\langle\phi^2\right\rangle,
\end{equation}
that sets the distribution of field values at the time of the transition.  The physics of interest demand that each lattice site be assigned the average value of the field contained in one Hubble volume.  We denote by $\langle \phi \rangle = \phi_v$, this average is done over one Hubble volume.  Since each lattice site is causally disconnected, we select the field value at each lattice site randomly from a probability distribution of possible field values.  Determining the distribution is done by using
\beq
P(\phi_v = \phi') =  \int D\phi \CP [\phi] \delta(\phi_v - \phi') = \langle \delta(\phi_v- \phi')\rangle.
\label{fullprob}
\eeq
Equation (\ref{fullprob}) finds the probability that $\phi_v$ has the value $\phi'$ by summing the probabilities of all the possible field configurations that for which $\phi_v = \phi'$, where $\CP[\phi]$ is the probability functional of the field configuration $\phi$. The volume-averaged field is given as
\beq
\phi_v = \frac{1}{V}\int_V d^3x\,\phi(x)= \frac{1}{V}\int_{-\infty}^{\infty} d^3x\, I(x)\phi(x),
\eeq
where, for computational convenience, we have introduced a window function $I(x)$, which is a function which is 1 for $x$ within the volume of interest and 0 outside of it.  We can approximate (\ref{fullprob}) using the cumulant expansion technique, which gives a Gaussian approximation
\begin{equation*}
P(\phi_v = \phi') \approx \sqrt{\frac{1}{2\pi \sigma^2}}
\exp{\left({-\frac{(\phi'-\mu)^2}{2 \sigma^2}}\right)},
\end{equation*}
with $\mu = \langle\phi_v\rangle$ and $\sigma^2 =
\langle\phi_v^2\rangle-\langle\phi_v\rangle^2$.  The problem is now to compute
the moments of the field given by
\begin{eqnarray}
\langle\phi_v\rangle &=& \int D\phi \mathcal{P}[\phi]\left(\frac{1}{V}\int_V
d^3x\, \phi(x) \right) \ , \nonumber \\
\langle\phi_v^2\rangle &=& \int
D\phi \mathcal{P}[\phi] \left(\frac{1}{V^2}\int_V
d^3xd^3y\, \phi(x)\phi(y) \right) \ .
\label{secondmoment} 
\end{eqnarray}
The full probability functional $\CP$, can be found in, \cite{Hindmarsh:1993mb},  but we consider only the leading order temperature dependence. We saw in (\ref{tpot}) that leading order temperature dependence modifies the potential by replacing $m$ with a temperature dependent term, $m_{\rm eff}(\beta)$.  For temperatures at and above $T_c$ we are in a symmetric phase of the effective potential.  We then have 
\beq
\CP[\phi] = \frac{e^{-\beta H_{\rm eff}[\phi]}}{Z} = \frac{1}{Z} \exp \left[- \beta \int d^3 x \lpa \frac{1}{2} \p_x \phi^2 + V_{\rm eff} \rpa \right]
\eeq
Note that the normalization $Z$ is given as
\beq
Z = \int \CD \phi \ e^{-\beta H_{\rm eff}[\phi]} \ .
\eeq
We can make use of a more general expression, known as a generating functional, that is related to the moments of the field by the expression
\beqa
W[J] &=&\frac{1}{Z} \int \CD \phi \ \exp \left[ -\beta \int d^3 x \lpa \frac{1}{2} \p_x \phi^2 + V_{\rm eff}  + J \phi \rpa \right] \ , \nonumber \\
&\approx&  \exp \left[ -\frac{ 1}{2\beta}J \cdot \Delta \cdot J \right] \ ,
\label{eq:GenFunct}
\eeqa
where in the last line we are taking using only the zeroth-order terms involving $m_{\rm eff}$; that is, we are neglecting the term $\lambda  \phi^4$ as it will produce higher-order corrections.  As well, in (\ref{eq:GenFunct}) $\Delta$ is given as 
\beqa
(-\p_x^2 +2 m_{\rm eff}^2)\Delta({\bf x},{\bf x'}) &=& \delta^{(3)}( {\bf x}-{\bf x'}) \ , \nonumber \\
\Delta({\bf x},{\bf x'}) &=& \int \frac{d^3 {\bf k} }{(2\pi)^3} \frac{e^{-i \bf k \cdot (x-x')}}{{\bf k}^2 +2 m_{\rm eff}^2} \ .
\eeqa
With the use of (\ref{eq:GenFunct}) the moments of the field can now be expressed as
\begin{eqnarray*}
\langle\phi_v\rangle &=&  \frac{1}{V}\int_{-\infty}^{\infty} d^3x\,
I(x)\frac{\delta}{\delta J(x)}W[J]\Bigr|_{J=0} \ , \\
\langle\phi_v^2\rangle &=&  \frac{1}{V}\int_{-\infty}^{\infty} d^3xd^3y\,
I(x)I(y)\nonumber\\
&\phantom{=}&
\phantom{\frac{1}{V}\int_{-\infty}^{\infty}}\times\frac{\delta}{\delta
J(x)}\frac{\delta}{\delta J(y)}W[J]\Bigr|_{J=0}.
\label{eq:Moments}
\end{eqnarray*}
We choose a Gaussian window function for which the integral $\int d^3x\, I(x)=4\pi R^3/3 = 4\pi/3 H_c^3$ is one Hubble volume at the time $T = T_c$. Evaluating the integrals of (\ref{eq:Moments}) at the point $T = T_c$ leads to
\beq
\langle\phi_v\rangle = 0 \ , \ \langle\phi_v^2\rangle = \frac{H_c T_c}{4\pi^{3/2}} \ .
\eeq
Putting this all together, we have
\begin{equation} 
P(\phi_v = \phi') = \sqrt{\frac{H_c T_c}{ 8 \pi^{5/2}}} \exp{\left(-\frac{\phi'^2}{4 \pi^{3/2}} H_c T_c\right)} \ ,
\end{equation}
from which we draw the initial values of the initial field values.
  
\subsection{Results of the Simulation}

 The simulation takes advantage of a $4^{\rm th}$ order RK algorithm to calculate the metric perturbation, and assumes  $g_S({\rm today})/g_S (t_f)=1/100$ for use in the transfer functions. The first major difference between the form of the gravitational-wave spectrum from self-ordering of the fields from domain interactions and that predicted by inflation is the lack of power at high-frequencies.  This cut-off feature exists because second order phase transition fluctuations occur on larger than Hubble length scales.
%----------------Figure------------------
\begin{figure}[tpb]
\begin{center}
\postscript{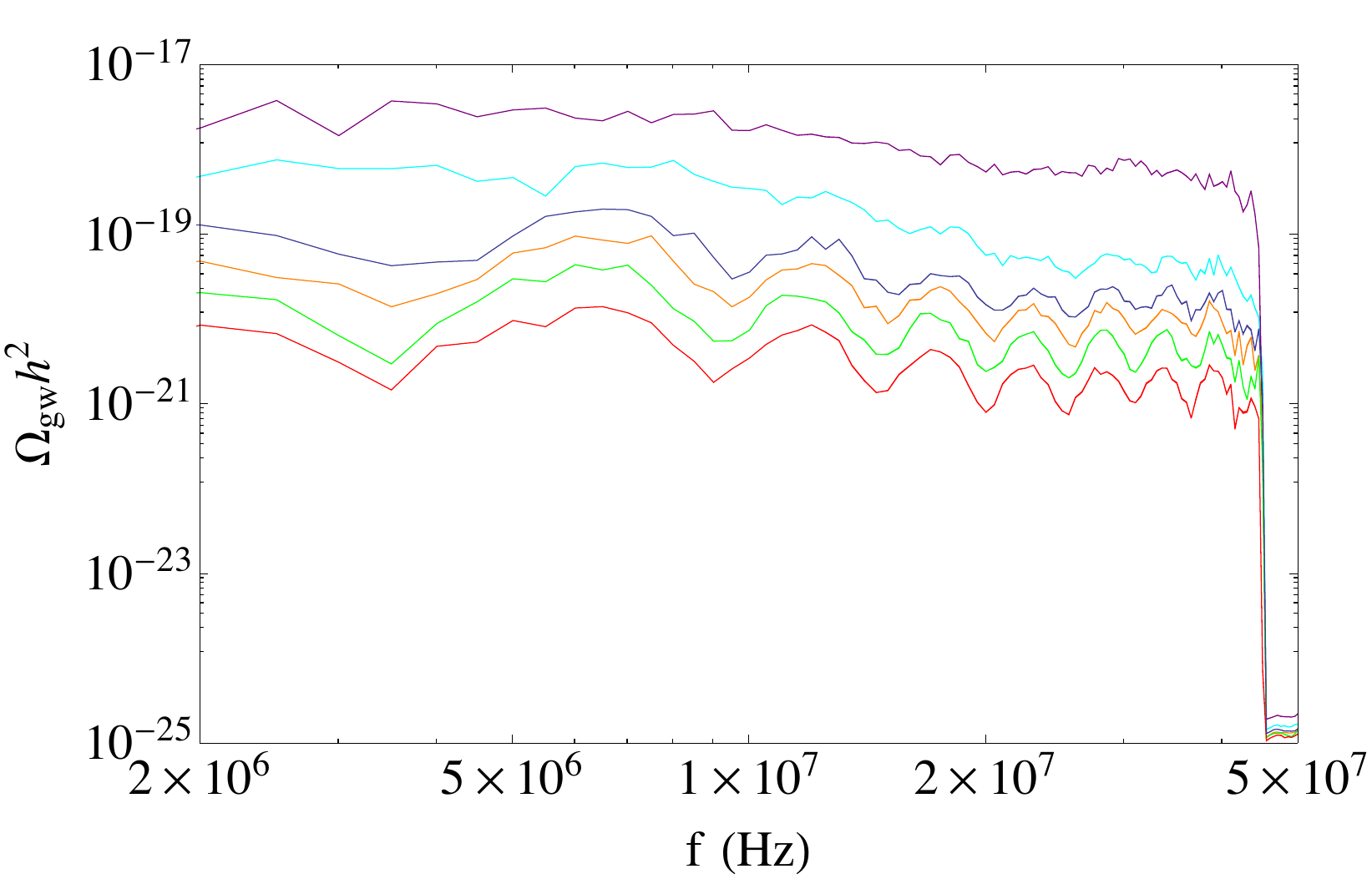}{0.8}
\caption[Grav-Wave Spectrum from Self Ordering for $N$ fields]{ \sglspc \small The figure displays the present-day gravitational wave spectrum as calculated from the lattice simulation, from self-ordering fields on larger than Hubble volume scales. The figure displays, from top to bottom, the spectrum for $N$ scalar fields which undergo breaking the $SO(N)$ symmetry with, $N=2,3,4,5,8,16$}
\label{fig:ncomparison}
\end{center}
\end{figure}
%-----------------------------------------
%----------------Figure------------------
\begin{figure}[tpb]
\begin{center}
\postscript{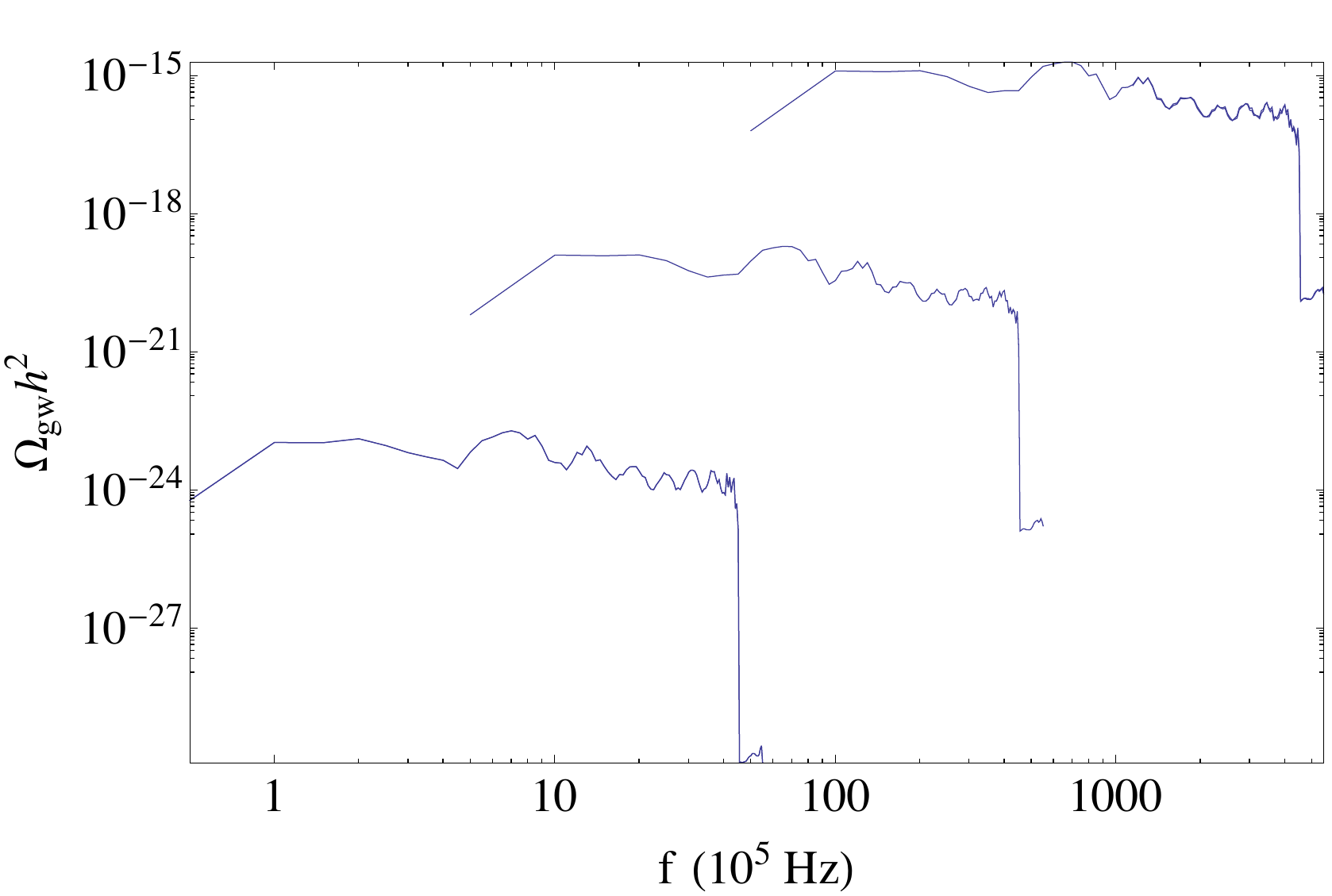}{0.8}
\caption[Grav-Wave Spectrum from Self Ordering Transition Energies Variation]{ \sglspc \small The present-day gravitational wave spectrum from self-ordering.  From top (rightmost) to bottom (leftmost), $\rho_T^{1/4}=10^{-3}\,M_{\rm pl}$, $\rho_T^{1/4}=10^{-4}\,M_{\rm pl}$, $\rho_T^{1/4}=10^{-5}\,M_{\rm pl}$.}
\label{fig:soucomparison}
\end{center}
\end{figure}
%-----------------------------------------

In \cite{Easther:2006gt} the cut-off frequency is shown to be related to the Hubble length at the time when the gravitational wave is generated $H$, the result of which is
\beq
f_{\rm peak} = 6\times 10^{-10} \frac{k}{\sqrt{M_{pl} H}}\,\rm{Hz}.
\eeq
where $k = H$ the largest possible frequency for this model. The first Friedmann equation,
\begin{equation}
H_c^2 = \frac{8\pi}{3}\frac{\rho_T}{M_{pl}^2},
\end{equation}
then implies that the cut-off should appear at
\begin{equation}
\label{peakeq}
f_{\rm peak} \simeq 10^{11} \frac{\rho_T^{1/4}}{M_{pl}}\,{\rm Hz}.
\end{equation}
For example, with $\rho_T^{1/4} = 10^{-4} M_{pl}$ we expect the cutoff to be at $f \simeq 10^{7}\, {\rm Hz}$, which agrees with the results of the simulation shown in Fig. \ref{fig:ncomparison}.  

This model provides a probe of the energy scale of some symmetry breaking on BSM physics, as the sharp Hubble-length wall cut off can probe the era at which the symmetry breaking occurs.  A flat signal as shown in Fig.  \ref{fig:ncomparison} could be misinterpreted as the gravitational radiation from primordial quantum fluctuations. It might, however, be possible to distinguish the two models at very high frequencies due to the Hubble length wall.  The behavior of the gravitational waves can be understood as follows. Initially at the Hubble-length the gradients due to differing field values sources gravitational waves at this scale. The waves generated continue to redshift to lower frequencies. After the fields have settled from self ordering, the overall configuration of the field is the same as the initial configuration, where differing regions of the the field have different domains of vacuum values, now at a larger Hubble-length scale.  The process repeats and sources gravitational waves at the scale of the new domains, and so on until the simulation ends.

The amplitude and the essentially flat nature of the spectrum were expected from the analytical approach of \cite{JonesSmith:2007ne,Krauss:2010df} and \cite{Fenu:2009qf}. It is interesting to note that the spectrum is flat for all the cases shown in Fig. \ref{fig:ncomparison}, including those for $N=2,3$, as the analytical methods employed by \cite{Fenu:2009qf} assumed that $N$ is large.

In Fig.~\ref{fig:soucomparison} we observe two important scaling effects.  First, we recover the fact that the high-frequency cutoff given by equation~(\ref{peakeq}) scales with $\rho_T^{1/4}$.  We can also see that the amplitude of the spectrum is proportional to the energy density of the universe at the time of the transition, $\Omega_{\rm gw}(k)h^2 \propto \rho_T$. This scaling was shown in \cite{JonesSmith:2007ne,Krauss:2010df} and \cite{Fenu:2009qf}.

 We can compare the numerical results we obtained from simulation with the analytic arguments of \cite{JonesSmith:2007ne,Krauss:2010df} and \cite{Fenu:2009qf}, where in both cases the authors use the model presented here to estimate the gravitational wave signal from the domain interactions.  The two sets of authors use slightly different parameterizations of the model; however, all authors arrive at the conclusion that there should be a flat gravitational wave spectrum from this transition.

In \cite{JonesSmith:2007ne} and ~\cite{Krauss:2010df} (JKM), the authors estimated the power in
gravitational waves from field reordering to be (given by equation~(10) in \cite{Krauss:2010df} with some modification outlined in~\cite{Giblin}) as
\begin{equation}
\label{Jestimate}
\Omega^{\rm JKM}_{\rm gw}h^2= \frac{99}{N}\Omega_{\rm rad}h^2
\left(\frac{v^4}{4 N M_{pl}^4}\right) \ .
\end{equation}
This can be reduced further by using $\Omega_{\rm rad}h^2\approx 2\times10^{-5}$ and estimating the total energy density at the phase transition by
\beq
\langle \rho_\phi \rangle \simeq \frac{\lambda v^4}{32} = \alpha \rho = \alpha \frac{3 M_{pl}^2}{8 \pi} H_c^2 \rightarrow \frac{v^2}{2} =  \sqrt{\frac{3 \alpha}{ \lambda \pi}} M_{pl} H_c.
\label{defofvev}
\eeq
From Eq. \ref{defofvev} and the relation $H_c^2 = 8\pi \rho_T / 3 M_{pl}^2$, Eq. (\ref{Jestimate}) becomes
\beq
\Omega^{\rm JKM}_{\rm gw}h^2 =  \frac{0.016}{N} \frac{\alpha}{\lambda}\frac{\rho_T}{M_{pl}^4}.
\eeq

Furthermore, in \cite{Fenu:2009qf} (FFDG), the authors predict a scale-invariant power spectrum (Eq.~(5.2) of \cite{Fenu:2009qf}) 
\begin{equation}
\label{fenuresult}
\Omega^{\rm FFDG}_{\rm gw}h^2 \simeq \frac{511}{N}\Omega_{\rm rad}h^2 \left(\frac{v}{\sqrt{2} M_{pl}}\right)^4,
\end{equation}
Using Eq. (\ref{defofvev}) along with $\Omega_{\rm rad}h^2 \approx 2\times10^{-5}$, the expression in (\ref{fenuresult}) reduces to
\beq
\Omega^{\rm FFDG}_{\rm gw}h^2 \simeq \frac{0.082}{N}\frac{\alpha}{\lambda} \frac{\rho_T}{M_{pl}^4}.
\label{Festimate}
\eeq

Both of these estimates assume the universe is comprised only of the scalar fields.  However, we diluted the source by a factor of $\alpha$, to preserve a radiation-dominated phase during and after the phase transition; this dilutes the analytic estimates~(\ref{Jestimate},\ref{Festimate}) by a factor of $\alpha^2$.

It is worth pointing out that some of the phase transitions we have simulated 
result in the production of global topological defects: global strings for $N=2$, 
global monopoles $N=3$, and global textures for $N>3$. 
Surprisingly, we find that the gravitational radiation produced 
is consistent with the large $N$ approximation even for low values 
of $N$, where the approximation is not valid (see the
analytic estimates above).
\sglspc
%--------------Table--------------
\begin{table}
\center
\caption[Grav-Wave Spectral Amplitude for $SO(N)$ Model with Varying $N$]{Spectral amplitudes as a
function of number of fields for simulations with
$\left(\rho_T^{1/4}=10^{-4} M_{pl}, \alpha = \lambda = 0.1 \right)$.  The
numerical values, $\Omega^{\rm SIM}_{\rm gw}h^2$, are an average value along the spectrum taken from the simulations, while the values in the second two columns are obtained from (\ref{Jestimate}) or (\ref{Festimate}).} 
\ \\
 \begin{tabular}{|cccc|}    
\hline
$N$ & $\Omega^{\rm SIM}_{\rm gw} h^2$ & $\alpha^2 \Omega^{\rm JKM}_{\rm gw} h^2$ & $\alpha^2 \Omega^{\rm FFDG}_{\rm gw}h^2$ \\
\hline
\hline
   2 & $1.0\times10^{-18}$ &$9.0\times10^{-21}$ & $4.1\times10^{-20}$ \\
   4 & $3.8\times10^{-20}$ & $4.0\times10^{-21}$& $2.1\times10^{-20}$ \\
   8 & $8.3\times10^{-21}$ & $2.0\times10^{-21}$& $1.0\times10^{-20}$ \\
  16 & $3.1\times10^{-21}$ & $1.0\times10^{-21}$& $5.1\times10^{-21}$ \\
  \hline
 \end{tabular} 
  \label{tab:specN} 
\end{table}
%------------------------------------
%--------------Table--------------
\begin{table}
\center
\caption[Grav-Wave Spectral Amplitude for $SO(N)$ Model with Varying $\rho_T$]{\label{tab:specr} Spectral amplitudes as a
function of $\rho_T$ for simulations with
$N=4, \alpha = \lambda = 0.1$.  The
numerical values $\Omega^{\rm SIM}_{\rm gw}h^2$ are an average of the spectral values taken from the simulations, while the values in the second two columns are obtained from (\ref{Jestimate}) or (\ref{Festimate}).}
\ \\ 
\ \\
 \begin{tabular}{|cccc|}    
\hline
$\rho_T^{1/4}(M_{pl})$ & $\Omega^{\rm SIM}_{\rm gw} h^2$& $\alpha^2 \Omega^{\rm JKM}_{\rm gw} h^2$ & $\alpha^2\Omega^{\rm FFDG}_{\rm gw} h^2$    \\
\hline
\hline
   $10^{-3}$ & $4.7\times10^{-16}$ & $4.0\times10^{-17}$ & $2.1\times10^{-16}$ \\
   $10^{-4}$ & $3.8\times10^{-20}$ & $4.0\times10^{-21}$& $2.1\times10^{-20}$ \\
   $10^{-5}$ & $4.0\times10^{-24}$ & $4.0\times10^{-25}$& $2.1\times10^{-24}$ \\
   \hline
 \end{tabular}
    \label{tab:specN2}  
\end{table}
%------------------------------------
\dblspc
The results of the estimates and simulation data are summarized in Tables~\ref{tab:specN} and~\ref{tab:specN2}.   The numerical results suggest a large value of $N$ is not needed to
make a scale invariant spectrum.  Since the results of \cite{JonesSmith:2007ne,Krauss:2010df,Fenu:2009qf}
are derived using a large $N$ approximation for the amplitude, one does not expect
these estimates to accurately approximate the amplitude of the gravitational waves in the low-$N$ limit. It can be seen the simulations differ from analytic estimates by an order of magnitude
and are more accurate at large $N$, consistent with the fact that analytic methods are derived
from a large $N$ expansion.
    
\section{\label{sec:AGFM} Adaptive Green's Function Method}

In Sec. \ref{sec:GWSPT}, the simulation of the metric perturbation was done via a $4^{\rm th}$ order RK algorithm (RK4). In Sec. \ref{sec:CompStability} however, we saw that in general explicit numerical differential equation solving methods suffer from stability issues, including the RK4 method.  As the system is iterated, the instability of the solution can typically be suppressed at the step where it occurs by increasing the accuracy of the method; in particular, changing from a $2^{\rm nd}$-order method to a $4^{\rm th}$-order method can move the instability by a power of $2$ in the number of iterations.  However, we saw that using implicit methods can remove stability issues, so by choosing a different algorithm the instability in numerical calculations can disappear entirely. 

\begin{figure}[tpb]
\begin{center}
\postscript{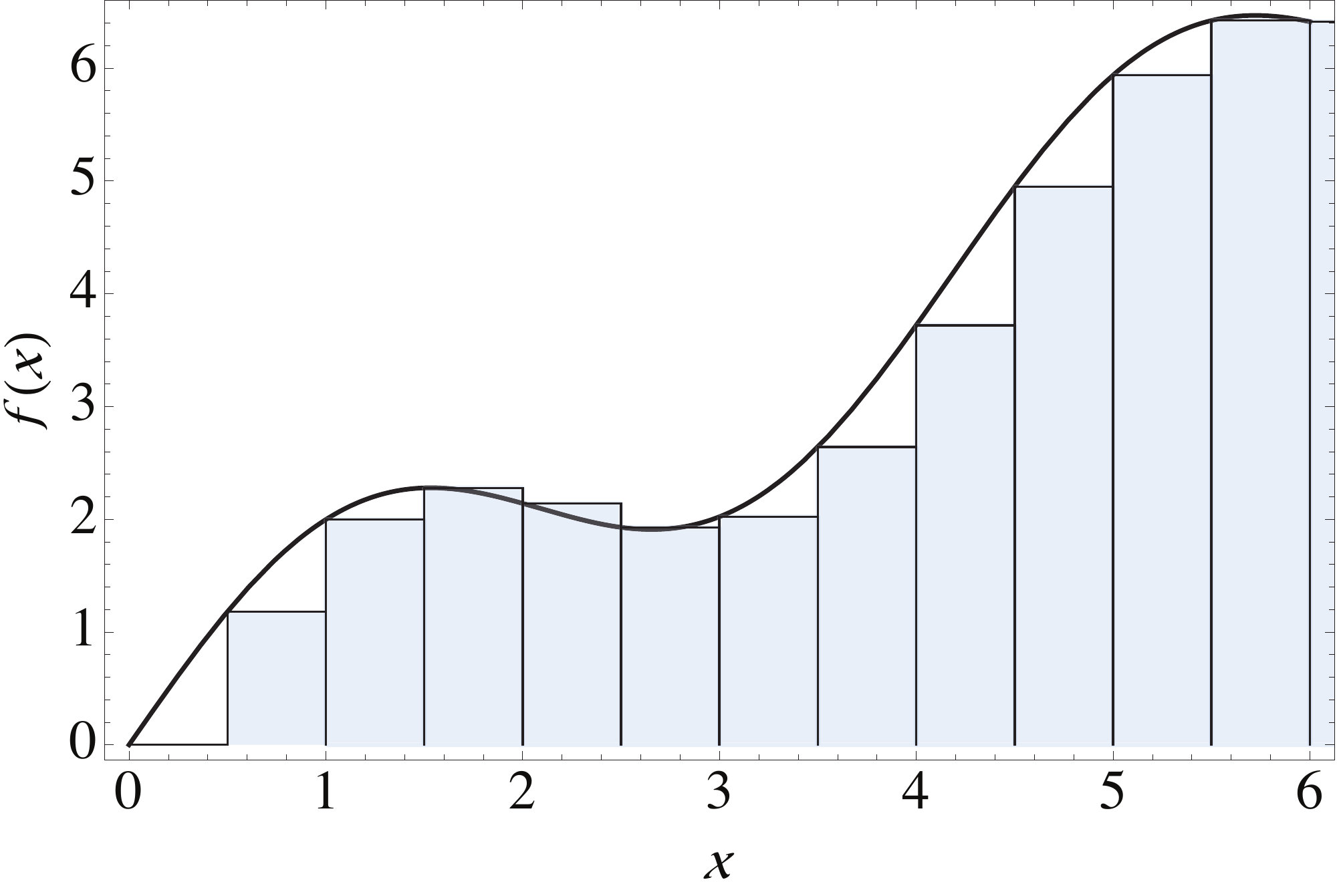}{0.7}
\caption[Rectangular Numerical Integration Method]{\sglspc \small The integral $\int_0^6 f(x) dx$ can be approximated in the rectangular sum method by summing the areas of the rectangular regions under the curve as seen in the figure above. }
\label{fig:RecMethod}
\end{center}
\end{figure}

An easy fix to an instability associated with the method of solving the metric perturbation is to use a method that numerically integrates an exact expression that only depends on the stability of the source.  For example we saw in Eq. (\ref{eq:GreensFuncs}) that the metric perturbation has an exact solution in terms of Green's functions (repeated here for connivence)
\beq
h^{\rm TT}_{ij} = \frac{16\pi}{M_{pl}^2}  \frac{k}{\eta^{n-1}} \int_{\eta_i}^{\bar \eta} d \eta' \ (\eta')^{n+1} \lpa j_{n-1}(k \eta') y_{n-1}(k \eta) - j_{n-1}(k \eta) y_{n-1}(k \eta') \rpa \ T^{\rm TT}_{ij}(\eta', {\bf k}),
\eeq
where the scale factor is assumed to have the form $a(\eta) = \alpha \ \eta^n$ and again $\eta$ is the conformal time coordinate.  To understand the computational advantage of this expression, let us consider the solution with $n=1$ appropriate for radiation dominated expansion.  Under this assumption Eq. (\ref{eq:GreensFuncs}) takes the form
\beq
h^{\rm TT}_{ij}(\eta, {\bf k}) = \frac{16\pi}{M_{pl}^2}  \int_{\eta_i}^{\bar \eta} d \eta'  \ \frac{\eta' \sin\lsb k(\eta - \eta')   \rsb }{k \eta} \ T^{\rm TT}_{ij}(\eta', {\bf k}),
\label{eq:GWrad}
\eeq
This expression can be approximately evaluated by any number of numerical integration methods, the simplest being the rectangular sum method (Riemann sum), which simply cuts the integration region into $N$ segments of  length $\Delta \eta$ and approximates the area under the curve as a sum of the values of integrand evaluated at $n \Delta \eta$ multiplied by the width $\Delta \eta$, where $n$ ranges from  $0$ to $N$, thus approximating the integration as a sum of the areas of rectangles making the area under the curve as seen in Fig. \ref{fig:RecMethod}
%----------------Figure------------------
%-----------------------------------------
In such a simple approximation the error associated with this method can be extracted by examining the integration over one of the rectangular regions
\beqa
\int_a^{a+\Delta \eta} f(\eta) d\eta &\approx&  \int_a^{a+\Delta \eta} \lpa f(a) + f'(a) (\eta - a) + \dots \rpa d\eta \nonumber \\
&=& f(a) \Delta \eta + f'(a) \frac{\Delta \eta^2}{2} +\dots
\eeqa
adding up all the rectangular regions takes the form
\beq
\int_a^b f(\eta) d\eta \approx \sum_{n=0}^{N-1} f(a + n \Delta \eta) \Delta \eta + \CO \lpa \Delta \eta \rpa \ ,
\eeq
thus the rectangular method is a $1^{\rm st}$ order approximation.  The equation  (\ref{eq:GWrad}) under this method takes the form
\beq
h^{\rm TT}_{ij}(\eta, {\bf k}) \approx \frac{16\pi  \Delta \eta}{M_{pl}^2}  \sum_{n=0}^{N-1} \ \frac{\eta_n \sin\lsb k(\eta - \eta_n)   \rsb }{k \eta} \ T^{\rm TT}_{ij}(\eta_n, {\bf k}) \  , \ \eta_n = \eta_i + n \Delta \eta.
\eeq
We can see the advantage of this form as the only source of instability comes from $T^{\rm TT}_{ij}(\eta_n, {\bf k})$, so long as $T^{\rm TT}_{ij}(\eta_n, {\bf k})$ is within machine sized numbers the method is absolutely stable, and furthermore much more computationally rapid.  This method was employed by~\cite{Price:2008hq} to calculate the spectrum associated with pre-heating mechanisms.  

\begin{figure}[tpb]
\begin{center}
\postscript{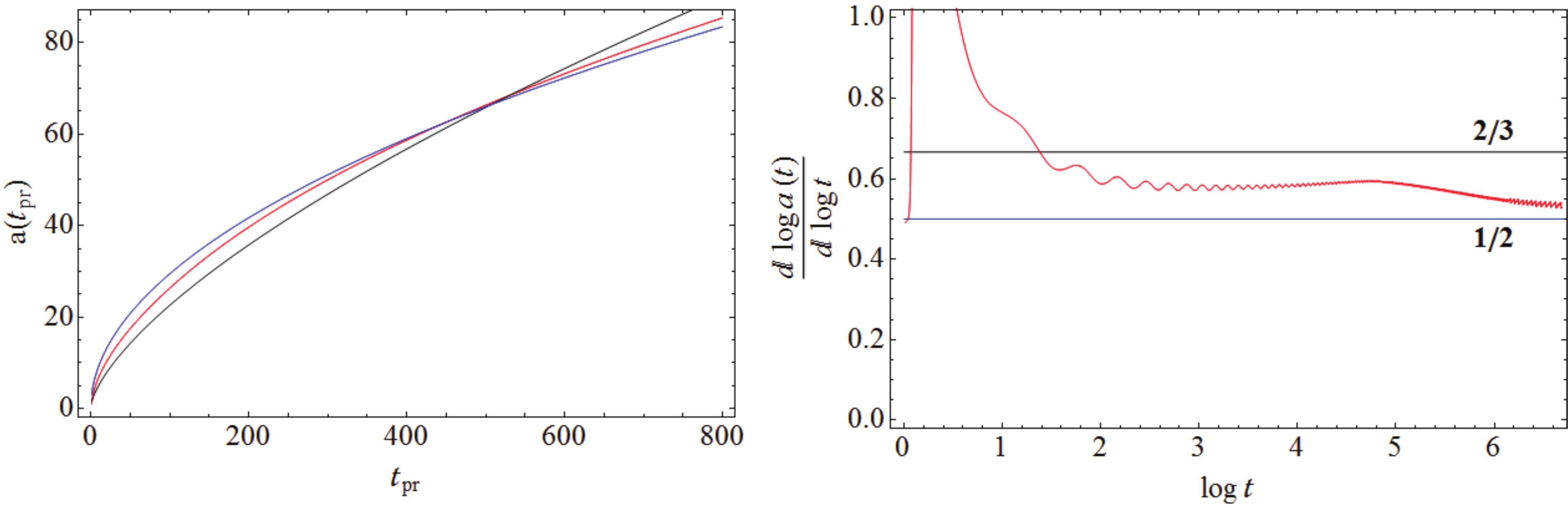}{1.0} 
\caption[Scale Factor Transitions from Semi-Matter to Radiation Dominated]{\sglspc \small (Left) The scale factor (red) initially starts in a state of close to that of a matter dominated form $a \propto t^{2/3}$ (black) but as the system evolves, the scale factor tends toward the scale factor for a radiation dominated universe $a \propto t^{1/2}$ (blue). (Right) The expression $d \log a / d \log t$ gives the power of the scale factor for a form $a = \alpha t^n \rightarrow \log a = \log \alpha + n \log t$.  In red is the value for the scale factor $d \log a / d \log t$ under self consistant evolution.  We can see the scale factor tends toward the form for matter domination $n = 2/3$ after which the scale factor falls towards a radiation dominated universe $n = 1/2$.  }
\label{fig:Rad2Mat}
\end{center}
\end{figure}

Unfortunately we've given up the flexibility of using the exact form of $a(\eta)$ for an approximate $a(\eta) = \alpha \eta^n$ form. This can present issues with this formulation, such as the case when the potential for pre-heating cases is taken as
\beq
V(\phi,\chi) = \frac{1}{2} m^2 \phi^2 + \frac{g^2}{2} \chi^2 \phi^2 \ ,
\label{eq:QuadPot}
\eeq 
where $\phi$ is the inflaton and $\chi$ is the field that will undergo pre-heating.  In this quadratic form of the potential {\sc LatticeEasy} can self consistently evolve the fields and scale factor.  Interestingly from this model the scale factor undergoes a transition, from an initially matter dominated scale factor (in comoving coordinates $a \propto t^{2/3}$) to a radiation dominated system (in comoving coordinates $a \propto t^{1/2}$), as is seen in Fig.  \ref{fig:Rad2Mat}.  
%----------------Figure------------------
%-----------------------------------------
In such a case it is not clear when to choose one scale factor form and exclude the other.  Unfortunately we have to make some approximation which naively the differential equation solving methods do not.  We do not want to sacrifice the numerical stability associated with the analytic method, but want to make as few approximations as possible. 

\subsection{Stability With No Scale Factor Assumption}

\begin{figure}[tpb]
\begin{center}
\postscript{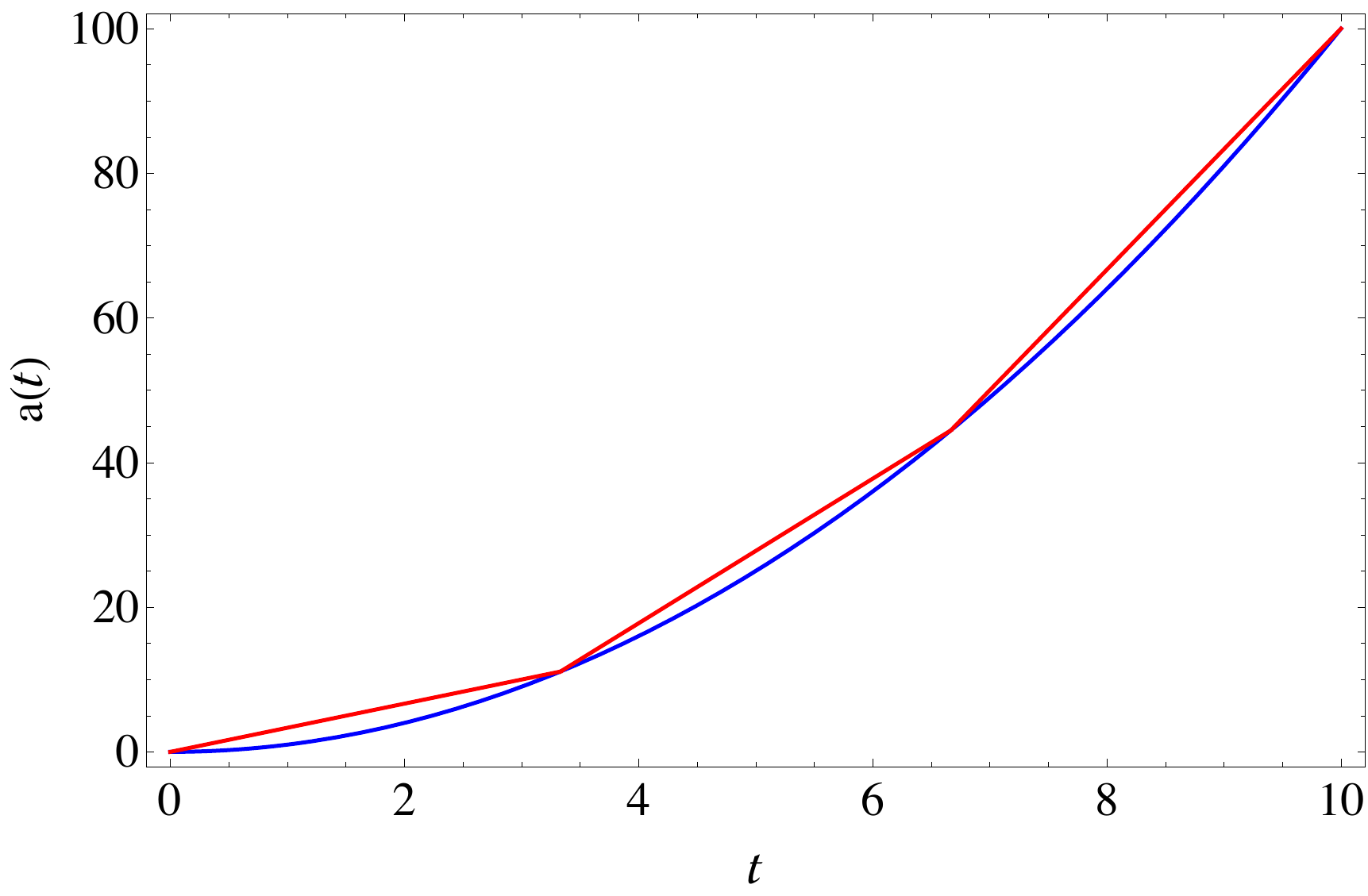}{0.6}
\caption[Approximate a Scale Factor as a Piece-wise Linear Segments Function]{ \sglspc \small The scale factor $a(t)$ as seen in blue, can be approximated as a piece-wise function of linear segments, as seen in red.  The approximation becomes better and better as the line segments become shorter and shorter.}
\label{fig:LineSegs}
\end{center}
\end{figure}

We retain both the stability of the numerical integration methods and have the flexibility of an arbitrary scale factor, by examining the form of the scale factor on the scale $\eta, \ \eta+ \Delta \eta$.  On this scale every scale factor can be approximated as linear segment. In fact, because all numerical methods must calculate the scale factor at discrete times in self consistent simulation; in numerical methods the scale factor is exactly a series of linear segments. To develop the method we take any scale factor and approximate its form as a piece wise function of linear segments as seen in Fig. \ref{fig:LineSegs}.
%----------------Figure------------------
%-----------------------------------------
With the explicit form
\beq
a(\eta) = \left\{ \begin{array}{l l}  \alpha_1 \eta + \beta_1 & \quad \eta_i \leq \eta < \eta_i + \Delta \eta \\ 
\alpha_2 \eta + \beta_2 & \quad \eta_i + \Delta \eta \leq \eta < \eta_i + 2\Delta \eta \\ 
\vdots \\ 
\alpha_N \eta + \beta_N & \quad \eta_i + (N-1) \Delta \eta \leq \eta < \eta_f \ .
\end{array} \right.
\eeq
Continuity of the scale factor is enforced by the conditions
\beq
\alpha_n \eta_{n-1} + \beta_n = \alpha_{n-1}\eta_{n-1} + \beta_{n-1} \ , \ \eta_m = \eta_i + m \Delta \eta \ .
\eeq
We can build up a solution of the metric perturbation $h_{ij}^{\rm TT}$.  To accomplish this, first assume that there is no source of gravitational waves before the time $\eta_i$, that $h_{ij}^{\rm TT}(\eta_i,{\bf k}) = 0$.  From the time $\eta_i$ to $\eta_i + \Delta \eta$ the scale factor is $\alpha_1 \eta + \beta_1$. The solution to the field equations (\ref{eq:DifEQGrav}) with this form of a scale factor is given by
\beqa
h_{ij}^{\rm TT} = & &\frac{1}{\alpha_1 \eta + \beta_1} \lpa A_1(k) \cos (k  \eta) + B_1(k) \sin (k \eta) \rpa \nonumber
\\
&+& \frac{16\pi}{M_{pl}^2 \ k (\alpha_1 \eta + \beta_1)} \int_{\eta_i}^{\bar \eta} d\eta' (\alpha_1 \eta' + \beta_1) \sin[ k (\eta - \eta')] T^{\rm TT}_{ij}(\eta',{\bf k}) \ , \nonumber \\
\label{eq:GenSol}
\eeqa
where
\beq
\bar \eta= {\rm Min}\left\{ \eta_1, \eta \right\}, \ \eta_i \leq \eta \leq \eta_1. 
\eeq
Examining (\ref{eq:GenSol}) we see we have a purely radiative part, which corresponds to the free propagation solution $T^{\rm TT}_{ij} = 0$, where $A_1, B_1$ are in general complex values that depend on $k$. We also we have a source term that generates the gravitational waves in this era.  To solve for the values $A_1$ and $B_1$ we use the initial conditions of the previous era.  For this particular era of $\eta_i \leq \eta < \eta_i + \Delta \eta$, the conditions are $h^{\rm TT}_{ij}(\eta_i, {\bf k} ) = 0$ and $\p_\eta h^{\rm TT}_{ij}(\eta_i, {\bf k} ) = 0$. 
  Computationally we are only interested in evaluating $h^{\rm TT}_{ij}$ at the endpoints of each era, {\it e.g.} $\eta_i + \Delta \eta, \ \eta_i + 2 \Delta \eta$ and so on. We need only to express (\ref{eq:GenSol}) at these points where the scale factor for era $n$ (conformal time given by $\eta_{n-1} \leq \eta < \eta_n $) is $a_n=\alpha_n \eta + \beta_n$.  With this notation the general solution for era $n$ of the metric perturbation becomes  
\beqa
h_{ij}^{\rm TT}(\eta_n,{\bf k}) = & &\frac{1}{a_n} \lpa A_n(k) \cos (k  \eta_n) + B_n(k) \sin (k \eta_n) \rpa \nonumber
\\
&+& \frac{16\pi}{M_{pl}^2 \ k a_n} \int_{\eta_{n-1}}^{\eta_n} d\eta' (\alpha_n \eta' + \beta_n) \sin[ k (\eta_n - \eta')] T^{\rm TT}_{ij}(\eta',{\bf k}) \ , \nonumber \\
\label{eq:GenSol2}
\eeqa
where continuity of the metric perturbation is used to solve for $A_n$ and $B_n$ as described for the first era.  To make this clear, we adopt a more compact notation denoting $h_n(\eta)$ as the metric perturbation evaluated at the time $\eta$ during the era $n$.  By demanding the continuity of the metric perturbation via
\beqa
h_{n-1}(\eta_{n-1}) &=& h_{n}(\eta_{n-1}) \ , \nonumber \\
\p_\eta h_{n-1}(\eta_{n-1}) &=& \p_\eta h_{n}(\eta_{n-1}) \ ,
\eeqa
we can solve for $A_{n+1}$ and $B_{n+1}$ in terms of the values of the generated gravitational waves generated from the previous era.  This can be understood as the generation of gravitational waves which then freely propagate from the source under the expansion of the universe, while the source generates new gravitational waves as well. 

By evaluating (\ref{eq:GenSol2}) and its derivative with respect to $\eta$ with the rectangular method, we have
\beqa
h_n(\eta_n) &\approx&  \frac{1}{a_n} \lpa A_n(k) \cos [k  \eta_n] + B_n(k) \sin [k \eta_n] \rpa \ , \nonumber \\
h'_n(\eta_n) &\approx& - \frac{\alpha_n}{a_n^2} \lpa A_n(k) \cos [k  \eta_n] + B_n(k) \sin [k \eta_n] \rpa \nonumber \\ & & +\frac{k}{a_n} \lpa B_n(k) \cos [k \eta_n] - A_n(k) \sin [k \eta_n] \rpa \nonumber
\\ & &+ \frac{16 \pi}{M_{pl}^2} T^{\rm TT}_{ij}(\eta_n, {\bf k}) \ .
\eeqa
The solution of the continuity conditions in this approximation give
\beqa
B_n &=& \frac{a_{n-1}}{k \sec (k \eta_{n-1})} \lsb h'_{n-1}(\eta_{n-1}) + \lpa\frac{\alpha_n}{a_{n-1}} + k \tan(k \eta_{n-1}) \rpa h_{n-1}(\eta_{n-1})  \rsb \ ,\nonumber \\
A_n &=& \frac{a_{n-1}}{\cos ( k \eta_{n-1} ) } h_{n-1}(\eta_{n-1}) - B_n \tan ( k \eta_{n-1} ) \ . 
\eeqa   
With this method we can build up the solution of $h'_n$ for each step throughout the program, for any scale factor evolution, while retaining the stability inherent in the numerical integration method. Since this method uses a Green's function approach to solving the field equations, while adapting the solution for differing regions of time it is dubbed the \emph{adaptive Green's function approach} (AGF).

\subsection{Example Use of Adaptive Green's Functions}

To show the effectiveness of the AGF approach we will simulate the motivating problem for a pre-heating simulation with the quadratic inflaton potential (\ref{eq:QuadPot}), and will compare the results of the AGF method with that of the Green's function methods assuming scale factors of definite matter and radiation form.  We will then compare the AGF method with a trapezoidal implicit numerical differential equation solver as a secondary check of the validity of the method.

The inherent scaling of the variables in {\sc LatticeEasy} for this particular potential enforces the condition $t_{pr} \propto t$. Since the source term is a function of the $t_{pr}$ the AGF method must be re-derived in terms of the solution to the field equation in comoving time, which results in a similar set of equations. 

\begin{figure}[tpb]
\begin{center}
\postscript{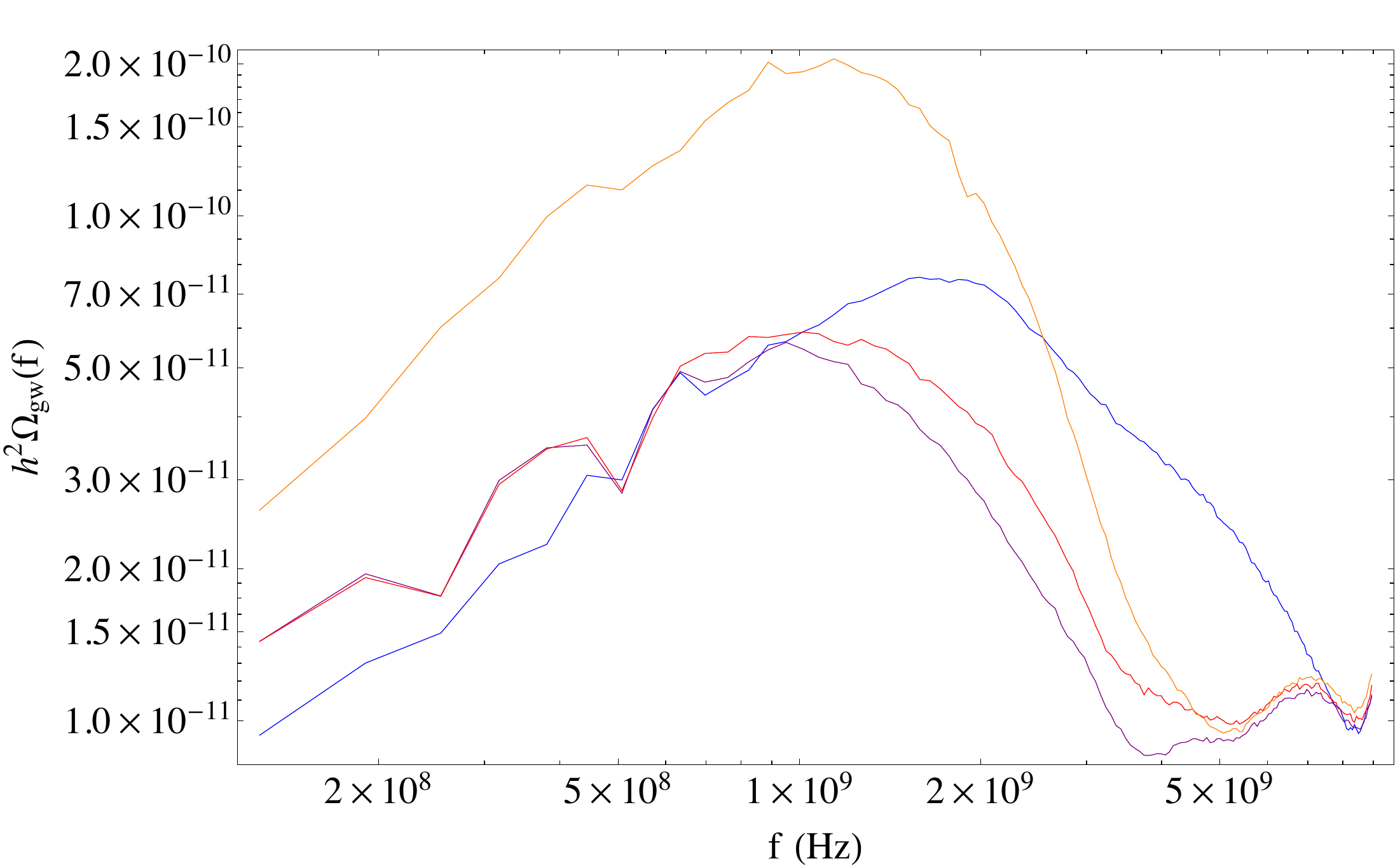}{0.852}
\caption[Comparing Numerical Methods for Power Spectrum of Grav-Waves]{ \sglspc \small The gravitational wave spectrum from pre-heating calculated from four different methods.  The orange curve is calculated from numerical integration for the Green's function method assuming a radiation dominated scale factor.  The Blue curve is calculated from numerical integration for the Green's function method assuming a matter dominated scale factor.  The red curve is the spectrum calculated from the AGF method, and finally the purple curve is calculated from the trapezoidal method.   It is easy to see that the AGF method and that of the trapezoidal method agree very well, while the matter and radiation methods tend to over estimate the spectrum. }
\label{fig:CompareMethods}
\end{center}
\end{figure}

We use a 3-dimensional lattice of size $N = 256$ points along each edge, along with a value $L_{pr} = 2.75$ which simulates a physical size of $L_{ph}=  2.8 \times 10^6 \ l_{pl}$.  The final program time is taken to be $t_f = 800$ and is the time in which the spectrum no longer receives large additions to its spectrum.  As was seen in Sec. \ref{sec:CompStability} the use of FFT causes the simulations to have many operations to perform.  In an attempt to decrease the computation time, we use parallel processing. specifically we used Open MP~\cite{OpenMP} to split the lattice in multiple segments to perform any operation on the lattice points in parallel so long as the operation did not depend on values being updated simultaneously in other lattice segments.  Even with this added improvement and attempts at optimization of calculations to avoid unnecessary repetitive calculations, the AGF calculation took 7.96 days on a computer with specifications: Intel¨ Xeon¨ X5680 12 core CPU at 3.33Ghz and 96 GB of 1333MHz RAM.  The simulations were computed with a metric perturbation step size of $\Delta t = 0.025 $ and field step size of $dt = 0.005$ to satisfy Courant conditions. The length of the calculation can be attributed to the fact that in comoving time the AGF method becomes twice as complicated, as the complex nature of the constants $A_n, B_n$ must be carefully handled.   The method of Green's functions with the same parameters for matter and radiation dominated scale factor forms took 4.81 days and 2.67 days respectively.  To make a comparison of the AGF with a numerical differential equation solving method we used an implicit trapezoidal method to solve the scaled equation 
\beq
\CH^{\rm TT}_{ij} (t, {\bf k})= a^r h^{\rm TT}_{ij} (t, {\bf k}) ,  \ s = 2 r- 3 \ , \ r = 3/2  \ , \nonumber
\eeq
\beq
\ddot{\CH}^{\rm TT}_{ij} (t, {\bf k}) + \lsb \frac{k^2}{a^{2(s+1)}} - r \frac{\ddot{a}}{a} + r(1-r) \lpa \frac{\dot{a}}{a} \rpa^2 \rsb \CH^{\rm TT}_{ij} (t, {\bf k}) = \frac{16\pi}{M_{pl}^2 a^{2(s+1)-r}} T^{\rm TT}_{ij} (t,{\bf k}) \ .
\eeq
The scaling is necessary as many numerical methods are unstable without the scaling procedure. In fact, 3 other possible methods were attempted without scaling (Implicit Euler, $2^{\rm nd}$ order RK , and Leap-frog Method) all of which developed instabilities under the evolution of the system.  After the simulation ends, we return to the metric perturbation via the relation
\beq
\dot{h}^{\rm TT}_{ij} (t, {\bf k}) = a^{-r} \dot{\CH}^{\rm TT}_{ij} (t, {\bf k}) - r a^{-(r+1)} \dot{a} \CH^{\rm TT}_{ij} (t, {\bf k}) \ .
\eeq
The trapezoidal method with the same parameters took 3.04 days to complete.  In Fig. \ref{fig:CompareMethods} the comparison of the four different methods is shown to demonstrates that the AGF gives the same value as the implicit trapezoidal method, while the radiation and matter scale factor Green's function methods over estimate the result. 
%----------------Figure------------------
%-----------------------------------------

The requirement for re-deriving the AGF equations for each unique potential because of the scaling of the time coordinate in {\sc LatticeEasy} is unfortunate as a general algorithm would be preferable.  The problem lies in the fact that $T^{\rm TT}_{ij}$ is calculated internally as function of $t_{pr}$.  In the metric perturbation equation $h^{\rm TT}_{ij} \propto \int d\eta' \dots T^{\rm TT}_{ij}(t_{pr})$ we need to numerically integrate a function which we do not have the form of in the appropriate coordinate system.  In order to circumvent this we can attempt to either transform $T^{\rm TT}_{ij}(t_{pr}) \rightarrow T^{\rm TT}_{ij}(\eta')$ at each step, which at this time has been left unexplored, or attempt to derive the AGF for $h^{\rm TT}_{ij}(t_{pr})$ that is a function of program time $t_{pr}$.  In program time, the field equations (\ref{eq:DifEQGrav}) take the form
\beq
\ddot{h}^{\rm TT}_{ij}(t_{pr},{\bf k}) + (3 + s) \frac{\dot{a}(t_{pr})}{a(t_{pr})} \dot{h}^{\rm TT}_{ij}(t_{pr},{\bf k}) + \frac{k^2}{a(t_{pr})^{2(s+1)} B^2} h^{\rm TT}_{ij}(t_{pr},{\bf k})  = \frac{16 \pi}{M_{pl}^2} \frac{T^{\rm TT}_{ij} (t_{pr},{\bf k})}{B^2 a(t_{pr})^{2 s} } \ ,
\eeq  
where the over dot indicates a derivative with respect to $t_{pr}$.  Taking the AGF approximation of linear segments gives
\beqa
&\ddot{h}^{\rm TT}_{ij}(t_{pr},{\bf k})& + \frac{(3 + s)  \alpha_n}{(\alpha_n t_{pr} + \beta_n)} \dot{h}^{\rm TT}_{ij}(t_{pr},{\bf k}) + \frac{k^2}{(\alpha_n t_{pr} + \beta_n)^{2(s+1)} B^2} h^{\rm TT}_{ij}(t_{pr},{\bf k}) \nonumber \\
  &=& \frac{16 \pi}{M_{pl}^2} \frac{T^{\rm TT}_{ij} (t_{pr},{\bf k})}{B^2 a(t_{pr})^{2 s} } \ .
\label{eq:GenAGF}
\eeqa
Unfortunately there seems to be no known general solution to this equation due to the $a^{2(s+1)}$ factor .  Even a change of variable of the form $h \rightarrow A a^r h$ does not help finding a general solution.    
%This admits solutions when $s \in R$  given by
%\beqa
%& &h^{\rm TT}_{ij}(t_{pr}) = \frac{\sqrt{k^2 + \alpha_n^2 a_n(t_{pr})^2}}{a_n(t_{pr})} \lpa C_{1,n} \cos \lsb \frac{k}{\alpha_n a_n(t_{pr})} + \tan^{-1}\lpa \frac{\alpha_n a_n(t_{pr})}{k} \rpa \rsb \right. \nonumber \\
%& +& \left. C_{2,n} \sin \lsb \frac{k}{\alpha_n a_n(t_{pr})} + \tan^{-1}\lpa \frac{\alpha_n a_n(t_{pr})}{k} \rpa \rsb \rpa %\nonumber \\
%&+& \frac{16\pi^2}{M_{pl}^2} \frac{\sqrt{k^2 + \alpha_n^2 a_n(t_{pr})^2}}{k^3 a_n(t_{pr})} \int_{t_{pr,n-1}}^{\bar t_{pr}} dt' \  a_n(t')^3 \lsb \alpha_n a_n(t')  \cos \varphi(t') - k \sin \varphi(t') \rsb T^{\rm TT}_{ij}(t',{\bf k}) \ , \nonumber \\
%\eeqa
%where $C_{1,n}, \ C_{2,n}$ are complex constants that do not depend on $t_{pr}$ or $k$ and
%\beqa
%\phi(t') &=&  \frac{k (t'-t_{pr})}{a_n(t_{pr}) a_n(t') } + \tan^{-1} \lpa \frac{ \alpha_n a_n(t_{pr})}{k} \rpa \ , \nonumber \\
%\bar t_{pr} &=& {\rm Min}\left\{ t_{pr,n}, t_{pr} \right\} \ , \ t_{pr,n-1} \leq t_{pr} < t_{pr,n} \ .
%\eeqa
Therefore a general AGF method at this point is not possible to derive by using {\sc LatticeEasy}, though the method of transforming the stress-energy tensor may result in some usefulness. The AGF method may still be viable if one uses a different method to solve the field equations rather than the method implemented in {\sc LatticeEasy} which forces $s = 2 r -3$, If another method is used to solve the field equations where we can set $s = -1$ permanently, then the AGF method should be able to be used with out problem.  The AGF method can certainly be many times slower than an alternative method, it does offer a stable method of calculating stochastic gravitational wave backgrounds that other numerical methods cannot.

\section{Summary of the Results and Conclusions}

The study of stochastic backgrounds reveals that it is possible to extract information on the scales and types of processes we associate with originating from the post inflationary era.  The sources studied here:  pre-heating and global second order phase transitions generate energy in gravitational waves on the scales $h^2 \Omega_{gw}\simeq 10^{-11}, \ 10^{-19}$, respectively. For pre-heating and second order phase transitions, the frequency of gravitational waves ranges of order $f\simeq 10^9, \ 10^6$ Hz,  respectively.   In studying global second order phase transitions flat spectra are produced with a Hubble wall like feature.  This Hubble-wall feature can be used to determine the scales at which the self ordering of the fields takes place, thus giving us knowledge of where we should expect new physics.  It was also investigated how the use of Green's functions methods can hope to improve the task of computing such stochastic backgrounds.  The use of numerical integration results in a more stable method of calculating the backgrounds, but it requires prior knowledge of the scale factor evolution.    Even in a more accurate self consistent solution, where the scale factor evolution is numerically solved for, the scale factor will be a series of linear segments due to the finite nature of numerical methods.  Motivated by the finite nature of the numerical process, one may assume that the scale factor truly is a linear piece-wise function.   This allows one to generate an algorithm for calculation of a stochastic background that makes use of the Green's functions and does not require prior knowledge of the form of the scale factor. We name this algorithm the adaptive Green's function method.  This method generates a stochastic background for the troublesome process of a pre-heating with one additional scalar field and a quadratic inflaton potential, that agrees well with the result of the numerical differential equation solving methods.  The use of {\sc LatticeEasy} is not straight forward when combined with the AGF method for general potentials. For this reason we are forced to abandon the use of {\sc LatticeEasy}'s field evolution methods and require further study into methods more compatible with the AGF method.

\newcommand{\teq}{t_{\rm EQ}}

% End of macros

%--------------------------------------------Begin New Section----------------------------

\newpage
\

\

\

\noindent\textbf{\Huge Part III:}

\

%---------------------------------Ch 4-----------------------------------------
\noindent\textbf{\huge Dark Sector Physics}
\addcontentsline{toc}{chapter}{Part III- Dark Sector Physics}
\ \\
\ \\
\noindent\textbf{\huge $N_{\rm eff} = 3.62 \pm 0.25 \ , \ \bar {\rm f} {\rm f} \rightarrow H^*/h^* \rightarrow \bar w w$}

\newpage

\thispagestyle{fancy}
\chapter{Dark Sector Physics}
\thispagestyle{fancy}
\pagestyle{fancy}

%---------------------------------Ch 4: Section 1-----------------------------------------

Beyond standard model (SM) physics models to be probed at the Large
Hadron Collider (LHC) often include the concept of a hidden sector,
consisting of $SU(3) \times SU(2) \times U(1)$ singlet
fields. Independent of any model, the standard sector and the hidden
sector are coupled by interactions of gauge-invariant operators which
illuminate the path for exploring structures in the hidden sector by
observing phenomena in the visible standard sector. A tantalizing
realization of this idea is provided by the Higgs portal, which
connects the Higgs fields in the two sectors by an elementary quartic
interaction~\cite{Schabinger:2005ei,Patt:2006fw,Barger:2007im,Barger:2008jx,Andreas:2010dz,Logan:2010nw,Bock:2010nz,Englert:2011yb}. Such
a construct moves a precision study of the Higgs sector into a central
position of new physics searches at the LHC. Likewise, astrophysical
observations open the gates for complementary information to further
test the Higgs portal hypothesis and to improve our understanding of
the physics in the hidden sector.

\section{\label{sec:DM} Extra Relativistic Degrees of Freedom}

Precision cosmology has been primarily driven by measurements of the CMB temperature anisotropies~\cite{Dodelson:2003ft}. The anisotropies can be decomposed into spherical harmonics
via
\beq
\delta T(\hat n) =  T(\hat n) - T_{\rm CMB} = \sum_{l,m} a_{lm} Y^m_l(\hat n) \ , 
\eeq
where $T_{\rm CMB}$ is the average temperature across the sky observed today as  $T_{\rm CMB}
\simeq 2.7255(6)~{\rm K}$~\cite{Beringer:1900zz}, and $\hat n$ a unit vector towards some point in the sky parameterized by polar and azimuthal angle.   From this decomposition, the angular power spectrum 
may be constructed in terms of the variance, $C_l$, of the CMB fluctuations as a function of multipole
number $l$ given by
\beq
\langle \frac{\delta T}{T}(\hat n) \frac{\delta T}{T} ( \hat n' )  \rangle_{\hat n \cdot \hat n' = \cos \theta} = \frac{1}{4\pi} \sum_{l=0}^\infty  (2l+1) C_l P_l(\cos \theta),
\eeq
where $P_l(x)$ are the Legendre polynomials, $\theta$ is the angle between two different directions in the sky $\hat n, \ \hat n'$, and $\la \dots \ra$ denote an average across the entire sky.  The most recent observational result for $C_l$ is given in Fig. \ref{fig:CSMTab}. The cause of the peaks and troughs of the CMB power spectrum is 
the compression due to gravitational potentials and resistance to this compression due to
pressure gradients in the early universe. The angular scales, $\theta_s$, of the acoustic
peaks are highly sensitive to the angular size of the sound horizon (the distance pressure waves of speed $c_s$ can travel in the early universe plasma since the big bang), and are given by
$\theta_s = r_s/D$, where the distance to the last scattering surface of the CMB is $D = \int_{\rm last \ scattering}^{\rm today} dt/a(t)$. The angular size $\theta_s$ is very precisely determined by the data, however, $D$ cannot be precisely determined because the density of the dark energy as a function
of the scale factor is unknown. Therefore, another angular scale, which can be measured, must be used to eliminate $D$. 

\begin{figure}[tbp]
\postscript{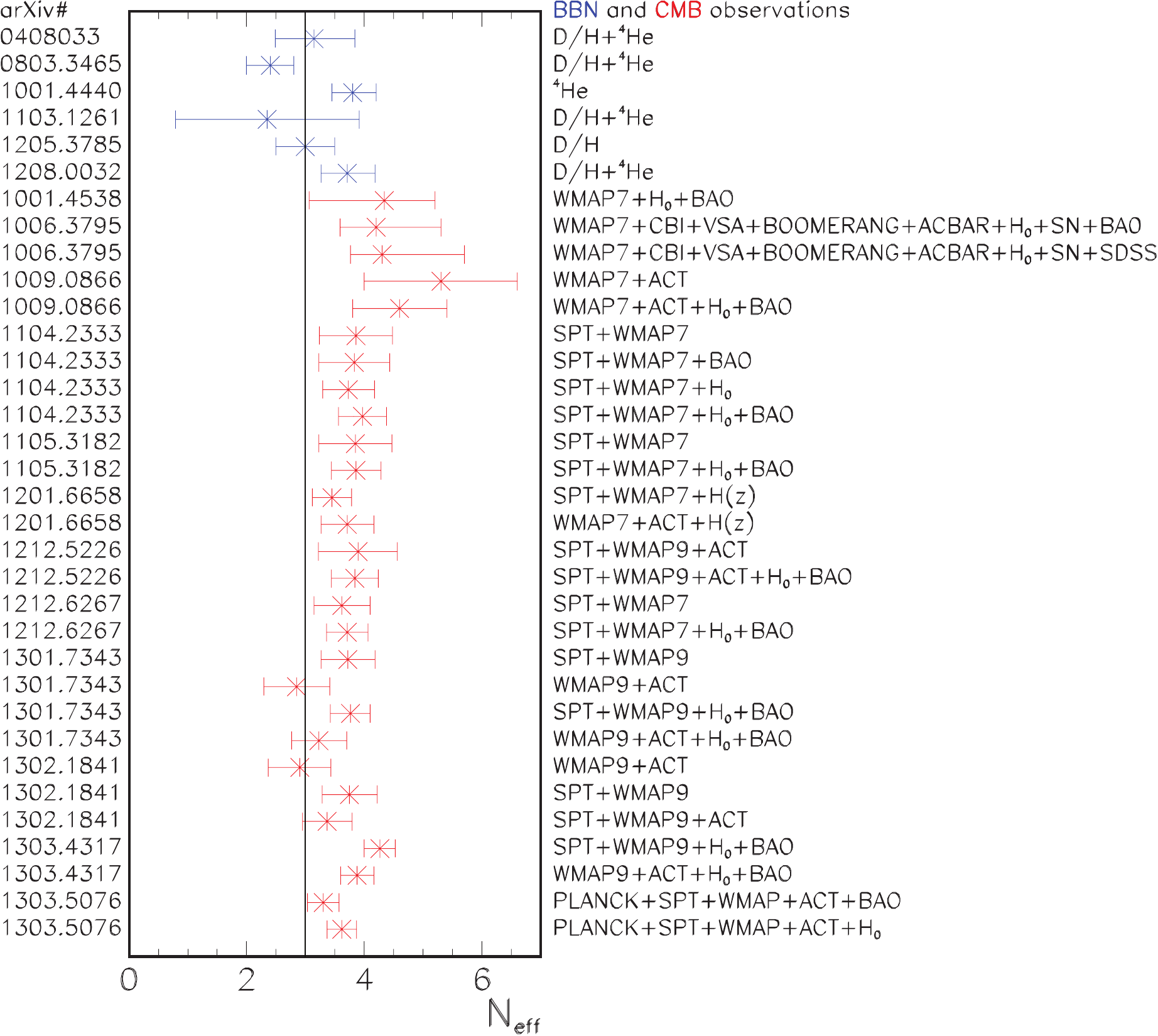}{1.0} 
\caption[Collection of Cosmological Measurements of $N_{\rm eff}$]{\sglspc \small A selection of the most recent cosmological $N_{\rm eff}$
  measurements and the $1\sigma$
  confidence  intervals from  various combinations of models and data
  sets.  The first five measurements correspond to BBN
  observations~\cite{Cyburt:2004yc,Simha:2008zj,Izotov:2010ca,Mangano:2011ar,Pettini:2012ph,Steigman:2012ve}. The
  other intervals result from a combination of the latest distance
  measurements from the baryon acoustic oscillations (BAO) in the
  distribution of galaxies~\cite{Percival:2009xn}, precise measurements of $H_0$ by
  the Hubble Space Telescope~\cite{Riess:2011yx}, and CMB data collected by the Wilkinson Microwave Anisotropy Probe
(WMAP), the South Pole Telescope (SPT), the Atacama Cosmology
Telescope (ACT), the Cosmic Background Imager (CBI), the Very Small
Array (VSA), BOOMERANG, the Arcminute Cosmology Bolometer Array
Receiver (ACBAR), and the Planck
mission~\cite{Komatsu:2010fb,GonzalezGarcia:2010un,Dunkley:2010ge,Hou:2011ec,Keisler:2011aw,Moresco:2012jh,Hinshaw:2012fq,Hou:2012xq,DiValentino:2013mt,Calabrese:2013jyk,Benetti:2013wla,Ade:2013lta}.  Image from~\cite{Anchordoqui1} .}
\label{fig:Neff} 
\end{figure}

The diffusion angular scale can be used for this purpose.  Peaks of temperature anisotropy on scales smaller than the photon diffusion length become Silk dampened, which is associated with random walks of photons on the small scale in the plasma.  This blurs out (weakens) correlations amongst small scale anisotropies. Diffusion causes a drop in power toward high $l$ and makes the CMB power spectrum sensitive to the
angular scale of the diffusion length $r_{\rm d}$ given by $\theta_{\rm d} = r_{\rm
  d}/D$. 
  For a random walk process, the diffusion distance increases as the square root of time, thus
\beq
t^{1/2} \propto a \rightarrow t  \propto \frac{1}{H} \rightarrow r_d \propto H^{-1/2}.
\eeq
Since the sound horizon $r_s = c_s \ t = c_s/2H$, we can form the relation 
\beq
\frac{\theta_{\rm d}}{\theta_{\rm s}} =\frac{r_d}{r_s} \propto H^{1/2} \ ,
\label{eq:HorizonsRatio}
\eeq 
 where $H$ is the Hubble parameter at the time the angular scales $\theta_{\rm d}$ and $\theta_{\rm s}$ froze into the CMB. At this time $t\simeq t_{\rm EQ} $, from which we can assume the energy density is dominated equally by $\rho_{\rm rad}$ and $\rho_{\rm mat}$. The Friedmann equation then allows 
 \beqa
 H^{1/2} = \lpa \frac{8 \pi}{3 M_{pl}^2} (\rho_{\rm rad} + \rho_{\rm mat}) \rpa^{1/4} &=& \lpa \frac{8 \pi}{3 M_{pl}^2} \frac{2 \pi^2}{30} g_\rho(T_{\rm EQ}) T_{\rm EQ}^4    \rpa^{1/4}\ , \nonumber \\
  g_\rho(T_{\rm EQ}) &=& 2 +\frac{7}{8} \cdot 2 \cdot N_{\rm eff} \lpa \frac{4}{11} \rpa^{4/3}
 \label{eq:HintoNeff}
 \eeqa 
 where $g_\rho$ is the effective degrees of relativistic species at the time of the formation of the angular scales.  The effective degrees from the concordance model of cosmology would suggest that $g_\rho$ consists of a photon with 2 spin degeneracies, and $N_{\rm eff} = 3$ left-handed neutrinos with 2 spin degeneracies each.  The factor $7/8$ reflects the fermi statistical nature of the neutrinos and $(4/11)^{4/3}$ reflects the difference in temperatures of the neutrinos to the photon plasma and can be derived from conservation of entropy. 
  The precise measurements of (\ref{eq:HorizonsRatio}) along with (\ref{eq:HintoNeff}) should give a consistency check of $N_{\rm eff}\approx 3$ for the effective number of relativistic species normalized to that of a left-handed neutrino at the time $\teq$.  Currently, high-resolution observations of the CMB temperature
anisotropy are providing a precise measurement of the damping tail of CMB power spectrum, shedding light on $N_{\rm eff}$, and over the past few years evidence has been accumulating for a possible
excess on the number of ``equivalent'' light neutrino species above SM
expectation, $N_{\rm eff} \approx 3 + \Delta N > 3$. A selection of the most
recent cosmological $N_{\rm eff}$ measurements and the $1\sigma$
confidence intervals from various combinations of models and data sets
are shown in Fig.~\ref{fig:Neff}.

At the time of this writing, the Planck spacecraft has measured the CMB to an
unprecedented precision~\cite{Ade:2013lta}. One of the most striking
results of the mission is that the best-fit Hubble constant
has the value $h= 0.674 \pm 0.012$, with a dark energy density
parameter $\Omega_\Lambda = 0.686 \pm 0.020$, and matter density
parameter $\Omega_{\rm M} = 0.307 \pm 0.019$. This result is at more
than 2.3$\sigma$ deviations of the value obtained with the Hubble
Space Telescope, $h = 0.738 \pm 0.024$~\cite{Riess:2011yx}.  The
impact of the new $h$ determination is particularly complex in the
investigation of the effective number of relativistic degrees of
freedom (as the times which the angular scales $\theta_s, \theta_r$ change with $h$). From the CMB data alone, the Planck Collaboration reported
$N_{\rm eff} = 3.36 \pm 0.34$. Adding baryon acoustic oscillation (BAO) data yields $N_{\rm eff} =
3.30 \pm 0.27$. Both of these values are consistent with the SM value
of 3.046 (slightly larger than 3 as the neutrino decoupling from photons is not an abrupt process~\cite{Mangano:2005cc}). Adding the $H_0$ measurement to the CMB data gives $N_{\rm
  eff} =3.62 \pm 0.25$ and relieves the tension between the CMB data
and $H_0$ at the expense of new neutrino-like physics (at around the
$2.3 \sigma$ level). In other words, it is possible to alleviate the
tensions between the CMB, BAO, and $H_0$ data by invoking an increase
in $N_{\rm eff}$. It should be noted, however, that any preference for
new physics comes almost entirely from the astrophysical data
sets. 

The CMB data can also be complemented by big bang nucleosynthesis (BBN) data, which offers the deepest reliable probe of the early universe, being based on well-understood SM
physics~\cite{Sarkar:1995dd,Olive:1999ij,Steigman:2007xt}.  The
expansion rate of the universe at early times increases with the
number of relativistic particle species in thermal equilibrium, and
this in turn sets timescales for BBN. One can then use predictions of
the abundances of light nuclei (D, $^3$He, $^4$He, $^7$Li) synthesized
at the end of the ``first three minutes'' to constrain the number of
light species. For instance, in SM cosmology the
neutron-proton interconversion rate $n \leftrightharpoons p$ drops out
of equilibrium at $T \sim 1~{\rm MeV}$.  Nearly all the surviving
neutrons when nucleosynthesis begins end up bound in the most stable
light element $^4$He. Therefore, the primordial mass fraction of
$^4$He, conventionally referred to as $Y_{\rm p}$, can be estimated by
the simple counting argument: $Y_{\rm p} = 2 n/p (1+ n/p)^{-1}$. If
$N_{\rm eff} >3$, the expansion rate at fixed temperature is increased and
the neutron-proton ratio freezes out at a higher temperature, leaving
more free neutrons, and thus a larger primordial abundance of
$^{4}$He.  By measuring $N_{\rm eff}$ we may be able to detect some new physics BSM.

\subsection{Adding New Species}

To account for $N_{\rm eff} > 3$, we must add some particles to the concordance model of cosmology that predicts the evolution of a spatially-flat expanding universe filled with dark energy $(\Lambda)$,
dark matter (DM), baryons (b), photons $(\gamma$), and and three
flavors of left-handed ({\it i.e.} one chiral state $\nu_L$)
neutrinos (along with their right-handed, antineutrinos $\bar
\nu_R$). The Hubble parameter $H$ is
determined by the total energy density
\beq
  H^2 (t) = \lpa \frac{\dot a(t)}{a(t)} \rpa^2 =  \frac{8 \pi}{3 M_{pl}^2} \
  [\rho_\Lambda (t) + \rho_{\rm DM} (t) + \rho_{\rm b}(t) + \rho_\gamma (t) + 3 \rho_{\nu_L} (t)] \, .
\label{uno}
\eeq
The quantities of importance can be expressed in units of $h$ to avoid uncertainties in $H$,  {\it
  e.g.} $\Omega_{M} = \Omega_{\rm DM} + \Omega_{\rm b} \simeq [0.111(6) +
0.0226(6)] h^{-2}$~\cite{Beringer:1900zz}.

In the early universe, the energy density is
dominated by radiation from extremely relativistic particles. When the temperature
 drops below the electron mass, the standard model of particle physics
constrains the relativistic particle content to photons and
neutrinos. As a result, the radiation energy density then is $\rho_{\rm R}
= \rho_\gamma + 3 \, \rho_{\nu_L}$.  To accommodate new physics in the
form of extra relativistic degrees of freedom, it is convenient to
account for the extra contribution to the SM energy density by
normalizing it to that of an ``equivalent'' neutrino
species. The number of ``equivalent'' light
neutrino species can then be expressed as
\begin{equation}
N_{\rm eff} \equiv \frac{\rho_{\rm R} - \rho_\gamma}{\rho_{\nu_L}} \,,
\label{eq:NeffMasterEQ}
\end{equation}
which quantifies the total ``dark'' relativistic energy density (including
the three left-handed SM neutrinos) in units of the energy density of a
single Weyl neutrino~\cite{Steigman:1977kc}. Any relativistic degree of freedom originating
from BSM physics is then included in $N_{\rm eff}$.

If neutrinos are fully decoupled prior to $e^\pm$ annihilation, they
do not share in the energy transferred from annihilating $e^\pm$ pairs
to photons. In this very good approximation, the
photons are hotter than the neutrinos in the post-$e^\pm$ annihilation
universe by a factor $T_\gamma/T_{\nu_L} = (11/4)^{1/3}$ and so
\begin{equation}
\rho_{\nu_L}  =   \frac{7}{8} \left(\frac{4}{11} \right)^{4/3} \rho_\gamma \,,
\label{cuatro}
\end{equation}
yielding
\begin{equation}
N_{\rm eff} = \frac{8}{7} \left(\frac{11}{4}\right)^{4/3} \frac{\rho_{\rm R} - \rho_\gamma}{\rho_\gamma} \, .  
\end{equation}
Since the temperature of the CMB is $T_{\rm CMB}
\simeq 2.7255(6)~{\rm K}$~\cite{Beringer:1900zz}, we can determine the energy density in photons as measured today to be
\begin{equation}
\Omega_{\gamma}  = 
\frac{\pi^2 (k_B T_{\rm CMB} )^4}{15 \hslash^2 c} \frac{8 \pi G}{3\times 10^4 \
   ({\rm km} \ {\rm s}^{-1} \ {\rm Mpc}^{-1})^2}\frac{1}{h^2} \simeq 2.471 \times 10^{-5} \ h^{-2}
\, .
\label{omegagamma}
\end{equation}
 On the other hand,  the energy density of the cosmic neutrino background
(C$\nu$B) is found to be
\begin{equation}
\Omega_{\nu_L} = \frac{3 \rho_{\nu_L}}{\rho_c} = \frac{\sum_i
  m_{\nu_i}}{93.14 \, h^2~{\rm eV}} \ ,
\label{tresmeses}
\end{equation}
where now we have included the fact that neutrinos are massive and $3 \rho_{\nu_L} \approx \sum_i m_{\mu_i}$, indicating the neutrino species are non-relativistic today. We know from neutrino oscillation experiments
that at least two of the neutrino states are non-relativistic today
because both $(\Delta m^2_{31})^{1/2} \simeq 0.05~{\rm eV}$ and
$(\Delta m^2_{21})^{1/2} \simeq 0.009~{\rm eV}$~\cite{GonzalezGarcia:2012sz} are larger than the
temperature $T_{\rm C \nu B} \simeq 1.96~{\rm K} \simeq 1.7 \times
10^{-4}~{\rm eV}$. If the third neutrino state
is very light and still relativistic, its relative contribution to $\Omega_{\nu_L}$
is negligible and (\ref{tresmeses}) remains an excellent approximation of the
total neutrino energy density. 

One finds that $\Omega_{\nu_L}$ is restricted to the approximate
range  $0.000637 (0.001078) < \Omega_{\nu_L} h^2 < 0.0637$, and therefore, the radiation energy density today, $\Omega_{\rm rad} \approx \Omega_\gamma$, is completely negligible.  It follows from (\ref{uno}) that new physics contributions to $\rho_{\rm rad}$ alter cosmology through the effect on the scale factor and since $\rho_{\rm rad}$ is negligible today, the early universe
becomes the sole laboratory in which one can measure $N_{\rm eff}$.

Several explanations have been proposed to explain a possible
$\Delta N$ excess. These include: \\
{\it (i)} models based on milli-weak
interactions of right-handed partners of three Dirac
neutrinos such as the SM$^{++}$ model among others~\cite{Anchordoqui:2011nh,SolagurenBeascoa:2012cz,Anchordoqui:2012qu}. \\
{\it (ii)} models based on active-sterile mixing of neutrinos in a
heat bath~\cite{Hamann:2010bk,Krauss:2010xg}. \\
{\it (iii)} models in
which the extra relativistic degrees of freedom are related to
possible dark matter candidates produced via decay of heavy
relics~\cite{Ichikawa:2007jv,Hasenkamp:2011em,Menestrina:2011mz,Feng:2011in,Hooper:2011aj,Bjaelde:2012wi,Hasenkamp:2012ii,DiBari:2013dna}. 

An interesting consequence of such a non-thermal DM scenario is that if the
lifetime of the decaying particle $X$ is longer than about $10^3$ seconds,
the expansion history of the universe during the era of BBN will not
have $\Delta N_X$ contributions to number of ``equivalent'' light
neutrino species. Moreover, if there is a light DM particle that
annihilates to photons after the $\nu_L$ have decoupled, the photons
are heated beyond their usual heating from $e^\pm$ annihilation,
reducing the late time ratio of neutrino and photon temperatures (and
number densities), leading to $N_{\rm eff} < 3$~\cite{Ho:2012ug,Ho:2012br,Steigman:2013yua}. This opens the window for the addition of one or more $\nu_R$ neutrino flavors while remaining consistent
with $N_{\rm eff} = 3$. A thorough study of the various possibilities listed above has been pursued in ref~\cite{Brust:2013ova}.

We will use the possibility that both right-handed Dirac
neutrinos and non-thermal dark matter particles can contribute to the
number of ``equivalent'' light neutrino species, {\it i.e.} $\Delta N
= \Delta N_\nu + \Delta N_X$.

\subsection{Non-Thermal Dark Matter}

A series of independent observations involving galaxies and clusters of
galaxies as well as the CMB seem to indicate that the most successful
structure-formation models have been those in which most of the mass
in the universe comes in the form of cold dark matter (CDM), {\it
  i.e.}  particles that were moving non-relativisticly in the early universe
~\cite{Jungman:1995df,Bertone:2004pz,Feng:2010gw}. A mixture of about
80\% CDM and 20\% hot dark matter would only reproduce the data on
nearby galaxies and clusters if the average density of matter in the
universe were at or close to the critical density, $\Omega_{\rm M} =
1$. However, like all such critical-density models, cold plus hot dark
matter models require that galaxies and clusters must have formed
fairly recently, which disagrees with
observations. The evidence increasingly favors $\Lambda$CDM models, in
which CDM and baryons make up about a third of the critical density,
with a cosmological constant or some other form of dark energy
contributing the remainder.  We now address adding non-thermal matter to the standard history of the universe.

\subsubsection{\underline{Relativistic Constraint on $X$}}
First we will address adding in non-thermal dark matter to the concordance model of cosmology by assuming that the total DM in existence today is conceived of a small fraction of particles of type $X$, which have cooled due to
expansion of the universe and are non-relativistic today,  but were relativistic at the CMB epoch, and a larger fraction of type $\chi$ that constitute the primordial CDM, which is cold today and has always been cold. We assume the particles of type $X$ are produced via decay of a heavy relic particle of type $X'$, which allows the particle $X$ to be relativistic at the CMB epoch so that these particles can add to $N_{\rm eff}$.  However, to do this we must ensure that during the CMB epoch the $X$ particles are relativistic.  We can set a limit on this by 
assuming the progenitor particle $X'$ is initially non-relativistic and unstable with a 
lifetime $\tau$. When $X'$ decay to $X' \rightarrow X + \gamma$ (at present, we do not
consider the more complicated scenario in which high energy neutrinos
are among the decay products~\cite{GonzalezGarcia:2012yq}) in the
center-of-mass frame of $X'$ (this should also be a good approximation
of any frame, as we assume that $X'$ is non-relativistic so its mass
energy dominates) we have from the conservation of energy
\begin{equation}
M_{X'}= E_\tau + p_\tau = \sqrt{p_\tau^2 + M_X^2} + p_\tau \Rightarrow p_\tau =
\frac{{M_{X'}}^2- M_X^2}{2 M_{X'}} \,,
\end{equation}
where $E_\tau = M_X \gamma_\tau, \ p_\tau$ is the energy and momentum, respectively, of the particle $X$ at
the time $\tau$ when $X'$, with mass $M_{X'}$ decays. The Lorentz boost factor $\gamma_\tau$ at time $\tau$ can be expressed as 
\begin{equation}
\gamma_\tau = \frac{E_\tau}{M_X} = \frac{M_{X'}}{2 M_X}+\frac{M_X}{2M_{X'}} \, .
\end{equation}
As the universe expands and cools, the $X$ momentum and thus energy will red shift, giving~\cite{Kolb:1990vq}  
\beq
E^2 (t) = p_\tau^2 \lpa \frac{a(\tau)}{a(t)} \rpa^2 + M_X^2  = M_X^2 \gamma(t)^2 \ .
\eeq
Using the redshift, we can express the Lorentz boost factor $\gamma$ as a function of time as
\beq
  \gamma^2(t)   =  1 + \left( \frac{p_\tau \ a(\tau)}{M_X \ a(t)}\right)^2 
   =   1 + \lpa \frac{a (\tau)}{a(t)} \right)^2 \lpa \frac{E_\tau^2 -
    M_X^2}{M_X^2}\rpa 
    =  1 +  \lpa \frac{a (\tau)}{a(t)} \rpa^2 \lpa \gamma_\tau^2 -
  1\rpa  \, .
\eeq
For $\left[ a(t) / a(\tau) \right]^2 \left( \gamma_\tau^2-1 \right) \ll 1$, the Lorentz boost factor takes the approximate form
\beq 
\gamma (t) \approx 1 + \frac{1}{2} \lpa \frac{a (\tau)}{a(t)} \rpa^2 \lpa \gamma_\tau^2 -1 \rpa - \frac{1}{8} \lpa
\frac{a (\tau)}{a(t)} \rpa^4 \lpa \gamma_\tau^2 -1 \rpa^2 \ + \cdots \, .  
\eeq
From this we can determine the non-relativistic limit by demanding the magnitude of
the second term in the expansion to be greater than the third term to consider the $X$ particle non-relativistic, 
which results in
\beq
 \lpa \frac{a (\tau)}{a(t)} \rpa^2 \lpa \gamma_\tau^2 -1 \rpa < 4 \, . 
\eeq 
Therefore, by this criteria the particle $X$ is relativistic if
$\gamma (t) > \sqrt{5}$, and so if this condition is met during the CMB epoch, then the $X$ particles can contribute to $N_{\rm eff}$.

\subsubsection{\underline{Adding the Relativistic DM to $\rho_{\rm rad}$ }}

Continuing, we must add the proper dark matter evolution into the Friedmann equations, so we now examine the time dependence of the DM energy density  (the components being the type $\chi$ and the type $X$). At any time after
the decay of $X'$ the total energy density in DM is
\begin{equation}
\rho_{\rm DM}(t) = \rho_X(t) + \rho_{\chi}(t) = \gamma(t) M_X n_X(t) + \rho_{\chi}(t) \,,
\label{nueve}
\end{equation} 
where $\rho_X(t)$ is the energy density in the DM particles of type $X$ (and may or may not be cold) and 
$\rho_\chi (t)$ is the energy density in the primordial CDM.  The second equality comes from the fact that we take $n_X(t)$ to be the number density of particles of type $X$ and each particle
has energy $M_X \gamma(t)$.

Now, since $\rho_{\chi}$ is made entirely of dust-like (cold) particles, it becomes diluted as
\begin{equation}
 \rho_\chi(t) = \frac{\rho_\chi(\rm today)}{a^3(t)} \, ,
\label{diez} 
\end{equation}
 where we have used the convention $a({\rm today}) = 1$.  Likewise,
 any number density of particles in an expanding universe will scale
 as $a^{-3}(t)$ and so we can write
\begin{equation}
 \rho_X(t) = \frac{M_X n_X({\rm today})}{a^3 (t)} \gamma(t).
\label{once} 
\end{equation}
At this point we make an assumption that the total dark matter today is entirely in the form of cold dark matter (CDM),
\begin{equation}
\rho_{\rm DM}(\rm today) = \rho_{\rm CDM} (\rm today) \,,
\end{equation}
and so the energy density for particles of type $X$ make up some fraction $f$ of the total
CDM today
\begin{equation}
\rho_X({\rm today}) = f \rho_{\rm CDM}({\rm today}) \, .
\label{ocho}
\end{equation}
By evaluating (\ref{once}) at $t = {\rm today}$  along with (\ref{ocho}), and (\ref{once}) we obtain
\beq
M_X \ n_X({\rm today}) \ \gamma({\rm today}) = f \ \rho_{{\rm CDM}}({\rm today}) \rightarrow \rho_X(t) = f \ \frac{\rho_{\rm CDM}({\rm today})}{a^3(t)} \ \gamma(t) \ .
\label{eq:DMradEn}
\eeq
Giving the full expression for the dark matter as a function of time as
\begin{equation}
\rho_{\rm DM}(t) = f \ \frac{\rho_{\rm CDM}({\rm today})}{a^3(t)} \ \gamma(t) + \rho_\chi(t).
\label{quince}
\end{equation}
This can be reduced even further by evaluating (\ref{quince}) at $t ={\rm today}$ to obtain
\beq
 \rho_{\rm CDM}({\rm today}) =  f \
\rho_{\rm CDM}({\rm today})  + \rho_\chi({\rm today}) \rightarrow \rho_\chi({\rm today})  =  (1-f) \rho_{\rm CDM}({\rm today}) \ .
\label{diesiseis}
\eeq
Finally, the dark matter content as a function of time is given as
\begin{equation}
\rho_{\rm DM}(t) = f \frac{\rho_{\rm CDM}({\rm today})}{a^3(t)}
\gamma(t) + (1-f)\frac{\rho_{\rm CDM}({\rm today})}{a^3(t)}.
\label{diesiocho}
\end{equation}

\subsubsection{\underline{Calculating $N_{\rm eff}$ from $\rho_{\rm rad}$}}

Now that we can identify a potentially relativistic part of the DM energy density, we can add its contribution to $\rho_{\rm rad}$ at this point we can also add in the right chiral neutrino states and their assumed relativistic contribution to the energy density of the universe. Taking the dominate energy density component at $t_{\rm EQ}$ gives
\begin{equation}
\rho_{\rm R}(\teq) = \rho_\gamma(\teq) + 3 \ \rho_{\nu_L}(\teq) + 3\ \rho_{\nu_R}(\teq) + \rho_X(\teq) + \rho_{\rm s} (\teq) \, ,
\label{veinte}
\end{equation} 
where $\rho_{\nu_R}(\teq)$ is the energy density contained in one
flavor of right chiral neutrinos and $\rho_{\rm s}$ is the energy
density in sterile neutrinos. From Eq. (\ref{veinte}) we can
calculate the effective number of neutrinos at the time of
radiation-matter equality from (\ref{eq:NeffMasterEQ}) as
\begin{equation}
N_{\rm eff} =  3 + 3 \frac{ \rho_{\nu_R} }{ \rho_{\nu_L} } +\frac{ \rho_X }{ \rho_{\nu_L} } + \frac{ \rho_{\rm s} }{ \rho_{\nu_L} } =   3 + \Delta N = 3 + \Delta N_\nu + \Delta N_X  \,,
\label{veintiuno}
\end{equation}
where we take all the energy densities evaluated at $\teq$ and,
unless otherwise specified for this point, all the quantities are evaluated
at $\teq$.  

To separate the effect of the additional particles we use $\Delta N_\nu = (3 \
\rho_{\nu_R} + \rho_{\rm s})/\rho_{\nu_L}$ and $\Delta N_X = \rho_X /\rho_{\nu_L}$ for the additional effect from right chiral neutrinos and $X$ particles, respectively.

To remove $\rho_{\nu_L}$ from $\Delta N$, we
make use of (\ref{cuatro}) to obtain
\begin{equation}
\Delta N_X = \frac{8}{7} \lpa \frac{11}{4} \rpa^{4/3} \frac{ \rho_X }{\rho_\gamma} \,,
\label{veinticinco}
\end{equation}
where we have assumed that particles of type $X$ decouple from the
plasma prior to $\nu_L$ decoupling, conserving the ratio
$T_\gamma/T_{\nu_L}$ from SM cosmology. Substituting
(\ref{eq:DMradEn}) into (\ref{veinticinco}) we finally obtain
\begin{equation}
\Delta N_X  =   \frac{8}{7} \ \lpa \frac{11}{4} \rpa^{4/3} \ \frac{\Omega_{\rm CDM}}{  \Omega_\gamma} \ a(\teq) \ \gamma(\teq) \ f \, ,
\label{vientisiete}
\end{equation}
where we used the standard relation $\rho_\gamma(\teq) = \rho_c \
\Omega_\gamma/a^4(\teq)$. The contribution from the right chiral neutrinos will be left to the next subsection.

\subsubsection{\underline{Scale Factor Consistant Reaction}}

Because we add new particles to the thermal history of the universe, the history of $a(t)$ and thus $\gamma(t)$ also changes.  To make corrections to the scale factor, we will neglect the accelerating phase of the scale factor and assume that we are still in a matter dominated universe. The matter phase was preceded by the phase of radiation domination, and for our calculations we will make the approximation of
instantaneous phase change.  That is, we assume at the time of
radiation-matter equality $\teq$ the scale factor instantaneously
changes from a radiation dominated phase to a matter dominated phase,
with continuity ensured.  With these considerations, the scale factor
can be expressed in a piece-wise form as
\begin{equation}
a(t)  = \left\{ \begin{array}{ll} \Big( \frac{3}{2} \ H_0 \ t \Big)^{2/3} & \ \ \ {\rm if} \ t > t_{\rm EQ}  \\
 \left( \frac{3}{2} \ H_0 \ t_{\rm EQ}^{1/4} \right)^{2/3} \ t^{1/2}  & \ \ \ {\rm if} \  t_{\rm EQ} > t  \end{array} \right. \ ,
\label{ochoteabrocho}
\end{equation}
where $H_0$ is the Hubble parameter today.  
Substituting (\ref{ochoteabrocho}) in (\ref{vientisiete}), we obtain
 \begin{equation}
 \Delta N_X  =  \frac{8}{7} \left( \frac{11}{4} \right)^{4/3} \
   \frac{\Omega_{\rm CDM}}{\Omega_{\gamma}} \ 
   \gamma(t_{\rm EQ}) \  \left( \frac{300~{\rm km} \ {\rm s}^{-1} \ {\rm Mpc}^{-1}}{2}
 \right)^{2/3} (h t_{\rm EQ})^{2/3} \ f.
\label{yaves} 
\end{equation}  
Next, we calculate the radiation-matter equality time $t_{EQ}$, {\it i.e.}, the time in which
 \begin{equation}
 \rho_{\rm M}(t_{EQ}) = \rho_{\rm R}(t_{EQ}) \Rightarrow \frac{\rho_{\rm R}}{\rho_{\rm M}} = 
   \frac{\rho_\gamma(t_{\rm EQ}) \left\{1 +
       \frac{7}{8}\left(\frac{4}{11}\right)^{4/3} \left[ 3 + \Delta
         N_\nu +\Delta N _X (t_{\rm EQ})\right] \right\}}{\rho_{M}
     (t_{\rm EQ})} =1 \, ,
\label{hey} 
\end{equation} 
and by scaling the quantities with their respective powers of the scale factor, we arrive at
 \begin{equation}
 \frac{1}{a(t_{\rm EQ})} \frac{\Omega_\gamma}{\Omega_{\rm M}} \left[1 + \frac{7}{8}\left(\frac{4}{11}\right)^{4/3} \left( 3 + \Delta N_\nu +\Delta N _X \right) \right] = 1 \, .
 \end{equation} 
Using the piece-wise form of the scale factor from (\ref{ochoteabrocho}), we can determine the radiation-matter equality time as
\begin{equation}
 h t_{EQ}= \frac{1}{150}  \left( \frac{\Omega_{\gamma}}{\Omega_{\rm M}} \right)^{3/2} \left[1 + \frac{7}{8}\left(\frac{4}{11}\right)^{4/3} \left( 3 + \Delta N_\nu +\Delta N _X \right) \right ]^{3/2}~{\rm km}^{-1} \, {\rm s} \, {\rm Mpc} \, .
\label{yaves2}
\end{equation}  
Note that the combination $h t_{\rm EQ}$ is independent of $h$.

 As a constancy check, we can set $\Delta N_\nu = \Delta N_X =0$, make use
of (\ref{omegagamma}), and take the central values recently
reported by the Planck Collaboration: $\Omega_{\rm M} \simeq 0.315$ and
$h \simeq 0.673$~\cite{Ade:2013lta}, which gives $t_{\rm EQ} = 1.19 \times
10^{12}~{\rm s}$, which is in very good agreement with the concordance
model of cosmology.  

Proceeding to insert (\ref{yaves2}) into (\ref{yaves}) then gives
  \begin{equation}
 \Delta N_X  =  \frac{8}{7} \left( \frac{11}{4} \right)^{4/3} \frac{
   \Omega_{\rm CDM} }{\Omega_{\rm M} } \ \gamma(t_{\rm EQ}) \  \left[1 + \frac{7}{8}\left(\frac{4}{11}\right)^{4/3} \left( 3 + \Delta N_\nu +\Delta N _X \right) \right] f \ ,
\label{treintaitres}
\end{equation}  
for which we must solve for $\Delta N_X$, however at this point it is worth exploring the quantity $\gamma(t_{\rm EQ})$ and its dependence on $h$.  By taking $R = M_{X'}/M_X$ and $\tau$ occurring in the radiation dominated era, $\gamma(t)$ becomes
\begin{equation}
\gamma(t_{\rm EQ}) = \sqrt{1 + \left(\frac{\tau}{t_{\rm EQ}}\right)
  \frac{\left( R^2 - 1\right)^2}{4 R^2}} \, ,
\label{treintaicinco}
\end{equation} 
which does have some $h$ dependence, $t_{\rm EQ} \propto h^{-1}$. To
make our analysis completely independent of $h$ (which we have seen has
large systematic uncertainties), we can rewrite the decay time as a
fraction of the time of the radiation matter equality, ${\cal T} =
\tau/t_{\rm EQ}$, {\it e.g.}, for $\mathcal{T} = 0.1$, the $X$ particle
is produced at a time which is 1\% of $t_{EQ}$. This new variable
allows expressing (\ref{treintaitres}) independently of $h$:
\begin{equation}
 \Delta N_X   = \frac{  \Omega_{\rm CDM}}{\Omega_{\rm M} }  \sqrt{1 + \mathcal{T} \frac{\left( R^2 - 1\right)^2}{4 R^2}} \  \left[
\frac{8}{7} \left( \frac{11}{4}
 \right)^{4/3} +   3 + \Delta N_\nu +\Delta N _X  \right] f.
 \label{trentaisiete}
 \end{equation}  
Note that (\ref{trentaisiete}) scales with the ratio $\Omega_{\rm CDM}/\Omega_{\rm M} \simeq 1.19$, which does not depend on the Hubble parameter. Solving for $\Delta N_X$ gives
\begin{equation}
 \Delta N_X = \left[ \frac{8}{7} \left(\frac{11}{4}\right)^{4/3} + 3 +
   \Delta N_\nu \right] \mathscr{Y}  \left(\frac{\Omega_{\rm M}}{\Omega_{\rm CDM} \ f} -  \mathscr{Y} \right)^{-1} , 
 \label{trentaiocho}
 \end{equation}
where 
\begin{equation}
\mathscr{Y} = \sqrt{1 + \mathcal{T} \frac{\left( R^2 - 1\right)^2}{4 R^2}} \, .
\end{equation} 
Note that (\ref{trentaiocho}) develops a pole if
\begin{equation}
\rho_{\rm M}(\teq)  = \frac{\rho_{\rm M} (\rm today) }{a^3(\teq)} =
\frac{\rho_{\rm CDM}(\rm today)}{a^3(\teq)}  \gamma(\teq) f =
\rho_X(\teq) \, ,
\label{cuarenta}
\end{equation}   
which implies $\rho_{\rm R} (\teq) = \rho_X (\teq)$. This saturates the 
regime for validity of (\ref{treintaitres}).

\subsubsection{\underline{Results}}

\begin{figure}[tpb]
\begin{center}
\postscript{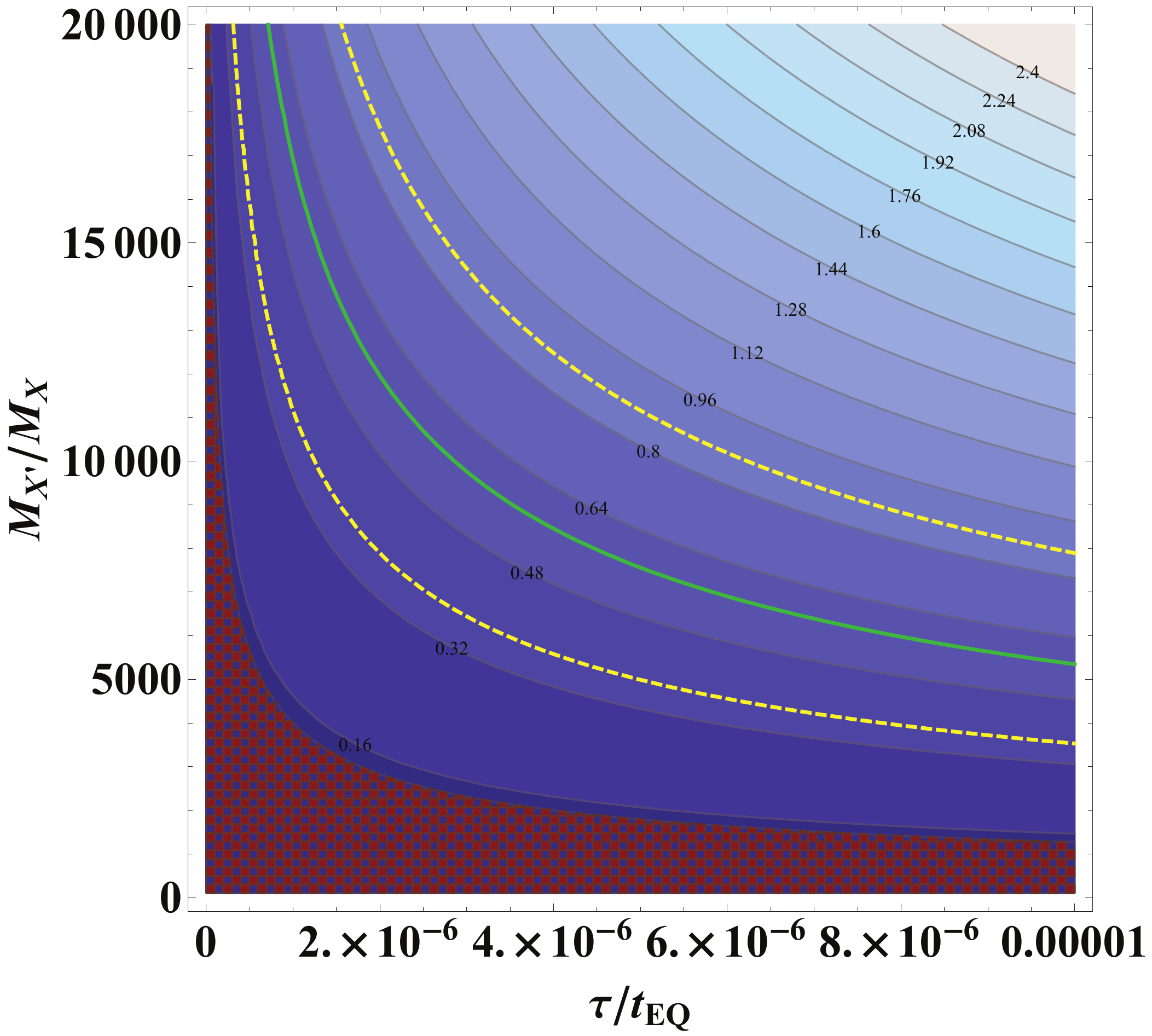}{0.7}
\postscript{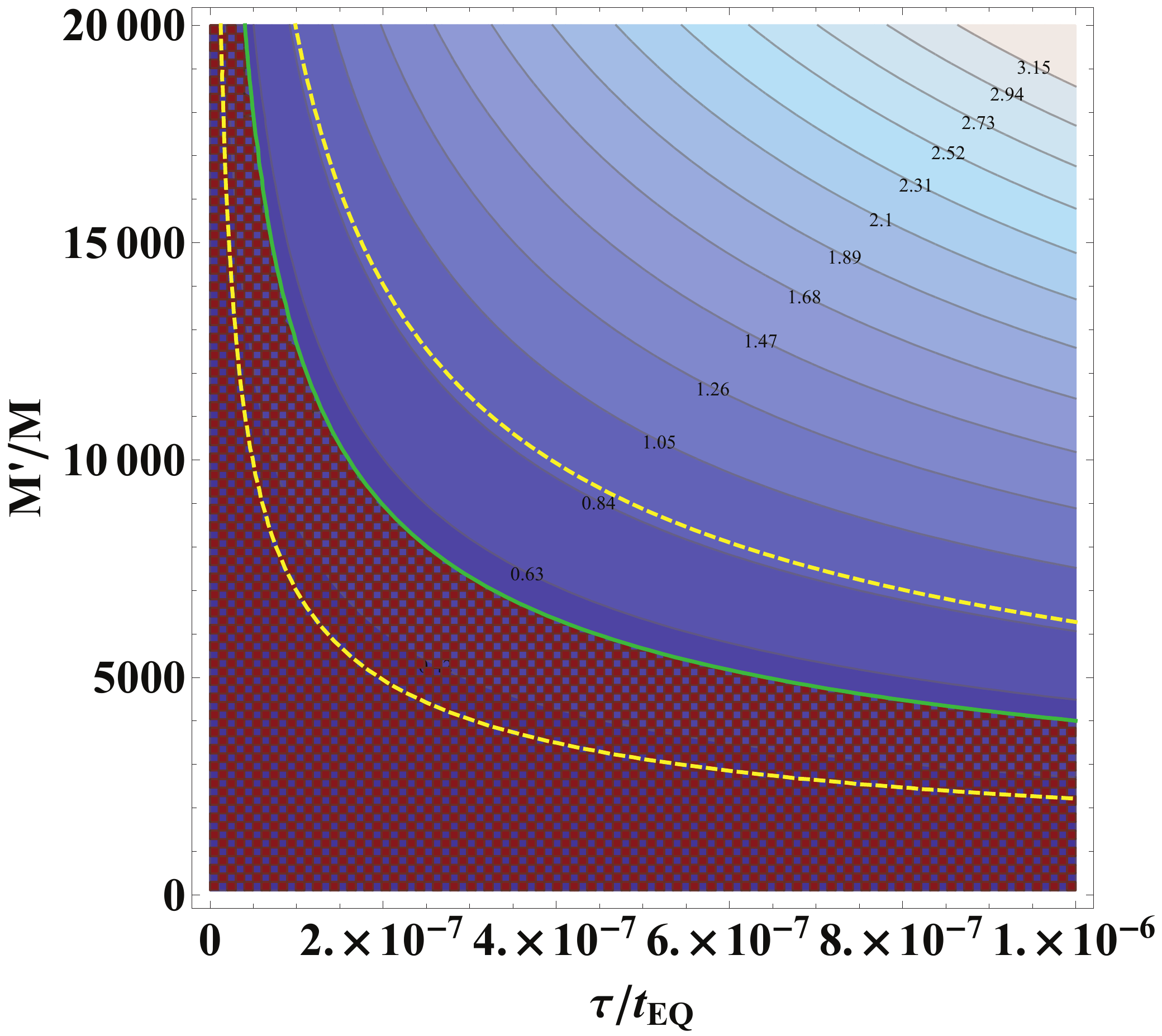}{0.725}
\caption[Contours of Constant $\Delta N_X$]{ \sglspc \small Contours of constant $\Delta N$ in the
$R$ {\it vs.} ${\cal T}$ plane, with $f = 1\%$ (top) and $f=3.8\%$ (bottom), for the case in which $\Delta N_\nu = 0$. The (green) solid line indicates the upper limit on $\Delta N$ (with $h \simeq 0.674$) as reported by the Planck Collaboration. The band between the dashed (yellow) lines corresponds to the allowed $\Delta N$ region reported by Planck Collaboration using $h \simeq 0.738$. The crosshatched area pertains to the region of the parameter space for which the $X$ particles are not relativistic at the CMB epoch, and should therefore not be taken under consideration.}
\end{center}
\label{fig:dos}
\end{figure}

In Fig.~\ref{fig:dos} we can see contours of constant $\Delta N_X$ in the
$R$ {\it vs.} ${\cal T}$ plane, for the case in which $\Delta N_\nu = 0$. As expected, to produce a 
given $\Delta N_X$ contribution, the required ratio of masses diminishes with increasing lifetime.
We can see that, for $h=0.647$, a fraction of $X$ particles larger than 3.8\% yields a
contribution to $N_{\rm eff}$ that is outside the 1$\sigma$ region allowed by Planck data. 
%----------------Figure------------------
%-----------------------------------------
The preceding discussion can be easily generalized to more complex
models endowed with a dynamical dark sector. In a fashion similar to~\cite{Dienes:2011ja}, we assume there are $n$ different decay
possibilities/channels for DM particles of the type $X'_j \rightarrow
X_i + \gamma$.  We further assume that each decay process occurs
instantaneously at a time $\tau_i$ and that of the total fraction of
dark matter particles coming from each decay process is $f_i$. For $N$
particle species, this gives the relation 
\beq f \rho_{\rm CDM}({\rm today}) =
\sum^N_{i=1} \rho_{X_i}({\rm today}) \Rightarrow \frac{\rho_{X_i}({\rm  today})}{ \rho_{\rm CDM}({\rm today})} = f_i \ ,
\label{One}  
\eeq with $\sum_{i=1}^N f_i = f$. Provided each particle in the ensemble $X_i$
meets the criteria for being relativistic, it will contribute 
\beq
\Delta N_{X_i} = \frac{\rho_{X_i}}{\rho_{\nu_L}} = \frac{8}{7} \lpa \frac{11}{4} \rpa^{4/3} \ \frac{\Omega_{\rm CDM}}{\Omega_\gamma} \ a(t_{\rm EQ}) \ \gamma_i  \  f_i \ ,
\label{Two}
\eeq where $\rho_{X_i}$ and $\rho_{\nu_L}$ are again taken at the time
of matter-radiation equality, and $\gamma_i (t)$ is the Lorentz factor
of each particle $X_i$.  As stated earlier, each particle may decay
from some particle $X'_j$ at a time $\tau_i$ such that each $\gamma_i$
is given by
\beq
\gamma_i= \sqrt{1 + \mathcal{T}_i \frac{\lpa R_i^2 - 1 \rpa}{4 R_i^2}},
\label{Three}
\eeq where $\mathcal{T}_i = \tau_i / t_{\rm EQ}$ and $R_i = M'_j /
M_{X_i}$. Once more we assume that the decays occur prior to the time
of matter-radiation equality. We denote the total $X_i$ contribution to the
effective number of neutrinos by
$\Delta N_{\Sigma X} = \sum_{i = 1}^N \Delta N_{X_i}$, so that
$\Delta N = \Delta N_\nu + \Delta N_{\Sigma X}$, with
\beq
\Delta N_{X_i} = \frac{8}{7} \lpa \frac{11}{4} \rpa^{4/3} \frac{\Omega_{\rm CDM}}{\Omega_\gamma}  \ \gamma_i (t_{\rm EQ}) \ \lpa \frac{300~{\rm km} \ {\rm s}^{-1} \ {\rm  Mpc}^{-1}}{2} \rpa^{2/3} \ \lpa h t_{\rm EQ} \rpa^{2/3} \ f_i \ .
\label{Six}
\eeq
Duplicating the procedure for a single $X$ particle, with the change of $\Delta N_X \rightarrow \Delta N_{\Sigma X}$ we obtain
\beq
h t_{\rm EQ} = \frac{1}{150} \ \lpa \frac{\Omega_\gamma}{\Omega_M}\rpa^{3/2} \ \left[ 1 + \frac{7}{8} \lpa \frac{4}{11} \rpa^{4/3} \lpa 3 + \Delta N_\nu + \Delta N_{\Sigma X} \rpa \right] \ .
\label{Seven}
\eeq
All in all the total contribution to the number of ``equivalent'' light
neutrino species is
\beq
\Delta N_{\Sigma X} = \left[\frac{8}{7} \lpa \frac{11}{4} \rpa^{4/3} + 3 + \Delta N_\nu \right] \left(\sum^N_{i=1} f_i \gamma_i \right) \left(\frac{\Omega_M }{\Omega_{\rm CDM}} - \sum^N_{i=1} f_i \gamma_i \right)^{-1} \, .
\label{Eight}
\eeq
It should be noted that the parent particle $X'_j$ can be the same for multiple particles $X_i$, as is the case for multiple decay channels.

\subsection{Right-Handed Neutrinos with Milli-Weak Interactions}

In addition to the $(2.984 \pm 0.009 )\nu_ L$ species measured from
the width for invisible decays of the $Z_0$ boson~\cite{:2005ema}, there
could also exist $\nu_R$ states that are sterile, {\it i.e.}  singlets
of the SM gauge group, and therefore insensitive to weak
interactions. Such sterile states are predicted in models involving
additional TeV-scale $Z'$ gauge bosons, which allow for milli-weak
interactions of the $\nu_R$. If the $\nu_R$ carry a non-zero $U(1)'$
charge, then the $U(1)'$ symmetry forbids them from obtaining a
Majorana mass much larger than the $U(1)'$-breaking scale. Therefore,
in most of these models there are no Majorana mass terms and the
$\nu_R$ states, which are almost massless, become the Dirac partners
of the SM $\nu_L$ species.

Here we will add in the right-handed neutrino species to study the
expected increase in $N_{\rm eff}$ due to the presence
of such light Dirac neutrinos with ongoing searches of $Z'$ gauge
bosons at the LHC. A critical input for such
an analysis is the relation between the relativistic degrees of
freedom (r.d.o.f.)  and the temperature of the primordial plasma. This
relation is complicated because the temperature which is of interest
for right-handed neutrino decoupling from the heat bath may lay in the
vicinity of the quark-hadron cross-over transition, which offers problems as the exact form
of $g_S(T)$ is in a non-perturbative regime of QCD and thus requires
high statistics lattice simulations of a QCD (LQCD) plasma in this phase.

We begin by first establishing $\Delta N_\nu$ as a function of the $\nu_R$ decoupling temperature.  
By taking into account the conservation of entropy of the rest of the plasma between the temperature
from $\nu_R$ decoupling, $T_{\nu_R}^{\rm dec}$, and the $\nu_L$ decoupling temperature, $T_{\nu_L}^{\rm dec}$,  we can arrive at an equation for $\Delta N_\nu$. First we express the total entropy of the universe at the $\nu_R$ decoupling as
\beq
S(T_{\nu_R}^{\rm dec}) \propto a(t_{\nu_R})^3 g_S( T_{\nu_R}^{\rm dec}) \lpa T_{\nu_R}^{\rm dec} \rpa^3 = {\rm const} \ ,
\eeq
where $t_{\nu_R}$ is the $\nu_R$ decoupling time
after which particles may annihilate and thus add energy to the photon matter plasma, which corresponds to an increase in temperature and change in $g_S(T)$.  The particles that decoupled have their thermal distribution temperature simply scale with $a^{-1}$; because of this, after $\nu_R$ decoupling the effective entropy density is
\beq
a(t_{\nu_R})^3 g_S(T_{\nu_R}^{\rm dec}) \lpa T_{\nu_R}^{\rm dec} \rpa^3   =  a(t)^3 g_S(T) T^3 + a(t)^3  g_{S,\nu_R} \lpa \frac{ T_{\nu_R}^{\rm dec} a(t_{\nu_R})}{a(t)} \rpa^3 \ ,
\eeq
where $g_{S,\nu_R}$ is the entropy degrees of freedom from $\nu_R$, which is now freely streaming.  Simplifying this expression results in
\beq
\lpa \frac{a(t_{\nu_R}) T_{\nu_R}^{\rm dec} }{a(t)} \rpa^3 \ \lpa g_S(T_{\nu_R}^{\rm dec})- g_{S,\nu_R} \rpa   = g_S(T) T^3   \ ,
\eeq
but since $T_{\nu_R}(t) = a(t_{\nu_R}) T_{\nu_R}^{\rm dec} / a(t)$ for $t > t_{\nu_R}$, this expression simplifies to
\beq
\lpa g_S(T_{\nu_R}^{\rm dec})- g_{S,\nu_R} \rpa T_{\nu_R}(t)^3 = g_S(T) T^3. 
\label{eq:ConsofS}
\eeq 
By redefining $g_S(T^{\rm dec})$ at decoupling temperatures to not include the particles which are decoupling, we can drop the $- g_{S,\nu_R}$ from above.  With Eq. (\ref{eq:ConsofS}), we can construct the contribution to $N_{\rm eff}$ by Eq. (\ref{veintiuno}) where we have the relation
\begin{equation}
\Delta N_\nu = 3 \frac{\rho_{\nu_R}}{\rho_{\nu_L}} =  3 \lpa \frac{T_{\nu_R}(\teq)}{T_{\nu_L}(\teq)}\rpa^4= 3 \left(\frac{g_S(T_{\nu_L}^{\rm dec})}{g_S(T_{\nu_R}^{\rm
      dec})} \right)^{4/3} \, ,
\label{7}
\end{equation}
where standard cosmology gives $g_s(T_{\nu_L}^{\rm
  dec}) = 43/4$~\cite{Kolb:1990vq}. For the particle content of the
SM, there is a maximum of $g_s(T_{\nu_R}^{\rm dec}) = 427/4$ (with
$T_{\nu_R}^{\rm dec} > m_{\rm top}$), which corresponds to a minimum
value of $\Delta N_\nu = 0.14$.  

If $T^{\rm dec}_{\nu_R}$ occurs during the QCD phase transition, then the value of $g_S(T_{\nu_L}^{\rm dec})$
is very complicated as the degrees of freedom are undergoing rapid changes.  At energies above the de-confinement transition towards the quark gluon plasma, quarks and gluons are the relevant degrees of freedom
for the QCD sector, such that the total number of SM r.d.o.f.  consisting of $\lpa \gamma, \ 8G, \ 3 \nu_L, \ e_{L,R}, \ \mu_{L,R}, \ u_{L,R}^{r,g,b}, d_{L,R}^{r,g,b},s_{L,R}^{r,g,b} \rpa$ 
\beq
g_S = 2(1 + 8) + \frac{7}{8} \lsb 3\cdot 2 + 2\cdot 2 + 2\cdot 2 +3(2\cdot 2+2\cdot 2+2\cdot 2) \rsb = 61.75 \ .
\label{eq:PartContent}
\eeq  
As the universe cools down, the SM plasma transitions to a
regime where mesons and baryons are the pertinent degrees of
freedom confining the quarks and gluons within. Precisely, the relevant hadrons present in this energy regime
are pions and charged kaons, such that $g_S =19.25$~\cite{Brust:2013ova}. This significant reduction in the degrees of freedom results from the rapid annihilation or decay of any more
massive hadrons which may have formed during the transition. The
quark-hadron crossover transition therefore corresponds to a large
redistribution of entropy into the remaining degrees of
freedom. We express the effective number of interacting r.d.o.f. in
the plasma at temperature $T$ by
\begin{equation}
g_S (T) \simeq  r (T) \left( \sum_{\rm bosons} g_{\rm b}+ \frac{7}{8} \sum_{\rm fermions} g_{\rm f} \right) \,,
\label{sietebravo}
\end{equation}
with $g_{\rm b} = 2$ for each real vector field and $g_{\rm f} = 2$
for each spin-$\frac{1}{2}$ Weyl field. The coefficient $r (T)$ is
unity for the lepton and photon contributions and is the ratio $s(T
)/s_{\rm SB}$ for the quark-gluon plasma, where $s(T)$ is the actual
entropy and $s_{\rm SB}$ is the ideal Stefan-Bolzmann entropy. We must now
examine LQCD simulations and radiative correction analyses to obtain the function $r(T)$.

\subsubsection{\underline{QCD Confinement Phase Transition Fitting}}

Two complementary approaches that describe high
temperature QCD phenomena result in similar expressions for $r(T)$. We will compare their
predictions during the de-confinement transition. 
  
\begin{figure}[tpb]
\begin{center}
\postscript{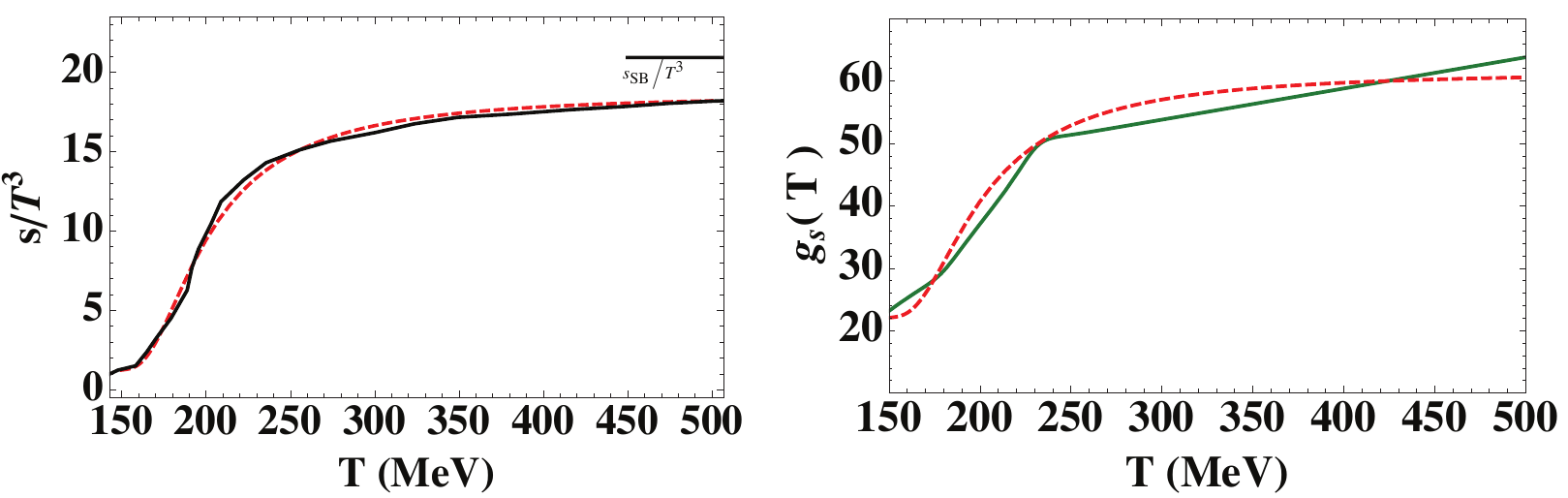}{0.99}
\caption[Entropy Fitting and Comparison with LQCD Simulations]{ \sglspc \small Left: The (black) solid line is the entropy density as
  obtained from high statistics lattice
  simulations~\cite{Bazavov:2009zn} (we used the results obtained
  using the p4-action~\cite{Heller:1999xz}). The (red) dashed line is
  the fit given in Eq. (\ref{asqtad}). Right: The effective
  number of interacting relativistic degrees of freedom, $g_S$, as a
  function of the temperature for $150~{\rm MeV} \leq T \leq 500~{\rm
    MeV}$. The solid line is obtained from the NLO correction result given in~\cite{Steigman:2012nb} (adapted
 from~\cite{Laine:2006cp}). The red dashed curve is the result from the fit of LQCD calculations. }
\label{fig:siete}
\end{center}
\end{figure}

The first approach relies on next-to-leading order (NLO) radiative
corrections to the non-interacting Stefan-Boltzmann law within the
$\overline{\rm MS}$ scheme~\cite{Laine:2006cp}. The second approach
is based on high statistics lattice simulations of the QCD plasma
during the changeover phase~\cite{Bazavov:2009zn}. In either case, all thermodynamic quantities can be obtained from the QCD partition function $Z_p(T,V)$.  The need for LQCD comes from the fact that perturbative calculations are no longer valid as the strong coupling becomes large at these temperatures.  In LQCD the basic outline is to calculate $Z_p$ via the path integration method we saw in Sec.~\ref{sec:Thermal} on a lattice.  The idea is that from the form
\beq
Z_p \propto \int \CD \phi \ e^{-S_E[\phi]} \ ,
\eeq
a field configuration $\phi$ can be discretized on a lattice (just like in Sec.~\ref{sec:CompStability}), and expectation values $\la \phi(x_1) \dots \phi(x_n) \ra$  can be calculated by treating $\exp(-S_E[\phi])$ as the relative probability of a field configuration with many degrees of freedom.  The many degrees of freedom offers a computational challenge that is typically solved by a Markov-Chain Monte Carlo (MCMC) method~\cite{MCMC}.  Lattice field theories are further complicated by the use of Weyl fields and gauge symmetry on a lattice, which is the subject of Wilson loops~\cite{Srednicki:2007qs} and a discrete lattice spacing RG analysis.  However the partition function is calculated it can be used for various quantities.
 For instance, its logarithm defines the free energy density,
\begin{equation}
\CF  = - \frac{T}{V} \ln~Z_p \, .
\label{freeenergy}
\end{equation}
The energy density and pressure are derivatives of $\ln~Z_p$ with respect to $T$ and
$V$, respectively,
\begin{equation}
\rho =  \frac{T^2}{V} \ \, \frac{\partial \ln~Z_p}{\partial T} \quad \quad {\rm and} \quad \quad
p = T \ \, {\partial  \ln~Z_p \over \partial V} \, .
\end{equation}
However, for sufficiently large volumes, $T$ is the only intensive
parameter controlling the thermodynamics and the pressure can  be
directly derived from the free energy density,
\begin{equation}
p = -\CF \, .   
\label{pressure}
\end{equation}
To obtain the entropy density, recall that from (\ref{eq:Sixty}) we can express the entropy density in an isotropic, homogenous universe as
\begin{equation}
s = \rho + p   =   T \ \frac{\partial p}{\partial T}. 
\label{entropy} 
\end{equation}
In practice, the quantity most convenient to calculate on the lattice is the trace
anomaly in units of the fourth power of the temperature $\Theta^{\mu
  \mu}/T^4$, where the QCD trace anomaly is given by
\begin{equation}
\Theta^{\mu \mu} \equiv \rho- 3p   =  T^5 \ \frac{\partial (p/T^4)}{\partial T} \, .
\label{theta}
\end{equation}
Using (\ref{theta}), the pressure is obtained by integrating
$\Theta^{\mu \mu}/T^5$ over the temperature
\begin{equation}
\frac{p(T)}{T^4} - \frac{p(T_0)}{T_0^4} = \int_{T_0}^T dT'
\frac{\Theta^{\mu\mu}(T')}{{T'}^5} \,,
\end{equation}
where $T_0$ is an arbitrary temperature that is generally chosen in
the low temperature regime where the pressure and other
thermodynamical quantities are suppressed exponentially by Boltzmann
factors associated with the lightest hadronic states; the convenient extrapolation 
$T_0 \to 0$ yields $p/T_0^4 \to 0$.  After $p/T^4$ is obtained, we can calculate the entropy density.
The increasing entropy curve from LQCD can be fit by
\begin{equation}
 \frac{s}{T^3} \simeq 18.62 \ \lpa \frac{175.41}{T_{\rm MeV} -148.46} \rpa^2 \ \frac{e^{ 175.41/(T_{\rm MeV} - 148.46)}}{\left[ e^{175.41/(T_{\rm MeV} -148.46)}-1 \right]^2} + \frac{42.82}{\sqrt{392 \ \pi}} \ e^{-\frac{\lpa  T_{\rm MeV} - 169.88\rpa^2}{392}} \, ,
\label{asqtad}
\end{equation}
where $T_{\rm MeV}$ is the temperature in units of MeV and $T_{\rm MeV}$ satisfies $150~{\rm MeV} < T_{\rm MeV} < 1~{\rm GeV}$. 
We compare the rise of $g_S(T)$ as given in (\ref{gdet}) with the LQCD result shown in Fig.~\ref{fig:siete}, as well as 
a comparison of $g_S(T)$ as obtained using LQCD and the NLO approach is shown.  Finally, we obtain the relevant degrees of entropy as
\begin{equation}
 g_S(T) \simeq 47.5 \ r(T) + 19.25 \,  ,
\label{gdet}
\end{equation}
which reflects the particle content in (\ref{eq:PartContent}) as well as 3 species of right-handed neutrinos along with
$r(T)$ determined from (\ref{asqtad}).

  If relativistic particles are present that have decoupled from the photons, it is necessary to
distinguish between two kinds of r.d.o.f.: those associated
with the total energy density $g_\rho$, and those associated with the
total entropy density $g_S$. Since the quark-gluon energy density in the
plasma has a similar $T$ dependence to that of the entropy (see Fig. 7
in~\cite{Bazavov:2009zn}), we take $g_\rho (T) \approx g_S(T)$.
%----------------Figure------------------
%-----------------------------------------

\subsubsection{\underline{Calculating the $\nu_R$ Decoupling Temperature}}

We can solve for the right-handed neutrino decoupling temperature by reason that
the right-handed neutrino decouples  from the plasma when its mean
free path becomes greater than the Hubble radius at that time.  
To do this, we first calculate the $\nu_R$ interaction rate given by
\begin{equation}
\Gamma (T) = {\cal K} \ \frac{1}{8} \ \left(\frac{\overline  g}{M_{Z'}}
\right)^4 \ T^5 \ \sum_{i=1}^6 {\cal N}_i \,,
\label{Gamma}
\end{equation}
where ${\cal N}_i$ is the number of chiral states that are available to scatter with $\nu_R$, with groupings as given in Table \ref{tab:P1ChiFerSpect}, such that $\CN_1$ is the number of chiral states in $u_R$ ($\sum_{i=1}^6 = 28$). The effective coupling $\bar g$ is
\begin{equation}
\overline g \equiv 
\left(\frac{\sum_{i =1}^6  {\cal N}_i g_i^2 g_6^2}{\sum_{i =i}^6  {\cal
    N}_i} \right)^{1/4}\, ,
\label{gbarra}
\end{equation}
with $g_i$ the chiral couplings of the $Z'$ gauge boson for species $i$, and the
constant ${\cal K}= 0.5 \ (2.5)$ for annihilation (annihilation + scattering)~\cite {Anchordoqui:2011nh}.

To illustrate, we calculate $\overline g$ for two candidate models. The
first is a set of variations on D-brane constructions which do not
have coupling constant unification. The second are two $U(1)$ models,
$U(1)_\psi$ and $U(1)\chi$, which are embedded in a grand unified
exceptional $E_6$ group, with breaking pattern
\begin{equation}
E_6 \to SO(10) \times
U(1)_\psi \to SU(5) \times U(1)_\psi \times U(1)_\chi \, .
\end{equation}
The latter two are interesting because they have long been suspected
to contribute to $N_{\rm eff}$~\cite{Ellis:1985fp,GonzalezGarcia:1989py,Lopez:1989dh,Barger:2003zh} and provide a test basis for $Z'$ searches at
ATLAS~\cite{Aad:2012hf} and CMS~\cite{Chatrchyan:2012it,Chatrchyan:2012oaa}. For each of the
$E_6$ models we may write $g_i$ in (\ref{gbarra}) as $g_i = g_0 Q_i$,
where in conformity with grand unification we
follow~\cite{Barger:2003zh} and choose
\begin{equation}
g_{0} = \sqrt{\frac{5}{3}} \ g_2  \ \tan \theta_W \sim 0.46 \,,
\end{equation}
with $g_2$ the $SU(2)_L$ coupling.  The charges $Q_i$ for the
different fermions in this model are given in Table~\ref{tablauno}~\cite{Barger:2003zh}.
\sglspc
\begin{table}
\begin{center}
\caption[Charges of Example Models for $\bar g$ Calculation]{\centering The charges of the $U(1)_\chi$ and the $U(1)_\psi$.\label{tablauno}}
\begin{tabular}{cccc}
\hline
\hline
~~~~~~~~ bin ~~~~~~~~ & ~~~~~~~~  Fields ~~~~~~~~ & ~~~~~~~~ $Q_\chi$ ~~~~~~~~ & ~~~~~~~~ $Q_\psi$ ~~~~~~~~  \\ \hline 
1 & $U_r$ & $\phantom{-} 1/(2 \sqrt{10})$ & $- 1/(2 \sqrt{6})$ \\
2& $D_R$ & $- 3/(2 \sqrt{10})$ &  $- 1/(2 \sqrt{6})$ \\
3 & $L_L$ & $\phantom{-} 3/(2 \sqrt{10})$ & $\phantom{-} 1/(2 \sqrt{6})$ \\
4 & $E_R$ & $\phantom{-} 1/(2 \sqrt{10})$ & $- 1/(2 \sqrt{6})$ \\
5 & $Q_L$ & $- 1/(2 \sqrt{10})$ & $\phantom{-} 1/(2 \sqrt{6})$ \\
6 & $N_R$ & $\phantom{-} 5/(2 \sqrt{10})$ & $- 1/(2 \sqrt{6})$ \\
\hline
\hline
\end{tabular}
 \end{center}
\end{table}
\dblspc
In the D-brane construction of the SM$^{++}$ outlined in Sec.~\ref{part:1} or any D-brane construct the Weyl fermions live at the brane intersections of a particular 4-stack quiver configuration: $U(3)_C
\times SU(2)_L \times U(1)_{I_R} \times U(1)_L$~\cite{Cremades:2003qj}. We consider here two possibilities in which the $Z''$ gauge boson of the SM$^{++}$ is taken at TeV scales and has branching ratio that is  mostly into $I_R$ or mostly $B-L$ (the specific definition is found in the section~\ref{sec:Bbound}). The chiral couplings ($g_i$) of these gauge bosons are given in Tables~\ref{tab:2} and \ref{newtable}~\cite{luisnMe1, Anchordoqui:2011eg}.

With (\ref{Gamma}) we can determine $T_{\nu_R}^{\rm dec}$  via
\begin{equation}
 \Gamma(T_{\nu_R}^{\rm dec}) = H(T_{\nu_R}^{\rm dec}) \, ,
\label{haim1}
\end{equation}
where $H$ can be retrieved from the Friedmann equations in the era of radiation dominance
\begin{equation}
H(T^{\rm dec}_{\nu_R}) = 1.66 \sqrt{g_\rho(T^{\rm dec}_{\nu_L}) } \
\frac{(T^{\rm dec}_{\nu_R})^2}{M_{\rm
    Pl}} \ \left( \frac{3}{\Delta
  N_\nu} \right)^{3/8} \,,
\label{Hubble}
\end{equation}
where above we have used (\ref{7}) to replace $g_\rho (T_{\nu_R}^{\rm dec})$ for $g_\rho(T_{\nu_L}^{\rm dec})$.  Solving (\ref{haim1}) results in
\begin{eqnarray}
 \Delta N_\nu & = & 3 \
\left(\frac{13.28 \ \sqrt{g(T_{\nu_L}^{\rm dec})} \ M_{Z''}^4}{M_{\rm Pl} \ {\cal
      K} \ (T_{\nu_R}^{\rm dec})^3 \ \overline g^4 \ \sum_{i=1}^6 {\cal N}_i} \right)^{8/3} \nonumber \\
& = & \left[\frac{5.39 \times 10^{-6}}{{\cal K} \sum_{i=1}^6 \mathcal{N}_i} \lpa  \frac{M_{Z''}/{\rm TeV}}{\bar{g}} \rpa^4 \ \frac{1}{(T_{\nu_R}/{\rm GeV})^3} \right]^{8/3}  \, .
\label{hitichi}
\end{eqnarray}
%----------------Figure------------------
\begin{figure}[tpb]
\begin{center}
\postscript{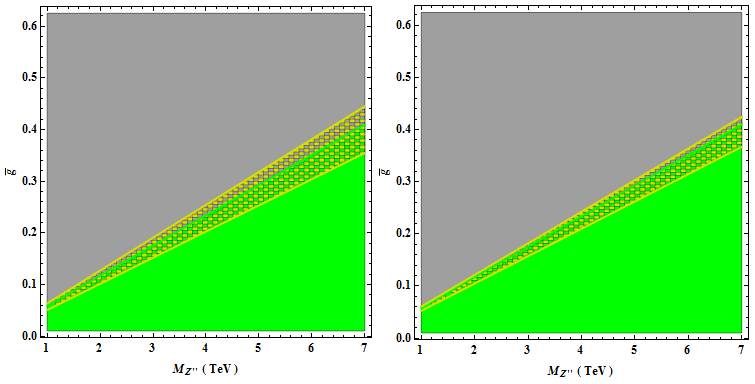}{0.99} %NuPlots
\caption[$N_{\rm eff}$ Allowed Regions for $\nu_R$ Contributions]{ \sglspc \small The green area shows the region allowed from
  decoupling requirements to accommodate $3.03 < N_{\rm eff} < 3.57$, 
and the yellow cross-hatched area shows the region allowed from
  decoupling requirements to accommodate $3.37 < \Delta
  N_{\rm eff} < 3.87$, respectively. We have taken ${\cal K} =0.5$
  (left) and ${\cal K} = 2.5$ (right).}
\label{fig:tost}
\end{center}
\end{figure}
%-----------------------------------------
In Fig.~\ref{fig:tost} we show the region of the parameter space allowed
from the decoupling Eq. (\ref{hitichi})  to accommodate contributions of $\Delta
N_\nu$ within the $1\sigma$ region of Planck data. 

\begin{figure}[tpb]
\begin{center}
\postscript{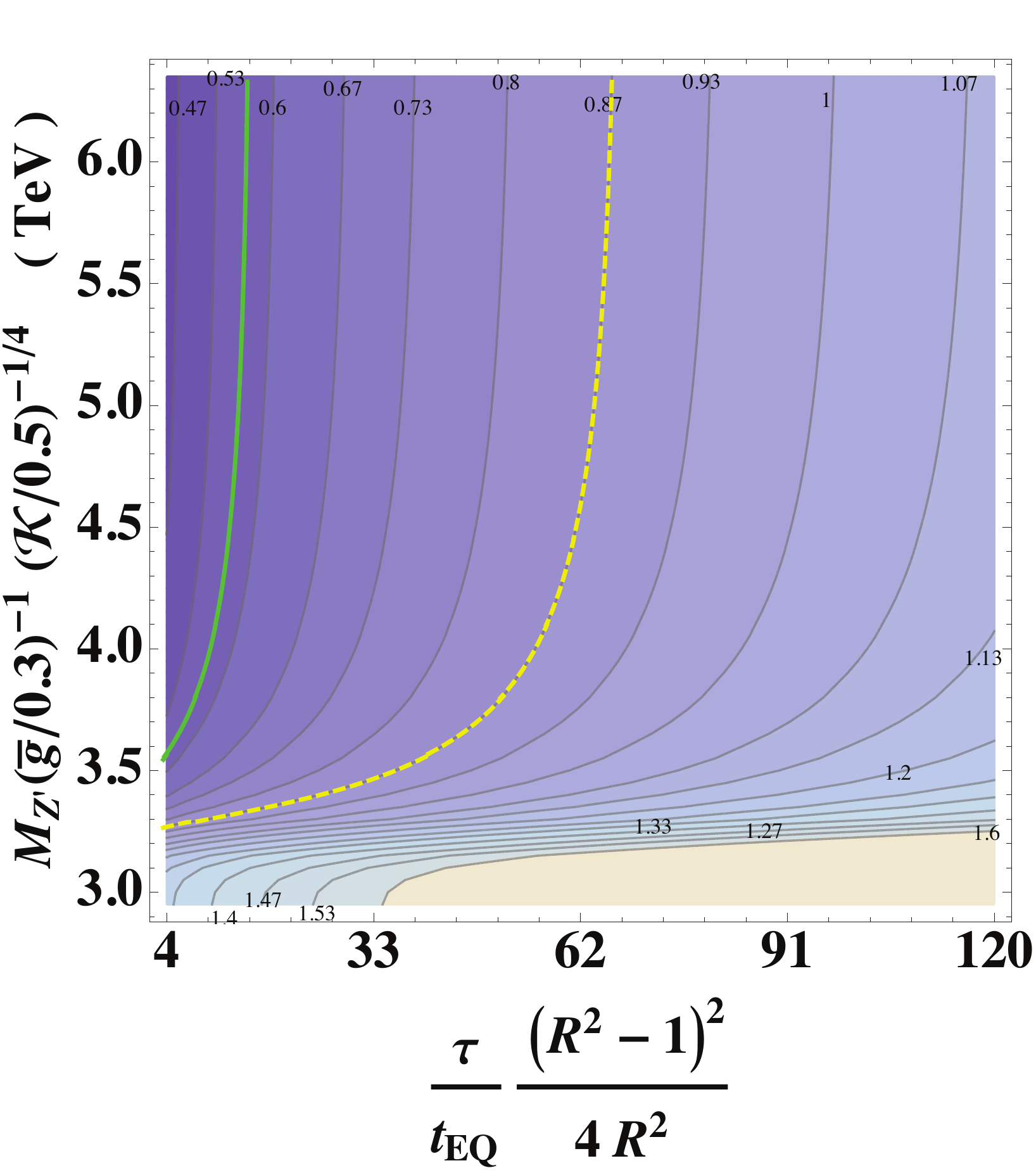}{0.9}
\caption[Contours for constant $\Delta N$ from $\nu_R$ and $X$ Contributions]{ \sglspc \small Contours of constant $\Delta N$ in the $M_{Z'}/ (K\overline
  g)$ {\it vs.} ${\cal T} (R^2-1)^2/(2R)^2$ plane, with $f= 1\%$.  The
  (green) solid line indicates the upper limit on $\Delta N$ (with $h
  \simeq 0.674$) as reported by the Planck Collaboration. The band
  between the dashed (yellow) lines corresponds to the allowed $\Delta
  N$ region reported by Planck Collaboration using $h \simeq
  0.738$.}
\label{fig:Both}
\end{center}
\end{figure}

Substituting (\ref{hitichi}) into (\ref{trentaiocho}), we obtain an expression for
$\Delta N$ for cases with a single non-thermal DM particle and right-handed neutrinos adding to the contribution of $N_{\rm eff}$. In Fig.~\ref{fig:Both} we show contours of constant
$\Delta N$ in the $M_{Z''}/ (K\overline g)$ {\it vs.} ${\cal T}
(R^2-1)^2/(2R)^2$ plane, with $f= 1\%$.  For all cases in Fig.~\ref{fig:Both},  $T_{\nu_R}^{\rm dec} \ll
m_{\rm charm}$, so our use of (\ref{gdet}) is validated.

For $\overline g \simeq 0.3$, there is a region of the
parameter space inside the $1\sigma$ interval of Planck data (with $h
\simeq 0.674$) that can accommodate contributions to $\Delta N$ from
both DM and $\nu_R$ with a $Z''$ gauge boson\footnote{In the general cases where we are not referring to the SM$^{++}$ model, one can interchange $Z'$ and $Z''$ as the TeV scale gauge boson.} within the LHC discovery
reach.  It is important to stress that for the $E_6$ $Z'_\psi$ model,
the LHC experimental limits on $M_{Z'}$ for null signals for
enhancements in dilepton searches entail $M_{Z'} > 2.3~{\rm TeV}$ at the
95~\%CL~\cite{Aad:2012hf,Chatrchyan:2012it,Chatrchyan:2012oaa}.
In some of the models with $\overline g \approx 0.3$, the $Z'$ may have
large couplings to quarks (e.g., $I_R$ and $B-L$ models) and the LHC
experimental limits are dominated by dijet final states which imply
$M_{Z'} < 4~{\rm TeV}$ at 95\% CL~\cite{ATLAS:2012pu,CMS:2012yf,Chatrchyan:2013qha}.  For all the
cases there is a region of the parameter space inside the $1\sigma$ interval of Planck data (with $h
\simeq 0.674$) that can accommodate contributions to $\Delta N$ from
both DM and $\nu_R$, and be in agreement with LHC limits.
%----------------Figure------------------
%-----------------------------------------

\subsection{Big Bang Nucleosynthesis Limits}

\begin{figure}[tpb]
\begin{center}
\postscript{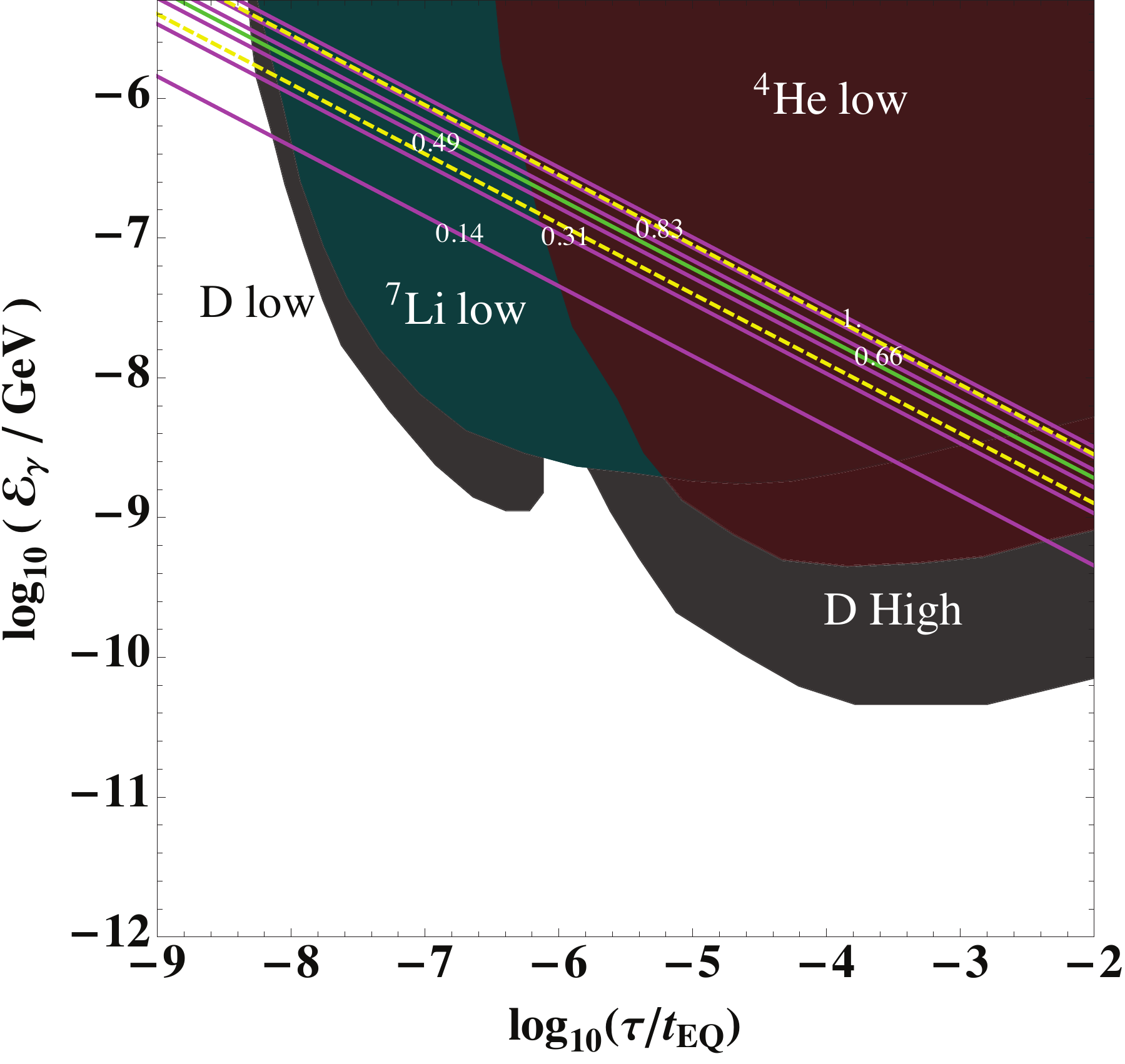}{0.9}
\caption[BBN Constraints on Non-Thermally Produced Dark Matter]{ \sglspc \small The released electromagnetic energy are represented via the lines labeled by $\Delta N_X$ varying from $0.14- 1.0$ as a function of $\log_{10}(\tau/\teq)$.  In order to avoid destroying the BBN results, these lines must not intersect any regions where any elemental abundance has been established.  Generically, the only parameter space available is for $\log_{10}(\tau/\teq) < -9$.  The
  (green) solid line indicates the upper limit on $\Delta N$ (with $h
  \simeq 0.674$) as reported by the Planck Collaboration. The band
  between the dashed (yellow) lines corresponds to the allowed $\Delta
  N$ region reported by Planck Collaboration using $h \simeq
  0.738$.}
\label{fig:BBNFig}
\end{center}
\end{figure}

 As a final constraint we now verify that $X' \to X + \gamma$ does not drastically alter any
 of the light elemental abundances synthesized during
 BBN. By following~\cite{Cyburt:2002uv}, we can assume that the photons injected
 into the plasma rapidly redistribute their energy through scattering
 with background photons and through inverse Compton scattering. As a
 consequence, the constraints from BBN are (almost) independent of the
 initial energy distribution of the injected photons and are only
 sensitive to the total energy released in the decay process. In the
 spirit of~\cite{Feng:2003uy}, we conveniently write the
 electromagnetic energy release as $\varepsilon_\gamma \equiv E_\gamma
 Y_{X'}$, where $E_{\gamma} = ({M_{X'}}^2 - M_X^2)/(2 M_{X'})$ is the initial
 electromagnetic energy release in each $X'$ decay and $Y_{X'} \equiv
 n_{X'}/n_\gamma^{\rm BG}$ is the number density of $X'$ before the
 decay, normalized to the number density of background photons
 $n_\gamma^{\rm BG} = 2 \ \zeta(2) \ T_\gamma^3/ \pi^2$.  For
 $Y_{X'}$, each $X'$ decay produces one $X$, and so the $X'$ abundance
 may be expressed in terms of the present $X$ abundance through
\begin{eqnarray}
Y_{X'} (\tau)  \! \! \!\! & = & \! \! \!\! Y_{X,  \tau} = Y_{X, {\rm today}} = \frac{\Omega_{X} \rho_c}{M_X \ n_{\gamma}^{\rm BG} ({\rm today})} \nonumber \\
    & \simeq & \! \! \! \! 2.26 \times 10^{-14} \ \lpa \frac{\rm
        TeV}{M_X} \rpa \ \frac{\Omega_{\rm CDM} h^2}{0.1199} \
      \frac{f}{0.01} \ ,
\label{Yx}
\end{eqnarray}
yielding
\begin{displaymath}
\varepsilon_\gamma  =  1.13 \times 10^{-11} \ \
  \frac{\Omega_{\rm CDM} h^2}{0.1199} \ \  \frac{f}{0.01} \ \left( \frac{M_{X'}}{M_X} - \frac{M_X}{M_{X'}} \right)~{\rm GeV} \, .
\end{displaymath}
The thorough analysis of electromagnetic cascades reported
in~\cite{Cyburt:2002uv} reveals that the shaded regions of
Fig.~\ref{fig:BBNFig} are ruled out by considerations of light
elemental abundances produced during BBN. The various regions are
disfavored by the following conservative criteria: {\it (i)}~D/H $<
10^{-4.9}$ (low); {\it (ii)}~D/H $> 10^{-4.3}$ (high);  {\it (iii)}~$^7$Li/H $< 10^{-10.05}$; {\it
  (iv)}~primordial $^4$He abundance $< 0.227$. The straight lines represent several
combinations of $R$ and $\tau/t_{\rm EQ}$ producing the $\Delta N_X$
indicated in the labels.  All straight lines intersect the BBN bounds
at about ${\rm log}_{10}(\tau/t_{EQ}) = -8.2$. The constraints from BBN are weak for early
decays because at early times the universe is hot and thus the $X'$
secondary photon spectrum is rapidly thermalized, leaving just a few
extra high-energy photons that cannot alter the light elemental
abundances. However, for $\tau/t_{\rm EQ} > 10^{-8.2}$, BBN excludes
most of the relevant parameter space.
%----------------Figure------------------
%-----------------------------------------

\newpage

\section{\label{sec:WWIMP} Higgs Portal to the Dark Sector and WIMPS}

Recent dark matter direct detection research has found that hints of a WIMP of mass $10$ GeV has been
seen by various collaborations reported by the
DAMA/LIBRA~\cite{Bernabei:2010mq},
CoGeNT~\cite{Aalseth:2010vx,Aalseth:2011wp},
CRESST~\cite{Angloher:2011uu}, and CDMS~\cite{Agnese:2013rvf}
Collaborations, each of which report signals consistent with a dark
matter particle of similar mass. These four experiments make use of
different technologies, target materials, and detection strategies,
but each reports results that are not compatible with known
backgrounds, but which can be accommodated by a dark matter
particle with a mass of about 10 GeV and an elastic scattering cross
section with nucleons of $1 - 2 \times 10^{-41}~{\rm
  cm}^2$~\cite{Fitzpatrick:2010em,Chang:2010yk,Hooper:2010uy,Buckley:2010ve}.

  Furthermore it is observed that around the Galactic Center (GC), there exists a bright and spatially
extended source of \mbox{$\gamma$-ray} emission peaking at energies of a few
GeV. The spectrum and morphology of this signal is consistent with one 
originating from dark matter
annihilations~\cite{Goodenough:2009gk,Hooper:2010mq,Hooper:2011ti,Abazajian:2012pn}. Very
recently, evidence of this signal has been found from regions outside
of the GC~\cite{Hooper:2013rwa} in the directions of the sky
coincident with the \emph{Fermi} Bubbles: 
%----------------Figure------------------
\begin{figure}[h]
\begin{center}
\postscript{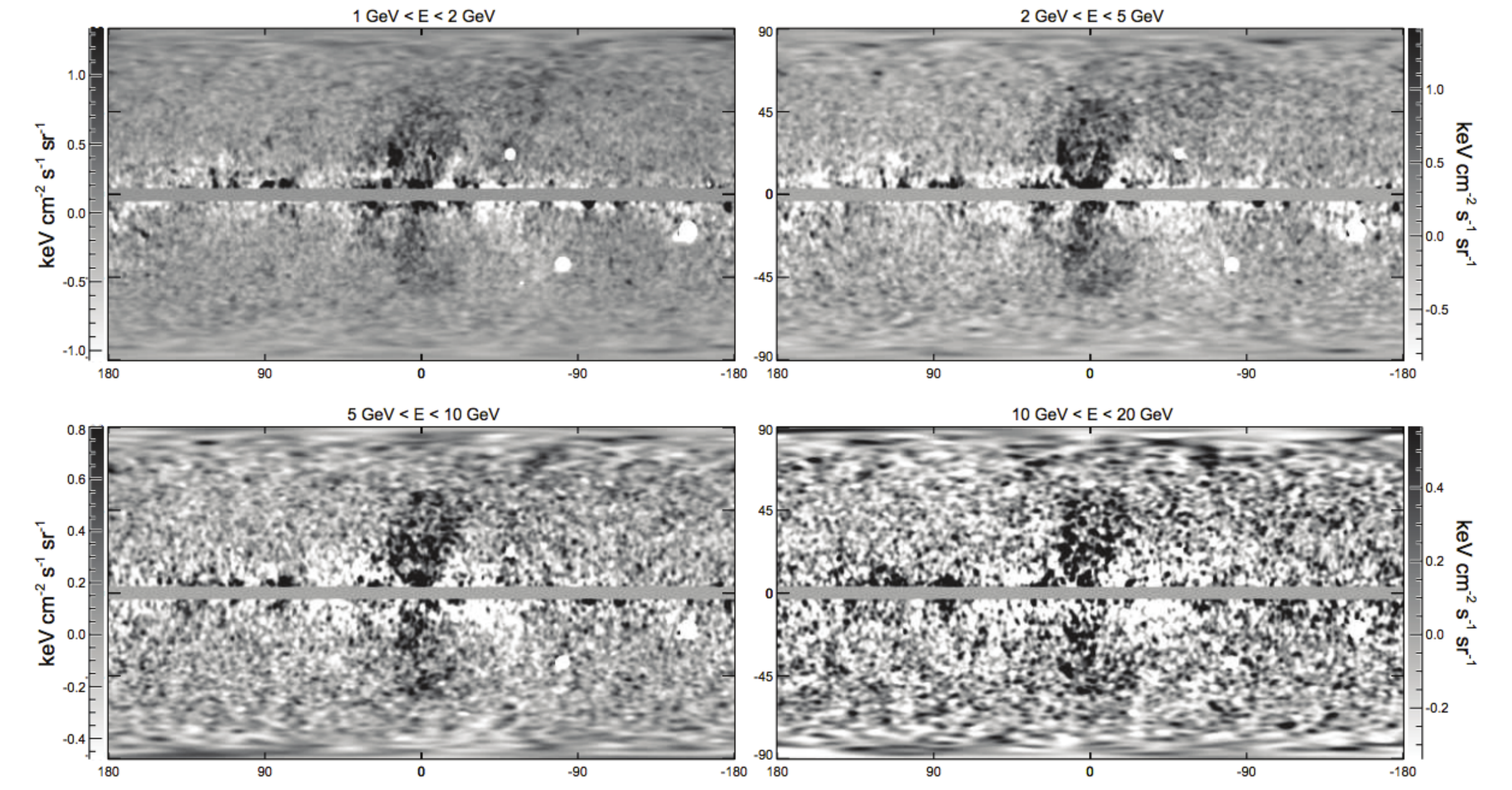}{0.9}
\caption[\emph{Fermi} Bubbles Appearing Above and Below Galactic Disk]{ \sglspc \small After subtracting the away the \emph{Fermi} diffuse galactic model from 1.6 years of LAT data two nearly symmetric bubble structures appear extending out from the galactic center know as \emph{Fermi} bubbles (image adopted from \cite{Su:2010qj}).}
\label{fig:FermiBubs}
\end{center}
\end{figure}
%-----------------------------------
two bilateral $\gamma$-ray
lobes centered at the core of the Galaxy and extending to around
$50^\circ$ above and below the Galactic plane ({\it i.e.}, $r =
\pm 10~{\rm kpc}$, where $r$ is the distance from the
GC)~\cite{Dobler:2009xz,Su:2010qj} as seen in Fig. \ref{fig:FermiBubs}. At lower Galactic latitudes these
structures are coincident with a non-thermal microwave `haze' found in
WMAP 23-33~GHz data~\cite{Finkbeiner:2003im} (confirmed recently by
the Planck space mission~\cite{:2012rta}) and the thermal $X$-ray
emission seen by ROSAT~\cite{Snowden:1997ze}.

Far from the Galactic plane ($|b| \geq 30^{\circ}$), the observed
energy-weighted $\gamma$-ray spectrum is nearly invariant with
latitude and fairly flat ($d\Phi_\gamma/dE_\gamma \propto E_\gamma^{-2}$) over
the energy range observed by \emph{Fermi}. The correlation found in
the multi-wavelength observations seems to indicate that the Bubbles
(measured in the range of $E_\gamma \sim 1 - 100~{\rm GeV}$) are
produced by a population of ${\rm GeV}-{\rm TeV}$ electrons (with an
approximately power-law spectrum $d\Phi_e/dE_e \propto E_e^{-3}$) via
inverse Compton scattering of ambient low-energy photons, as the same
electrons can also simultaneously produce radio synchrotron radiations
in the presence of magnetic
fields~\cite{Dobler:2009xz,Mertsch:2011es}. The transparency of
  this elementary and self-consistent framework provides strong
  support for a leptonic origin of the high-latitude emission from the
  \emph{Fermi} Bubbles.

Conversely, at latitudes closer to the disk ($|b| \geq 20^\circ$), the
spectrum of the emission correlated with the Bubbles possesses a
pronounced spectral feature in $E_\gamma^2d\Phi_\gamma/dE_\gamma$ peaking at
$E_\gamma \sim 1-4~{\rm GeV}$, which cannot be produced by any
realistic spectrum of electrons~\cite{Hooper:2013rwa}. This implies
that a second (non-inverse-Compton) emission mechanism must be
responsible for the bulk of the low-energy, low-latitude emission. The
spectral shape of this second component is similar to the one reported
from the GC.  The intrinsic non-inverse-Compton emission appears
spatially consistent with a luminosity per volume falling
approximately as $r^{-2.4} - r^{-2.8}$.  As a consequence, the
spectral feature visible in the low-latitude Bubbles is most likely
the extended counterpart of the GC excess, now detected out to at
least $r \sim 2-3~{\rm kpc}$. Even though millisecond pulsars possess
a spectral cutoff at approximately the required energy, these sources
exhibit a spectral shape that is much too soft at sub-GeV energies to
accommodate this signal~\cite{Hooper:2013nhl}.

The spectrum and angular distribution of the signal is broadly
consistent with one predicted from $\sim 10~{\rm GeV}$ dark matter
particles annihilating to leptons, or from $\sim 50~{\rm GeV}$ dark
matter particles annihilating to quarks, following a distribution
similar to, but slightly steeper than, the canonical
Navarro-Frenk-White profile. In either case, the morphology of the
$\gamma$-ray signal requires a dark matter distribution that scales
approximately as $\rho_{\rm DM} \propto r^{-1.2} - r^{-1.4}$, that is
the annihilation rate per volume is proportional to the square of the
dark matter density. Such a dark matter distribution is in good 
agreement with current observational constraints~\cite{Iocco:2011jz}.

For the 10~GeV dark matter candidate, the normalization of the
observed signal requires a velocity average annihilation cross section
on the order of
\begin{equation}
\langle \sigma_\tau v \rangle
\sim 2 \times 10^{-27}~{\rm cm}^{3}/{\rm s} = 1.7 \times 10^{-10}~{\rm GeV}^{-2} \,,
\label{fermibubble}
\end{equation}
up to overall uncertainties in the normalization of the halo
profile~\cite{Hooper:2012ft}.  

 Moreover dark matter particles can also elastically scatter with nuclei in the Sun,
leading to their gravitational capture and subsequent
annihilation. Electrons and muons produced in such annihilations
quickly lose their energy to the Solar medium and produce no
observable effects. Annihilations to taus, on the other hand, produce
neutrinos which, for a 10~GeV, can be observed by
Super-Kamiokande. For the required branching into $\tau^+ \tau^-$ of
about 10\% -- as given by (\ref{fermibubble}) -- existing data
constrain the dark matter spin-independent elastic scattering cross
section with protons to be less than $4 \times 10^{-41}~{\rm
  cm}^{2}$~\cite{Hooper:2008cf,Kappl:2011kz}.

For the 50~GeV dark matter particle, the normalization of the
observed signal requires a velocity average annihilation cross section
on the order of
\begin{equation}
\langle \sigma_b v \rangle
\sim 8 \times 10^{-27}~{\rm cm}^{3}/{\rm s} = 6.7 \times 10^{-10}~{\rm GeV}^{-2} \, .
\label{fermibubble2}
\end{equation}
The XENON-100 Collaboration reported a 90\%~CL bound on the elastic
scattering cross section with nuclei of ${\cal O} (10^{-44}~{\rm
  cm}^2)$~\cite{Aprile:2011hi}. A later analysis has arrived at
alternative conclusions allowing for a signal of two events with a
favored mass of 12~GeV and large error contour extending to about 50~GeV~\cite{Hooper:2013cwa}.

It is worthwhile to point out that the bounds from the combined
analysis of 10 dwarf
spheroidals~\cite{Ackermann:2011wa,GeringerSameth:2011iw}, galaxy
clusters~\cite{Ackermann:2010rg}, or diffuse $\gamma$-ray
emission~\cite{Abazajian:2010sq,Abdo:2010dk} are not sensitive enough to
probe the velocity average annihilation cross sections
(\ref{fermibubble}) and (\ref{fermibubble2}).

With this recent (at the time of this writing) observations a minimal hidden sector Higgs
portal model has been suggested by Weinberg~\cite{Weinberg:2013kea}. 

It has attractive qualities to explain these observations, and we will examine to what extent its free parameters can be adjusted to fit, experimental searches for new physics at the LHC,  constraints from cosmological observations,  constraints from direct detection searches, and a possible explanation of the low-latitude
$\gamma$-ray emission from the \emph{Fermi} Bubbles.

\subsection{Constructing Weinberg's WIMP\lowercase{s}}

Let us start by examining the Weinberg Higgs portal model.  It is based on a broken global $U(1)$
symmetry associated with the dark matter charge $W$: the number of
weakly-interacting massive particles (WIMPs) minus the number of their
antiparticles.  The  hidden sector contains a
Dirac field $\psi$ (carrying WIMP quantum number $W = +1$) and a
complex scalar field (with $W = 2$, so that its expectation value
leaves an unbroken reflection symmetry $\psi \to - \psi$). All SM
fields are assumed to have $W=0$. 

The scalar potential consists of the SM component $[s]$, the
isomorphic component in the hidden sector $[h]$, and the quartic
interaction coupling between the two sectors with strength
$\eta_\chi$. The Lagrangian density for the scalar sector reads
\beqa 
\mathscr{L} = | \p \Phi_h |^2 + | \p \Phi_s |^2 +
\mu_h^2 |\Phi_h|^2 - \lambda_h |\Phi_h|^4 +
\mu_s^2 |\Phi_s|^2 - \lambda_s |\Phi_s|^4 -
\eta_\chi |\Phi_h|^2|\Phi_s|^2  \,, \nonumber \\
\label{eq:one}
\eeqa 
where $\Phi_s$ is the SM scalar doublet and $\Phi_h$ is
a complex scalar field. We separate a massless Goldstone boson field
$\alpha(x)$ and a massive radial field $r(x)$ by defining
\begin{equation}
\Phi_h(x) =  \frac{1}{\sqrt{2}} \ r(x) \ e^{i\, 2 \alpha(x)} \,, 
\end{equation}
where $r(x)$ and $\alpha(x)$ are real, with the phase of $\Phi_h(x)$
adjusted to make the vacuum expectation value (VEV) of $\alpha (x)$
zero. The $SU(2) \times U(1)$ symmetry of the SM is (of course) broken
by a non-vanishing VEV of the neutral component $\phi$
of the scalar doublet,
\begin{equation}
\Phi_s = \frac{1}{\sqrt{2}} \ \left( \begin{array}{c} G^\pm \\
v_\phi + \phi' + iG^0 \end{array} \right) ,
\end{equation}
where $v_\phi \simeq 246~{\rm GeV}$. The $G$ fields are the familiar Goldstone
bosons, which are eaten by the vector bosons ({\it i.e.} the $G^\pm$
become the longitudinal components of the charged $W$-boson and $G^0$
becomes the longitudinal component of the $Z$-boson). In terms of real
fields the Lagrangian density  (\ref{eq:one}) takes the form
\begin{equation}
\mathscr{L} = \frac{1}{2} \p r^2 +  \frac{1}{2} \p \phi^2 + 2 r^2 \p \alpha^2 +
\frac{\mu_h^2}{2} r^2 - \frac{\lambda_h}{4}r^4 + \mu_s^2
|\phi|^2 - \lambda_s |\phi|^4 - \frac{\eta_\chi}{2} r^2 |\phi|^2
 \, .
\label{eq:three}
\end{equation}
The $U(1)$ symmetry of $W$ conservation is also broken and $r$ gets a VEV
 \begin{equation}
r (x)  = v_r  + r' (x) \,,  
\label{eq:four}
\end{equation}
with $v_r$ real and non-negative.  We demand the scalar
potential obtains its minimum value at
\beq
{\cal V} =   -\frac{\mu_h^2}{2} v_r^2 + \frac{\lambda_h}{4} v_r^4 - \frac{\mu_s^2}{2}  v_\phi ^2
+ \frac{\lambda_s}{4} v_\phi^4 + \frac{\eta_\chi}{4} v_r^2 v_\phi^2 \, . 
\label{eq:five}
\eeq
Physically, the most interesting solutions to the minimization of
 (\ref{eq:five}), 
\begin{equation}
\p_{v_r} {\cal V} = - \mu_h^2 v_r +  \lambda_h v_r^3 +  \frac{\eta_\chi}{2} v_r  v_\phi^2 =0 
\end{equation}
 and
\begin{equation}
\p_{v_\phi} {\cal V}  =  - \mu_s^2 v_\phi  +  \lambda_s v_\phi^3 
+ \frac{\eta_\chi}{2}  v_r ^2  v_\phi =0 \,, 
\end{equation}
are obtained for $v_r $ and $v_\phi$ both non-vanishing
\begin{equation}
v_\phi^2  = \frac{1}{\lambda_s}\left(\mu_s^2 - \frac{\eta_\chi v_r^2}{2}  \right)
\end{equation}
and
\begin{equation}
v_r^2  =
\frac{1}{\lambda_h}\left(\mu_h^2 - \frac{\eta_\chi v_\phi^2}{2} \right) \,,
\label{eq:six}
\end{equation}
respectively. To compute the scalar masses, we must expand the
potential  around the minima 
 \beqa 
 \mathscr{L} & = &\frac{1}{2}
(\p r')^2 \nonumber
\\
& + & 2 v_r^2 \p \alpha^2 + 4 v_r r' \p \alpha^2 +  2 r'^2 \p \alpha^2 \nonumber
\\
& - & \lambda_h v_r^2 r'^2 - \lambda_s v_\phi^2 \phi^{'2} -
\eta_\chi v_r v_\phi r' \phi' + \cdots \,, 
\label{eq:seven}
\eeqa 
where the dots indicate 3-point and 4-point interactions, as well as
the SM interactions.  There is a mixing term present for $r'$ and
$\phi'$. We find the fields of definite mass by diagonalizing the mass
matrix for $r'$ and $\phi'$.  We denote by $H$ and $h$ the scalar
fields of definite masses, $m_H = 125~{\rm GeV}$ and $m_h$, respectively. After a
bit of algebra, the explicit expressions for the scalar mass
eigenvalues and eigenvectors are given by 
\begin{equation}
m^2_h =  \lambda _h v_r^2 +  \lambda_s v_\phi^2 - \sqrt{(\lambda_s v_\phi^2 - \lambda _h
  v_r^2)^2 + (\eta_\chi v_r  v_\phi )^2} 
\end{equation}
and
\begin{equation}
m^2_H =  \lambda _h v_r^2 + \lambda _s v_\phi^2 +\sqrt{(\lambda_s v_\phi^2 - \lambda _h
  v_r^2)^2 + (\eta_\chi v_r  v_\phi )^2} \,,
\label{eq:eight}
\end{equation}
with
\begin{equation}
\left( \begin{array}{c} h \\ H \end{array}\right) =
\left( \begin{array}{cc} \cos{\chi}& -\sin{\chi}\\ \sin{\chi}& \phantom{-}\cos{\chi}
\end{array}\right) \left( \begin{array}{c} r' \\ \phi' \end{array}\right) \, ,
\label{eq:nine}
\end{equation}
where $\chi \in [-\pi/2,\pi/2]$ also fulfills
\begin{equation}
\sin 2\chi = \frac{ \eta_\chi v_\phi v_r}{\sqrt{(\lambda_s v_\phi^2 - \lambda _h
  v_r^2)^2 + (\eta_\chi v_r  v_\phi )^2} } = \frac{2 \eta_\chi v_\phi v_r}{m_H^2 - m_h^2},
\label{correa}
\end{equation}
and
\begin{equation}
\cos 2\chi = \frac{ \lambda _s  v_\phi^2 - \lambda_h v_r^2}{\sqrt{(\lambda_s v_\phi^2 - \lambda _h
  v_r^2)^2 + (\eta_\chi v_r  v_\phi )^2} } \ , 
\end{equation}
yielding
\begin{equation}
\tan 2\chi = \frac{ \eta_\chi v_r  v_\phi}{ \lambda_s v_\phi^2 -  \lambda_h v_r^2} \, .
\label{eq:ten}
\end{equation}
The Goldstone boson in  (\ref{eq:seven}) has to be be re-normalized so
that it resumes the standard canonical form. This is achieved through
scaling $\alpha \rightarrow \alpha' = 2 v_r \alpha$, giving
\beq
2 v_r^2 \p \alpha^2 + 4 v_r r' \p \alpha^2 +  2 r'^2 \p \alpha^2 \rightarrow \frac{1}{2} \p \alpha'^2 + \frac{1}{v_r} r' \p \alpha'^2 + \frac{1}{2 v_r^2} r'^2 \p \alpha'^2.
\label{eq:eleven} 
\eeq 
Adding in the dark matter sector requires at least one Dirac field
\beq
\mathscr{L}_\psi = i \bar{\psi}\gamma \cdot \p \psi - m_\psi \bar{\psi} \psi 
- \frac{f}{\sqrt{2}} \bar{\psi^c} \psi \Phi_h^\dagger  - \frac{f^*}{\sqrt{2}} \bar{\psi} \psi^c \Phi_h  .
\label{eq:fortyone}
\eeq We assign $\psi$ a charge $W = 1$, so that the Lagrangian is
invariant under the global transformation $e^{i W \alpha}$. Applying a phase change allows us to express $\psi$ as 
\beq
\psi(x) = \psi'(x) e^{i \alpha(x)}.
\label{eq:fortytwo}
\eeq
We can now rewrite  (\ref{eq:fortyone}) in terms of $\psi',$ $\alpha,$ and $r$
\beq
\mathscr{L}_\psi = i \bar{\psi}'\gamma \cdot \p \psi' - (\bar{\psi}' \gamma \psi' )  \cdot \p \alpha - m_\psi \bar{\psi}' \psi' - \frac{f}{2} \bar{\psi'}^{c} \psi' r  - \frac{f}{2}  \bar{\psi}' {{\psi'}^{c}} r  \,,
\label{eq:fortythree}
\eeq where we have taken $f$ to be real. Once $r$ achieves a VEV we
can expand the dark matter sector to get \beqa \mathscr{L}_\psi &=&
\frac{i}{2}\left(\bar{\psi}'\gamma \cdot \p \psi' + \bar{\psi'}^{c}
  \gamma \cdot \p \psi^{c'} \right), \nonumber
\\
&-& \frac{m_\psi}{2} \left( \bar{\psi}' \psi' + \bar{\psi'}^{c}
  {\psi'}^{c} \right)-\frac{f v_r}{2} \bar{\psi'}^{c} \psi' - \frac{f
  v_r}{2} \bar{\psi}' {\psi'}^{c} , \nonumber
\\
&-& \frac{1}{2} (\bar{\psi}' \gamma \psi' - \bar{\psi'}^{c} \gamma
{\psi '}^{c} ) \cdot \p \alpha , \nonumber
\\
&-& \frac{f}{2} r' \left( \bar{\psi'}^{c}\psi' + \bar{\psi}' {\psi'}^{c}
\right).
\label{eq:fortyfour}
\eeqa
Note that we have made the Lagrangian explicitly symmetric via relations like
\begin{eqnarray}
\psi^c &=& C \bar{\psi}^T  \\
\label{eq:fourtyfive}
\bar{\psi^c} \psi^c &=& ( - \psi^T C^{-1} C \bar{\psi}^T) = \bar{\psi} \psi \\
\bar{\psi}_c \gamma \cdot \p \psi_c &=& - \psi^T C^{-1} \gamma C \cdot \p \bar{\psi}^T \nonumber
\\
\label{eq:fourtysix}
&=& \psi^T \gamma^T \cdot \p \bar{\psi}^T = - (\p \bar{\psi} \cdot \gamma \psi) \rightarrow \bar{\psi} \gamma \cdot \p \psi \, .
\end{eqnarray} 
In  (\ref{eq:fourtyfive}) we used the Grassman nature of the spinor
fields; in the second line of  (\ref{eq:fourtysix}) we used integration
by parts to transfer the derivative onto the $\psi$ field. Similar
results can be found for the other expressions.

Diagonalization of the $\psi'$ mass matrix  generates the mass eigenvalues,
\begin{equation}
m_\pm =  m_\psi \pm  f  v_r, 
\end{equation}
for the two mass eigenstates
\begin{equation}
\psi_- = \frac{i}{\sqrt{2}} \lpa \psi'^c - \psi'  \rpa
  \quad {\rm and}   \quad \psi_+ = \frac{1}{\sqrt{2}}\lpa \psi'^c+\psi' \rpa  \, .
\label{eq:fourtyseven}
\end{equation}
In this basis, the act of charge conjugation on $\psi_\pm$ results in
\beq
\psi^c_\pm =  \psi_\pm.
\label{eq:fourtyeight}
\eeq This tells us that the fields $\psi_\pm$ are Majorana
fermions. The Lagrangian is found to be 
\beqa \mathscr{L}_\psi
&=&\frac{i}{2}\bar{\psi_+}\gamma \cdot \p \psi_+ +
\frac{i}{2}\bar{\psi_-}\gamma \cdot \p \psi_- - \frac{1}{2} m_+
\bar{\psi}_+ \psi_+ - \frac{1}{2}m_- \bar{\psi}_- \psi_- , \nonumber
\\
&-&\frac{i}{4 v_r} (\bar{\psi}_+ \gamma \psi_- - \bar{\psi}_-
\gamma \psi_+) \cdot \p \alpha' , \nonumber
\\
& -& \frac{f}{2} r' (\bar{\psi}_+\psi_+ - \bar{\psi}_- \psi_-).
\label{eq:fortynine} 
\eeqa
We must now put $r'$ into its massive field representation, for which the interactions of interest are
\beq
-\frac{f \sin \chi}{2} H (\bar{\psi}_+\psi_+ -  \bar{\psi}_-
\psi_-) - \frac{f  \cos \chi}{2} h (\bar{\psi}_+\psi_+ -  \bar{\psi}_- \psi_-). 
\label{eq:fifty}
\eeq This leads to 3-point interactions between the $W$-WIMPs and the
Higgs boson of the SM.

In summary, instead of one Dirac $W$-WIMP, there are two Majorana
$W$-WIMPs of different masses. However, the heavier $W$-WIMP will
decay into the lighter one by emitting a Goldstone boson, while the
lighter one is kept stable by an unbroken reflection
symmetry. Therefore in this model we can expect that the universe
today will contain only one type of Majorana $W$-WIMP, the lighter one
$w$, with mass $m_w$ equal to the smaller of $m_\pm$. Throughout,
$\Delta m = |m_+ - m_-| = 2 | f v_r|$ denotes the mass splitting of
the $W$-WIMP states.  (The most common variables used in this discussion
are summarized in Table~\ref{table:0}.)

A cautionary note is worth taking on board at this juncture. It has
long been known that the spontaneous breaking of a global $U(1)$
symmetry have several disconnected and degenerate vacua (the phase of
the vacuum expectation value $\langle 0 | \Phi_h |0 \rangle$ can be
different in different regions of space, and actually we expect it to
be different in casually disconnected regions), leading to
catastrophic domain-wall structure in the early
universe~\cite{Sikivie:1982qv,Vilenkin:1982ks}. In the spirit
of~\cite{Sikivie:1982qv}, it may be possible to introduce a small
explicit breaking of the symmetry, such that the domain walls
disappear before dominating the matter density of the universe, while
leaving \mbox{(pseudo-)Goldstone} bosons and the same dark matter
phenomenology.\footnote{ \sglspc \small Other approaches, if exceedingly fine-tuned,
  may offer alternative
  solutions~\cite{Hill:1989sd,Linde:1990yj,Turner:1990uz,Dvali:1991ka,Hiramatsu:2012sc}.}

The absence of new physics signals at the LHC place constraints on the
model. We discuss this next.
\begin{table}
\begin{center}
\caption[Common Variables Table for W-WIMP Model]{\centering Definition of Most Common Variables of W-WIMP Model. \label{table:0}}
\ \\
\begin{tabular}{| ll |}
\hline
\hline
$\Phi_s$  & ~~SM Scalar Doublet \\
\hline
$\Phi_h$  & ~~Complex Scalar Field \\
\hline
$\phi$ & ~~Neutral Component of the Scalar Doublet \\
\hline
$r$ & ~~Massive Radial Field \\
\hline
$\alpha$ & ~~Goldstone Boson \\
\hline
$v_\phi$ & ~~Vacuum Expectation Value of $\phi$ \\
\hline
$v_r$ & ~~Vacuum Expectation Value of $r$ \\
\hline
$H$ & ~~SM Higgs Boson \\
\hline
$h$ & ~~Hidden Scalar \\
\hline
$\eta_\chi$ & ~~Quartic Interaction Coupling Between SM and Hidden Sectors  \\
\hline
$\chi$ & ~~$H$-$h$ Mixing Angle \\
\hline
$w$ & ~~lightest $W$-WIMP \\
\hline
$\Delta m$ & ~~$W$-WIMP Mass Splitting\\
\hline
$f$ & ~~Coupling Between Hidden Majorana Fermions and Complex Scalar Field\\
\hline
\hline
\end{tabular}
\end{center}
\end{table}

\newpage

\subsection{\label{sec:Bbound} Constraints from Collider Experiments}

As was stated previously the recent discovery~\cite{ATLASnew,Chatrchyan:2012ufa} SM Higgs boson like particle suggests that with measurements of branching ratios
in various channels, a study of the properties of the
Higgs-like state has the potential for revealing
whether or not the Higgs sector is as simple as envisioned in the
SM. Since invisible decays reduce the branching fraction to the
(visible) SM final states, it is to be expected that ${\cal B} (H \to
\, {\rm invisible})$ is strongly constrained. Indeed ${\cal B} (H \to
\, {\rm invisible})$ is known to be less than about 19\% at
95\%CL~\cite{Espinosa:2012vu,Giardino:2013bma,Ellis:2013lra}. Thus,
the mixing of the SM with the hidden sector must be weak. Note that
for $\eta_\chi \ll 1$  the relations between masses and angles then 
becomes
\begin{equation}
m^2_h \approx 2 \lambda _h v_r^2, \quad \quad m^2_H \approx  2 \lambda_s
v_\phi^2, \quad  \quad
\tan 2\chi \approx  \frac{2 \eta_\chi v_r  v_\phi}{m_H^2 - m_h^2} \, ,
\label{eq:ten-1}
\end{equation}
where we have assumed $ \lambda _s v_\phi^2 > \lambda _h v_r^2$. For a
Higgs width of about 4~MeV, the partial width for decay into 
unobserved particles is found to be
\begin{equation}
\Gamma_{H \to \, {\rm invisible}} < 0.8~{\rm MeV} \, . 
\label{LHCwidth}
\end{equation}

The phenomenology of a Higgs portal to the hidden sector depends on
whether the SM Higgs particle is lighter or heavier than the new
companion. In this study we take $m_H > m_h$. The decay rate into
invisible stuff, $\Gamma_{H \to \, {\rm invisible}}$, has two distinct
contributions: $\Gamma_{H \to \, {\rm invisible}}^{\rm SM}$ and
$\Gamma_{H \to \, {\rm hidden}}$. The former is dominated by $H \to 2
Z \to 4 \nu$, with an invisible $Z$ branching ratio of 4\%. The $4
\nu$ rate can be predicted from observed decays $H \to 2Z \to 4l$. For
the sake of simplicity, hereafter we will omit the contribution of
$\Gamma_{H \to \, {\rm invisible}}^{\rm SM}$. Unless expressly stated
otherwise herein, we assume $m_w + \Delta m > m_H/2$ and thus $H$
decays (invisibly) into the hidden sector via three channels: $H \to 2
\alpha'$, ${H \to 2 h}$, and $H \to 2 w$. From the event rates for
visible Higgs production and decay channels we could derive upper
bounds on non-SM admixtures in the wave-function of the Higgs boson
and on the new three invisible decay channels. To this end we now
compute the decay rates for these three processes.

\subsubsection{$\underline{\bf{\Gamma_{H\to 2 \alpha'}}}$}

Substituting in  (\ref{eq:eleven}) $r'$ by the field of definite mass,
$r' = h \cos \chi + H \sin \chi$, we can write the Higgs--Goldstone
boson interaction term as \beq \frac{1}{v_r} r' \p \alpha'^2
\rightarrow \frac{\sin \chi}{v_r} H (\p \alpha')^2 + \frac{\cos
  \chi}{v_r} h (\p \alpha')^2 \, .
\label{eq:twelve}
\eeq
Using  (\ref{eq:twelve}) we write the Feynman rule for
        interactions of the type $H, \ \alpha', \ \alpha'$ as 
\beq
-i\frac{2 \sin \chi}{v_r} k \cdot k' ,
\label{eq:thirteen}
\eeq where $k \, (k')$ is the 4-momentum of the incoming (outgoing)
$\alpha'$ particle, and the factor of $2$ is a symmetry factor, as one
can exchange incoming-outgoing $\alpha'$ twice.  From this 3-point
interaction we can calculate the decay width of the SM Higgs $H$
into 2 Goldstone bosons $\alpha'$. In the rest frame of the Higgs, the
differential decay probability per unit time is given by 
\beq
d\Gamma_{H \rightarrow 2 \alpha'} = \frac{1}{2 m_H} \left(\frac{2 \sin \chi}{v_r} k_1 \cdot k_2\right)^2 \, d\mathcal{Q} \,,
\label{eq:fourteenW}
\eeq
where 
\beqa
d\mathcal{Q} &=& \frac{1}{2!} \ \frac{d^3 k_1}{(2\pi)^3 2 k_1}\frac{d^3 k_2}{(2\pi)^3 2 k_2}
(2\pi)^4 \delta(m_H - k_1 - k_2)\delta^{(3)}(\textbf{k}_1 +
\textbf{k}_2)  \nonumber
\\
&=& \left. \frac{1}{16} \frac{d\Omega_{k_1}}{(2 \pi)^2} \right|_{k_1 = m_H/2} 
\eeqa
is the phase space for a two-body final state (the factor of $1/2!$
is included because of  identical particles in the final state).  After some
algebra  (\ref{eq:fourteenW}) can be re-written as
\begin{equation}
d\Gamma_{H \rightarrow 2 \alpha'}  =  \frac{d \Omega_{k_1}}{128~\pi^2
  m_H} \left[\frac{2 \, \sin\chi}{v_r} \, 2 \, \left( \frac{m_H}{2} \right)^2 \right]^2 \, .
\end{equation}
The partial decay width can now be expressed as
\begin{equation}
\Gamma_{H \rightarrow 2 \alpha'}  = \frac{1}{32 \pi}
\left(\frac{\sin \chi}{v_r}\right)^2 m_H^3 \, .
\end{equation}
For $m_H \gg m_h$ and $m_H^2 \gg 2 \, \eta_\chi \, v_r  v_\phi$,
we can use the small angle approximation
\begin{equation}
\sin \chi \approx \chi =  \eta_\chi \, v_r v_\phi
/(m_H^2-m_h^2) \, . 
\end{equation}
In this very good approximation
the decay width becomes
\beq
\Gamma_{H \rightarrow 2 \alpha'} =\frac{1}{32 \pi}
\left(\frac{\eta_\chi \, v_\phi }{m_H^2-m_h^2}\right)^2 m_H^3.
\label{eq:fifteen}
\eeq

\subsubsection{$\underline{\bf{\Gamma_{H\to 2 h}}}$}

We begin by expanding the scalar potential  $\CV$ around the VEVs of $r$
and $\phi$ after which we diagonalize the mass matrix. Together this
requires that we expand around the fields \beqa r(x) &=& v_r + h \cos
\chi + H \sin \chi \ , \nonumber
\\
\phi(x) &=& v_\phi + H \cos \chi -h \sin \chi \ , \eeqa 
 which puts $\CV$  in the form 
\begin{eqnarray}
\CV &=& \frac{1}{2}
m_H^2 H^2 + \frac{1}{2} m_h^2 h^2 \nonumber \\
& - &  \frac{1}{16} \lpa \eta_\chi + 3 (\lambda_h+\lambda_s)+3(\eta_\chi
-\lambda_h-\lambda_s)\cos 4\chi \rpa H^2 h^2 \nonumber
\\
&- &\frac{1}{4}   v_\phi \cos \chi [6 \lambda_s - \eta_\chi +
3(\eta_\chi - 2 \lambda_s)\cos 2\chi]  H h^2  \nonumber \\
& + & \frac{1}{4} v_r \sin \chi [\eta_\chi - 6 \lambda_h +
3 (\eta_\chi - 2 \lambda_h) \cos 2 \chi ]  H h^2 \nonumber
\\
&- &\frac{1}{4} \lpa \lambda_s \cos^4 \chi + \eta_\chi \cos^2 \chi
\sin^2 \chi + \lambda_h \sin^4 \chi \rpa H^4 \nonumber
\\
&- &\frac{1}{4} \lpa \lambda_h \cos^4 \chi + \eta_\chi \cos^2 \chi
\sin^2 \chi + \lambda_s \sin^4 \chi \rpa h^4 \nonumber
\\
&+&\frac{1}{2} \lpa v_\phi \sin \chi ( \eta_\chi \cos^2 \chi + 2
\lambda_s \sin^2 \chi) - v_r (2 \lambda_h \cos^3 \chi + \eta_\chi \cos
\chi \sin^2 \chi) \rpa h^3 \nonumber
\\
&- &\frac{1}{2} \lpa v_r \sin \chi (\eta_\chi \cos^2 \chi + 2 \lambda_h
\sin^2 \chi) + v_\phi (2\lambda_s \cos^3 \chi + \eta_\chi \cos \chi
\sin^2 \chi)\rpa H^3 \nonumber
\\
&+ &\frac{1}{4} \lpa \lambda_h - \lambda_s +(\lambda_s + \lambda_h -
\eta_\chi) \cos 2\chi \rpa \sin 2\chi \ H h^3 \nonumber
\\
&+ & \frac{1}{4} \lpa \lambda_s - \lambda_h +(\lambda_s + \lambda_h -
\eta_\chi) \cos 2\chi \rpa \sin 2\chi \ H^3 h \nonumber
\\
&+ &\frac{1}{2}  v_\phi \sin \chi [2 (3 \lambda_s - \eta_\chi)
\cos^2 \chi + \eta_\chi \sin^2 \chi] H^2 h  \nonumber \\
&+ & \frac{1}{2}v_r (\eta_\chi \sin \chi \sin 2
\chi - 6 \lambda_h \cos \chi \sin^2 \chi - \eta_\chi \cos^3 \chi) 
H^2 h \, .
\end{eqnarray} 
Since $\chi < 1$,  we first
expand the potential  around $\chi=0$, and then using (\ref{correa})  we further expand around
$\eta_\chi = 0$  retaining only the terms of first order in
$\eta_\chi$;  this results in \beqa \CV &\approx& \frac{1}{2} m_H^2 H^2
+ \frac{1}{2} m_h^2 h^2 \nonumber
\\
&-&\frac{\eta_\chi }{4} H^2 h^2 - \frac{\lambda_h}{4} h^4 -
\frac{\lambda_s}{4} H^4 - \frac{\eta_\chi \lambda_h v_r v_\phi}{m_H^2
  -m_h^2} H h^3 + \frac{\eta_\chi \lambda_s v_r v_\phi}{m_H^2 -m_h^2}
H^3 h \nonumber
\\
&-& \lambda_h v_r h^3 - \lambda_s v_\phi H^3 -\frac{\eta_\chi v_\phi}{2}
\lpa \frac{6 \lambda_h v_r^2}{m_H^2 - m_h^2} +1 \rpa H h^2 \nonumber \\
&+& \frac{\eta_\chi v_r}{2} \lpa \frac{6 \lambda_s v_\phi^2}{m_H^2 -
  m_h^2} -1\rpa H^2 h \ . \nonumber 
\eeqa 
Using  (\ref{eq:ten-1}) we can manipulate this expression to write the scalar
potential as 
\beqa \CV &\approx&\frac{1}{2} m_H^2 H^2 + \frac{1}{2}
m_h^2 h^2 \nonumber
\\
&-&\frac{\eta_\chi }{4} H^2 h^2 - \frac{\lambda_h}{4} h^4 -
\frac{\lambda_s}{4} H^4 - \frac{\eta_\chi \lambda_h v_r v_\phi}{m_H^2
  -m_h^2} H h^3 + \frac{\eta_\chi \lambda_s v_r v_\phi}{m_H^2 -m_h^2}
H^3 h \nonumber
\\
&-& \frac{m_h^2}{2 v_r} h^3 - \frac{m_h^2}{2 v_\phi}H^3 -\frac{\eta_\chi
  v_\phi}{2} \lpa \frac{m_H^2 + 2 m_h^2}{m_H^2 - m_h^2} \rpa H h^2 \nonumber
\\
&+& \frac{\eta_\chi v_r}{2} \lpa \frac{2 m_H^2 + m_h^2}{m_H^2 - m_h^2}
\rpa H^2 h \ . 
\eeqa
Under the approximations
taken previously, $m_H \gg m_h$ and $m_H^2 \gg 2 \eta_\chi v_r v_\phi$,
the relevant $Hhh$ interaction term results in 
\beq -\frac{\eta_\chi v_\phi}{2} \lpa \frac{m_H^2 + 2
  m_h^2}{m_H^2-m_h^2} \rpa H h^2. \eeq
The differential decay probability  per unit time is given by
\beqa
&d\Gamma_{H \to 2h}& = \nonumber \\
&\frac{1}{2 m_H}& \left( \frac{\eta_\chi
  v_\phi}{m_H^2 - m_h^2} \right)^2 \ \left(m_H^2 + 2 m_h^2 \right)^2
\frac{1}{2!} \frac{k^2 \, dk \, d \Omega}{(2 \pi)^2 4  E_k}
\frac{1}{2k} \ \delta \left( m_h - 2 \sqrt{k^2 + m_H^2} \right) \, . \nonumber \\
\eeqa
The partial $H \to 2 h$ decay width can now be expressed as
\begin{equation}
\Gamma_{H \to 2h} =  \frac{1}{32 \, \pi \, m_H^2}  \left( \frac{\eta_\chi
  v_\phi}{m_H^2 - m_h^2} \right)^2 \  \left(m_H^2 + 2 m_h^2 \right)^2
\ \sqrt{m_H^2 - 4 m_h^2} \, .
\end{equation}
In the limit $m_H \gg m_h$ we obtain
\begin{equation}
\Gamma_{H \to 2h} = \frac{1}{32 \pi} \left( \frac{\eta_\chi
  v_\phi}{m_H^2 - m_h^2} \right)^2  \ m_H^3 \, .
\end{equation}

\subsubsection{$\underline{ \bf{\Gamma_{H\to 2 w}}}$}

For $m_w < m_H/2$, the $r-\phi$ mixing allows the Higgs boson to decay
into pairs of the lightest $W$-WIMP.  We obtain the invariant
amplitude for this process (a description of Feynman rules for
Majorana fermions can be found in Ref.~\cite{Srednicki:2007qs}),
\begin{equation} 
i \CM = i f \sin \chi \bar{u}(p) v(p') \,, 
\end{equation}
where $u(p)$ and $v(p)$ are Dirac spinors. The spin average rate is
given by
\begin{equation}
\sum_s | \CM|^2 = 4 f^2 \sin^2 \chi( p\cdot p' - m_w^2).  
\end{equation}
The
partial $H$-decay rate into $2w$ is
\begin{eqnarray}
d\Gamma_{H \to 2 w} &=& \frac{|\CM|^2}{2 m_H}\frac{d^3 p'}{(2\pi)^3 2 E_{p'}} \frac{d^3 p}{(2\pi)^3 2 E_p} (2\pi)^4 \delta^{(3)}(\textbf{p}'+\textbf{p}) \delta(m_H - p' - p) , \nonumber
\\
&=& \frac{1}{2!} \frac{d\Omega}{64 \pi^2 m_H^2} \sqrt{m_H^2-4m_w^2} |\CM|_{\textbf{p}'=-\textbf{p}, \ p = \sqrt{(m_H/2)^2-m_w^2}}^2 \, ,
\end{eqnarray}
and so the partial width for this decay is given by
\beq
\Gamma_{H \to 2 w} = \frac{ 2( m_H^2- 4 m_w^2)  }{32 \pi m_H^2} \left
  ( \frac{f \eta_\chi v_r v_\phi}{m_H^2-m_h^2} \right)^2
\sqrt{m_H^2-4m_w^2} \, .
\label{eq:fiftyone}
\eeq
For  $m_H \gg 2 m_w$,  (\ref{eq:fiftyone}) becomes 
\beq
\Gamma_{H \rightarrow 2 w} = \frac{1}{16 \pi} \left( \frac{f
    \eta_\chi v_r   v_\phi }{m_H^2 - m_h^2}\right)^2
\sqrt{m_H^2-4m_w^2}.
\label{godoy}
\eeq

\subsubsection{$\underline{\bf{\Gamma_{H\to \ {\rm hidden}}}}$}

All in all, the decay width of the Higgs into the
hidden sector  is given by \beq \Gamma_{H
  \rightarrow {\rm hidden}}  =  \frac{1}{16 \pi}
\left(\frac{\eta_\chi \, v_\phi }{m_H^2-m_h^2}\right)^2 m_H^3 +
\frac{1}{16 \pi} \left( \frac{f \, \eta_\chi \ v_r \
 \ v_\phi }{m_H^2 - m_h^2}\right)^2
\sqrt{m_H^2-4m_w^2}.
\label{eq:invis}
\eeq
Assuming $m_H \gg m_h$, this decay width is 
\beq
\Gamma_{H \rightarrow {\rm hidden}}  = \frac{\eta_\chi^2
  v_\phi^2}{16 \pi m_H } + \frac{\eta_\chi^2 \Delta m^2 v_\phi^2}{64
  \pi m_H^3} .
\label{eq:fifteen-1}
\eeq 
Comparing (\ref{LHCwidth}) and  (\ref{eq:fifteen-1}) we obtain
\begin{equation}
|\Delta m| >  2 m_H \ \sqrt{ \frac{8.3 \times 10^{-5}}{\eta_\chi^2 } - 1} \, ,
 \label{eq:DeltaMlim}
\end{equation}
which is satisfied if $|\eta_\chi| < 0.009$.

\subsection{Constraints from Direct Detection Experiments}

Direct detection experiments attempt to observe the recoil from the
elastic scattering of dark matter particles interacting with nuclei in
the detector. Since the late 90's the \mbox{DAMA/NaI}
Collaboration~\cite{Bernabei:1998fta} has been claiming to observe the
expected annual modulation of the dark matter induced nuclear recoil
rate due to the rotation of the Earth around the
Sun~\cite{Drukier:1986tm, Freese:1987wu}. The upgraded DAMA/LIBRA
detector confirmed~\cite{Bernabei:2008yi} the earlier result adding
many more statistics, and it has reached a significance of $8.9\sigma$
for the cumulative exposure~\cite{Bernabei:2010mq}.  In 2010, the
CoGeNT Collaboration reported an irreducible excess in the counting
rate~\cite{Aalseth:2010vx}, which may also be ascribed to a dark
matter signal. One year later, the same collaboration reported further
data analyses showing that the time-series of their rate is actually
compatible with an annual modulation effect~\cite{Aalseth:2011wp}. In
CoGeNT data the evidence for the annual modulation is at the
$2.8\sigma$ level. In the summer of 2011, the CRESST Collaboration
also reported an excess of low energy events that are not consistent
with known backgrounds~\cite{Angloher:2011uu}. In particular, 67
counts were found in the dark matter acceptance region and the
estimated background from leakage of $e/\gamma$ events, neutrons,
$\alpha$ particles, and recoiling nuclei in $\alpha$ decays is not
sufficient to account for all the observed events. The CRESST
Collaboration rejected the background-only hypothesis at more than $4
\sigma$.  Of particular interest here, the DAMA (after including
the effect of channeling in the NaI crystal
scintillators~\cite{Petriello:2008jj}) and CoGeNT results appear to be
compatible with a relatively light dark matter particle, in the few
GeV to tens of GeV mass range, with a scattering cross section against
nucleons of about $7 \times 10^{-41}~{\rm
  cm}^2$~\cite{Fitzpatrick:2010em,Chang:2010yk,Hooper:2010uy,Buckley:2010ve}.
The central value favored by CRESST data points to somewhat larger
dark matter masses, but it is still compatible at the $1\sigma$ level
with the range determined by the other two experiments.

Very recently, CDMS II Collaboration reported three candidate events
with an expected background of 0.7 events~\cite{Agnese:2013rvf}.  If interpreted as a signal
of elastically scattering dark matter, the central value of the
likelihood analysis of the measured recoil energies favors a mass of
$8.6~{\rm GeV}$ and a scattering cross section on nucleons of 
\begin{equation}
\sigma_{w N}^{m_w \approx 10~{\rm GeV}} \approx 1.9
\times 10^{-41}~{\rm cm}^{2} \, . 
\label{CDMS}
\end{equation}
The 68\% confidence band is somewhat large and overlaps with previous signal
claims.

Alongside these ``signals'' stands the series of null results from the
XENON-100~\cite{Aprile:2011hi} and XENON-10~\cite{Angle:2011th}
experiments, which at present have the world's strongest exclusion
limit.  Some authors have pointed out that uncertainties in the
response of liquid xenon to low energy nuclear recoil may be
significant, particularly in the mass region of
interest~\cite{Collar:2010ht,Collar:2011wq}. In light of these
suspicions, a recent re-analysis of XENON data suggests candidates in
fact may have been observed~\cite{Hooper:2013cwa}. The data favor a
mass of 12~GeV, though the 90\% error contours extend from 7 to 30 GeV
with the cross section varying between $6\times 10^{-41}~{\rm cm}^2$
and $4 \times 10^{-45}~{\rm cm}^2$. Taken together, these
different arguments suggest that the existing data set is not
inconsistent with a dark matter candidate of about 10~GeV.

The $wN$ cross section for elastic scattering is given by
\begin{equation}
\sigma_{wN} = \frac{4}{\pi} \frac{m_w^2 m_N^2}{(m_w + m_N)^2} \ \frac{f_p^2 + f_n^2 }{2} \,,
\label{sigmawN}
\end{equation}
where $N \equiv \frac{1}{2} (n+p)$ is an isoscalar nucleon, in the
renormalization group-improved parton
model~\cite{Ellis:2000ds,Beltran:2008xg}. The effective couplings to
protons $f_p$ and neutrons $f_n$ are given by \beq f_{p,n} = \sum_{q =
  u,d,s} \frac{G_q}{\sqrt{2}} f_{Tq}^{(p,n)} \frac{m_{p,n}}{m_q} +
\frac{2}{27} f_{TG}^{(p,n)} \sum_{q = c,b,t} \frac{G_q}{\sqrt{2}}
\frac{m_{p,n}}{m_q},
\label{effective-coup}
\eeq where $G_q$ is the $W$-WIMP's effective Fermi coupling for a
given quark species, \beq \mathscr{L} = \frac{G_{q}}{\sqrt{2}} \bar
\psi_- \psi_- \bar \psi_q \psi_q \,, \eeq with $\psi_q$ the SM quark
field of flavor $q$. The first term in (\ref{effective-coup}) reflects
scattering with light quarks, whereas the second term accounts for
interaction with gluons through a heavy quark loop. The terms
$f_{Tq}^{(p,n)}$ are proportional to the matrix element, $\la \bar q q \ra$, of quarks in a nucleon,
and are given by \beqa
f^{p}_{Tu} = 0.020 \pm 0.004, \quad \ f^{p}_{Td} = 0.026 \pm 0.005, \quad f^{p}_{Ts} = 0.118 \pm 0.062, \nonumber  \\
f^{n}_{Tu} = 0.014 \pm 0.003, \quad f^{n}_{Td} = 0.036 \pm 0.008,
\quad f^{n}_{Ts} = 0.118 \pm 0.062 \, .  \eeqa We also have
$f^{(p,n)}_{TG} = 1 - \sum_{u,d,s} f^{(p,n)}_{Tq}$, which is  $f^{p}_{TG} \approx 0.84$ and $f^{n}_{TG} \approx 0.83$~\cite{Ellis:2000ds}.

To establish the value of $G_{q}/m_q$ we look back at
 (\ref{eq:fifty})  along with the SM Yukawa interaction term, which involves the mixing
of both scalar fields, $H$ and $h$. For interactions of $W$-WIMPs with SM quarks, the
relevant terms are \beq \mathscr{L} = \frac{m_q \cos \chi}{v_\phi} H \bar
\psi_q \psi_q - \frac{m_q \sin \chi}{v_\phi} h \bar \psi_q \psi_q +
\dots +\frac{f \sin \chi}{2} H \bar \psi_- \psi_- + \frac{f \cos
  \chi}{2} h \bar \psi_- \psi_-.  \eeq The scattering of a $w$
particle off a quark then gives \beqa \CM &=& i \frac{f m_q \sin \chi
  \cos \chi}{v_\phi} \bar u_q (p') u_q(p) \lpa \frac{1}{t-m_H^2} -
\frac{1}{t-m_h^2} \rpa \bar u(k') u(k)  \nonumber
\\
&\approx& i \frac{f m_q \eta_\chi v_r }{m_H^2 m_h^2} \bar u_q (p')
u_q(p) \bar u(k') u(k)  \nonumber
\\
&\approx& i \frac{m_q \eta_\chi \Delta m }{2 m_H^2 m_h^2} \bar u_q
(p') u_q(p) \bar u(k') u(k)  .  \eeqa This leads to the identification
of the effective coupling \beq  \frac{2 G_{q}}{\sqrt{2}} = \frac{m_q \eta_\chi
  \Delta m }{2 m_H^2 m_h^2} \Rightarrow \frac{G_{q}}{m_q} =
\frac{\eta_\chi \Delta m }{2\sqrt{2} \ m_H^2 m_h^2}.
\label{eq:effectiveCoup}
\eeq 
Insertion of   (\ref{eq:effectiveCoup}) and (\ref{effective-coup}) into
(\ref{sigmawN}) yields \beq
\sigma_{wN} \approx  3 \times 10^{-7} \,  \left[
  \frac{226.27 \ \eta_\chi \Delta m \ {\rm GeV}}{m_h^2} \right]^2~{\rm
  pb} \, .
\label{NOB}
\eeq
 
Combining (\ref{NOB}) with the signals/bounds on elastic scattering of dark matter particles on nucleons we obtain a constraining relation for $\eta_\chi \Delta m$. For $m_w = 10~{\rm GeV}$, we use the cross section reported by CDMS Collaboration (\ref{CDMS}) to obtain 
\beq
\eta_\chi \Delta m = \frac{3.5 \times 10^{-2}}{\rm GeV} \   m_h^2    \, .
 \label{eq:DetAcdms}
\eeq 
For $m_w = 50~{\rm GeV}$, we adopt the 90\% CL upper limit reported by
the XENON-100 Collaboration~\cite{Aprile:2011hi} to obtain 
\begin{equation}
   \eta_\chi \Delta m <  \frac{3.6 \times 10^{-4}}{\rm GeV} \   m_h^2
   \, .
 \label{eq:DetAxenon}
\end{equation} 

\subsection{Constraints from Cosmological Observations}

 As noted in~\cite{Weinberg:2013kea} the Goldstone boson $\alpha$ is a
 natural candidate for an impostor equivalent neutrino. The
 contribution of $\alpha$ to $N_{\rm eff}$ is $\Delta N =
 \rho_\alpha/\rho_{\nu_L}$,  which can also be expressed as
\begin{equation}
\Delta N = \frac{4}{7} \left( \frac{g(T^{\rm dec}_{\nu_L})}{g(T^{\rm dec}_\alpha)} \right)^{4/3} \,,
\end{equation}
where $g(T) = g_S (T) \approx g_\rho(T)$ is the effective number of interacting (thermally
coupled) relativistic degrees of  freedom at temperature $T$~\cite{Kolb:1990vq}.

We now turn to calculating the interaction rate for Goldstone bosons,
\beq
\Gamma  (T) = \sum_{\rm fermions} n_{\rm f}(T) \langle \sigma v \rangle \,,
\label{eq:P3twenty}
\eeq where 
\beq
n_{\rm f}(T) = \frac{ g_{\rm f} }{2 \pi^2} \int_0^\infty \frac{k^2}{e^{\beta \sqrt{k^2+m_{\rm f}^2}}+1} dk
\label{eq:twentyone}
\eeq is the number density of an interacting fermion of type f (with
mass $m_{\rm f}$) in thermal equilibrium with, $\beta = (k_B T)^{-1}$, and $g_{\rm f}$, the number
of chiral states.  The average in (\ref{eq:P3twenty}), indicates an average over the
statistical distribution for a given temperature. For $T \gg m_{\rm
  f}$, we obtain \beq n_{\rm f}(T) \approx g_{\rm f}
\frac{3 \zeta(3)}{4 \pi^2} \left( \frac{k_B T}{\hbar c} \right)^3.
\label{eq:twentytwo}
\eeq
This results in a simplification of  (\ref{eq:P3twenty}) 
\beq
\Gamma (T) \approx   \frac{3 \zeta(3)}{4 \pi^2} \left( \frac{k_B T}{\hbar c} \right)^3 \sum_{\rm fermions} g_{\rm f} \langle \sigma v \rangle.
\label{eq:twentythree}
\eeq

Since the Goldstone boson only interacts with the SM fields via the
Higgs, we can have scatterings of the type $\alpha \psi \to \alpha \psi$,
with $\psi$ a generic SM fermion. The $\alpha$ scattering off fermions is 
described by SM Yukawa interaction terms that can be written  as
\beqa Y_{\rm f} \phi \bar{\psi} \psi & \rightarrow & Y_{\rm f} v_\phi \bar{\psi} \psi +
Y_{\rm f} \phi' \bar{\psi} \psi , \nonumber
\\
&=& m_{\rm f} \bar{\psi} \psi + \frac{m_{\rm f}}{v_\phi}  H \bar{\psi}
\psi  \, \cos \chi - \frac{m_{\rm f}}{v_\phi} h \bar{\psi}
\psi  \, \sin \chi \,,
\label{eq:fourteen}
\eeqa where $Y_{\rm f}$ is the Yukawa coupling  of the fermion in question. 

We proceed to calculate the scattering cross section.  The invariant
amplitude follows from the Feynman rules
\beqa
i \CM &=& \frac{2  m_f \sin \chi \cos \chi}{v_r v_\phi} (k\cdot k') \frac{i}{t
  - m_H^2} \bar{u}(p') u(p) \nonumber \\ 
  &-&\frac{2 m_f \cos \chi \sin \chi}{v_r v_\phi} (k\cdot k') \frac{i}{t
  - m_h^2} \bar{u}(p') u(p)  . 
\eeqa
The momenta  of   incoming
and outgoing (outgoing primed) particles are defined by
\beqa
p^\mu &=& \left(p, p \sin \varphi, 0, -p \cos \varphi \right) \nonumber
\\
k^\mu &=& \left(k, 0, 0, k \right) \nonumber
\\
k'^\mu &=& \left(k', k' \sin \vartheta, 0, k' \cos\vartheta \right) \nonumber
\\
p'^\mu &=& \left(p', -p' \sin \vartheta', 0, - p' \cos \vartheta' \right) \,,
\label{eq:twentysix}
\eeqa 
with  $t = p'-p$.  To obtain the (unpolarized) cross section, we have
to take the square of the modulus of ${\cal M}$ and then carry out the
spin and color (if appropriate) sums
\begin{equation}
\frac{1}{2} \sum_{\rm spins,\;  colors} |\CM|^2 = 8 N_c \lpa \frac{m_f \sin \chi \cos \chi}{v_r v_\phi} \rpa^2 \lpa \frac{m_H^2-m_h^2}{(t-m_H^2)(t-m_h^2)} \rpa^2 (k\cdot k')^2 (p\cdot p' + m_f^2) ,
\label{eq:seventeenW}
\end{equation}
where $N_c =3$ for quarks and $N_c =1$ for leptons. 
The cross section in the center-of-mass frame (c.m.) in the highly relativistic approximation is given by
\beq
\frac{d\sigma}{d \Omega} \approx \frac{N_c}{8 \pi^2 s}\lpa \frac{m_f \eta_\chi}{(t-m_H^2)(t-m_h^2)} \rpa^2 (k\cdot k')^2 (p\cdot p') \,,
\label{eq:seventeen}
\eeq where $s = (k + p)^2 \approx 4 k^2$ and finally $\eta_\chi \ll
1$. To make progress on this problem we take the effective coupling
form \beq \sigma (s) \approx \frac{N_c}{64 \pi} \lpa \frac{m_f
  \eta_\chi}{m_H^2 m_h^2} \rpa^2 s^2 \, . \eeq

Non-equilibrium thermal physics tells us that the way to do thermal
averaging within Boltzmann's  approximation is
\begin{eqnarray}
\la \sigma v \ra & = & \int d\Pi_{p'} d\Pi_{k'} d\Pi_{k} d\Pi_{p}
|\CM(k+p\rightarrow k' + p')|^2 f_{\rm f}(p, T) f_\alpha(k, T) \nonumber \\
& \times & (2\pi)^4 \delta^{(4)} (p+k-p'-k'),
\label{assman}
\end{eqnarray} 
with $d\Pi_{p} = d^3 p'/[(2\pi)^3 2 E_{p'}]$ and likewise for the
other parameters.  Here, $f_{\rm f}$ and $f_\alpha$ are Fermi and Bose
equilibrium normalized distributions, corresponding to the ${\rm f}$
fermion and $\alpha$ boson, respectively.  The expression from
non-equilibrium thermal physics [Eq.~(\ref{assman})] is approximated
by \beq \la \sigma v \ra \approx \int \frac{d^3 k}{(2\pi)^3}\frac{d^3
  p}{(2\pi)^3} f_{\rm f}(p, T) f_\alpha(k, T) \ v_M \ \sigma(s) \, ,
\eeq where $v_M \approx k \cdot p/(pk) = 2 (1+\cos \varphi)$ is the
M\"oller velocity in the ultra-relativistic
limit~\cite{Gondolo:1990dk,Weiler:2013hh} and $s = 2 k p (1+\cos
\varphi)$ is the c.m. energy of two interacting particles with initial
momenta not necessarily co-linear.  The velocity average cross section
then is found to be \beqa \la \sigma v \ra &\approx& \frac{1}{8 \pi^4}
\int_0^\infty p^2 dp \int_0^\infty k^2 dk \int_0^\pi \sin \varphi d
\varphi \ f_{\rm f}(p,T) f_\alpha(k,T) \nonumber \\ & \times & 2 (1+
\cos\varphi) \ \, \sigma_{\rm c.m.} [2 k p (1+\cos \varphi)],
\nonumber
\\
&=& N_c \frac{15 \zeta^2(5)}{\pi \zeta^2(3)} \lpa \frac{m_f
  \eta_\chi}{m_H^2 m_h^2} \rpa^2 (k_B T)^4, \nonumber
\\
&\approx& 3.55 N_c \lpa \frac{m_f \eta_\chi}{m_H^2 m_h^2} \rpa^2 (k_B
T)^4.  \eeqa Putting this all together, we obtain \beq \Gamma (T)
\approx 0.32 \lpa \frac{\eta_\chi}{m_H^2 m_h^2} \rpa^2 (k_B T)^7
\sum_{\rm fermions} g_{\rm f} \, N_{\rm c} \, m_{\rm f}^2.  \eeq

Now, since we can approximate the energy density (at high temperatures) by
including only particles species $i$ with $T \gg m_i$, it follows that
\begin{equation}
\rho_{\rm R} = \left(\sum_{\rm bosons} g_{\rm b} + \frac{7}{8} \sum_{\rm fermions} g_{\rm f} \right) \frac{\pi^2}{30} (k_BT)^4 = \frac{\pi^2}{30} g(T) (k_BT)^4
\end{equation}
and therefore the Hubble parameter (\ref{Hubble}) becomes
\begin{equation}
  H(T)  \simeq \frac{1.66}{M_{\rm Pl}}
  \sqrt{g(T)} \ (k_B T)^2  \, ,
\end{equation}
where $g_{\rm b  (f)}$ is the number of degrees of freedom of each boson (fermion) and the sum runs over all boson and fermion states with $T
\gg m_i$. The factor of 7/8 is due to the difference between the Fermi
and Bose integrals.

The Goldstone boson decouples from the plasma when its mean free path
becomes greater than the Hubble radius at that time
\begin{equation}
\Gamma (T_\alpha^{\rm dec}) =  H(T_\alpha^{\rm dec}) \, .
\end{equation}
The most interesting thermodynamics originates if $\alpha$ goes out of
thermal equilibrium while $T$ is still above the mass of the
muons but below the mass of all other particles of the SM, a time when
neutrinos are still in thermal equilibrium. For instance, with $\eta_\chi = 0.005$ and $m_h \approx 500~{\rm MeV}$ we obtain~\cite{Weinberg:2013kea}
\begin{equation}
\Delta N
= (4/7) (43/57)^{4/3} = 0.39 \, .
\end{equation}
This corresponds to a number of equivalent light neutrino species that
is consistent at the $1\sigma$ level with both the estimate of
$N_{\rm eff}$ using Planck + BAO data as well as the estimate using Planck +
$H_0$ data.

However, of particular interest here is the case where the mass of the
Goldstone boson companion field is $m_h \approx 98~{\rm GeV}$ and
$\eta_\chi = 0.0003$. For such set of parameters, $\alpha$ decouples
when 
\beq 0.32 \lpa \frac{\eta_\chi}{m_H^2 m_h^2} \rpa^2 \lpa k_B T
\rpa^7  12 \ m_b^2 = \frac{1.66}{M_{\rm
    Pl}}\sqrt{86.25}\lpa k_B T \rpa^2,
\label{eq:Nfixer}
\eeq 
where we have approximated $\sum_{\rm fermions} N_c \ g_{\rm f} 
\ m_{\rm f}^2 \approx 12 m_b^2$.  This gives $T \approx 5~{\rm GeV}$, and so the
$\alpha$ contribution to $N_{\rm eff}$ is found to be \beq \Delta N
\approx 0.036. \eeq The corresponding value of $N_{\rm eff}$ is within
the $1\sigma$ interval of the value reported by the Planck
Collaboration using Planck + BAO data, but far out from the value
derived using Planck + $H_0$ data. Should future data point towards
the Planck + $H_0$ value, one should find a different origin to
explain the extra relativistic degrees of freedom (if $m_h \approx
98~{\rm GeV}$). One interesting possibility is to include the
right-handed partners of the three left-handed, SM neutrinos. It was
shown
elsewhere~\cite{Anchordoqui:2011nh,luisnMe1,SolagurenBeascoa:2012cz,Anchordoqui:2012qu,Anchordoqui1}
that milli-weak interactions of these Dirac states (through their
coupling to a TeV-scale $Z'$ gauge boson) may allow the $\nu_R$'s to
decouple during the course of the quark-hadron crossover transition,
just so that they are partially reheated compared to the $\nu_L$'s.
Remarkably, the required mass for the $Z'$ gauge boson is within the
range of discovery of LHC.

\subsection{Fitting Fermi data and the Observed Dark Matter Density}

Next, in line with our stated plan, we use \emph{Fermi} data and the
observed relic density to determine the free parameters of the
model. To this end we first calculate the annihilation rate into SM
fermions and Goldstone bosons.

\subsection{$\bf{W}$-WIMP Annihilation into SM Fermions}

The $W$-WIMP can annihilate into SM fermions  via $\bar{\psi}_- \psi_-
\rightarrow \phi^*/r^* \rightarrow \bar{\psi}{\psi}$, with an $s$-channel
Higgs or $h$ mediator. The matrix element of this process is
given by \begin{equation} i \CM = i f \sin \chi \, \cos\chi \
  \bar{v}(p') u(p) \left(\frac{i }{s - m_H^2} -
    \frac{i}{s-m_h^2}\right) \frac{i m_{\rm f}}{v_\phi} \bar{u}(k')
  v(k) \, . \end{equation} The minus sign in the second propagator
is necessary because the $r$ couples with a negative sign to fermions compared
to the Higgs; see~ (\ref{eq:fourteen}). The spin-averaged invariant
amplitude reads \begin{equation} \frac{1}{4}\sum|\CM|^2 = 
  N_c \left( \frac{f m_{\rm f} \sin \chi \, \cos \chi}{v_\phi}
  \right)^2\frac{4 \, \left( m_h^2 - m_H^2 \right)^2
     (p \cdot p'-m_w^2)(k \cdot k' - m_{\rm f}^2)}{(s-m_h^2)^2(s-m_H^2)^2} \, .
\end{equation}
Now, let's calculate the cross section for ${\rm f} \bar {\rm f}$-pair production
\beqa
d\sigma  &=& \frac{1}{8 E_p |p| } |\CM|^2 \frac{d^3 k}{(2\pi)^3 2 E_k} \frac{d^3 k'}{(2\pi)^3 2 E_{k'}} (2\pi)^4 \delta^{(3)}(\textbf{k}'+\textbf{k}) \nonumber \\
& \times & \delta(2 E_p - E_k - E_{k'}), 
\end{eqnarray}
and so 
\begin{eqnarray}
  \sigma &=& \frac{|\CM|^2}{64 \pi } \frac{|k'|}{|p| E_p^2} \nonumber
  \\
  &=&  \frac{N_c}{16 \pi}  \, \left( \frac{f m_{\rm f} \sin \chi \ \cos \chi}{v_\phi} \right)^2\frac{|k'|}{|p|} \frac{ \left( m_h^2 - m_H^2\right)^2}{(s-m_h^2)^2(s-m_H^2)^2}  \frac{\lpa p \cdot p' -m_w^2 \rpa \lpa k \cdot k' - m_{\rm f}^2 \rpa}{ E_p^2}  \nonumber
  \\
  &\approx&   \frac{N_c}{16 \pi} \left( \frac{\eta_\chi m_{\rm f}  \Delta m }{2(s-m_h^2)(s-m_H^2)} \right)^2 \sqrt{\frac{|s-4m_{\rm f}^2|}{|s-4m_w^2|}} \frac{(s-4m_w^2)(s-4m_{\rm f}^2)}{s}\, .
\label{CDMx}
\end{eqnarray}  
In this case the out
state does not consist of identical particles. For phenomenological purposes, the $h$ pole needs to be softened to a Breit-Wigner form by obtaining and utilizing the correct total widths
$\Gamma_h$ of the resonance.  This is accomplished by modification of the $s$-channel propagator for $h$ via
\beq
\frac{i}{s -m_h^2} \rightarrow \frac{i}{s- m_h^2- i m_h\Gamma_h} \, .
\eeq
It should be noted that we could also do the same analysis on the $H$ pole which may also have some phenomenological interest as an independent analysis of the \emph{Fermi} bubbles resulted in a best fit annihilation of dark matter of a particle with mass $m_w \approx 61$ GeV into $b \bar b$~\cite{AUrbano}.  
After this is done, the contribution of the ${\rm f} \bar {\rm f}$ channel is as follows:  
\beqa \sigma &=& \frac{N_c}{16 \pi} \left( \frac{\eta_\chi
    m_{\rm f} \Delta m }{2 (m_H^2-m_h^2) (s-m_H^2)} \right)^2 \frac{
  \left( m_H^2 - m_h^2\right)^2 + m_h^2 \Gamma_h^2}{(s-m_h^2)^2 +
  m_h^2 \Gamma_h^2}  \sqrt{\frac{|s-4m_{\rm f}^2|}{|s-4m_w^2|}}
\nonumber \\
& \times & \frac{(s-4m_w^2)(s-4m_{\rm f}^2)}{s} \ , \nonumber \\
&\approx& \frac{N_c}{16 \pi} \left( \frac{\eta_\chi
    m_{\rm f} \Delta m }{2  (s-m_H^2)} \right)^2 \frac{
  1 }{(s-m_h^2)^2 +
  m_h^2 \Gamma_h^2}  \sqrt{\frac{|s-4m_{\rm f}^2|}{|s-4m_w^2|}}
\frac{(s-4m_w^2)(s-4m_{\rm f}^2)}{s}  . \nonumber \\
\eeqa 
For $\Delta m > m_H/2$, the decay channels of the $h$ field are:
$h \rightarrow {\rm f} \bar {\rm f}$, $h \rightarrow w \bar w$, and $h
\rightarrow 2 \alpha'$. The corresponding decay widths are given by
\beqa \Gamma_{h\rightarrow {\rm f} \bar {\rm f} } &=& \sum_{\rm
  fermions} \frac{N_c}{8 \pi m_h^2} \lpa \frac{m_{\rm f} \sin \chi}{v_\phi} \rpa^2 (m_h^2-4 m_{\rm f}^2)^{3/2} \nonumber
\\
&\approx& \sum_{\rm fermions}  \frac{N_c}{8 \pi m_h^2} \lpa \frac{m_{\rm f} \eta_\chi v_r}{m_H^2-m_h^2} \rpa^2 (m_h^2-4 m_{\rm f}^2)^{3/2} \nonumber
\\
&\approx& \sum_{\rm fermions}  \frac{N_c}{8 \pi m_h^2 f^2} \lpa \frac{m_{\rm f} \eta_\chi \Delta m}{2(m_H^2-m_h^2)} \rpa^2 (m_h^2-4 m_{\rm f}^2)^{3/2} \nonumber
\\
&\approx& \frac{3}{8 \pi m_h^2 f^2} \lpa \frac{m_{\rm b} \eta_\chi \Delta m}{2(m_H^2-m_h^2)} \rpa^2 (m_h^2-4 m_{\rm f}^2)^{3/2}
\eeqa (in the last line we have taken
$m_b < m_w < m_t$), \beqa \Gamma_{h\rightarrow w \bar w} &=& \frac{2
  \, ( m_h^2- 4 m_w^2) \, f^2 \cos^2 \chi }{32 \pi m_h^2}
\sqrt{m_h^2-4m_w^2} \nonumber
\\
&\approx& \frac{f^2}{16 \pi m_h^2} ( m_h^2- 4 m_w^2)^{3/2} \eeqa
(inclusion of this channel requires $2 m_w < m_h$), and \beqa
\Gamma_{h\rightarrow 2 \alpha'} &=& \frac{1}{32 \pi}\lpa \frac{\cos
  \chi}{v_r} \rpa^2m_h^3 \nonumber
\\
&\approx& \frac{f^2 }{8 \pi \Delta m^2} m_h^3 \, .  \eeqa The dominant 
terms of the total decay width come from the hidden sector. Hence, in what
follows we  neglect  terms accounting for $h$ decay into the visible
sector and consider $m_h < 2 m_w$  (so that the decay $h \rightarrow w 
\bar w $ is closed). Under these assumptions  the decay width takes a particularly simple form
\beq \Gamma_h = \frac{f^2 }{8 \pi \Delta m^2} m_h^3  \, .
\label{finalform}
\eeq

Next, we compute the averaged cross section for thermal interactions.
In the cosmic comoving frame (the frame where the gas is assumed to be
at rest as a whole) we have \beq \la \sigma v \ra = \frac{\int d^3 p
  d^3 p' f_w(p, T) f_w(p', T) \, \sigma v_M}{\int d^3 p d^3 p' f_w(p,
  T) f_w(p', T) }, 
\label{eq:ThermAve}
\eeq where $\bf{p}$ and $\bf{p}'$ are the
three-momenta of the colliding particles, whose equilibrium
distribution function at temperature $T$ is Maxwell-Boltzmann, \beq
f_w(p,T) \approx e^{- \beta \sqrt{p^2 + m_w^2}} \, , \eeq with $p =
|\bf{p}|$ and $p' = |\bf{p}'|$. The Maxwell-Boltzmann distribution
remains a good approximation
provided $3 \, m_w \, \beta > 1 $. The M\"oller velocity can be expressed as
\beq v_M = \frac{1}{E E'} \sqrt{(p\cdot p')^2 - m_w^4} = \frac{1}{2 E
  E'} \sqrt{s (s-4m_w^2)} \, ,
\label{nonrelmoller}
\eeq where $E$ and $E'$ are the energies of the scattering
particles. Note that in the c.m. frame the velocity of the colliding
$W$-WIMPs is half the M\"oller velocity, $v = \sqrt{1-4m_w^2/s} =
v_M/2$.

For $s \gg m_{\rm f}$, from (\ref{CDMx}) and (\ref{nonrelmoller}) we
obtain
\begin{equation}
\sigma v_M = \frac{N_c}{8 \pi} \left( \frac{\eta_\chi m_{\rm f}  \Delta m }{2(s-m_h^2)(s-m_H^2)} \right)^2 \,  (s - 4 m_w^2) \, .
\end{equation}
We evaluate  (\ref{eq:ThermAve}) by expanding $\sigma v_M$ around \beq
s = 4 E^2 = \frac{4 m_w^2}{1-v^2} \approx 4m_w^2 (1 + v^2 + \dots) \
 \eeq to obtain a series solution in powers of $v$ of which the
leading order term is \beq \la \sigma v \ra \approx \frac{N_c}{2\pi}
\lpa \frac{ \eta_\chi m_{\rm f} m_w \Delta
  m}{2(4m_w^2-m_h^2)(4m_w^2-m_H^2)} \rpa^2 \la v^2 \ra \,,\eeq where
$\la v^2 \ra$ is the $W$-WIMP thermally averaged velocity.  

All in all, the total average annihilation cross section
into SM particles (labelled by subindex $i$) is given by \beq
\sum_{\rm fermions} \la \sigma_i v \ra \approx \frac{ 3 }{ 2 \pi }
\left( \frac{\eta_\chi m_b m_w \Delta m }{2 (4 m_w^2-m_H^2)} \right)^2
\, \frac{\la v^2 \ra}{(4 m_w^2-m_h^2)^2 + m_h^2 \Gamma_h^2} \, ,
\label{eq:TEWII}
\eeq 
where we have assumed that the overwhelming contribution into $b \bar b$ dominates the process.

Provided the theory is not strongly coupled,  (\ref{eq:TEWII}) is
generally a good approximation for relativistic particles, but for low
velocities and in the presence of a long-range force (classically,
when the potential energy due to the long-range force is comparable to
the particles' kinetic energy), the perturbative approach breaks
down. In the non-relativistic limit, the question of how the long-range
potential modifies the cross section for short-range interactions can
be formulated as a scattering problem in quantum mechanics, with
significant modifications to the cross sections occurring when the
particle wavefunctions are no longer well approximated by plane waves
(so the Born expansion is not well-behaved). The deformation of the
wavefunctions due to a Coulomb potential was calculated by
Sommerfeld~\cite{Sommerfeld:1931}, yielding a $\sim 1/v$ enhancement
to the cross section for short-range interactions (where the
long-range behavior due to the potential can be factorized from the
relevant short-range behavior). Along these lines, for low-velocity
($v \sim 10^{-3}$) $W$-WIMPs in our Galactic halo, we expect
interactions with the $H$ and $h$ fields to enlarge the cross section,
as the attractive Yukawa potential \beq V_w(r) = -\frac{f^2
  \cos^2(\chi)}{4\pi} \frac{e^{-m_h r}}{r} - \frac{f^2
  \sin^2(\chi)}{4\pi} \frac{e^{-m_H r}}{r} \approx -\frac{f^2}{4\pi}
\frac{e^{-m_h r}}{r} \simeq -\frac{f^2}{4\pi} \frac{1}{r} \ .  \eeq
cause passing $W$-WIMPS to be drawn toward each
other~\cite{ArkaniHamed:2008qn}.
 %%%%%%%%%%%%%%%%%%%%%%%
\begin{figure}[tpb]
\begin{center}
\includegraphics[width=0.7\textwidth]{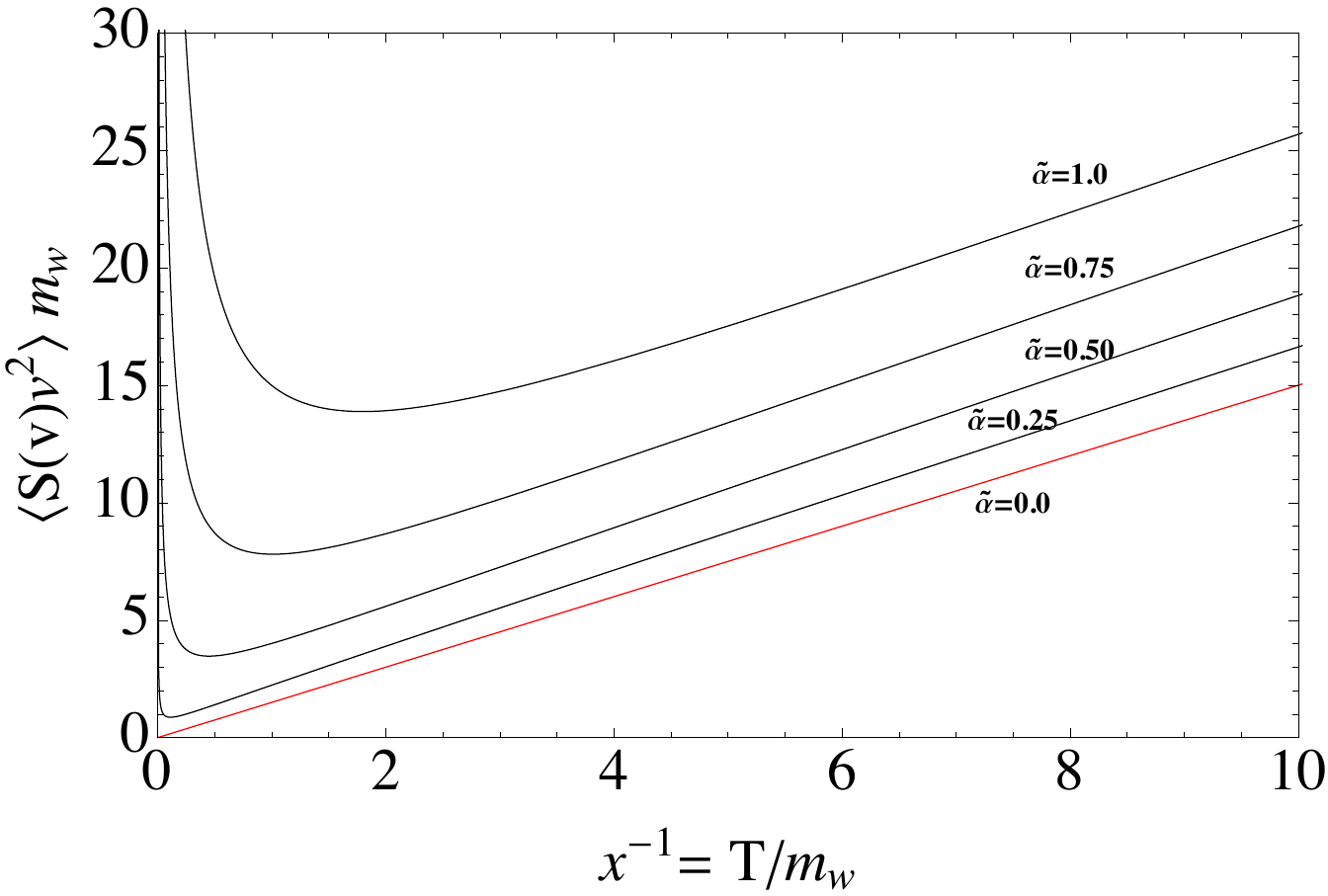}
\caption[Sommerfeld Enhancements Give Large Effect a low T]{ \sglspc \small Sommerfeld enhancements allow the thermally averaged cross section to increase at low temperatures T. The figure shows varying values of $\tilde \alpha$ and the red curve is the result with no Sommerfeld enhancement.}
\end{center}
\label{fig:P3Som}
\end{figure} 
%%%%%%%%%%%%%%%%%%%%%%%%%
For $p$-wave scattering, $ \la v^2
\ra \rightarrow \la S (v) v^2 \ra ,$ where
 \begin{equation}
S (v)  \approx   \frac{\pi \tilde{\alpha}}{v}\frac{1}{1-e^{- \pi
    \tilde{\alpha}/v}} \lpa 1+ \frac{\pi^2 \tilde{\alpha}^2}{4 v^2}
\rpa \ ,
 \label{eq:SomS1}
 \end{equation}
 is the Sommerfeld enhancement factor in the Coloumb approximation,
 with $\tilde \alpha = f^2/(4\pi)$~\cite{Cassel:2009wt}.
 Following~\cite{Feng:2010zp} we compute the thermally averaged
 Sommerfeld enhancement factor by approximating  $\lpa 1-e^{- \pi
   \tilde{\alpha}/v} \rpa^{-1}$  with $\tilde{\alpha} \ll 1$ \beq
 \la S (v) v^2 \ra \approx 6 x^{-1} + 4 \sqrt{\pi} \tilde{\alpha}
 x^{-1/2} + \frac{4 \pi^2 \tilde{\alpha}^2}{3} + \pi^{5/2}
 \tilde{\alpha}^3 x^{1/2} +\frac{\pi^4 \tilde{\alpha}^4}{6} x, \eeq
 where $x = m_w/T$.  Figure \ref{fig:P3Som} shows the effect of the Sommerfeld enhancements at low temperatures. For interactions in the Galactic halo (G.h.), we have
 $\la v^2 \ra \sim 10^{-6}$, and therefore the thermally average
 annihilation cross section into $b \bar b$ becomes \beqa \la \sigma_b
 v \ra & \approx & \frac{ 3 }{ 2 \pi } \left( \frac{\eta_\chi m_b m_w
     \Delta m }{2 (4 m_w^2-m_H^2)} \right)^2 \, \frac{1}{(4
   m_w^2-m_h^2)^2 + m_h^2 \Gamma_h^2} \nonumber
 \\
 &\times& \frac{1}{4} \lpa6 x_{\rm G.h.}^{-1} + 4 \sqrt{\pi}
 \tilde{\alpha} x_{\rm G.h.}^{-1/2} + \frac{4 \pi^2
   \tilde{\alpha}^2}{3} + \pi^{5/2} \tilde{\alpha}^3 x_{\rm
   G.h.}^{1/2} +\frac{\pi^4 \tilde{\alpha}^4}{6} x_{\rm G.h.} \rpa \,, \nonumber \\
 \eeqa with $x_{\rm G.h.}  \approx 3 \times 10^{6}$.

\subsection{$\bf{W}$-WIMP Annihilation into Pairs of Goldstone Bosons}

In addition to the annihilation into SM fermions we must consider the  $w \bar w
\rightarrow 2 \alpha'$ annihilation channel. The invariant amplitude for this process
is given by \beq i \CM = \frac{2 i f}{v_r} \bar{v}(p) u(p') \left(
  \frac{\sin^2 \chi}{s-m_H^2} - \frac{\cos^2 \chi}{s-m_h^2} \right) k
\cdot k'.  \eeq We then average over the in state spins to obtain
\beqa \frac{1}{4} \sum_{s,s'} |\CM|^2 &=& \frac{ f^2 s^2
  [(s-m_h^2)\sin^2 \chi - (s-m_H^2)\cos^2 \chi]^2}{2 v_r^2
  (s-m_h^2)^2(s-m_H^2)^2} (s - 4m_w^2) . \nonumber \eeqa The general
expression for the cross section reads \beq \sigma  =
\frac{1}{16 \pi \sqrt{s} \sqrt{|s-4m_w^2|} } \frac{f^4 s^2
  [(s-m_h^2)\sin^2 \chi - (s-m_H^2)\cos^2 \chi]^2}{\Delta m^2
  (s-m_h^2)^2(s-m_H^2)^2} (s - 4m_w^2) \, .  \eeq Using the small
angle approximation, {\it i.e.} $\cos \chi \approx 1$, we obtain \beq
\sigma  \approx \frac{f^2 s^2 \sqrt{|s-4m_w^2|}}{16 \pi
  \sqrt{s} (s-m_h^2)^2 } \lpa \frac{f^2}{\Delta m^2} +
\frac{(m_h^2+m_H^2-2 s)}{ 2 (s-m_H^2)^2}\frac{\eta_\chi^2
  v_\phi^2}{(m_H^2-m_h^2)^2} \rpa \ .  \eeq Taking a thermal average
gives \beq \la \sigma_{\alpha'} v \ra  \approx  \frac{2 f^4 m_w^4}{\pi \Delta m^2 [(m_h^2-4 m_w^2)^2 + m_h^2 \Gamma_h^2]} \la v^2 \ra \, .
\label{eq:siduerme}
\eeq
If the $W$-WIMPs are highly non-relativistic we have to correct  (\ref{eq:siduerme}) to account for the Sommerfeld enhancement
\beq \la \sigma_{\alpha'} v \ra  \approx  \frac{2 f^4 m_w^4}{
\pi \Delta m^2 [(m_h^2-4 m_w^2)^2 + m_h^2 \Gamma_h^2]} \la  S(v) v^2 \ra \, .
\eeq

\subsection{\label{sec:BFit} $\bf{W}$-WIMP Parameter Fits}

The total flux of $\gamma$-rays per solid angle from $W$-WIMP annihilation
into  SM particles (labelled by subindex $i$) is given by
\begin{equation}
\frac{d\Phi_\gamma}{dE_\gamma} = \sum_{\rm fermions} \frac{\langle
  \sigma_i v
  \rangle}{2} \ \frac{{\cal
  J}_{\Delta \Omega}}{J_0} \, \frac{1}{\Delta \Omega_{\rm obs} m_w^2} \,
\left. \frac{dN_\gamma}{dE_\gamma} \right|_i \,,
\end{equation}
where ${\cal J}/J_0$ is the normalized integral of mass density
squared of the dark matter in the line of sight, $dN_\gamma/dE_\gamma$ is the
$\gamma$-ray spectrum per annihilation into particle species $i$,
$\Delta \Omega_{\rm obs}$ is the observational solid angle in
steradians, and the sum runs over all possible annihilation
channels. It is noteworthy that $d\Phi_\gamma/dE_\gamma$ is the total photon
number flux per unit energy per unit steradian for a full sky
observation and, when compared to the total photon count of the
Fermi-LAT observation with $|b| > 10^\circ$, must be scaled to the field of
view of that observation, $\Delta \Omega_{\rm obs} = 10.4~{\rm sr}$.
%%%%%%%%%%%%%%%%%%%%%%%
\begin{figure}[h]
\begin{center}
\includegraphics[width=0.9\textwidth]{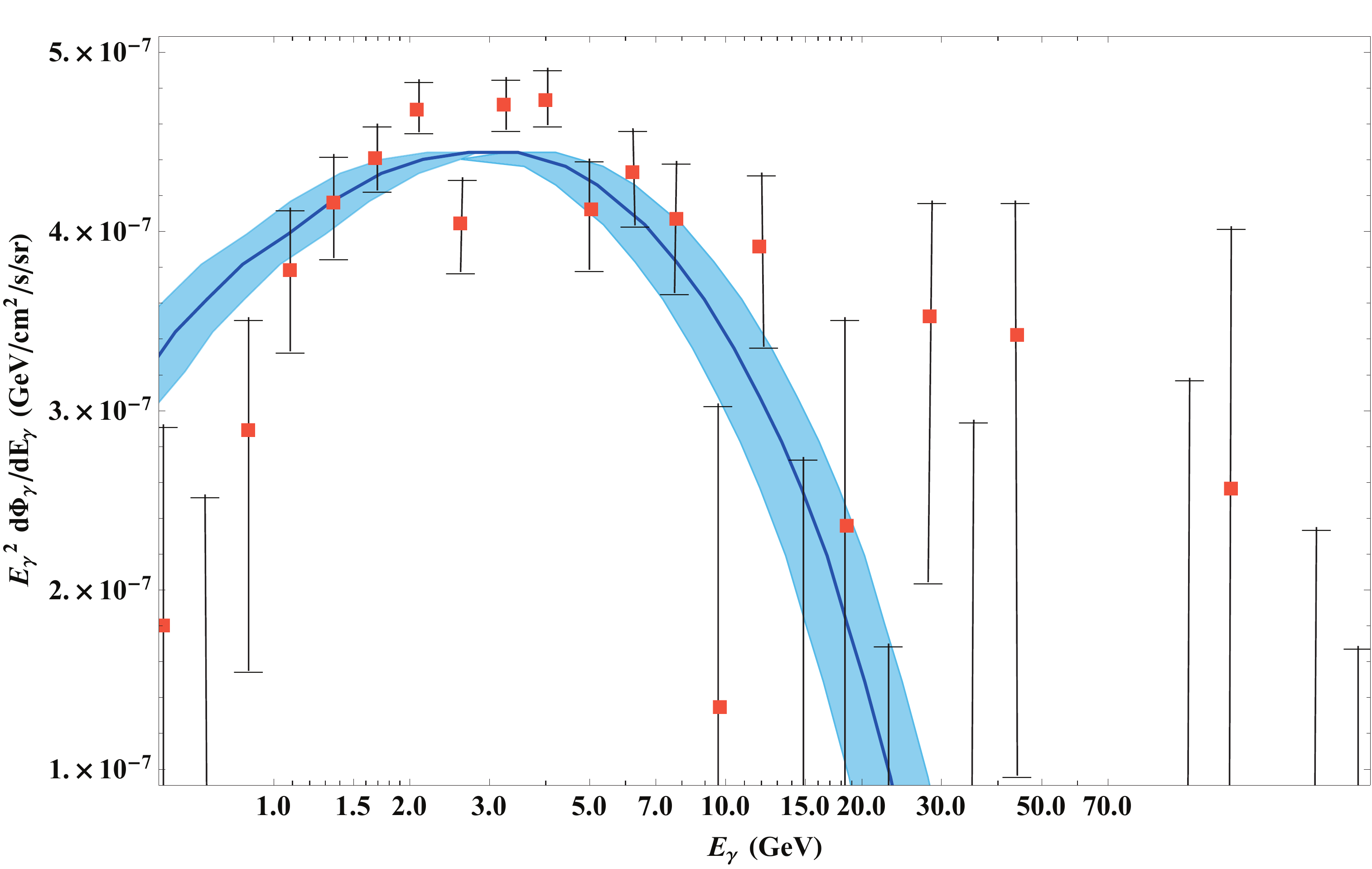}
\caption[Fermi Bubble Emission Curves with Dark Matter Annihilation Fit]{ \sglspc \small Comparisons of the observed $\gamma$-ray spectrum of the
  low-latitude ($|b| = 10^\circ - 20^\circ$) emission, after
  subtracting the contribution from inverse Compton scattering to that
  predicted from 50~GeV $W$-WIMPs annihilating to $b \bar b$. We have adopted a generalized NFW profile with an inner slope of $\gamma = 1.2$, and normalized the signal to a local density of $0.4~{\rm GeV}/{\rm cm}^3$ and an annihilation cross section of $\langle \sigma_b v \rangle = 8 \times 10^{-27}~{\rm cm}^3/{\rm s}$. The band shows the variation in the mass range $45~{\rm GeV} < m_w < 55~{\rm GeV}$ for the same normalization. Adapted from Fig.~14 of Ref.~\cite{Hooper:2013rwa}.}
\end{center}
\label{fig:P3dos}
\end{figure} 
%%%%%%%%%%%%%%%%%%%%%%%%%
{}From  (\ref{eq:TEWII}) we see that, for $10~{\rm GeV} < m_w <
50~{\rm GeV}$, the dominant annihilation channel is $b \bar
b$. Annihilation into $c \bar c$ and $\tau^+ \tau^-$ are suppressed by
about an order of magnitude. Hereafter we make the case for a $w$ with
a mass of about $50~{\rm GeV}$, which annihilates into $b \bar b$.
The photon flux expected from the \emph{Fermi} Bubbles is shown in
Fig.~\ref{fig:P3dos}. Comparing (\ref{fermibubble2}) and
 (\ref{eq:TEWII}) we obtain \beqa \la \sigma_b v \ra & \approx & \frac{
  3 }{ 2 \pi } \left( \frac{\eta_\chi m_b m_w \Delta m }{2 (4
    m_w^2-m_H^2)} \right)^2 \, \frac{1}{(4 m_w^2-m_h^2)^2 + m_h^2
  \Gamma_h^2} \, \nonumber \\
&\frac{1}{4}& \lpa6 x_{\rm G.h.}^{-1} + 4 \sqrt{\pi} \tilde{\alpha} x_{\rm G.h.}^{-1/2} + \frac{4 \pi^2  \tilde{\alpha}^2}{3}
+  \pi^{5/2} \tilde{\alpha}^3 x_{\rm G.h.}^{1/2} +  \frac{\pi^4 \tilde{\alpha}^4}{6} x_{\rm G.h.} \right) \nonumber \\
& =& 6.7 \times 10^{-10} \rm GeV^{-2} \, . \nonumber \\
 \label{eq:FermiBub}
\eeqa
To be produced thermally
in the early universe in an abundance equal to the measured dark
matter density,  $\Omega_{\rm DM} h^2 = 0.1120 \pm 0.0056$~\cite{Beringer:1900zz}, the 50 GeV $w$-particle must have an annihilation 
cross section of
\beq
\sum_{\rm all \, species} \langle \sigma_i v \rangle \sim 3
\times 10^{-26}~{\rm cm}^3/{\rm s} = 2.5 \times 10^{-9}~{\rm GeV}^{-2} , 
\label{gary}
\eeq when thermally averaged over the process of
freeze-out, $x_{\rm f.o.} \sim 20$~\cite{Scherrer:1985zt,Steigman:2012nb}. It is noteworthy that for $\tilde \alpha \lessapprox 0.01$ the effect of the Sommerfeld enhancement on the final relic particle abundance is negligible~\cite{Dent:2009bv,Chen:2013bi}. Herein we will work on the range of the coupling $\tilde \alpha$ over which Sommerfeld annihilation can be neglected in the calculation of relic densities.
Because  {\it a priori} we do not know whether $\langle
\sigma_{\alpha'} v\rangle$ or $\langle \sigma_b v\rangle$ dominates
the total annihilation cross section at freeze-out, we combine
 (\ref{eq:TEWII}) and  (\ref{eq:siduerme}) evaluated at $v(x_{\rm
  f.o.})$ together with (\ref{gary}) to obtain 
  \beq 
\left[ \frac{2 f^4 m_w^4}{ \pi \Delta m^2} +\frac{ 3 }{ 2 \pi } \left(
    \frac{\eta_\chi m_b m_w \Delta m }{2 (4 m_w^2-m_H^2)} \right)^2
\right] \, \frac{1}{(4 m_w^2-m_h^2)^2 + m_h^2 \Gamma_h^2}\frac{3}{2 x_{\rm f.o.}} \approx 2.5 \times 10^{-9} {\rm GeV}^{-2}.
\label{eq:RelD}
\eeq

To determine the allow region of the parameter space, for $m_w =
50~{\rm GeV}$, we solve  (\ref{eq:FermiBub}) and  (\ref{eq:RelD}) while
simultaneously demanding that $\tilde{\alpha} \lessapprox 0.01$, and that the
upper limit on the invisible decay width for the SM Higgs
(\ref{LHCwidth}) is not violated by  (\ref{eq:invis}). The best fit
parameters are given in Table~\ref{table:1}, for an example with
$\Delta m = 6000~{\rm GeV}$.  We can see that the
annihilation into pairs of Goldstone bosons is dominating the $w \bar
w$ interactions at freeze-out by a factor of about $9$. Precise
determination of the parameters is at present hampered by the large
uncertainties in the dark matter halo profile. Interestingly, the
$W$-WIMP-nucleon cross section is within the reach of the XENON1T
experiment~\cite{Aprile:2012zx}, providing a strong motivation for the ideas
discussed in this section.  Again I reiterate that the fine tuned nature of the dark sector may
be avoidable by the use of the result of $m_w = 61$ GeV~\cite{AUrbano} we may fit the result
to the Higgs pole.
\begin{table}
\begin{center}
\caption[Best Fit Parameters to Observations for $\Delta m = 6000$ GeV]{\centering Best fit parameters for $\Delta m = 6000$ GeV. \label{table:1}}
\begin{tabular}{| c  c | }
\hline
\hline
$\Delta m$ & $6000~{\rm GeV}$ \\
\hline
$m_h$ & $98.8\ {\rm GeV}$ \\
\hline
$f$ & $0.34$ \\
\hline
$\tilde{\alpha}$ & $0.009$ \\
\hline
$\eta_\chi$ & $1.8 \times 10^{-4}$ \\
\hline
$\chi$ & $0.049$  \\
\hline
$\Gamma_{H \rightarrow \, {\rm invisible}}$ & $0.65~{\rm MeV}$ \\
\hline
$\quad \la \sigma_{\alpha'} v (x_{\rm f.o.}) \ra \quad$ & $\quad 2.7 \times 10^{-26} \ {\rm cm^3 s^{-1}} \quad $ \\
\hline
~~~~~~~~~~~~~~~$\la \sigma_b v(x_{\rm f.o.}) \ra$ ~~~~~~~~~~~~~~& ~~~~~~~~~~~~~~$0.3 \times 10^{-26} \ {\rm cm^3 s^{-1}} $~~~~~~~~~~~~~~\\
\hline
$\quad \la \sigma_{\alpha'}  v (x_{\rm G.h.}) \ra \quad$ & $7.8 \times 10^{-26} \ {\rm cm^3 s^{-1}}$ \\
\hline
\hline
\end{tabular}
\end{center}
\end{table}

\begin{figure}[h]
\begin{center}
\includegraphics[width=0.8\textwidth]{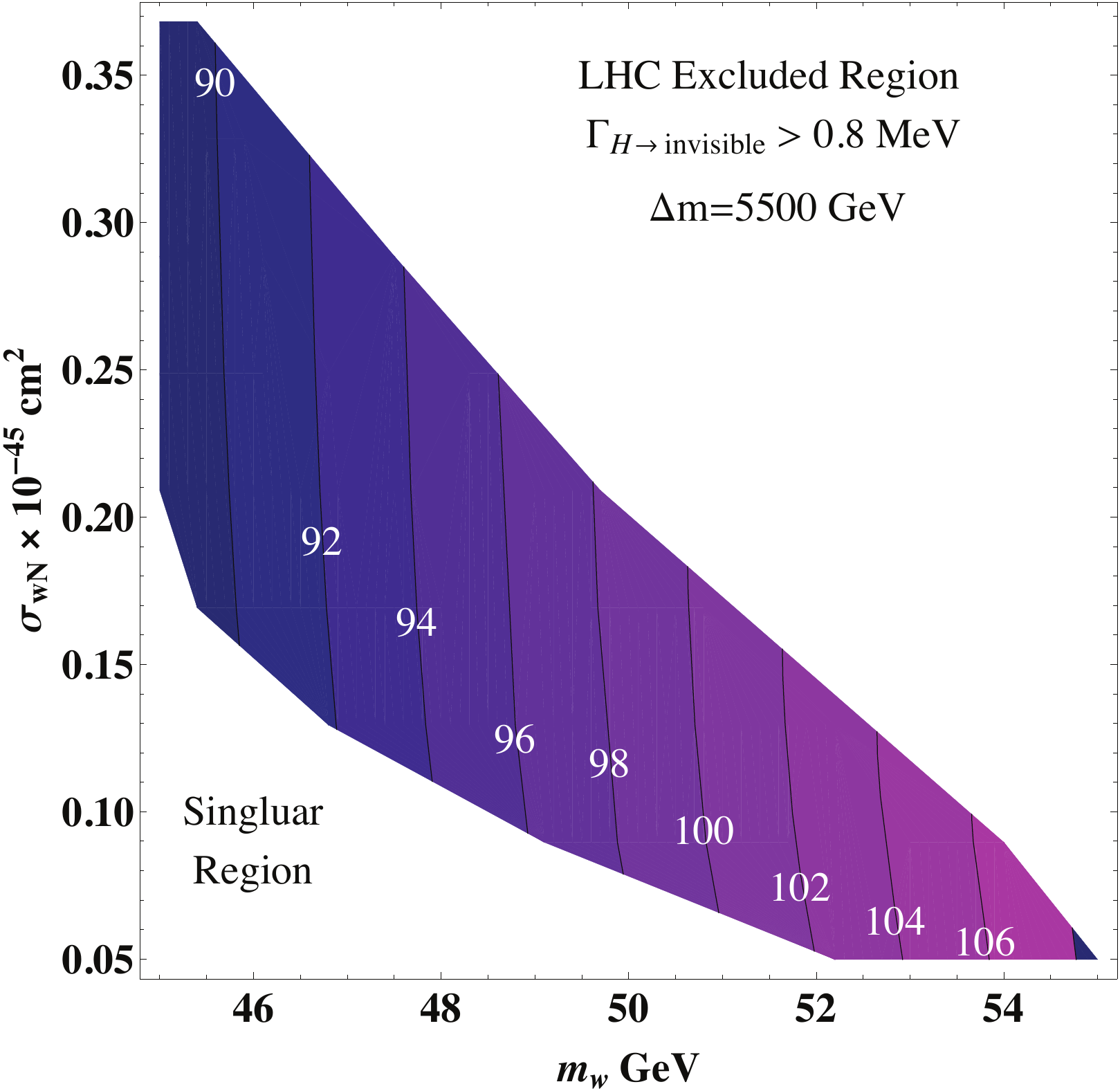}
\caption[Contours of Fixed $m_h$ in W-WIMP Model That Satisfy Observations]{ \sglspc \small Contours of constant $m_h/{\rm GeV}$ in the $\sigma_{wN} - m_w$
  plane. The contours satisfy \emph{Fermi} data, the relic density
  requirement, and the LHC bound ${\cal B} (H \to \ {\rm invisible})$.
  We have required $\tilde{\alpha}\lessapprox 0.01$ and taken $\Delta m = 5500~{\rm GeV}$.}
  \end{center}
\label{fig:tres}
\end{figure}

\begin{figure}[h]
 \begin{center}
\includegraphics[width=0.8\textwidth]{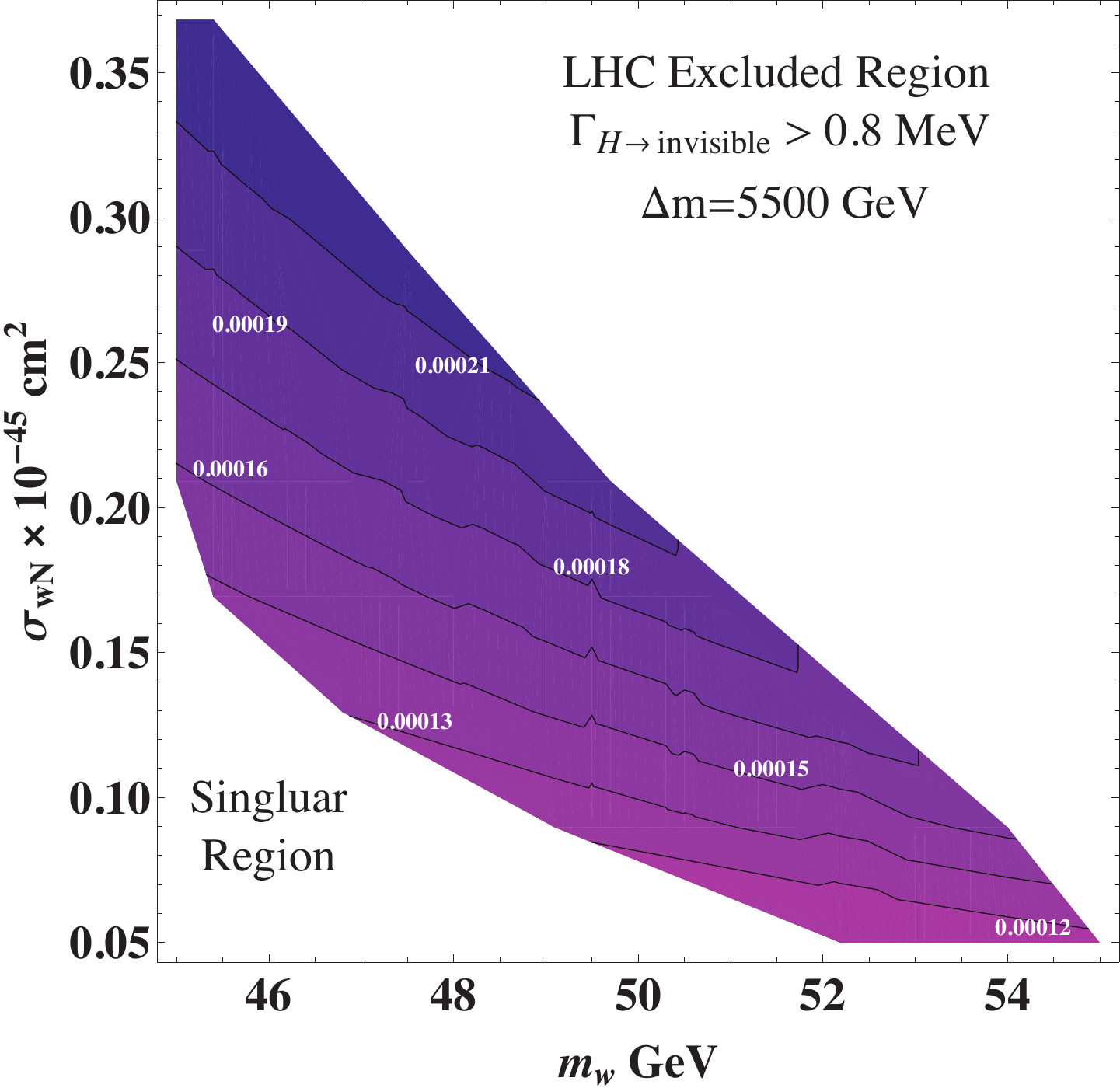}
\caption[Contours of Fixed $\eta_\chi$ in W-WIMP Model That Satisfy Observations]{ \sglspc \small Contours of constant $\eta_\chi$ in the $\sigma_{wN} - m_w$
  plane. Again the contours satisfy \emph{Fermi} data, the relic
  density requirement, and the LHC bound ${\cal B} (H \to \ {\rm
    invisible})$. We have required $\tilde{\alpha}\lessapprox 0.01$ and taken
  $\Delta m = 5500~{\rm GeV}$.}
  \end{center}
\label{fig:cuatro}
\end{figure}

\begin{figure}[h]
\begin{center}
\includegraphics[width=0.8\textwidth]{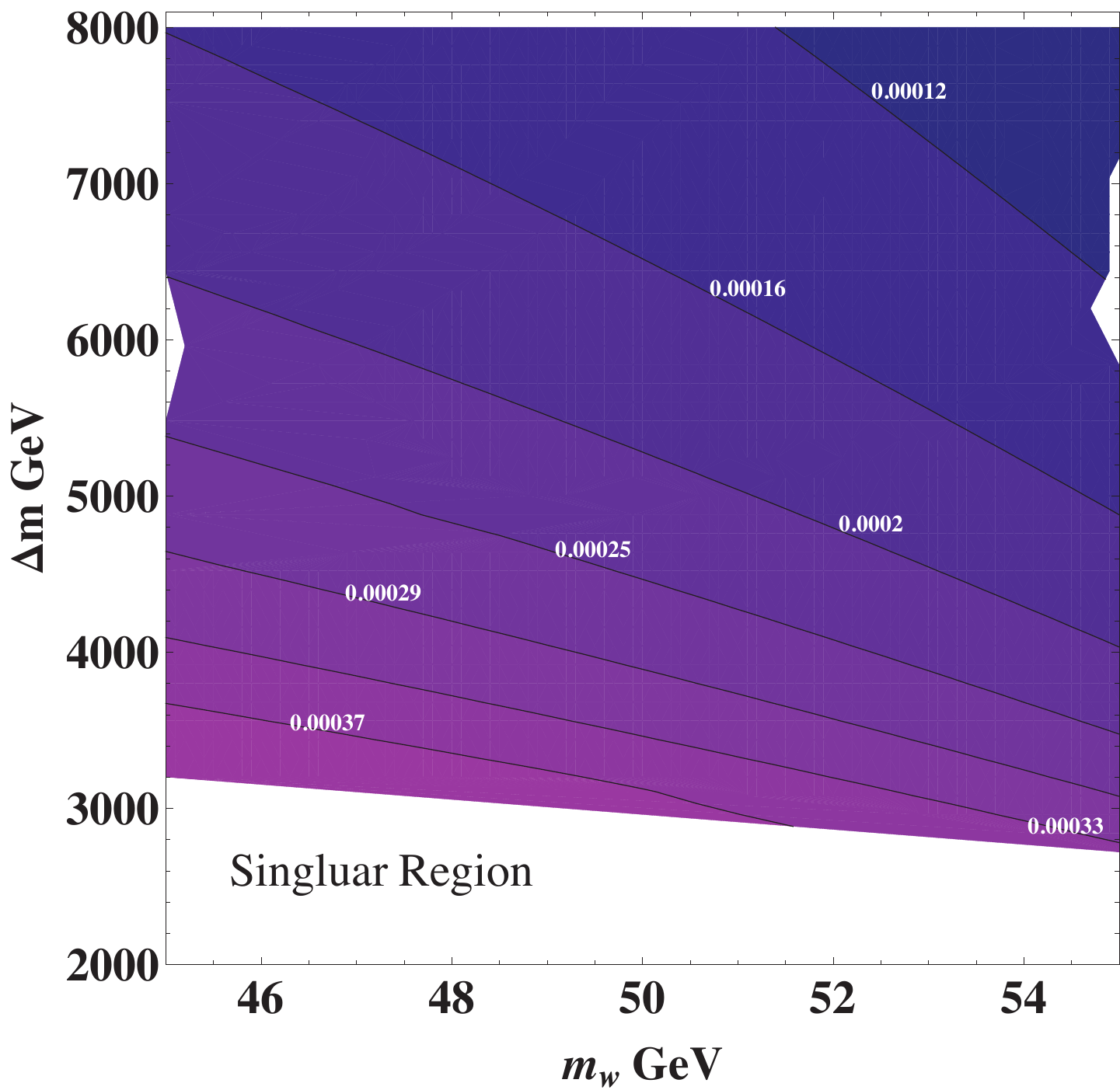}
\caption[Contours of Fixed $\eta_\chi$ in W-WIMP Model That Saturates $\Gamma_{H\rightarrow \rm invisibles}$]{\sglspc \small Contours of constant $\eta_\chi$ in the $\Delta m- m_w$
  plane. The contours satisfy \emph{Fermi} data, the relic density
  requirement, and saturates the LHC bound ${\cal B} (H \to \ {\rm
    invisible})$. We have required $\tilde{\alpha} \lessapprox 0.01$ and we have verified
that the XENON-100 upper limit~\cite{Aprile:2011hi} is not violated.}
\end{center}
\label{fig:cinco}
\end{figure}

Duplicating the procedure described above, we have scanned the mass
range of the parameter space that is consistent with Fermi data:
$45~{\rm GeV} < m_w < 55~{\rm GeV}$; see Fig.~\ref{fig:P3dos}. Our
results are encapsulated in Figs.~\ref{fig:tres}, \ref{fig:cuatro} and
\ref{fig:cinco}.  In particular, Fig.~\ref{fig:tres} and
\ref{fig:cuatro} display, for $\Delta m = 5500~{\rm GeV}$, the region
of the parameter space of $m_w$ {\it vs.} $\sigma_{wN}$ not yet
excluded by current direct detection experiments or the LHC.  Future
LHC data will either more tightly constrain this parameter space or
will turn up evidence for a signal.  Note that  the region
excluded by nonexistence of a solution ($\Gamma_{H \to \, {\rm
    invisible}} \approx 0.3~{\rm MeV}$) up to the current LHC bound
will be very tightly constrained after the LHC coming upgrade,
assuming no signal appears. In the case that a signal does appear, the
combination of relations shown in Figs.~\ref{fig:tres} and
\ref{fig:cuatro} will constrain model parameters providing the XENON1T
experiment with the specific cross section required to confirm this
model. As an illustration, in Fig.~\ref{fig:cinco} we show contours of
constant $\eta_\chi$ in the $\Delta m - m_w$ plane for the case in
which ${\cal B} (H \to\, {\rm invisible})$ saturates the current
limit, $\Gamma_{H \to \, {\rm invisible}} = 0.8~{\rm MeV}$. The direct
  detection cross section sampling this sub-region of the parameter
  space varies between $1.8 \times 10^{-46}~{\rm cm}^2$ and $2.2
  \times 10^{-46}~{\rm cm}^2$, with an average of $1.9 \times
  10^{-46}~{\rm cm}^2$.
%%%%%%%%%%%%%%%%%%%%%%%
%%%%%%%%%%%%%%%%%%%%%%%%%
%%%%%%%%%%%%%%%%%%%%%%%%%

\subsection{$\bf{W}$-WIMP Interpretation for Hints of Light Dark Matter}

Signals broadly compatible with $\sim 10~{\rm GeV}$ dark matter have
been observed in four direct detection experiments:
DAMA/LIBRA~\cite{Bernabei:2010mq},
CoGeNT~\cite{Aalseth:2010vx,Aalseth:2011wp},
CRESST~\cite{Angloher:2011uu}, and CDMS-II~\cite{Agnese:2013rvf}. In
this section we explore the compatibility with one particular region
of the $W$-WIMP parameter space.  The features of this region of the
parameter space has bearing on the evidence for extra-relativistic
degrees of freedom at the CMB epoch.

In order to elaborate on the case for $m_w \sim 10~{\rm GeV}$, we
consider $m_h \approx 500~{\rm MeV}$ and $\eta_\chi \approx
0.005$. Substituting these values  in  (\ref{eq:DetAcdms}), it is
straightforward to see that to comply with the elastic cross section
``signal'' reported by the CDMS Collaboration~\cite{Agnese:2013rvf},
we must set $\Delta m \approx 1.75~{\rm GeV}$. This in turn determines
via  (\ref{eq:TEWII}) a thermal average annihilation cross section into
quarks: $\langle \sigma_b v (x_{\rm G.h.}) \rangle \approx 1.3 \times
10^{-39}~{\rm cm}^3 \, {\rm s}^{-1}$. Note that this is more than ten
orders of magnitude smaller than current limits on light dark matter
from anti-proton data~\cite{Lavalle:2010yw,Evoli:2011id,Kappl:2011jw}.

\begin{figure}[h]
\begin{center}
\includegraphics[width=0.8\textwidth]{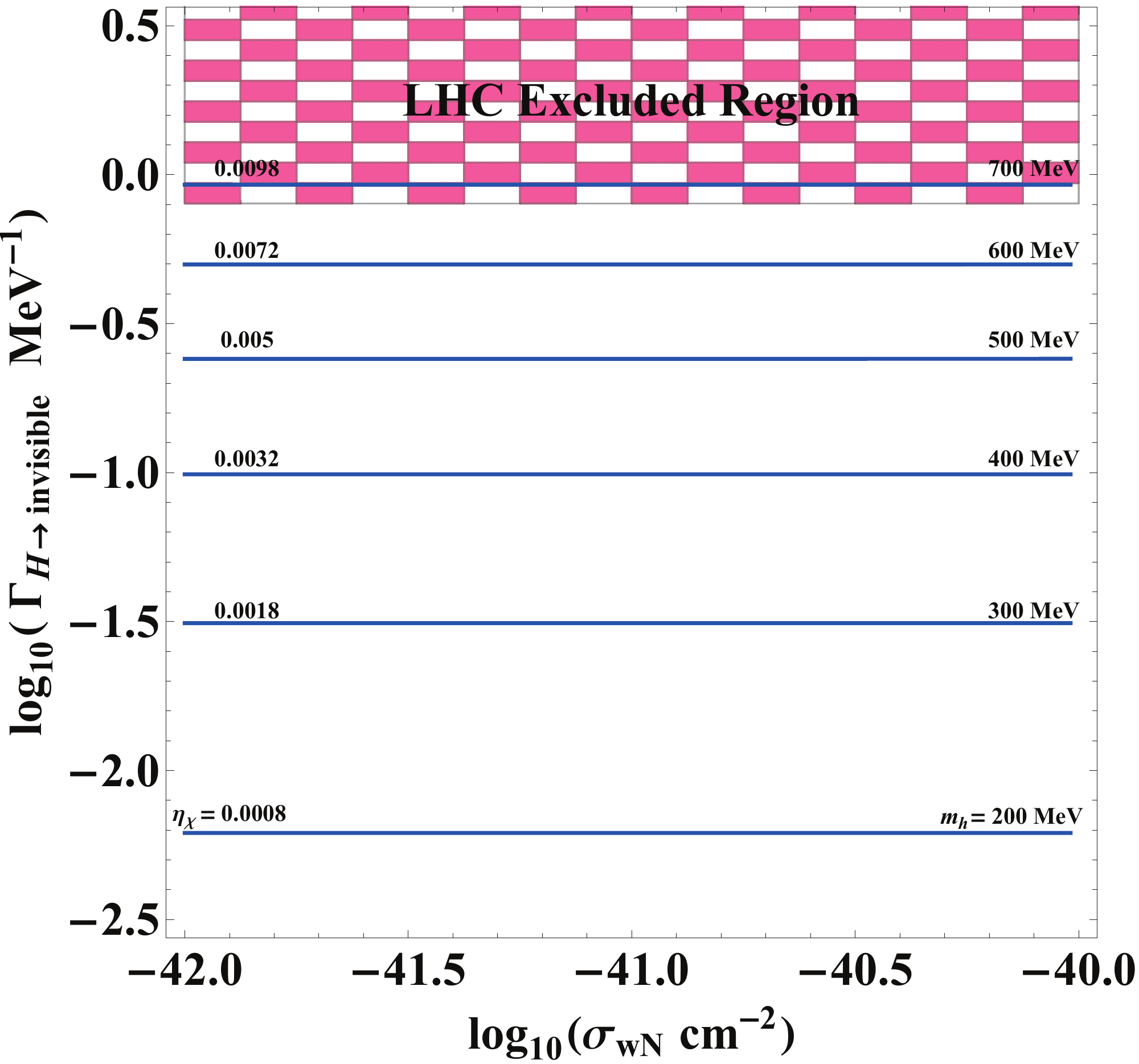}
\caption[LHC Predicted Observations For W-WIMP Model]{\sglspc \small $\Gamma_{H \rightarrow {\rm invisible}}$ for varying values
  of $\sigma_{wN}$. The plotted values are nearly constant as the
  terms from $\Gamma_{H \rightarrow \alpha, h}$ dominate the decay
  width, and thus there is weak dependence on the direct detection
  cross section. For varying values of $m_h$ we adjust the value of
  $\eta_\chi$ so that the Goldstone bosons decouple from the
  primordial plasma at $k_BT \approx m_\mu$, yielding $N_{\rm eff} =
  3.39$. For $200~{\rm MeV} \leq m_h \leq 700~{\rm MeV}$, the Higgs
  decay width into the hidden sector varies between $(0.006 -
  0.92)$~MeV. The constant $\eta_\chi/m_h$ contours shown here are
  independent of $m_w$, and therefore span the mass range $7~{\rm
    GeV} \lessapprox m_w/{\rm GeV} \lessapprox 10$.}
    \end{center}
\label{fig:seis}
\end{figure}

The observed dark matter density is obtained again through dominant $W$-WIMP
annihilation into the hidden sector.  To demonstrate this point, we
must first compute the $w \bar w  \rightarrow 2 h$ annihilation cross
section, as this channel is now open. We consider the relevant terms of  (\ref{eq:fifty}), \beq
\frac{f \cos \chi}{2} h \bar \psi_- \psi_- + \frac{f \sin \chi}{2} H
\bar \psi_- \psi_- \ , \eeq as well as the relevant terms of the
scalar potential \beq \CV \approx \dots - \frac{m_h^2}{2 v_r} h^3
-\frac{\eta_\chi v_\phi}{2} \lpa \frac{m_H^2 + 2 m_h^2}{m_H^2 - m_h^2}
\rpa H h^2 \ ; \eeq together this gives the total reaction matrix
element
\begin{eqnarray} \CM  & = & i f
\bar{v}(p) u(p') \frac{i}{s-m_h^2} \lpa \frac{-i 3! m_h^2}{2v_r}\rpa +
i f \frac{\eta_\chi v_r v_\phi}{m_H^2-m_h^2} \bar{v}(p)
u(p')\frac{i}{s-m_H^2}  \nonumber \\
& \times &  \lpa \frac{-i \eta_\chi v_\phi
  (m_H^2+2m_h^2)}{m_H^2-m_h^2} \rpa \ . 
\end{eqnarray}
Assuming $\eta_\chi \ll 1$ and $m_h \ll m_H$, we arrive at a
manageable form of the spin-averaged $w \bar w \to 2h$ amplitude
\begin{eqnarray}
\frac{1}{4} \sum_{\rm spins}|\CM|^2 & \approx &  f^2 \lpa \frac{9 m_h^4}{v_r^2 (s-m_h^2)^2} + \frac{6 m_h^2 \eta_\chi^2 v_\phi^2 (m_H^2+2m_h^2)}{(s-m_h^2)(s-m_H^2)}\rpa (p\cdot p' - m_w^2) \ , \nonumber
\\
& \approx &  f^2 \lpa  \frac{18 f^2 m_h^4}{\Delta m^2 (s-m_h^2)^2} + \frac{3 m_h^2 \eta_\chi^2 v_\phi^2 m_H^2}{(s-m_h^2)(s-m_H^2)}\rpa (s - 4m_w^2) \ , \nonumber \\
\end{eqnarray}
and the scattering cross section
\begin{equation}
\sigma \approx \frac{f^2}{32 \pi s }\sqrt{\frac{s-4m_h^2}{s-4m_w^2}} \lpa  \frac{18 f^2 m_h^4}{\Delta m^2 (s-m_h^2)^2} + \frac{3 m_h^2 \eta_\chi^2 v_\phi^2 m_H^2}{(s-m_h^2)(s-m_H^2)}\rpa (s - 4m_w^2) \ . 
\end{equation}
We take the thermal average in the low temperature limit, that is $T \ll m_w$,
\beq \la \sigma_h v \ra \approx \lpa \frac{9 f^4 m_h^4}{8 \pi \Delta m^2 (m_h^2-4m_w^2)^2}+\frac{3 f^2 v_\phi^2 \eta_\chi^2 m_h^2}{16 \pi (m_h^2-4 m_w^2)} \rpa \la v^2 \ra.
\eeq
By demanding the total annihilation cross section to comply with the
relic density requirement~\cite{Steigman:2012nb} we obtain \beq \la
\sigma_{\alpha'} v \ra_ + \la \sigma_h v \ra + \sum_{\rm fermions} \la
\sigma_i v \ra \sim 3 \times 10^{-26}~{\rm cm}^3/s \,,  \eeq and so
\begin{equation}
f \approx 0.070 \ ,
\label{eq:WeinLaster}
\end{equation} 
yielding $\chi \approx 10^{-3}$. The latter is consistent with the upper bound on the mixing angle 
$\chi < 10^{-2}$~\cite{Kingman} derived from the invisible Higgs search by the OPAL Collaboration~\cite{Abbiendi:2007ac}.  In addition, the production of pion pairs plus a large missing energy carried away by the Goldstone boson, $\alpha'$, could become a smoking gun at the LHC~\cite{Kingman}.
As a final check we ensure that the LHC
upper limit on the hidden decay width of the Higgs is satisfied;
taking note that the decay channel $H \rightarrow \bar \psi_+ \psi_+$
is now open, we have 
\beq \frac{\eta_\chi^2 v_\phi^2}{16\pi m_H} +
\frac{\eta_\chi^2 \Delta m^2 v_\phi^2}{32\pi m_H^3}=0.24~{\rm MeV} <
0.8~{\rm \rm MeV} \, . \eeq In Fig.~\ref{fig:seis} we exhibit the
range of parameters consistent with the 95\%~CL upper limit on ${\cal
  B} (H \to {\rm invisible})$~\cite{Espinosa:2012vu,Giardino:2013bma,Ellis:2013lra}
together with possible signal regions associated with data from
CDMS-II~\cite{Agnese:2013rvf}. For $m_w = 10~{\rm GeV}$, the best-fit
intervals at the 68\%~CL and the 90\%~CL are $3 \times 10^{-42} <
\sigma_{wN}/{\rm cm}^2 < 2.5 \times 10^{-41}$ and $2 \times 10^{-42} <
\sigma_{wN}/{\rm cm}^2 < 3 \times 10^{-41}$, respectively. The
horizontal lines preserve the constant $\eta_\chi/m_h$ ratio that
allows decoupling of $\alpha'$ at $T \approx m_\mu$, yielding $N_{\rm
  eff} = 3.39$. 
%%%%%%%%%%%%%%%%%%%%%%%%% 

In summary, we have shown that $W$-WIMPs of about 10~GeV can
simultaneously explain the observed relic density and the possible
signals observed by direct detection experiments, while avoiding
limits from indirect detection experiments. In the near future, the
Large Underground Xenon (LUX) dark matter
experiment~\cite{Akerib:2012ys} will collect enough statistics to
probe the $\sim 10~{\rm GeV}$ dark matter hypothesis. Concurrent with
LUX observations will be precise measurements of the Higgs branching
fractions by the LHC ATLAS and CMS experiments (operating at $\sqrt{s}
= 14~{\rm TeV}$). This new arsenal of data, when combined with
observations the Phased IceCube Next Generation Upgrade
(PINGU)~\cite{Koskinen:2011zz}, will have the potential to single out
this distinctive Higgs portal light dark matter model.\footnote{ \sglspc \small Since
  the annihilation rate into SM particles is largely suppressed
  compared to annihilations into the hidden sector, this particular
  model predicts null results at PINGU.}

\section{Summary of the Results and Conclusions}

Light-element abundances probing big bang nucleosynthesis and precision data from cosmology probing the CMB decoupling epoch have hinted at the presence of extra relativistic degrees of freedom. This is widely referred to as ``dark radiation'', suggesting the need for new light states in the UV completion of the SM. We provided a brief and concise overview of the current observational status of such dark radiation and we investigated the interplay between two possible interpretations of the extra light states: the right-handed partners of three Dirac neutrinos (which interact with all fermions through the exchange of a new heavy vector meson) and dark matter particles that were produced through a non-thermal mechanism, such us late time decays of massive relics. Interestingly, the first scenario ties together cosmological indications of the extra light states in SM$^{++}$ and the production of the heavy vector particle $Z''$ at the LHC. 

We have also studied the minimal hidden sector recently introduced by Weinberg, which communicates with the visible sector via the Higgs portal. We have re-examined the possibility that the Goldstone boson associated with the hidden scalar may be masquerading as a fractional cosmic neutrino.  The broken symmetry associated with this Goldstone boson could regulate the conservation of the particles in the dark matter sector. We have studied the implications of this model for direct and indirect detection experiments. In particular we have shown that $W$-WIMPs (with $m_w \approx 50~{\rm GeV}$) are capable of accommodating the desired effective annihilation into $b \bar b$ to reproduce the photon spectrum of the \emph{Fermi} Bubbles. We have also demonstrated that the thermal cross section required to account for the relic dark matter abundance can easily be obtained if $w \bar w \to 2 \alpha$ is the dominant annihilation channel. However, given that the Goldstone bosons would decouple at 5 GeV (i.e. in the very early universe), the contribution to the effective number of neutrinos for the described parameter space is negligible, and thus cannot explain the evidence for dark radiation.  In the near future, the upgraded LHC together with the new XENON1T experiment will further whittle down the parameter space, or else make a discovery. On the other hand, if $m_w \approx  10~{\rm GeV}$, Weinberg's hidden sector does not provide a viable explanation of the \emph{Fermi} Bubbles. However, there remains an interesting region of the parameter space which can account for the alleged signals recently reported by direct detection experiments. In this region, the Goldstone bosons decouple from the primordial plasma near the 100~MeV temperature, consistent with the two measurements of the effective number of neutrinos reported by the Planck Collaboration. In this region of the parameter space, $W$-WIMP annihilation into Goldstone bosons is also sufficient for consistency of the observed dark matter abundance. Furthermore, future LHC measurements will further constrain this sector of the Higgs portal (or better, find a signal), while LUX will close the deliberations on the alleged direct signals. 

\newpage

%---------------------------------Ch 4-----------------------------------------
\thispagestyle{fancy}
\chapter{Future Prospects on Beyond the Standard Model Physics}
\thispagestyle{fancy}
\pagestyle{fancy}

In 2012, the LHC entered its first long shutdown for upgrades.  Starting as early as 2015 the LHC beams will attain collisions at an energy $\sqrt{s} = 13~{\rm TeV}$~\cite{CMSnote}, in an effort to prepare for design collision energy of  $\sqrt{s} = 14~{\rm TeV}$ at a luminosity of $10^{34}~{\rm cm}^{-2} \, {\rm s}^{-1}$.  In addition to the 2015 run of the LHC, there are also plans for high luminosity-LHC (HL-LHC) in 2023, which has the prospective goal of accumulating $3000 \ {\rm pb^{-1}}$ of data.  With this exciting future nearly here, beyond the standard model physics may soon become experimental fact. 

\begin{figure}[h]
\begin{center}
\postscript{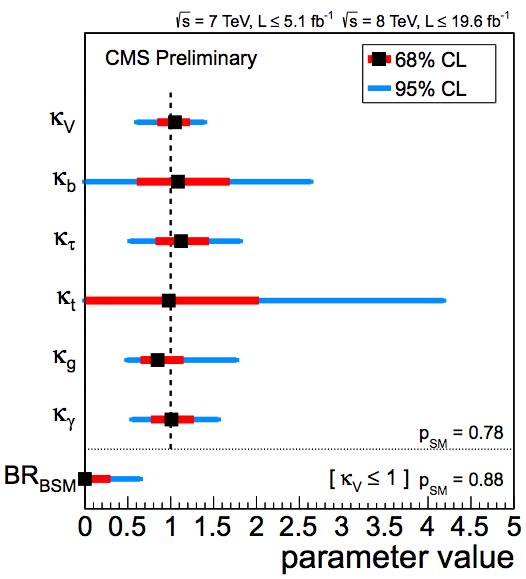}{0.5}
\caption[Higgs Best Fit Couplings for CMS Data]{ \sglspc \small The best fit to observation couplings for the Higgs-like boson detected at CMS, along with corresponding confidence intervals.  The couplings are for $\kappa_V, \kappa_{\rm b}, \kappa_\tau , \kappa_t, \kappa_{\rm g}, \kappa_\gamma$ for the vector bosons ($W_{\pm}, Z_0$), bottom, $\tau$, top, gluon, and photon couplings assuming no BSM physics. This results in a p-value of $p_{\rm SM} = 0.78$, allowing BSM physics while restricting the vector boson coupling to $\kappa_V \leq 1.0$ results in an upper limit on the branching ratio for the Higgs into invisibles (image from~\cite{CMSnote}).}
\label{fig:HiggsBRs}
\end{center}
\end{figure}

   The Higgs-like boson discovery~\cite{ATLASnew,CMSnew} of 2012 seems to be more and more likely the actual Higgs boson of the standard model, as evidenced in Fig. \ref{fig:HiggsBRs}.
%----------------Figure------------------
%-----------------------------------
As discussed in Sec.~\ref{part:1}, it may be possible in the future for the LHC to detect additional gauge bosons, which may reflect an underlying symmetry that can be supported by a D-brane construct~\cite{luisnMe1}.  The additional bosons may also help to solve the vacuum stability problem of the standard model as shown in Sec.~\ref{III}~\cite{luisnMe2}.  Furthermore, in Sec.~\ref{sec:DM}, it was shown that it may be possible to bring collider physics and cosmological observations together by explaining the possible additional relativistic degrees of freedom inferred from CMB anisotropy analysis, with the addition of extra gauge bosons and the right chiral components of neutrinos~\cite{Anchordoqui1}.  The latter also solves the neutrino oscillation problem outlined in Sec.~\ref{sec:NuOsc}.

Further investigation into the Higgs boson particle can result in a portal to BSM physics. In Sec.~\ref{sec:WWIMP} we have explored the minimal hidden sector recently introduced by Weinberg~\cite{Weinberg:2013kea}.  This model can explain the $\gamma$-ray emission from the low-latitude regions of the \emph{Fermi} Bubbles, for which the spectral
shape is consistent with an approximately $50~{\rm GeV}$ dark matter particle annihilating into $b \bar b$, with a normalization corresponding to $\langle \sigma_b v
\rangle \sim 8 \times 10^{-27}~{\rm cm}^3/{\rm s}$.  Moreover, in a separate region of the parameter space, $m_w \approx 10~{\rm GeV}$  the model has the potential to explain recent hints from dark matter direct detection experiments and at the same time it provides predictions for LHC~\cite{luisnMe4}. In the near future, the upgraded LHC, together with the new XENON1T experiment, will further whittle down the parameter space, or else make a discovery.

\begin{figure}[h]
\begin{center}
\postscript{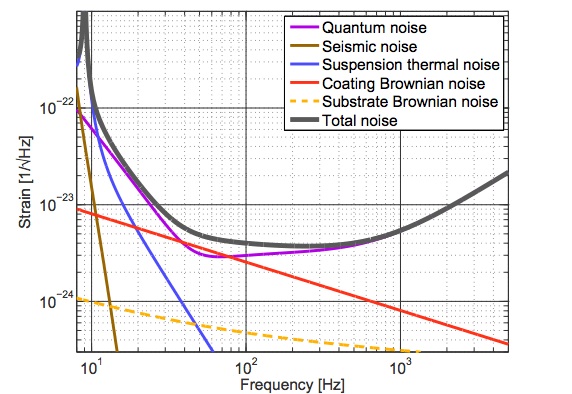}{0.7}
\caption[Advanced LIGO Prospective Noise Curve]{ \sglspc \small The prospective noise curve for Advanced LIGO results in a spectral strain sensitivity of order $\sqrt{{\rm Noise}} \sim 10^{-23} \ {\rm Hz^{-1/2}}$, allowing for better noise suppression and possible direct detection of gravitational waves. Image from~\cite{AdvLIGO}. }
\label{fig:NoiseCurve}
\end{center}
\end{figure}

Far from the particle physics experiments lies another method of discovery of BSM physics, that of gravitational wave astronomy.  As of 2011, the LIGO detectors at Hanford, WA, and Livingston, LA, are undergoing upgrades to Advanced LIGO, whose prospective noise curve can be seen in Fig.~\ref{fig:NoiseCurve}. 
%----------------Figure------------------
%----------------------------------- 
%----------------Figure------------------
\begin{figure}[h]
\begin{center}
\postscript{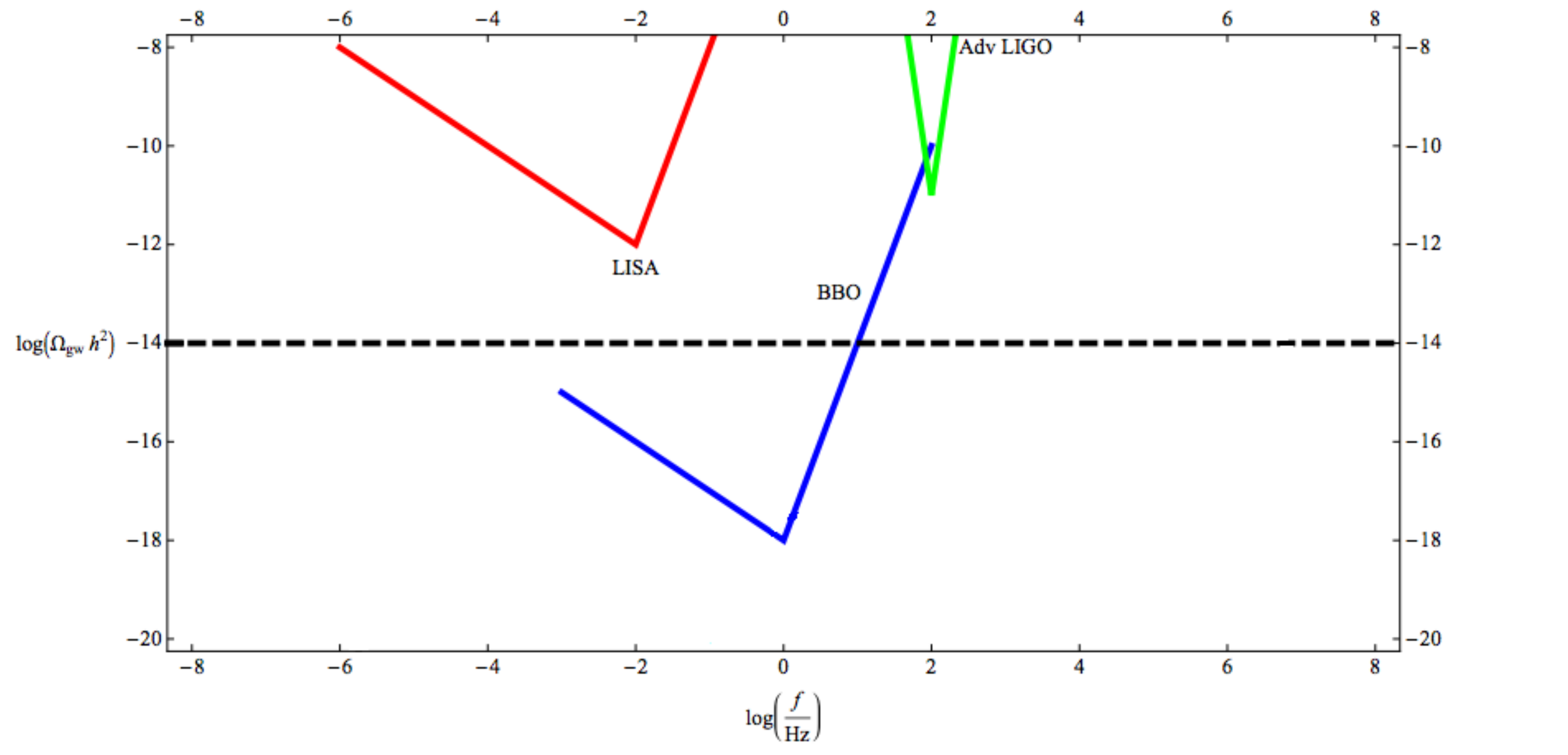}{1.0}
\caption[Grav-Wave Stochastic Background Projected Sensitivity]{ \sglspc \small The future detection of a gravitational wave stochastic background with the projected sensitivity curves of possible future detectors such as Advanced LIGO and BBO.  The spectra of gravitational waves from second order global phase transitions may be detected in the far future if BBO is actually completed, but in the near future seems to be out of reach for detection. Image modified from~\cite{Easther:2008sx}. }
\label{fig:GWBounds}
\end{center}
\end{figure}
%-----------------------------------
Advanced LIGO is expected to be completed in 2015, which should allow unprecedented detection ability of gravitational waves~\cite{AdvLIGO}.  The study of gravitational wave astronomy opens the door to probes of matter densities that are far from obtainable in a lab, such as that found in pulsars~\cite{Maggiore}.  It is a new set of eyes on the sky as everything we have received from the cosmos has been in the form of photons.  The field of gravitational wave astronomy may even reveal a pathway to probing physics of the GUT scale as discussed in Sec.~\ref{sec:GWSPT}, though detection of a signal of the strength $\Omega_{gw} h^2 \sim 10^{-24} - 10^{-15}$~\cite{Giblin}  lies out of reach of even possible future detectors except the possible future big bang observer (BBO)~\cite{Crowder:2005nr} (see Fig. \ref{fig:GWBounds}); the possibility is exciting that far in the future, it may be possible.

 Whatever lies beyond the SM whether it be SUSY, a Higgs portal, a GUT theory, or string theory,  it may soon be within reach with as the discovery of the Higgs, future upgrades of the LHC, and Advanced LIGO hold in store an exciting future for all.

%---------------------------------Bibliography-----------------------------------------

\thispagestyle{fancy}
\newpage
\thispagestyle{fancy}
\pagestyle{fancy}

 \newpage
 \sglspc
 %------------------------------------CV--------------------------------------------------------------------------------
 {\raggedleft  \LARGE{\textbf{Brian J. Vlcek}} \hfill \textit{Curriculum Vitae}}
\begin{center}
\line(1,0){420}
\end{center}
{\raggedleft \textbf{Work Address}} \\ 
University of Wisconsin-Milwaukee \\
Department of Physics
 \hfill e-mail: bvlcek@uwm.edu\\ 
1900 E. Kenwood Blvd.\\
Milwaukee, WI \ 53211 USA\\
\\
{\large \textbf{Research Interests}}
\begin{itemize}
\item Theoretical high energy particle physics, with interests of beyond the standard model physics.
\item High energy particle physics phenomenology.
\item Astroparticle physics, and cosmology.
\end{itemize}
{\large \textbf{Education}}\\
\\
\textbf{PhD:  University of Wisconsin-Milwaukee} \quad (Milwaukee, WI) \hfill Sept 2007 - Current\\
Graduation Date: August 2013 (GPA = 3.693/4.0)\\
Co-Advisors: Associate Professor Xavier Siemens and\\
Associate Professor Luis Anchordoqui.\\
\\
Dissertation Title: {``}Beyond the Standard Model: LHC Phenomenology, Cosmology from Post-Inflationary Sources, and Dark Matter.{"}\\
\\
\textbf{BSc: Illinois State University} \quad (Bloomington-Normal, IL)  \hfill Aug. 2004 - May 2007\\
Graduated May 2007 with focus on Computational Physics. (Final GPA = 3.52/4.0)\\
\\
{\large \textbf{Relavant Work Experience}}\\
\\
\textbf{Research Assistant} (UW-Milwaukee) \hfill Jan 2013- Aug 2013\\
Completed research with Luis Anchordoqui on projects concerning extensions of the Standard Model of particle physics.\\
\\
\textbf{Graduate Teaching Assistant} (UW-Milwaukee) \hfill Sept 2007 - Dec 2012\\
Taught various undergraduate level courses for the department of Physics. \\
11 courses taught, with an average evaluation score of $3.2/4.0$.\\
\\
\textbf{LIGO Collaboration Scientific Monitor} (Hanford, WA USA) \hfill Aug - Sept 2010\\
On-site observation of LIGO telescope for data quality monitoring.\\
\\
\textbf{Preparation Class Leader for Phd Written Qualifying Exam } (UW-Milwaukee) \hfill \\Jan '09, '10, '11\\
Invited/chosen by members of physics faculty to lead an open study session to assist 1st and 2nd year students in the department in preparation for taking the Spring semester qualifying exam.\\
\\
\textbf{Graduate Astro-club TA/{``}Arciebo@UWM{''} Trainer} (UW-Milwaukee) \hfill \\Sept '09 - Sept '11\\
Trained undergraduates on how to analyze pulsar data for proper identification of possible pulsars, as well as monitoring students during remote usage of the Arciebo radio telescope for pulsar search surveys. Additionally handled duties associated with hosting the UWM Astronomy Club.\\
\\
\textbf{Undergraduate Research Assistant} (Illinois State Univ.  Bloomington-Normal, IL)\hfill May '05 - Sept '07\\
Conducted research in computational simulation of chaotic systems under the supervision of Associate Professor Epaminondas Rosa Jr.\\
\\
\textbf{University Employed Tutor} (Illinois State Univ.  Bloomington-Normal, IL) \hfill May '05 - Sept '07\\
University employed tutor in math and physics (by appointment) available for all under-graduate students, including specific tutoring for student athletes. \\
\\
{\large \textbf{Skills}}

\begin{itemize}
\item 12+ years programing in C/C++

\item 6+ years programming in Mathematica.

\item Moderate level experience with UNIX systems.
\end{itemize}
{\large \textbf{Honors and Awards}}\\
\\
\textbf{Univ. of Wisconsin- Milwaukee {``}Chancellor's Graduate Student Award{"} and ``Research Excellence Award"} \\
Competitive award given in order to supplement TA/RA salaries. Received this award/support in numerous semesters during my PhD studies at UW-Milwaukee.\\
\\
\textbf{Skadron Computational Physics Award for Research \hfill 2007 Recipient}\\
Award for outstanding research involving simulation by undergraduate physics student at Illinois State University.\\
\\
\textbf{Skadron Prize Competition \hfill 1st prize recipient - 2007}\\
1st prize recipient, in collaboration with Michael P. Morrissey, on programing/simulation competition, hosted by the Physics Department at Illinois State University.\\
\\
{\large \textbf{Talks, Conferences and Workshops}}\\
\\
\textbf{33rd International Cosmic Ray Conference (poster presentation)} \hfill July 2013\\
Presented poster on non-thermal dark matter as explanation of excess relativistic degrees of freedom from Planck and HST data, hosted by Centro Brasileiro de Pesquisas F'sicas (CBPF) (Rio de Janeiro, RJ, Brazil)\\
\\
\textbf{Phenomenology 2010 Symposium (presenter)} \hfill May 2010\\
Presented on gravitational wave cosmology and its relation to high energy physics; hosted by University of Wisconsin-Madison (Madison, WI)\\
\\
\textbf{UW-Milwaukee, Center for Gravitation and Cosmology (presenter)} \hfill May 20, 2011.\\
Friday seminar series attended by members of UWM's ``Center for Gravitation and 
 Cosmology" (which includes graduate students, post docs and faculty).  Talk title was: ``Gravitational Waves from 
Pre-Heating and Phase Transitions,{"} which focused on my research of gravitational wave cosmology, specifically on computational methods that improve the standard ways of calculating a stochastic background of gravitational waves from a phase transition in the early universe. \\
\\
\textbf{Gravitational-wave Physics and Astronomy Workshop} {\bf (Attendee)}\hfill January 2011\\
Hosted by University Wisconsin-Milwaukee (Milwaukee, WI)\\
\\
\textbf{20th Annual Midwest Relativity Meeting} {\bf (Attendee)} \hfill November 2010\\
 Hosted by University of Guelph (Guelph, Ontario - Canada)\\
\\
{\large \textbf{Scholarly publications in refereed journals}}\\
\begin{itemize}
\item{``}W-WIMP Annihilation as a Source of the Fermi Bubbles,{"} \\
Luis~A.~Anchordoqui and {\bf Brian J. Vlcek}, \\ 
Phys.\ Rev.\ D {\bf 88}, 043513 (2013)  [arXiv:1305.4625 [hep-ph]]. \\
\item {``}LHC phenomenology and cosmology of string-inspired intersecting D-brane models,{"}\\
 Luis A. Anchordoqui, Ignatios Antoniadis, Haim Goldberg, Xing Huang,
Dieter L\"{u}st, Tomasz R. Taylor, and \textbf{Brian Vlcek}, \\
 Phys. Rev. D \textbf{86}, 6 066004 (2012) [arXiv:1206.2537 [hep-ph]].\\
\item {``}Vacuum Stability of SM++,{"} \\
Luis A. Anchordoqui, Ignatios Antoniadis, Haim Goldberg, Xing Huang, Dieter L\"{u}st, Tomasz R. Taylor, and \textbf{Brian Vlcek}, \\
JHEP {\bf 1302},  074 (2013) [arXiv:1208.2821[hep-ph]].\\
\item {``}Gravitational Waves from Global Second Order Phase Transitions.{"}\\
John. T. Giblin, Jr., Larry R. Price, Xavier Siemens, and \textbf{Brian Vlcek}, \\
JCAP  {\bf 11}, 006 (2012) [arXiv:1111.4014 [astro-ph]].
\end{itemize}
{\large \textbf{Publications in conference proceedings}}\\
\begin{itemize}
\item ``Tracing the Interplay between Non-Thermal Dark Matter and Right-Handed Dirac Neutrinos with LHC Data.{"} \\
Luis~A.~Anchordoqui, Haim~Goldberg and {\bf Brian Vlcek}; \\ 
in Proceedings of the 33rd International Cosmic Ray Conference (ICRC 2013), Rio de Janeiro, Brazil, 2 - 9 July 2013  [arXiv:1305.0146 [astro-ph.CO.]]. \\
\end{itemize}

\end{document}